THE UNIVERSITY OF NEW SOUTH WALES

SCHOOL OF PHYSICS

# Fractal Magneto-conductance Fluctuations in Mesoscopic Semiconductor Billiards

Adam P. Micolich

May, 2000

*A thesis submitted in fulfilment*
*of the requirements for the degree of*
*Doctor of Philosophy*

# Abstract

Epitaxially-grown $Al_xGa_{1-x}As$-GaAs heterostructures allow electron confinement to a thin sheet known as a two-dimensional electron gas (2DEG). Using negatively biased surface-gates to electrostatically deplete selected regions of the 2DEG, allows lateral confinement of the 2DEG to specific geometries smaller than the 2DEG electron mean free path. These confined regions are called billiards. Electron transport through billiards is ballistic. At millikelvin temperatures, the electron phase coherence length is sufficient that electron quantum interference effects produce reproducible magneto-conductance fluctuations (MCF). MCF act as a 'magneto-fingerprint' of the scattering dynamics of electrons traversing the billiard. It has been predicted and demonstrated experimentally that billiard MCF are fractal.

Fractal MCF in mesoscopic semiconductor billiards is investigated. The MCF of a Sinai billiard (circle at the centre of a square) displayed exact self-similarity (ESS). Self-similarity is a property of fractal behaviour taking one of two forms: exact where precise structural features are repeated at different scales, and statistical where structure at different scales is linked by the same scaling relationship. A correlation function analysis is introduced to quantify the presence of ESS. A model for the Sinai billiard MCF based on a Weierstrass function is presented. Using a bridging interconnect, a continuous transition between the Sinai and an empty square geometry is achieved. The removal of the circle induces a transition from ESS to statistical self-similarity (SSS), suggesting that ESS is due to the presence of an obstacle at the centre of the billiard. A mechanism for the transition between ESS and SSS is proposed. Fractal behaviour is observed over 3.7 orders of magnitude for ESS and 2.1 orders for SSS. The physical dependencies of SSS are investigated and produce variations in the fractal dimension, rather than the fractal scaling range. SSS obeys a unified picture where the fractal dimension depends only on the ratio between the average spacing and broadening of billiard energy levels, irrespective of other billiard parameters. The semiclassical origin of SSS is demonstrated and the suppression of SSS is observed in both the quantum and classical limits. The influence of soft-wall potential profile on fractal MCF is investigated using double-2DEG billiards.

# Acknowledgements

So many people to thank. I just hope I don't miss anyone. First and foremost I would like to thank (enormously) my supervisors A/Prof. Richard Newbury and Dr Richard Taylor. I could never have hoped for two better people to work with over the last three and a half years. Without fail, they have always been helpful and supportive and have given me a lot of freedom to chase my own ideas and do things my way. They have also persevered with my stubbornness and occasional bouts of stupidity in that time. Secondly, I'd like to thank Prof. Jon Bird for allowing me to work on his phase-breaking studies in the early days, for providing me with a large amount of experimental data and for patiently answering all of my questions along the way. Dr Mark Fromhold and his Ph.D. student Chris Tench have provided theoretical support for this project since the beginning. I'd like to thank Mark also for answering many questions along the way and allowing me to use his results, diagrams and simulations over this time. Dr Giles Davies and Dr Linda Macks have been instrumental in the double-2DEG billiards project. I'd like to thank them for providing these samples, helping out with the experiments and freely giving their advice. I'd also like to thank Linda for proofreading the fabrication chapter in this thesis. Without these seven people, this thesis would not be.

A large debt of gratitude must also go to René Wirtz. He has helped immensely with numerous discussions, suggestions and assistance both on physics and on experimental matters. René's constant cheerfulness has made many a hard day easier. Heiner Linke has been a good office-mate for the last year and has also helped out with advice, discussions and many suggestions. Thanks also to David Reilly (who also proofread this thesis) for helpful discussions on measurement techniques. Jack Cochrane, Bob Starrett, David Jonas and Andrei Skougarevski have worked as technical staff on this project. In particular, many thanks to Bob Starrett who has dealt with the fools at BOC over the years, stayed here late and come in on weekends (often at very short notice) to help out with the logistics of helium use. Jack Sandall (staff/student mechanical workshop) played an important part in the dilution refrigerator refit. Jack's skill and expertise is awe-inspiring. He is always helpful, friendly and full of good advice on tough jobs. I'd also like to thank the students who have worked with me on parts of this project – Mark Hallett, Tammy Humphrey, Alex Ehlert and Michelle Neilson.

During this project I have had the opportunity to learn the art of billiard fabrication during two visits to the Semiconductor Physics Group at the University of Cambridge. I am most grateful to Dr Charles Smith for looking after me on my first visit, getting me started in the clean-rooms, and teaching me the basics of electron beam lithography and device fabrication, to Dr Bill Tribe for his advice on the finer points of EBL and for teaching me other parts of the fabrication process also and to Prof. Mike Pepper for supporting my visits and allowing me to use his groups' facilities. Thanks also to Dr Jon Cooper (who also fabricated the RIKEN Sinai billiards), Dr Chris Ford, Dr Mark Graham, Dr Geb Jones, Dr Chi-te Liang, Dr Linda Macks, Dr James Nicholls Harry Clark, Masaya Kataoka and Graham Winiecki for offering advice and help along the way. Thank you to Dr Dave Ritchie for supplying the heterostructure material and arranging for me to stay at Robinson College on my second visit, Dr Giles Davies for letting me stay in his office at Selwyn college and to Mary Holland for her hospitality on my first visit to Cambridge. Thank you also to Tina Jost who dealt with a lot of the administrative hassles and lent me her pink bicycle on my first visit and Kurt Hasselwimmer who kept my social life going whilst in Cambridge and for partially inspiring the fridge refit.

I'd also like to thank Dr Carl Dettmann and A/Prof. Gary Morriss for helpful discussions regarding fractals and chaos, Dr Shiro Kawabata (ETL, Japan) for helpful discussions regarding the Weierstrass project, and Dr Andrew Dzurak and Dr Nancy Lumpkin for helping me out with the occasional re-bonding job in the SNF. Thanks also to all of the other students and staff at UNSW who have offered suggestions, support or friendship over the last few years, in particular (in alphabetical order): Saskia Besier, Rolf Brenner, Dr Simon Brown, Tilo Bühler, Jason Chaffey, Geoff Facer, Alex Hamilton, Rachael Heron, Mike Johnston, Garth Jones, Dr Emma Mitchell, Stuart Moin, Derek Munneke, Jeremy O'Brien, Eileen O'Hely, Dapsy Olatona, David Reilly, Kiyonori Suzuki and my bros. Damien McLachlan and Adam Tonitto. Thanks also to the Australian government for my APA scholarship (although I don't think we can thank Penfold and his mates for much else lately) and to my supervisors for topping me up from their ARC grants. I am also grateful to Dr Andy Sachrajda and Dr Roland

Ketzmerick for keeping the coals ‘obscurely’ burning in the boiler over the last three years.

Finally, I owe a great debt to my family, particularly Mum and Dad who looked after me so well for all those years and supported me in the numerous times when life got really tough, and my grandmothers who have also offered much support throughout the years. Thanks also to my second family (the Kays), particularly Aunty Gloria who has made many wonderful dinners and lunches and has also been very supportive over the years. And last, but by no means the least I would like to thank my wonderful girlfriend Merlinde Kay. Amongst other things she has proof-read this thesis, helped out with the French translations, helped me flesh out ideas over the last three years, lent a hand while I was working on the fridge, put up with my moody spells, bouts of whinging and complaining and lack of free time, and supported me all the way. Without her the last three years would not have been as enjoyable as they have been and I would probably no longer be sane.

OK. That’s it. Thanks to anyone I’ve missed and happy reading…

# Publications Arising from this Work

## Refereed Journal Publications

- "Fractal Conductance Fluctuations as a Probe of the Transition from Classical to Quantum Behaviour in Semiconductor Billiards"
  A.P. Micolich, R.P. Taylor, A. Ehlert, A.G. Davies, J.P. Bird, T.M. Fromhold, R. Newbury, H. Linke, L.D. Macks, W.R. Tribe, P.D. Rose, E.H. Linfield, D.A. Ritchie, M. Pepper, J. Cooper, Y. Aoyagi and T. Sugano
  Submitted to Physical Review Letters.

- "An Investigation of Weierstrass Self-similarity in a Semiconductor Billiard"
  A.P. Micolich, R.P. Taylor, R. Newbury, T.M. Fromhold and C.R. Tench
  Published: Europhysics Letters, **49**, 417, (2000).

- "Comment on Fractal Conductance Fluctuations in a Soft-wall Stadium and a Sinai Billiard"
  R.P. Taylor, A.P. Micolich, T.M. Fromhold and R. Newbury
  Published: Physical Review Letters, **83**, 1074, (1999).

- "Exact and Statistical Self-similarity in Magneto-conductance Fluctuations: A Unified Picture"
  R.P. Taylor, A.P. Micolich, R. Newbury, J.P. Bird, T.M. Fromhold, J. Cooper, Y. Aoyagi and T. Sugano
  Published: Physical Review B (Brief Reports), **58**, 11107, (1998).

- "Experimental and Theoretical Investigations of Electron Dynamics in a Semiconductor Sinai Billiard"
  A.P. Micolich, R.P. Taylor, R. Newbury, C.P. Dettmann and T.M. Fromhold
  Published: Australian Journal of Physics, **51**, 547, (1998).

- "Fractal Transistors"
  R.P. Taylor, A.P. Micolich, R. Newbury, C.P. Dettmann, and T.M. Fromhold
  Published: Semiconductor Science and Technology, **12**, 1459, (1997).

- "Correlation Analysis of Self-similarity in Semiconductor Billiards"
  R.P. Taylor, A.P. Micolich, R. Newbury and T.M. Fromhold
  Published: Physical Review B (Rapid Communications), **56**, R12733, (1997).

## Refereed Conference Proceedings

- "Semiconductor Billiards – A Controlled Environment for the Study of Fractals"
  R.P. Taylor, A.P. Micolich, A. Ehlert, A.G. Davies, R. Newbury, W.R. Tribe, E.H. Linfield, P.D. Rose, D.A. Ritchie, L.D. Macks, H. Linke, T.M. Fromhold, and C.R. Tench
  Invited Talk (RPT) at "Quantum Chaos Y2K", Sweden, June 13-17, 2000.
  To be published in Physica Scripta

- "Temperature and Size Dependence of Fractal MCF in Semiconductor Billiards"
  A.P. Micolich, R.P. Taylor, J.P. Bird, R. Newbury, H. Linke, Y. Aoyagi, and T. Sugano
  Presented at the 3rd International Conference on Low Dimensional Structures and Devices, Antalya, Turkey, 15-17th September 1999. (Poster – RPT)
  Published: Microelectronic Engineering, **51-52**, 241, (2000).

- "Chaotic Ray Dynamics and Fast Optical Switching in Micro-cavities with a Graded Refractive Index"
  P.B. Wilkinson, T.M. Fromhold, C.R. Tench, R.P. Taylor, A.P. Micolich, and R. Newbury
  Presented at the 11th International Conference on Hot Carriers in Semiconductors, Kyoto, Japan, 19-23rd July 1999. (Poster – PBW)
  Published: Physica B, **272**, 484, (1999).

- "A Physical Explanation for the Origin of Self-similar Magneto-conductance Fluctuations in Semiconductor Billiards"
  C.R. Tench, T.M. Fromhold, M.J. Carter, R.P. Taylor, A.P. Micolich, and R. Newbury
  Presented at the 9th International Conference on Modulated Semiconductor Structures, Fukuoka, Japan, 12-16th July 1999. (Poster – TMF)
  Accepted for Publication in Physica E.

- "Physical realisation of Weierstrass scaling using a quantum interferometer"
  A.P. Micolich, R.P. Taylor, T.M. Fromhold, C.R. Tench, and R. Newbury
  Proceedings of "The Conference on Optoelectronic and Microelectronic Materials and Devices", Perth, Australia, 1998. (Poster – APM)
  Published: IEEE, ISBN 0-7803-4513-4, 468, (1999).

- "The Temperature Dependent Fractal Dimension of Magneto-Conductance Fluctuations in Semiconductor Billiards"
  A.P. Micolich, R.P. Taylor, J.P. Bird, R. Newbury, Y. Aoyagi, and T. Sugano
  Proceedings of "The Conference on Optoelectronic and Microelectronic Materials and Devices", Perth, Australia, 1998. (Poster – APM)
  Published: IEEE, ISBN 0-7803-4513-4, 471, (1999).

- "Scale Factor Mapping of Self-similarity in Semiconductor Billiards"
  R.P. Taylor, A.P. Micolich, R. Newbury, T.M. Fromhold, C.R. Tench, J.P. Bird, and J. Cooper
  Proceedings of "The Conference on Optoelectronic and Microelectronic Materials and Devices", Perth, Australia, 1998. (Poster – RPT)
  Published: IEEE, ISBN 0-7803-4513-4, 475, (1999).

- "Observation of Fractal Conductance Fluctuations over Three Orders of Magnitude"
  R.P. Taylor, A.P. Micolich, R. Newbury, T.M. Fromhold, and C.R. Tench
  Proceedings of "The 8th Gordon Godfrey Workshop on Condensed Matter Physics", Sydney, Australia, 1998. (Talk – RPT)
  Published: Australian Journal of Physics, **52**, 887, (1999).

- “Temperature Dependence of the Fractal Dimension of Magneto-conductance Fluctuations in a Mesoscopic Semiconductor Billiard”
  A.P. Micolich, R.P. Taylor, J.P. Bird, R. Newbury, Y. Aoyagi and T. Sugano
  Proceedings of “The 11th International Conference on Superlattices, Microstructures and Microdevices”, Hurghada, Egypt, 1998. (Poster – APM)
  Published: Superlattices and Microstructures, **25**, 157, (1999).

- “Physical Realisation of Weierstrass Scaling using a Quantum Interferometer”
  A.P. Micolich, R.P. Taylor, T.M. Fromhold, C.R. Tench and R. Newbury
  Proceedings of “The 11th International Conference on Superlattices, Microstructures and Microdevices”, Hurghada, Egypt, 1998. (Talk – APM)
  Published: Superlattices and Microstructures, **25**, 207, (1999).

- “Exact and Statistical Self-similarity in Semiconductor Billiards”
  R.P. Taylor, A.P. Micolich, T.M. Fromhold, C.R. Tench, R. Newbury, J.P. Bird, J. Cooper, Y. Aoyagi and T. Sugano
  Proceedings of “The 24th International Conference on the Physics of Semiconductors”, Jerusalem, Israel, 1998. (Poster – RPT)
  Published: World Scientific, ICPS-24 CD, Chapter VII, A14, (1999).

- “Scale Factor Mapping of Statistical and Exact Self-similarity in Billiards”
  A.P. Micolich, R.P. Taylor, J.P. Bird, R. Newbury, T.M. Fromhold, J. Cooper, Y. Aoyagi and T. Sugano
  Proceedings of “The 2nd International Workshop on Surfaces and Interfaces of Mesoscopic Devices”, Hawaii, USA, 1997. (Talk – JPB)
  Published: Semiconductor Science and Technology, **13**, A41, (1998).

- “Geometry-induced Fractal Behaviour: Fractional Brownian Motion in a Ballistic Mesoscopic Billiard”
  A.P. Micolich, R.P. Taylor, R. Newbury, J.P. Bird, T.M. Fromhold, Y. Aoyagi and T. Sugano
  Proceedings of “The 12th International Conference on the Electronic Properties of Two Dimensional Systems”, Tokyo, Japan, 1997. (Poster – TMF)
  Published: Physica B, **249-251**, 343, (1998).

- “Effect of Wavefunction Scarring in the Magneto-transport of Quantum Dots”
  Y. Ochiai, Y. Okubo, N. Sasaki, J.P. Bird, K. Ishibashi, Y. Aoyagi, T. Sugano, A.P. Micolich, R.P. Taylor, R. Newbury, D. Vasileska, R. Akis and D.K. Ferry
  Proceedings of “The 12th International Conference on the Electronic Properties of Two Dimensional Systems”, Tokyo, Japan, 1997. (Poster – YO)
  Published: Physica B, **249-251**, 353, (1998).

- “Self-similar Conductance Fluctuations in a Sinai Billiard with a Mixed Chaotic Phase Space: Theory and Experiment”
  T.M. Fromhold, C.R. Tench, R.P. Taylor, A.P. Micolich and R. Newbury
  Proceedings of “The 12th International Conference on the Electronic Properties of Two Dimensional Systems”, Tokyo, Japan, 1997. (Poster – TMF)
  Published: Physica B, **249-251**, 334, (1998).

- "Experimental and Theoretical Investigations of Clusters in the Magneto-fingerprints of a Sinai Billiard"
  R.P. Taylor, A.P. Micolich, R. Newbury, T.M. Fromhold, C.P. Dettmann and C.R. Tench
  Proceedings of "The 2$^{nd}$ International Conference on Low Dimensional Structures and Devices", Lisbon, Portugal, 1997. (Talk – RPT)
  Published: Materials Science and Engineering B, **51**, 212, (1998).

## Articles in Scientific Magazines

- "Fractals and Self-Similarity in Mesoscopic Semiconductor Billiards"
  A.P. Micolich, R.P. Taylor, R. Newbury, T.M. Fromhold, J.P. Bird, J. Cooper, and C.R. Tench
  Published: Australian and New Zealand Physicist Magazine, **35**, 151, (1998).

## Unrefereed Publications

- "Fractal Conductance Fluctuations in Mesoscopic Billiards: An Observation Over Three Orders of Magnitude"
  R.P. Taylor, A.P. Micolich, R. Newbury, T.M. Fromhold and H. Linke
  Proceedings of "The American Physical Society Annual Meeting", Atlanta, Georgia, USA, 1999.
  Published: American Physical Society Bulletin, Centennial Meeting Program, DP01.185, (1999).

- "Fractal Magneto-conductance in Mesoscopic Billiards: Temperature and Size Dependence"
  A.P. Micolich, R.P. Taylor, J.P. Bird, R. Newbury, Y. Aoyagi and T. Sugano
  Proceedings of "The New Zealand and Australian Institutes of Physics Annual Condensed Matter Meeting" (Wagga'99), Wagga Wagga, Australia, 1999.
  Published: "Wagga'99" ISSN 1037-1214, Page TM8, (1999). (Talk – APM)

- "Physical Realisation of Weierstrass Scaling in a Soft-wall Antidot Billiard"
  A.P. Micolich, R.P. Taylor, T.M. Fromhold, C.R. Tench and R. Newbury
  Proceedings of "The New Zealand and Australian Institutes of Physics Annual Condensed Matter Meeting" (Wagga'99), Wagga Wagga, Australia, 1999.
  Published: "Wagga'99" ISSN 1037-1214, Page WP14, (1999). (Poster – APM)

- "Correlation Analysis of Statistical and Exact Self-similarity in Billiards"
  A.P. Micolich, R.P. Taylor, R. Newbury, J.P. Bird, T.M. Fromhold and J. Cooper
  Proceedings of "The New Zealand and Australian Institutes of Physics Annual Condensed Matter Meeting" (Wagga'98), Wagga Wagga, Australia, 1998.
  Published: "Wagga'98" ISSN 1037-1214, Page TM11, (1998). (Talk – APM)

- "Fractional Brownian Statistics of Magneto-conductance Fluctuations"
  A.P. Micolich, R.P. Taylor, R. Newbury and J.P. Bird
  Proceedings of "The New Zealand and Australian Institutes of Physics Annual Condensed Matter Meeting" (Wagga'98), Wagga Wagga, Australia, 1998.
  Published: "Wagga'98" ISSN 1037-1214, Page TP49, (1998). (Poster – APM)

- “Electron Behaviour in AlGaAs/GaAs Square Quantum Dots”
  A.P. Micolich, R.P. Taylor, J.P. Bird and R. Newbury
  Proceedings of “The New Zealand and Australian Institutes of Physics Annual Condensed Matter Meeting” (Wagga’97), Pakatoa Island, New Zealand, 1997
  Published: “Wagga’97” ISSN 1037-1214, Page TP20, (1997). (Poster – APM)

- “Temperature Saturation of Fractal Magneto-conductance Fluctuations”
  R.P. Taylor, A.P. Micolich, R. Newbury, J.P. Bird, T.M. Fromhold, H. Linke, Y. Aoyagi and T. Sugano
  Presented at the LT22 Satellite Conference on Electron Transport in Mesoscopic Systems, Göteburg, Sweden, 12-15$^{th}$ August 1999. (Talk – RPT)
  Published: Electron Transport in Mesoscopic Systems, Conference Program and Abstracts, 130, (1999).

- “Self-similar Magnetoconductance Fluctuations in Semiconductor Sinai Billiards”
  C.R. Tench, T.M. Fromhold, M.J. Carter, R.P. Taylor, A.P. Micolich and R. Newbury
  Proceedings of “The Condensed Matter and Materials Physics Conference”, Manchester, UK, 1998. (Talk – CRT)

- “Fractal Transistors”
  R.P. Taylor, A.P. Micolich and R. Newbury
  Book of Abstracts for “The International Centennial Symposium on the Electron” Cambridge University, UK, 1997.

- “Self-similar Conductance Fluctuations in a Sinai Billiard with a Mixed Chaotic Phase Space”
  C.R. Tench, T.M. Fromhold, R.P. Taylor, A.P. Micolich, and R. Newbury
  Proceedings of “The Condensed Matter and Material Physics Conference”
  Exeter, UK, 1997. (Poster – CRT)

## Colloquia and Other Presentations

- “Geometry-Induced Fractal Behaviour in a Semiconductor Billiard”
  Max-Planck-Institut für Strömungsforschung und Institut für Nichtlineare Dynamik
  Universität Göttingen, Germany – 5$^{th}$ October 1997

- “Physical Realisation of Weierstrass Scaling using a Quantum Interferometer”
  Semiconductor Physics Group, Cavendish Laboratory
  University of Cambridge, UK – 17$^{th}$ July 1998

- “The Quantum Billiard – A New Probe of Quantum Mechanics in Ultra-small Electronic Devices”
  Australian Institute of Physics (NSW) Awards for Postgraduate Excellence
  12$^{th}$ October 1999

# Table of Contents

# Chapter 1 – Introduction

At the end of the 1970s, Benoît Mandelbrot started a mathematical revolution with his essays *Fractals: Form, Chance and Dimension* [1] and *The Fractal Geometry of Nature* [2]. Whilst his work on fractals was an important advancement in mathematics, the ramifications of his work are equally significant, if not more so, in the natural sciences following his observation of the abundance of fractals in nature. In the words of Mandelbrot: "Clouds are not spheres, mountains are not cones and lightning does not travel in a straight line." It wasn't long before what began as a menagerie of mathematical monsters and pretty pictures rapidly spread into various areas of physics ranging from fluid dynamics and fracture surfaces to diffusion and dynamics. In particular however, fractal structures are commonly observed in association with the presence of chaos – the unpredictability of a system due to an exponential sensitivity to initial conditions [3].

Concurrently and independently, the progress in semiconductor physics continued relentlessly at an ever increasing rate, driven by the success of the transistor and the search for faster, smaller and cheaper computer and electronic technologies. Following the discovery of molecular beam epitaxy by Cho and Arthur in 1970 [4] and modulation doping by Dingle *et al.* in 1978 [5] it became possible to form a high mobility quasi-two-dimensional electron gas (2DEG) in an AlGaAs-GaAs heterostructure. With the development of electron-beam lithography [6] in the late 1970s and early 1980s it became possible to define sub-micron features on the surface of semiconductor materials. This allowed the lateral confinement of 2DEGs to widths comparable to the electron Fermi wavelength and much smaller than the electron elastic mean free path, and opened the way for studies of ballistic electron transport and electron quantum interference effects. This was highlighted by the discovery of quantised conductance through a narrow aperture (quantum point contact) by Wharam *et al.* [7] and van Wees *et al.* [8] in 1988. Billiards, where the 2DEG is confined to some chosen geometry (e.g. a square or a circle), provide an ideal system for the investigation of two-dimensional electron dynamics in the quantum mechanical limit. This is highlighted by the initial experiments of Marcus *et al.* [9], Berry *et al.* [10] and Chang *et al.* [11]. The dynamics of electrons in these billiards is investigated via small openings in the confinement geometry by measuring the electrical conductance through the billiard as a function of magnetic field. The magneto-conductance of billiards exhibits a set of reproducible fluctuations that act as a ‘magneto-fingerprint’ of the electron dynamics within the billiard.

In 1996, it was predicted theoretically [12], and established experimentally [13,14] that the magneto-conductance fluctuations observed in billiards exhibit fractal behaviour. This thesis follows on from this discovery to investigate fractal behaviour in the magneto-conductance fluctuations of semiconductor billiards, with the ultimate aim of understanding this phenomenon and with the view towards furthering the knowledge of both fractal behaviour in physical systems and electron transport in semiconductor billiards. The latter is doubly important since the knowledge gained from the study of electron transport in billiards may play a significant role in the semiconductor technologies of the future.

## 1.1 – Thesis Outline

Chapter 2 is an introductory review of semiconductor billiards and fractals. This Chapter commences with a discussion of the important concepts of low-dimensional semiconductor physics. Attention is then turned to billiards specifically, with a discussion of the research that led up to this thesis. A brief introduction to fractals is also presented in this Chapter.

Chapter 3 presents the various techniques involved in fabricating the billiards investigated in this thesis. This Chapter begins with a discussion of the fabrication of simple billiards, followed by the fabrication of Sinai billiards which require a multi-level gate architecture, and double-2DEG billiards which have been investigated for the first time as part of this work.

Electrical measurements on billiards are performed at low temperatures using low currents/biases to avoid electron heating which reduces electron phase coherence and suppresses quantum interference effects. Chapter 4 describes the experimental methods, both cryogenic and electrical, involved in investigating electron transport in billiards.

Methods for characterising the various important parameters that affect electron transport through the billiard are also discussed.

Chapter 5 presents the analysis of data obtained from the NRC Sinai billiard experiment performed by Taylor *et al.* [13] in 1995. The magneto-conductance fluctuations (MCF) exhibit structure on distinct scales (coarse and fine) that are related by constant scaling factors in magnetic field and conductance. These two structural levels bear a striking visual similarity and are strongly suggestive of the presence of exact self-similarity in the MCF of the Sinai billiard. This self-similarity is confirmed using a correlation analysis technique. Further inspection reveals the presence of further structural levels larger than the coarse (ultra-coarse) and smaller than the fine (ultra-fine). The fine and ultra-fine, and coarse and ultra-coarse levels are related by the same scaling factors that relate the coarse and fine. The MCF for the Sinai billiard is shown to be fractal over 3.7 orders of magnitude. A model for the observed exact self-similarity based on the Weierstrass function is also presented. Finally, it is shown that removal of the circle at the centre of the Sinai billiard, which transforms it into a square billiard, causes the removal of exact self-similarity from the MCF data.

Chapter 6 commences with a discussion of techniques that assess fractal behaviour in the absence of exact self-similarity. These techniques are applied to the MCF data of the square billiard and demonstrate that this data is also fractal, albeit over a shorter range of 2.1 orders of magnitude. An investigation of the scaling behaviour of this data demonstrates the presence of statistical self-similarity in this data. The techniques in this Chapter are used to demonstrate that statistical self-similarity is not observed in the Sinai billiard, contrary to the claims made by Sachrajda *et al.* [15]. Finally, the Sinai-square transition is addressed again following the findings earlier in this Chapter and a model is proposed to explain the transition between exact and statistical self-similarity observed in the Sinai-square transition.

Chapter 7 presents an investigation of the physical dependencies of the statistically self-similar fractal behaviour discussed in Chapter 5. It is shown that the exact self-similarity observed in the Sinai billiard is generated by the circular obstacle located at the centre of the billiard rather than merely the presence of curvature in the billiard geometry. Furthermore, it is shown that all 'empty' billiards examined to date exhibit statistical self-similarity. Whilst the range over which fractal behaviour is observed is found to remain constant, the fractal dimension $D_F$ varies under changes in the various billiard parameters. An extensive investigation of the fractal dimension as a function of the various parameters was conducted. It was found that the fractal dimension, when plotted against the ratio $Q$ of the average energy level spacing of the billiard to the average energy level broadening, lies on a single curve, irrespective of billiard geometry, entrance and exit port location, area, temperature, number of modes conducting in the ports and the electron mobility in the 2DEG. The MCF was found to become non-fractal in the fully classical ($Q \rightarrow 0$) and fully quantum ($Q >> 1$) limits for the billiard. The maximum $D_F$ was found at a ratio of $Q = 1$ demonstrating that fractal MCF is indeed a semiclassical effect as proposed by Ketzmerick [12]. Finally, the double-2DEG billiard was used to investigate the dependence of the fractal dimension upon the depth of the 2DEG beneath the heterostructure surface. The potential profileof the billiard is expected to vary with 2DEG depth. It is found that the form of the curve remains intact but has a different maximum $D_F$ value.

Chapter 8 discusses suggestions for further investigations of fractal MCF in semiconductor billiards based on the results and conclusions obtained from the research presented in this thesis.

Overall, this study demonstrates the presence of fractal behaviour in semiconductor billiards – exact self-similarity in the MCF of the Sinai billiard and statistical self-similarity in the MCF of other empty billiards. The properties and physical dependencies of the fractal MCF are investigated with the hope that the observed fractal behaviour may be used as a tool for investigating electron transport in billiards in the future. Furthermore, this study also presents the results of the first experiment on a double-2DEG semiconductor billiard – a device with potential for further investigations of electron transport through semiconductor billiards and examinations of the properties of coupled billiards, which are also topical.

# Chapter 2 – Background

This thesis investigates fractal behaviour in the magneto-conductance fluctuations of semiconductor billiards. Hence it draws on two separate fields of research – low dimensional semiconductors and fractals. This Chapter introduces the basic concepts of these two fields of research to a level sufficient to understand the results presented in this thesis. Further details and in-depth treatments may be found in the many references included in this Chapter.

For clarity, this Chapter is divided into four sections. §2.1 commences with a review of low-dimensional systems introducing the concepts required to realise a semiconductor billiard such as those studied in this thesis. The physics of billiards leading up to the discovery of fractal magneto-conductance fluctuations in the Sinai billiard is discussed in §2.2. In §2.3, fractals are discussed presenting the preliminary concepts of fractal behaviour required in this thesis. Finally §2.4 returns to the discussion of billiards, presenting recent developments with a particular focus on the study of fractal magneto-conductance fluctuations.

## 2.1 – Low-Dimensional Electron Systems: Preliminary Concepts

Semiconductor technologies have advanced at an amazing rate since the discovery of the transistor by Shockley, Bardeen and Brattain in 1947 [16,17]. More that 50 years on, semiconductor devices form the basis of the most important technologies on which modern society depends. The most significant achievement of the last 50 years of semiconductor research is miniaturisation, which has continued to such an extent that modern computer chips contain over 5 million transistors, and associated circuitry, built on a piece of semiconductor material, a small fraction of the size of the first transistor. Indeed, it is this drive for miniaturisation that has opened the field of low-dimensional semiconductor physics.

### 2.1.1 – The AlGaAs-GaAs Heterostructure: Forming a Two-Dimensional Electron System

$Al_xGa_{1-x}As$ is a ternary alloy with properties that can generally be described by some interpolation between those of the binary alloys GaAs and AlAs that occur for the trivial mole fraction values $x = 0$ and $x = 1$ respectively. The properties of these materials are discussed at length in [18]. GaAs is a Γ-direct band-gap semiconductor and for $x < 0.45$, $Al_xGa_{1-x}As$ is also Γ-direct. The parameters of interest here are the lattice parameter $a$ and the band-gap energy $E_g$. In terms of $x$, these are given by $a = (5.6533 + 0.0078x)$Å and $E_g = (1.424 + 1.247x)$eV for $0 \leq x \leq 0.45$ [18]. The mole-fraction $x = 0.33$ is commonly used in studies of low-dimensional systems (and throughout this thesis) in order to maintain the direct band-gap at the Γ-point in both materials forming the heterostructure, and as a compromise between a significant difference in band-gap energy and well-matched lattice parameters. With this $x$ value, the lattice parameters of AlGaAs and GaAs differ by less than 0.05%, ensuring an essentially strain free interface with few dislocations and lattice defects, whilst maintaining a significant band-gap difference of 0.41eV. Note, however, that it is not uncommon for different values of $x$ to be used instead.

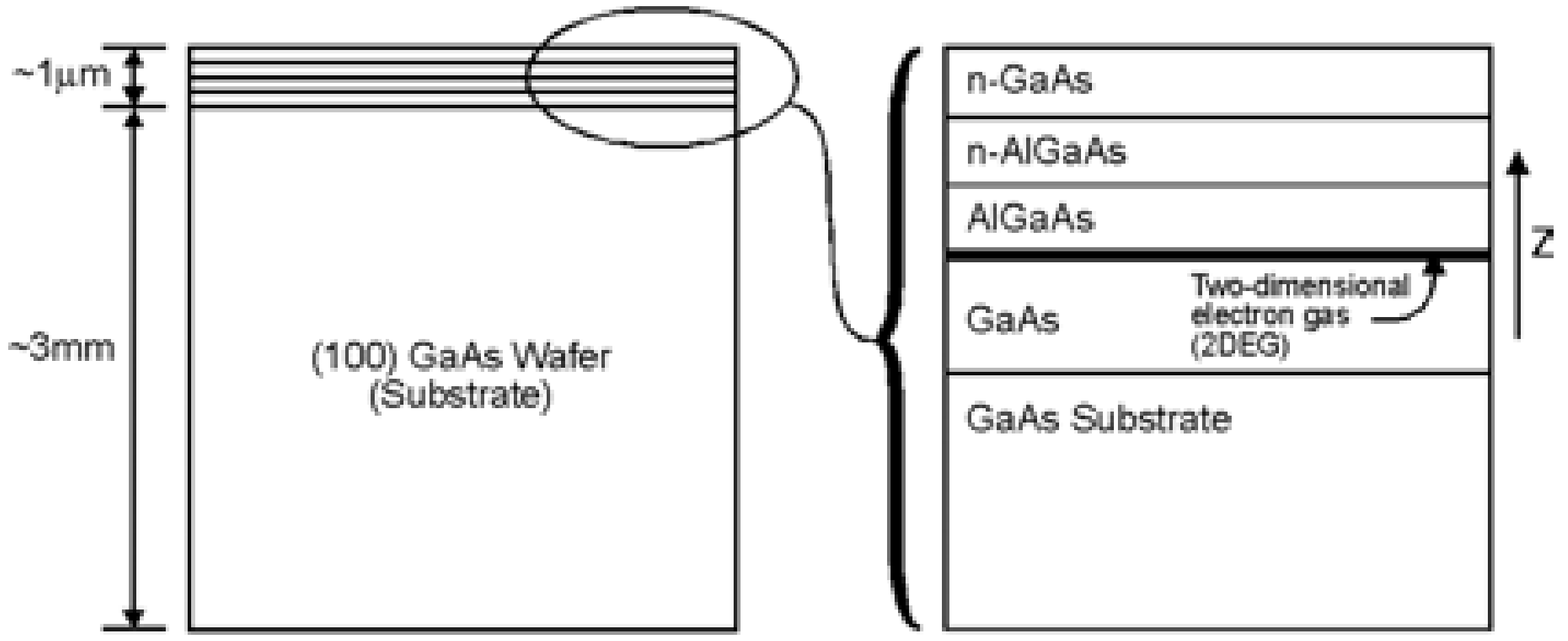

Figure 2.1: Schematic of an AlGaAs-GaAs heterostructure. Electrons are confined to a thin sheet (2DEG) located at the interface between AlGaAs and GaAs (see text).

Figure 2.1 shows a schematic of an AlGaAs-GaAs heterostructure. The heterostructure is grown on the (100) surface of a GaAs substrate wafer using a technique known as molecular beam epitaxy (see §3.1 for details). Starting from the substrate, a narrow GaAs layer is grown first, followed by a layer of undoped AlGaAs. The purpose of the GaAs layer is to even out any surface roughness in the substrate. The heterostructure interface (GaAs surface) needs to be atomically flat in order to obtain two-dimensional confinement of electrons at the AlGaAs-GaAs interface. At thermal equilibrium, the Fermi energy across the interface is constant due to the redistribution of free charge in the heterostructure. However, because both the AlGaAs and GaAs are intrinsic, there are few free electrons available. This situation is rectified by doping the AlGaAs with Si, which occupies some of the Al lattice positions in the AlGaAs. Si 'donor' atoms have an extra valence electron compared to Al. Thermal or optical excitation may free this valence electron from the Si atom, allowing it to travel about the heterostructure, thus increasing the number of free electrons. Instead of doping the AlGaAs layer immediately in contact with the GaAs however, it is better to grow an n-AlGaAs layer on top of the undoped AlGaAs 'spacer' layer. This technique is known as modulation doping [5,19] and increases the mobility of electrons confined at the AlGaAs-GaAs interface by increasing the distance between the ionised Si donors and the heterostructure interface.

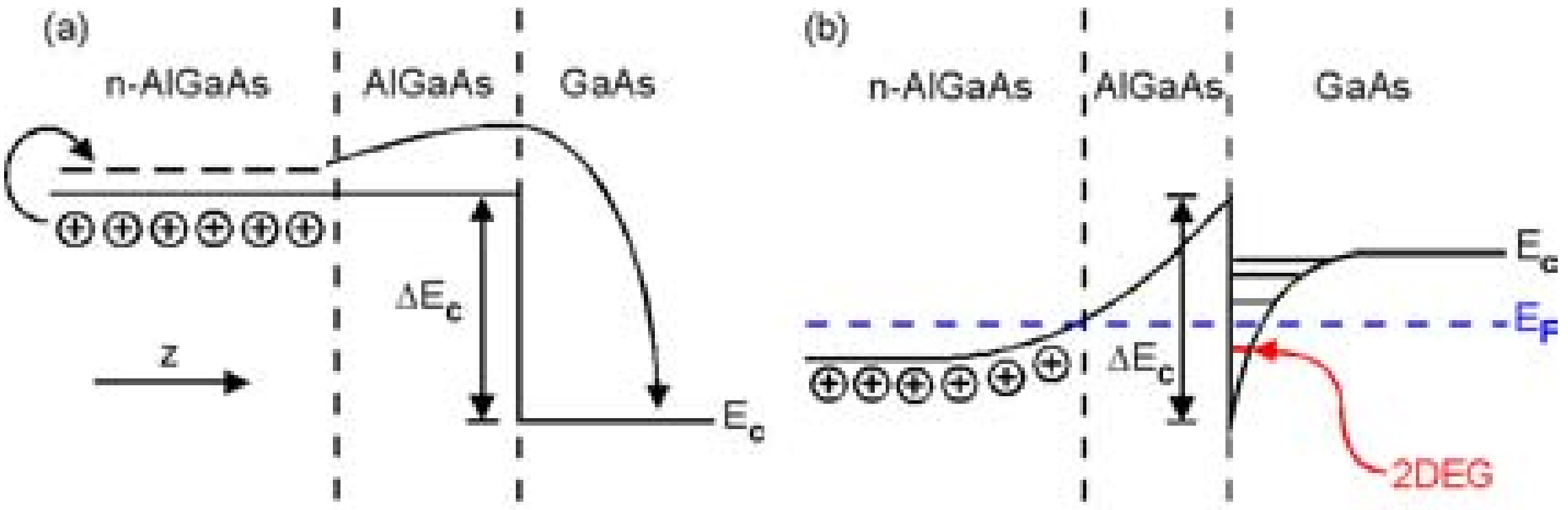

Figure 2.2: (a) Conduction band diagram of the n-AlGaAs/AlGaAs/GaAs layers at the instant of contact and (b) after free charge has redistributed. Electrons are confined to the ~10nm wide potential well located at the AlGaAs-GaAs interface to form a two-dimensional electron gas (2DEG).

The unanswered question at this point is how the two-dimensional confinement is achieved. Figure 2.2(a) shows the conduction band structure for the n-AlGaAs/AlGaAs/GaAs heterostructure at the instant of contact. The conduction band discontinuity is caused by the band-gap mismatch between AlGaAs and GaAs. At this point, the valence electrons remain with the Si atoms until some excitation (typically optical at low temperatures) frees them so that they can move up to the conduction band. These electrons are then free to move and they migrate to the lowest potential, that is, into the GaAs. As the electrons migrate into the GaAs they are unable to return to the n-AlGaAs and AlGaAs layers because they lack the energy to get back over the conduction band discontinuity $\Delta E_c$. This migration process leads to an accumulation of electrons in the GaAs, raising the potential in the GaAs, lowering the potential in the n-AlGaAs and modifying the shape of the conduction band. Electrostatic attraction between the ionised Si atoms and the free electrons keeps them within close proximity of the AlGaAs-GaAs interface. This results in the free electrons being held in an approximately triangular potential well in the GaAs close to the interface, as shown in Fig. 2.2(b). This well is ~10nm wide, comparable to the Fermi wavelength $\lambda_F$ of the electrons. Hence the energy spectrum of the electrons in the direction perpendicular to the interface is split into a set of discrete energy states as shown in Fig. 2.2(b). Typically only the ground state is occupied. As a result of this confinement, the electrons are only free to travel in the plane parallel to the interface, despite the fact that their wave-function has spatial extent in the direction perpendicular to the interface, forming a quasi-two-dimensional system known as a two-dimensional electron gas (2DEG). Since the 2DEG is located in the GaAs layer, the various material dependent parameters such as the electron effective mass $m^*$ will take the values normally found for electrons in bulk GaAs. An extensive discussion of the electronic properties of 2DEGs is presented by Ando, Fowler and Stern [20].

## 2.1.2 – Lateral Confinement of the 2DEG

Further studies of electron transport and quantum interference effects can be performed by restricting the 2DEG to a specific geometry in the plane of the 2DEG. This lateral restriction is commonly achieved in one of three ways: deep etching (Fig. 2.3(a)), shallow etching (Fig. 2.3(b)) or using negatively biased metal gates on the heterostructure surface (Fig. 2.3(c)). Etching and surface-gate fabrication techniques are discussed in detail in Chapter 3.

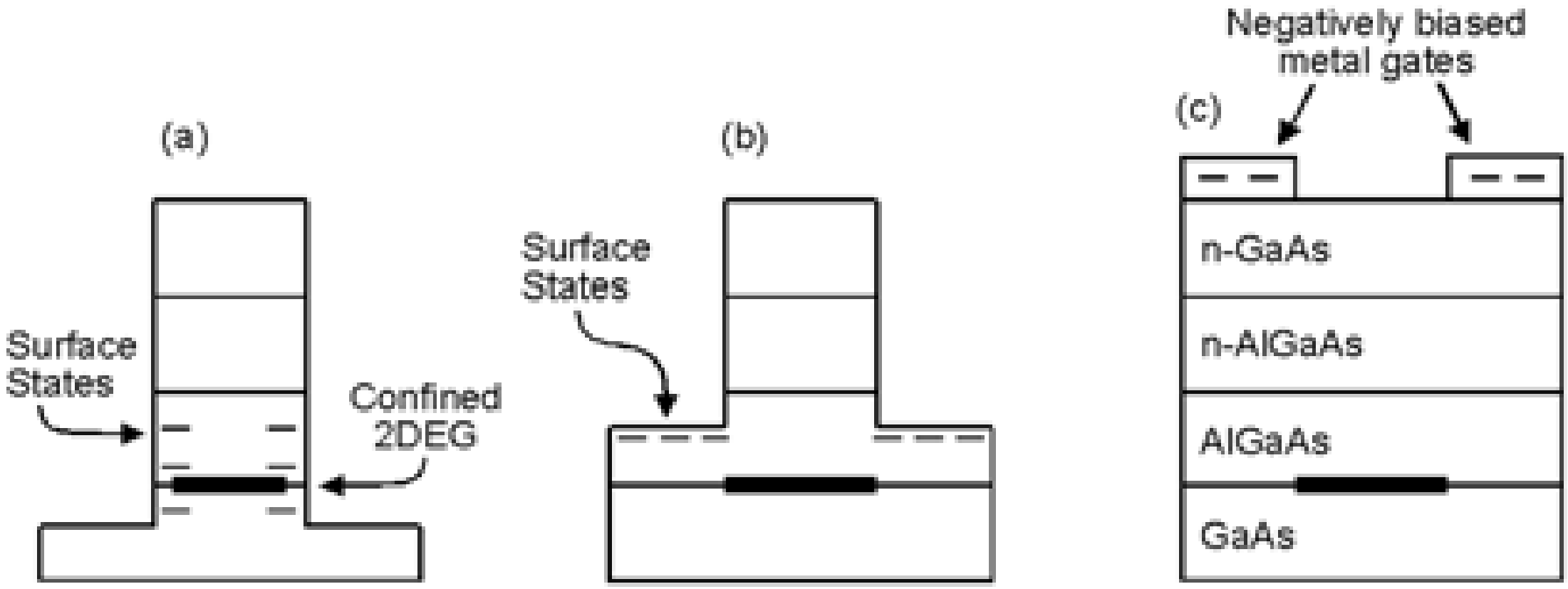

Figure 2.3: Methods for lateral confinement of a 2DEG: (a) deep etching, (b) shallow etching and (c) using negatively biased surface-gates.

Deep etching acts to directly confine the 2DEG by limiting the lateral extent of the AlGaAs-GaAs interface. In reality, however, the 2DEG is confined to a region slightly smaller than the remaining AlGaAs-GaAs interface. This is due to electrostatic repulsion by negatively charged surface states that form on the new surfaces created during the etching process. With shallow etching, rather than etching through the AlGaAs-GaAs interface, etching is stopped in the AlGaAs layer above the 2DEG. Again negatively charged surface states form at the new surface. These surface states electrostatically repel electrons in the 2DEG, leading to depletion of the 2DEG below the etched regions. The surface-gate technique also operates by electrostatic repulsion of the electrons in the 2DEG, this time using negatively biased metal gates on the heterostructure surface. This repulsion leads to depletion of the 2DEG below the surface-gates in a similar way to shallow etching. There are two disadvantages in using etching for confinement however. The first disadvantage is that the presence of surface states close to the AlGaAs-GaAs interface reduces the electron mobility in the 2DEG. The second disadvantage is that it is not possible to adjust the confinement width during an experiment. The surface-gate technique overcomes both of these problems because the heterostructure remains intact and the negative bias applied to the surface-gates can be adjusted easily allowing control over the electrostatic depletion of the 2DEG. Hence, all lateral confinement to form billiards in this thesis has been performed using surface-gate techniques.

## 2.1.3 – Ballistic Transport

An important parameter of the 2DEG in relation to the study of billiards is the elastic mean free path $l_{el}$. This parameter is defined as the average length that an electron will travel between elastic collisions with impurities and imperfections in the 2DEG. The technique used for assessing the value of $l_{el}$ is discussed in §4.2.1. The nature of electron transport through a laterally confined region is determined by the relationship between $l_{el}$ and the dimensions of the confined region. For ease of explanation, this is discussed further by considering electron transport through the simplest lateral confinement geometry – a narrow rectangular channel of width $W$ and length $L$ – shown in Fig. 2.4. Electron transport in such a channel may be classified into one of three regimes: diffusive, quasi-ballistic and ballistic[1]. The diffusive transport regime occurs when $l_{el}$ is smaller than both $W$ and $L$. In this regime there are a number of impurities in the channel and scattering from impurities dominates over scattering from the confinement boundary. As a result, the random impurity distribution rather than the confinement geometry determines the trajectories of electrons in the channel, as illustrated in Fig. 2.4(a). As $W$ (and $L$) are decreased or $l_{el}$ is increased, electron transport enters the quasi-ballistic regime where $W < l_{el} < L$. In the quasi-ballistic transport regime, impurity and boundary scattering are of roughly equal importance in determining electron trajectories through the channel, as illustrated in Fig. 2.4(b). Further decreases in $L$ (and $W$) or increases in $l_{el}$ allow both $L$ and $W$ to become smaller than $l_{el}$. Transport then enters the fully ballistic regime where there are, on average, no impurities in the channel and the electron trajectories are entirely determined by scattering from the channel boundary, as illustrated in Fig. 2.4(c). Note that by 'on average', I mean that it is still possible that one, two or even more impurities may be present within a particular channel, however taking an average channel randomly located in the 2DEG, the channel region is likely to be impurity-free. Quantitatively, 'on average' depends on how much larger $l_{el}$ is compared to $L$ and $W$. Where $l_{el}$ is very large compared to $L$ and $W$ it is very rare to find an impurity in a randomly located channel. However, if $l_{el}$ is only slightly larger than $L$ and $W$ the probability of finding an impurity in a randomly placed channel is higher.

---

1 The word ballistic means "of or pertaining to the motion of projectiles proceeding under no power…" [21] and is used because in the ballistic regime, electrons travel along straight trajectories between scattering events from the confinement boundary.

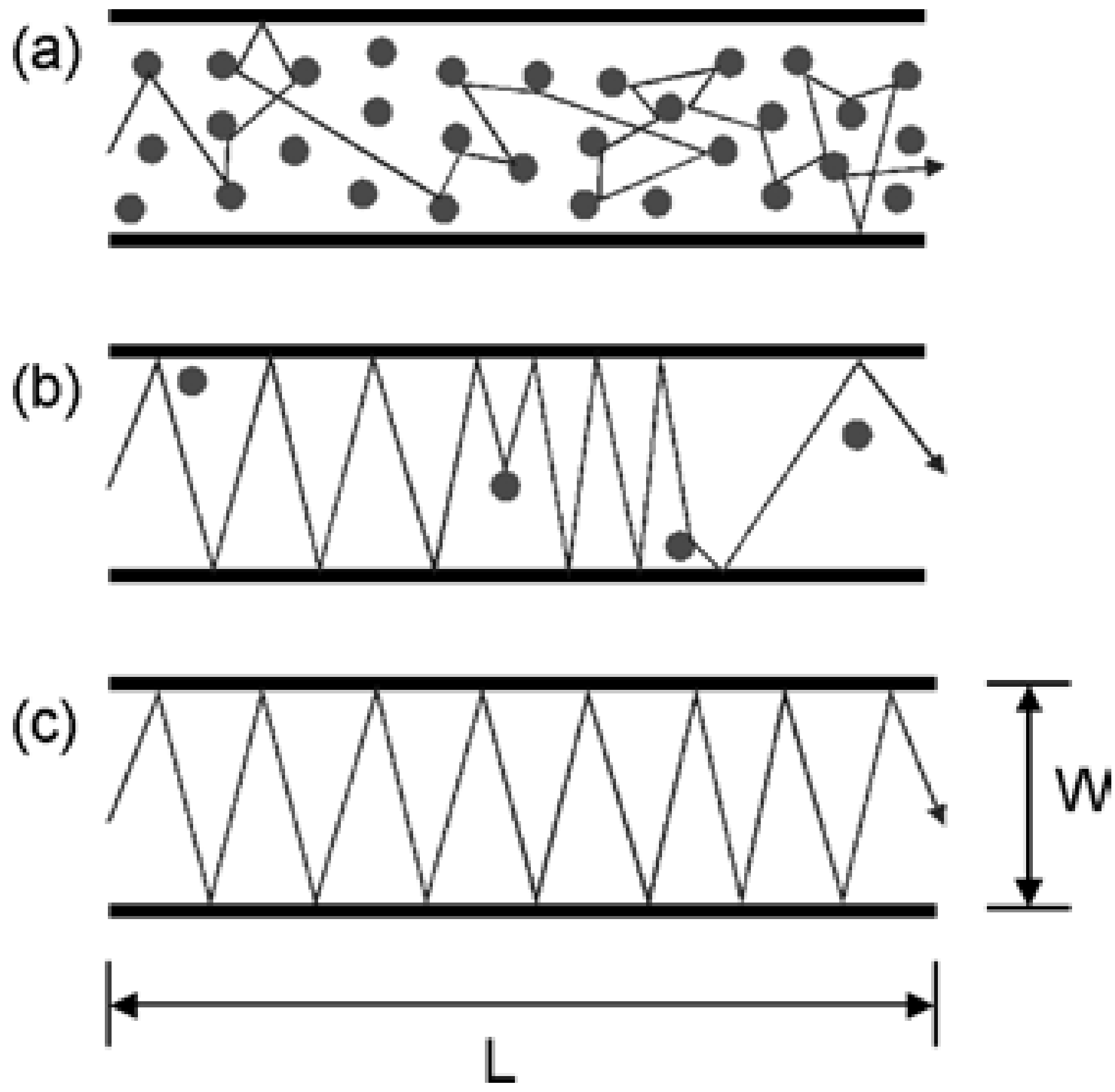

Figure 2.4: Schematic illustrating the three electron transport regimes: (a) diffusive where $l_{el} < W, L$; (b) quasi-ballistic where $W < l_{el} < L$ and (c) ballistic where $W, L < l_{el}$.

Electron transport through a laterally confined structure is investigated by measuring the electrical conductance through the structure, typically as a function of magnetic field. In the diffusive regime, the channel is the two-dimensional analogue of a macroscopic wire and electron transport follows from the Drude model of conduction [22]. In particular, the concept of conductivity $\sigma$ still holds so that the conductance through the channel is given simply by $G = (W/L)\sigma$. In the quasi-ballistic and ballistic regimes, the local conductivity is no longer defined. Instead it is necessary to view the conductance in terms of the transmission probability $T$ of electrons through the channel via the Landauer formula $G = (e^2/h)T$ [23]. The Landauer formula and its derivation is discussed in [22]. For the remainder of this thesis only ballistic transport will be discussed. The physics of diffusive and quasi-ballistic electron transport in laterally confined 2DEGs is discussed extensively in [22].

### 2.1.4 – Electron Quantum Interference Effects

Another important parameter of electron transport through a laterally confined region in a 2DEG is the electron phase coherence length $l_\phi$. If $l_\phi$ is larger than the minimum path length between the entrance and exit of the confined structure, it is possible to observe quantum interference effects in the measured electrical conductance through the structure. I will begin by discussing the simplest quantum interference scenario – namely the thought experiment proposed by Aharonov and Bohm in 1959 [24] – since this serves as the basis for understanding the more complicated quantum interference effects observed in billiards. The original experiment, proposed by Aharonov and Bohm, was to split a beam of electrons travelling in a vacuum so that the two halves of the beam travel around a region containing a magnetic flux and then recombine into a single beam. At the recombination point, the two halves of the beam undergo quantum interference[2] resulting in a recombined beam intensity that was expected to oscillate as a function of magnetic field. In 1960, an experiment demonstrating this effect was successfully performed by Chambers [25]. A laterally confined region of a 2DEG in the ballistic regime is analogous to the vacuum in the original Aharonov-Bohm experiment. Hence, by using the ring-shaped structure shown in Fig. 2.5(a), it is possible to reproduce this experiment in a semiconductor system as achieved, for example, by Timp *et al.* [26] and Ford *et al.* [27]. In the semiconductor version of the Aharonov-Bohm experiment, a current of electrons is injected into the ring geometry via the bottom lead as shown in Fig. 2.5(a). Each electron may take one of two possible paths – around the left arm or around the right arm of the ring. Partial-waves of each electron take both possible paths and interfere at the top of the ring. In the presence of an applied magnetic field $B$, a magnetic flux $\Phi = BA$ is enclosed by the two possible paths where $A$ is the area enclosed by the paths (indicated by the blue shading in Fig. 2.5(a)).

[2] Note that it is the magnetic vector potential **A** rather than the magnetic field $B$ that leads to a phase difference between the two paths in Aharonov and Bohm's work (see [24] for details).

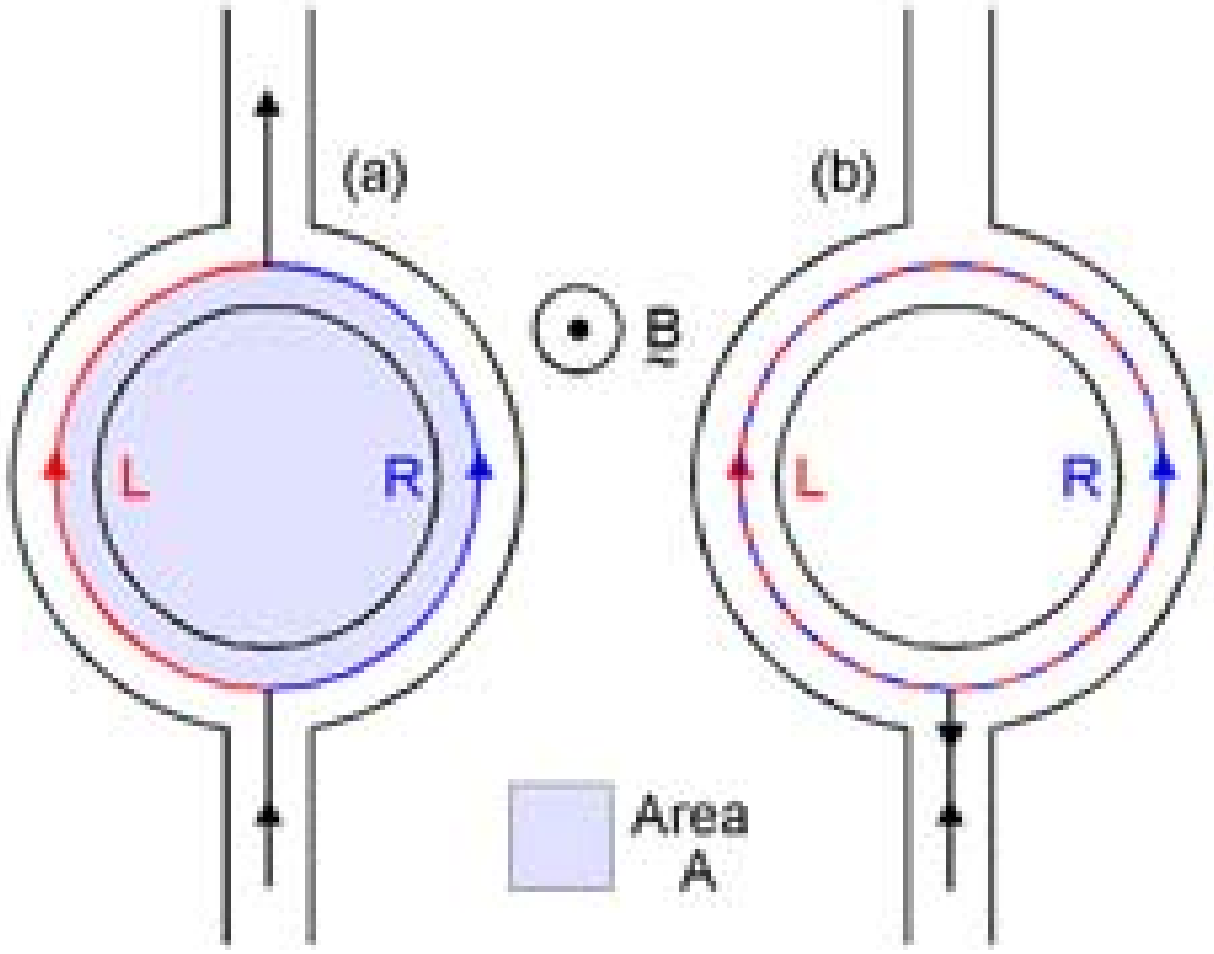

Figure 2.5: Schematics of the Aharonov-Bohm effect producing (a) $h/e$ and (b) $h/2e$ oscillations.

The magnetic flux affects the phase of the electron partial-waves along the two paths such that the phases at the recombination point (top of the ring) $\phi_L$ and $\phi_R$ are given by:

$$\phi_L = \phi_0 + \frac{1}{\hbar}\int_{\pi}^{0}(\mathbf{p} + e\mathbf{A}).(\hat{\theta})rd\theta \qquad \phi_R = \phi_0 + \frac{1}{\hbar}\int_{-\pi}^{0}(\mathbf{p} - e\mathbf{A}).(-\hat{\theta})rd\theta \tag{2.1}$$

where $\phi_0$ is the phase of the electron at the separation point, $\mathbf{A}$ = (0, $Bx$, 0) is the magnetic vector potential (Landau gauge – rectangular coordinates) and $\mathbf{p}$ is the electron momentum. Cylindrical coordinates have been used in the integrals over the electron paths and $\hat{\theta}$ is the tangential unit vector. This results in a phase difference at the recombination point of:

$$\delta\phi = \phi_L - \phi_R = \frac{e}{\hbar}\int_{-\pi}^{\pi}\mathbf{A}.\hat{\theta}rd\theta = \frac{2\pi BA}{\Phi_0} \tag{2.2}$$

where $\Phi_0 = h/e$ is the flux quantum. The phase difference $\delta\phi$ is dependent only upon the magnetic field $B$ and the area $A$ enclosed by the two possible paths and changes by $2\pi$ for each quantum of flux $\Phi_0$ threaded through the area $A$. Hence the quantum interference at the top of the ring modulates the conductance through the ring producing magneto-conductance oscillations with a period of $\Delta B = h/(eA)$. Note that it is also possible for the electron partial-waves to travel all the way around the ring before recombining, as shown in Fig. 2.5(b). This process constitutes coherent backscattering into the entrance lead. Such trajectories are time-reversed pairs, enclose twice as much area, and hence twice as much magnetic flux, producing oscillations with period $\Delta B = h/(2eA)$ known as Al'tshuler-Aronov-Spivak (AAS) oscillations [28]. These time-reversed trajectories are responsible for the weak localisation effect [29], which is discussed in relation to billiards in §2.2.3.

An important distinction can be made between the $h/e$ oscillations and the $h/2e$ oscillations. Suppose that one of the arms in the experiment discussed above is slightly longer than the other arm. In the case of the $h/e$ oscillations, there will be an added term in the phase difference $\delta\phi$ that is dependent upon the length difference between the two paths in the two arms. This means that $\delta\phi$ does not necessarily equal zero at $B = 0$ and hence the conductance oscillation due to the interference is not necessarily a maximum at $B = 0$ either. In contrast, for the $h/2e$ oscillations, the path length for both partial-waves is the same, irrespective of the fact that one arm is longer than the other, because they follow the same path in opposite directions. Hence $\delta\phi = 0$ at $B = 0$, leading to coherent backscattering into the entrance lead and a reduced conductance through the ring geometry. As a result the $h/2e$ magneto-conductance oscillations always have a minimum at $B = 0$. The importance of this distinction between $h/e$ and $h/2e$ oscillations will be seen in §2.2.3.

# 2.2 – Electron Transport in Semiconductor Billiards

## 2.2.1 – Introduction

The concept of a billiard originated in theoretical studies of classical particle dynamics and scattering in the 1800s [30]. The name 'billiard' however, was based on the direct analogy between these systems and the game of billiards that originated in England in the 15$^{th}$ century [31], as highlighted by the title of Coriolis' paper *Théorie*

*Mathematique des effets du jeu de billard* [3] [30]. To quote a precise definition for a billiard in terms of theoretical studies, "A billiard is a two-dimensional planar domain in which a point particle moves with constant velocity along straight line orbits between specular bounces from the boundary of the domain" [32]. Note that under these definitions the billiard is a ballistic system. The study of billiards returned to importance in the 1970s following the work of Lorenz in the early 1960s [33] from which 'chaos' was born[4] [34]. In particular, billiards were used as a tool for theoretical investigations of chaotic particle dynamics, with particular geometries exhibiting chaotic dynamics (e.g. stadium, Sinai billiard) and others exhibiting non-chaotic, integrable dynamics (e.g. square, circle) [32,35-37]. In the 1980s and 1990s, the theoretical study of billiards extended further to quantum chaos – the quantum mechanical limit of chaotic systems [37-40].

Experimental realisation of quantum-mechanical billiards is non-trivial. The traditional particles used in quantum mechanical investigations are the photon and the electron. The transport of these particles through the billiard must be assessed remotely[5]. This is achieved using an open billiard with entrance and exit ports and examining the transmission of particles through the billiard. To date this has been achieved using microwave billiards and ballistic semiconductor billiards [39]. An extensive discussion of the study of quantum chaos in microwave billiards is presented by Doron, Smilansky and Frenkel [41]. Since the focus of this thesis is mesoscopic semiconductor billiards, results obtained in microwave billiards are not discussed.

## 2.2.2 – Ballistic Semiconductor Billiards

Semiconductor billiards are formed by confining the 2DEG in an AlGaAs-GaAs heterostructure to a specific planar geometry that is much smaller than the elastic mean free path of electrons in the 2DEG. The billiard is connected to the rest of the 2DEG via narrow entrance and exit openings in the confinement geometry. In the high-mobility 2DEGs used for investigations of billiards in this thesis, $l_{el}$ is of order 10μm. Hence sub-micron confinement is required to ensure ballistic transport through the billiard. Confinement is achieved throughout this thesis using the surface-gate technique discussed in §3.3. Examples of surface-gates used to define a square and a circular billiard are shown in the scanning electron micrographs in Figs. 2.6(a) and (b) respectively. The various geometries investigated in this thesis are discussed in Chapters 3,5,6 and 7.

---

[3] Translation: Mathematical theory of the effects of the game of billiards

[4] Note, however, that the central idea of chaos, namely the sensitivity of some systems to their initial conditions, was understood by Poincaré at the turn of the 20$^{th}$ century, well before Lorenz's work [34].

[5] Even if it were possible to observe these particles travelling through the billiard, the act of observation would change the behaviour of the particle.

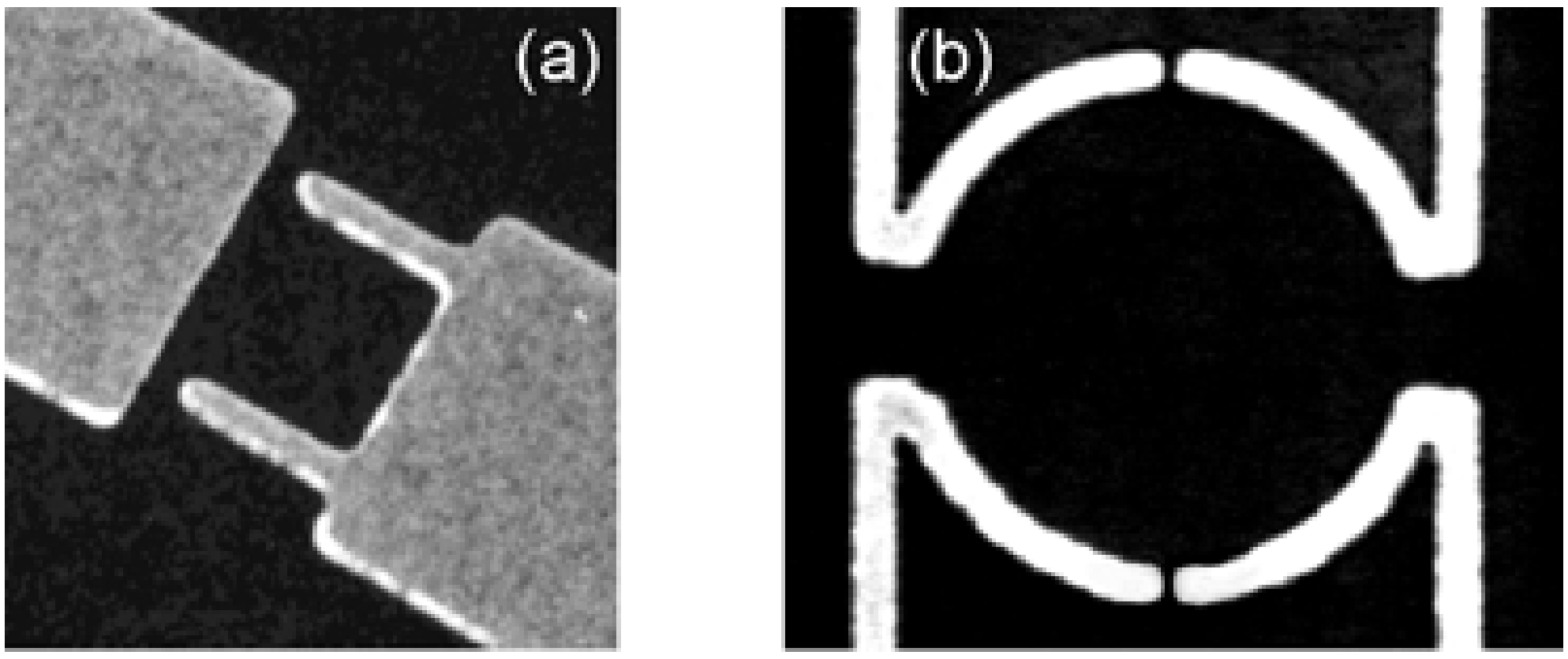

Figure 2.6: Scanning electron micrographs of surface-gates used to define (a) a 1μm square billiard and (b) a 1μm diameter circular billiard. The small openings at the top and bottom of the circle in (b) allow the four gates to be biased independently and are fully depleted during the experiment. The only entrance/exit for electrons is via the larger openings at the left and right-hand sides of the circle.

The entrance and exit ports to the billiard are typically less than 0.2μm wide. Increasing the negative bias applied to the surface-gates reduces the width of these ports in the 2DEG. Ultimately, these ports can be 'pinched-off' (i.e. become totally depleted) at larger negative biases. Given that the typical Fermi wavelength of an electron in such a billiard is approximately 40nm, the entrance and exit ports are strongly quantised and are known as quantum point contacts (QPCs). The properties of the QPC are discussed extensively in [42]. A QPC contains a finite number of conducting modes (also known as 1D-subbands) that correspond in energy to the condition that that an integer number of half wavelengths $\lambda = h/(2m^*E)^{1/2}$ is equal to the width of the QPC. An equal amount of current flows via each conducting mode and as a result the conductance of the QPC is quantised in integer multiples of $2e^2/h$ [7,8]. Hence, as the width of the QPC is adjusted via the surface-gate bias, the conductance changes in steps of $2e^2/h$. Characterisation of the QPCs used as entrance and exit ports in billiards is discussed further in §4.2.2. A discussion of the fabrication of semiconductor billiards and techniques for performing electrical measurements are presented in Chapter 3 and §4.1.2 respectively.

## 2.2.3 – Magneto-conductance Fluctuations

Returning to the discussion of §2.1.4, providing $l_\phi$ is longer than the minimum path length between the entrance and exit ports it is possible to observe electron quantum interference effects in the conductance through the billiard. In ballistic semiconductor billiards, $l_\phi$ values exceeding 50μm have been reported at millikelvin temperatures [43-47]. The minimum distance between entrance and exit in most billiards studied to date is of the order of (or less than) 1μm, which means that many possible trajectories traversing the billiard will still be phase coherent upon exit from the billiard. In §2.1.4, electron quantum interference was examined for the conductance through a ring-shaped geometry where the incident electron could take one of two possible paths through the ring. The conductance was found to oscillate as a function of magnetic field with period $\Delta B = h/eA$ where $A$ is the area enclosed by the two possible electron trajectories through the ring. In a billiard, however, the quantum interference processes are more complicated. The reason is that the injection properties of the entrance QPC allow a significantly greater number of possible electron trajectories between the entrance and the exit of the billiard. Both semiclassical theory [48,49] and quantum mechanical simulations [50-52] have shown that the injection distribution of the QPC is roughly collimated into narrow beams. The direction of these narrow beams depends upon the ratio of the QPC width to $\lambda_F$ and the precise geometry of the QPC [50-52].

It is possible to understand quantum interference in the ballistic semiconductor billiard using a semiclassical approach [53,54], where electron partial-waves follow all possible classical paths through the billiard, accumulating a quantum phase along those paths. This semiclassical approach is valid providing that the dimensions of the billiard are much larger than the Fermi wavelength. At the exit QPC, the various partial-waves interfere. Considering them in pairs, each pair encloses an area $A$ that leads to an oscillation in the magneto-conductance with period $\Delta B = h/eA$ (n.b. the magnetic field is applied perpendicular to the 2DEG as in §2.1.4). However, each pair encloses a different area $A$, contributing a magneto-conductance oscillation with a different period $\Delta B$. Furthermore, since the two paths in each pair are not necessarily the same length, the various magneto-conductance oscillations are not in phase either. Ultimately, the final transmission probability through the billiard is the sum of the various paired interference conditions. Hence the magneto-conductance through the billiard is the sum of the various magneto-conductance oscillations with different periods and relative phases, producing a set of reproducible magneto-conductance fluctuations (MCF) as shown in Fig. 2.7.

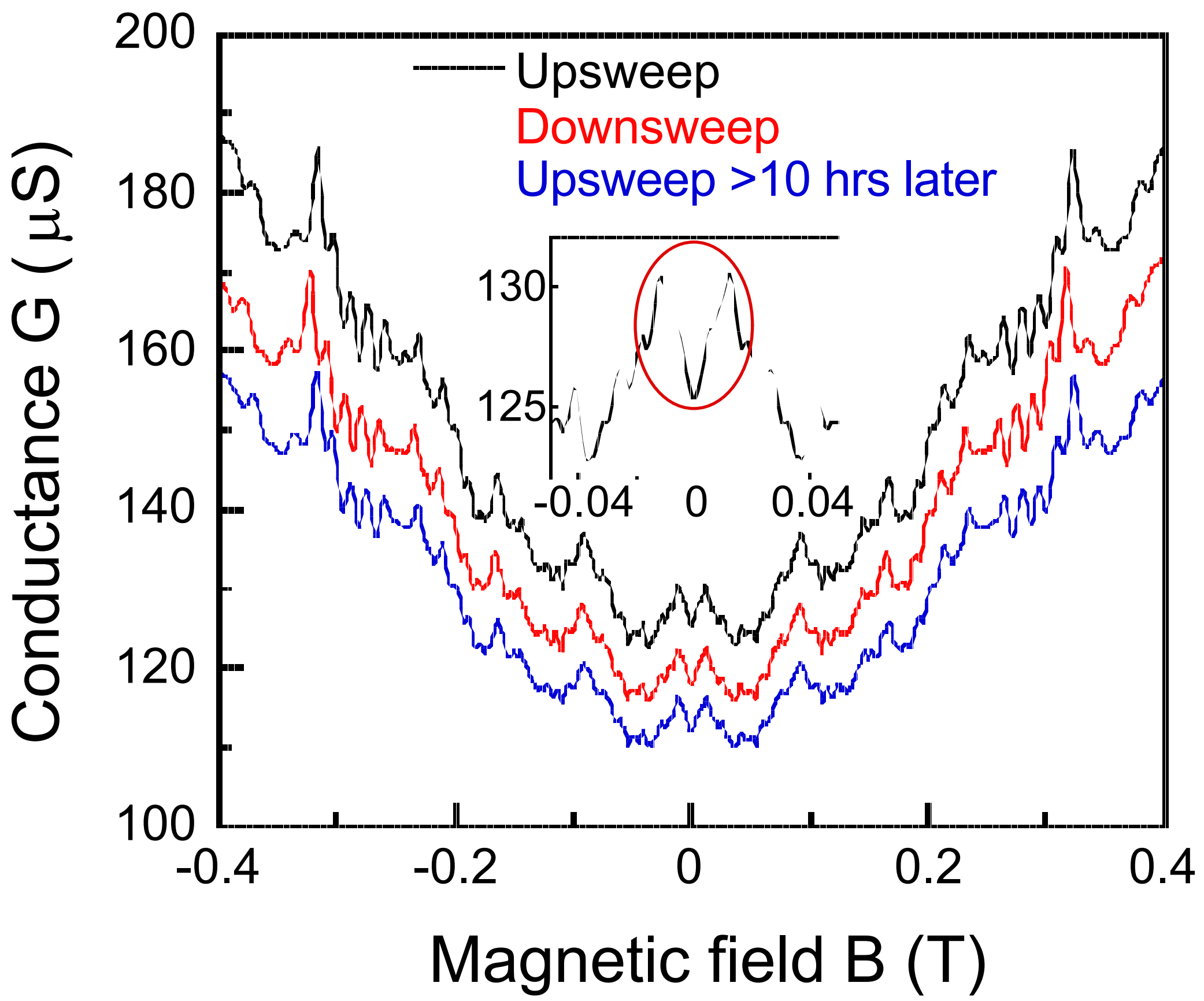

Figure 2.7: Reproducible magneto-conductance fluctuations (MCF) obtained from a ballistic semiconductor billiard investigated as part of this thesis. The inset is a close-up around $B = 0$ for the upsweep (black trace). The weak localisation (WL) feature is circled in red.

$h/2e$ oscillations due to time-reversed pairs, as discussed in §2.1.4, are also possible in billiards. In this case the two partial-waves follow the same trajectory in opposite directions back to the entrance QPC where they interfere. As with the $h/e$ oscillations, there are a number of possible time-reversed trajectories possible, each enclosing a different area $A$ and hence generating magneto-conductance oscillations with a different period $\Delta B = h/2eA$. However, unlike the $h/e$ oscillations, the time-reversed trajectory pairs have the same length and hence the various $h/2e$ oscillations are in-phase. Furthermore, at $B = 0$, the phase difference between the time-reversed trajectories is zero resulting in maximum coherent backscattering into the entrance QPC for each $h/2e$ oscillation and a conductance minimum at $B = 0$, as discussed in §2.1.4. The $h/2e$ oscillations due to each time-reversed pair add in the same way that the $h/e$ oscillations for each transmitted partial-wave pair did earlier. However, the time-reversal symmetry is gradually broken with the application of a magnetic field in billiards due to magnetically-induced curvature which acts to curve the oppositely directed trajectories away from one another. The removal of time-reversal symmetry by the applied magnetic field causes the superposition of the various $h/2e$ oscillations to appear as a conductance minimum centred at $B = 0$. This process is known as weak localisation [9,29,48]. Hence the conductance minimum at $B = 0$ is known as the weak localisation (WL) feature and is highlighted in Fig. 2.7(inset). Note that the MCF in Fig. 2.7 is superimposed upon a smoothly varying conductance background. By increasing the temperature sufficiently that inelastic scattering events reduce $l_\phi$ to less than the minimum distance between QPCs, the MCF and the WL feature are suppressed and only this smooth background remains. This indicates that the smooth background is due to classical magneto-transport through the billiard.

## 2.2.4 – Magneto-transport Regimes in the Billiard

As the magnetic field is increased, the electrons begin to follow curved paths of radius $r_{cyc} = \hbar k_F/eB$. Essentially, there are four magnetic field regimes for electron transport, as indicated schematically in Figs. 2.8(a)-(d). At $B = 0$ there is no magnetically-induced curvature and the electrons travel in straight lines between specular reflections from the billiard walls, as shown in Fig. 2.8(a). The situation shown in Fig. 2.8(a) also holds for very small magnetic fields where the magnetically-induced curvature has a negligible effect on the trajectories but induces a phase difference between possible electron trajectories through the billiard via the Aharonov-Bohm effect. Time-reversal symmetry also holds in this very low (quasi-zero) field regime, giving rise to coherent backscattering. In this regime, the semiclassical theory for MCF and WL [53,54] holds, providing the dimensions of the billiard are much larger than the Fermi wavelength. As the magnetic field is increased further, magnetically-induced curvature begins to become significant and the remaining three regimes are determined by the relationship between $r_{cyc}$ and the width of the billiard and QPCs.

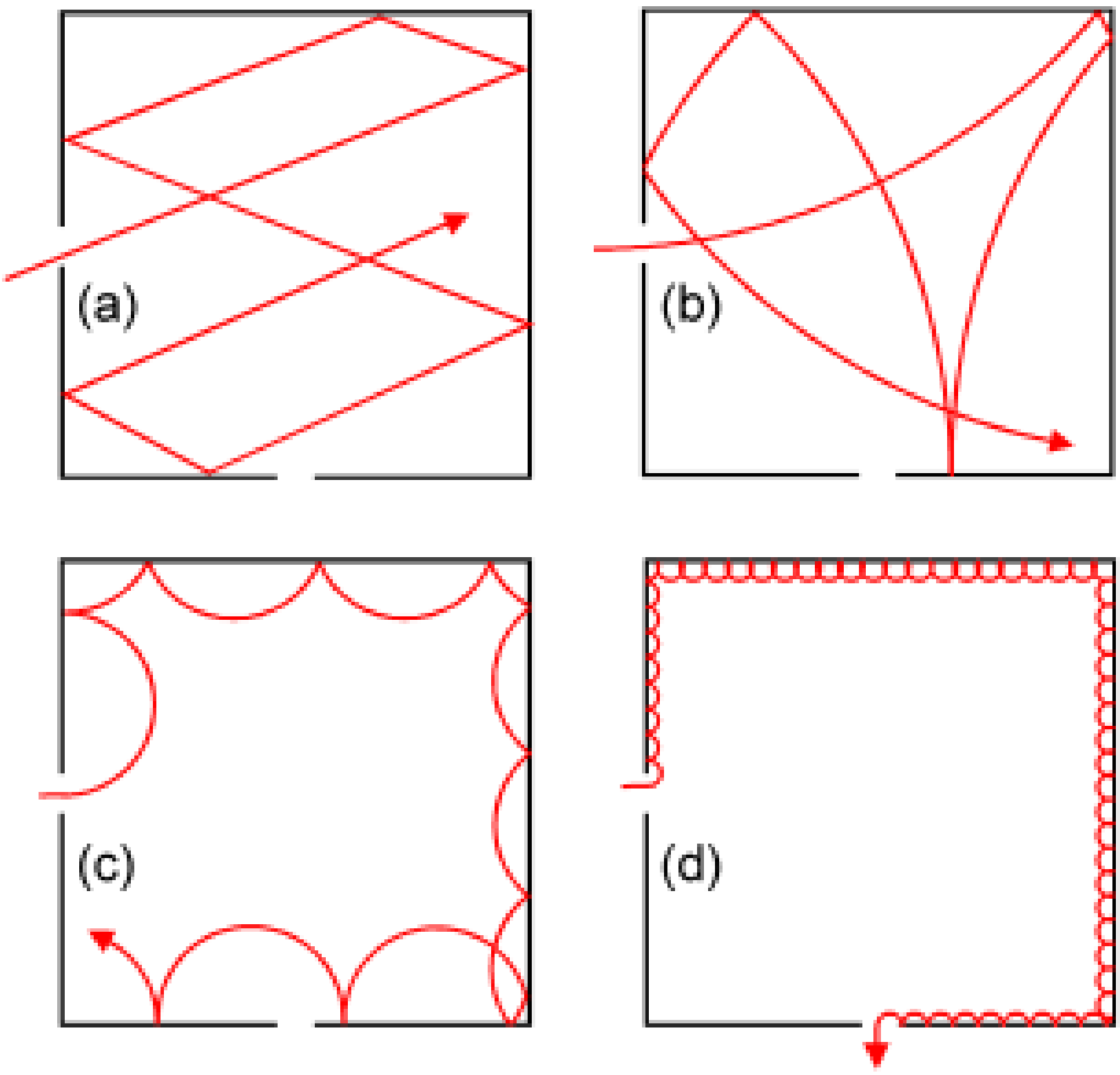

Figure 2.8: Schematic of magneto-transport regimes in a billiard: (a) very low (quasi-zero) field regime, (b) low field regime, (c) skipping orbit regime and (d) high-field regime.

Where $r_{cyc}$ is larger than the width of the billiard, the electrons follow a curved path but can still scatter from opposite walls in subsequent scattering events and still access all regions of the billiard, as shown in Fig. 2.8(b). This is known as the low field regime. The semiclassical theory of MCF [53,54] also still holds in this regime, however weak localisation is suppressed by the breaking of time-reversal symmetry by the magnetically-induced curvature. All of the investigations of fractal MCF in this thesis are conducted in the very low (quasi-zero) and low field regimes.

As the magnetic field is increased further, $r_{cyc}$ becomes small enough ($r_{cyc} < W/2$) that a cyclotron orbit can fit inside the billiard (but not the QPC). Once this occurs, electron transport takes place via skipping orbits along the billiard walls as shown in Fig. 2.8(c). This is known as the skipping orbit regime. In this regime the electron can no longer scatter from opposite walls of the billiard in subsequent scattering events and trajectories can no longer access the central region of the billiard. The character of the MCF is different in this regime and the semiclassical approach [53,54] is no longer expected to be valid. Fluctuations in this regime are used to calculate the quantum lifetime using the method discussed in §4.2.3.

Once $r_{cyc}$ becomes small enough for a cyclotron orbit to fit inside the QPCs, transport still occurs via skipping orbits, however the electron can only travel along the wall as far as the exit QPC, where is it guaranteed to exit, as shown in Fig. 2.8(d). Whilst this is often called the edge-state regime, the requirement for edge-state formation is Landau level quantisation (see §4.2.1) rather than $r_{cyc} < W_{qpc}/2$. Landau level quantisation occurs once $\omega_c \tau_Q << 1$ and $\hbar\omega_c << kT$, where $\omega_c = eB/m^*$, $\tau_Q$ is the quantum lifetime (see §4.2.3) and $T$ is the temperature [55]. Hence the regime illustrated schematically in Fig. 2.8(d) is known instead as the high-field regime. The area of the billiard $A_B$ is assessed in the high-field regime (at sufficiently high $B$ for edge-states) using the edge-state Aharonov-Bohm effect, as discussed in §4.2.4. Extensive discussions of skipping orbit and edge-state transport may be found in [22,56] and with reference to billiards in [57]. As a final note, the presence of Landau level quantisation in the bulk 2DEG surrounding the billiard can occur in the low-field regime for very high mobility 2DEGs. This is often indicated by the presence of bulk-2DEG Shubnikov-de Haas oscillations (see §4.2.1) 'contaminating' the billiard MCF. Investigations in this thesis are only conducted in the very low and low field regimes for magnetic fields where Landau levels are not formed in either the billiard or the bulk-2DEG. This issue is discussed further in Chapter 6.

## 2.2.5 – Geometry-induced Transport in Billiards

The spectral content of the MCF is directly dependent upon the distribution of enclosed trajectory areas, and this in turn is dependent upon the trajectories of the electrons through the billiard. As illustrated in Fig. 2.9 and demonstrated by classical dynamics simulations [32], the trajectories through the billiard, and more importantly the distribution of trajectory areas, depend upon the geometry of the billiard. Hence the MCF is often referred to as a 'magneto-fingerprint' of electron transport through the billiard.

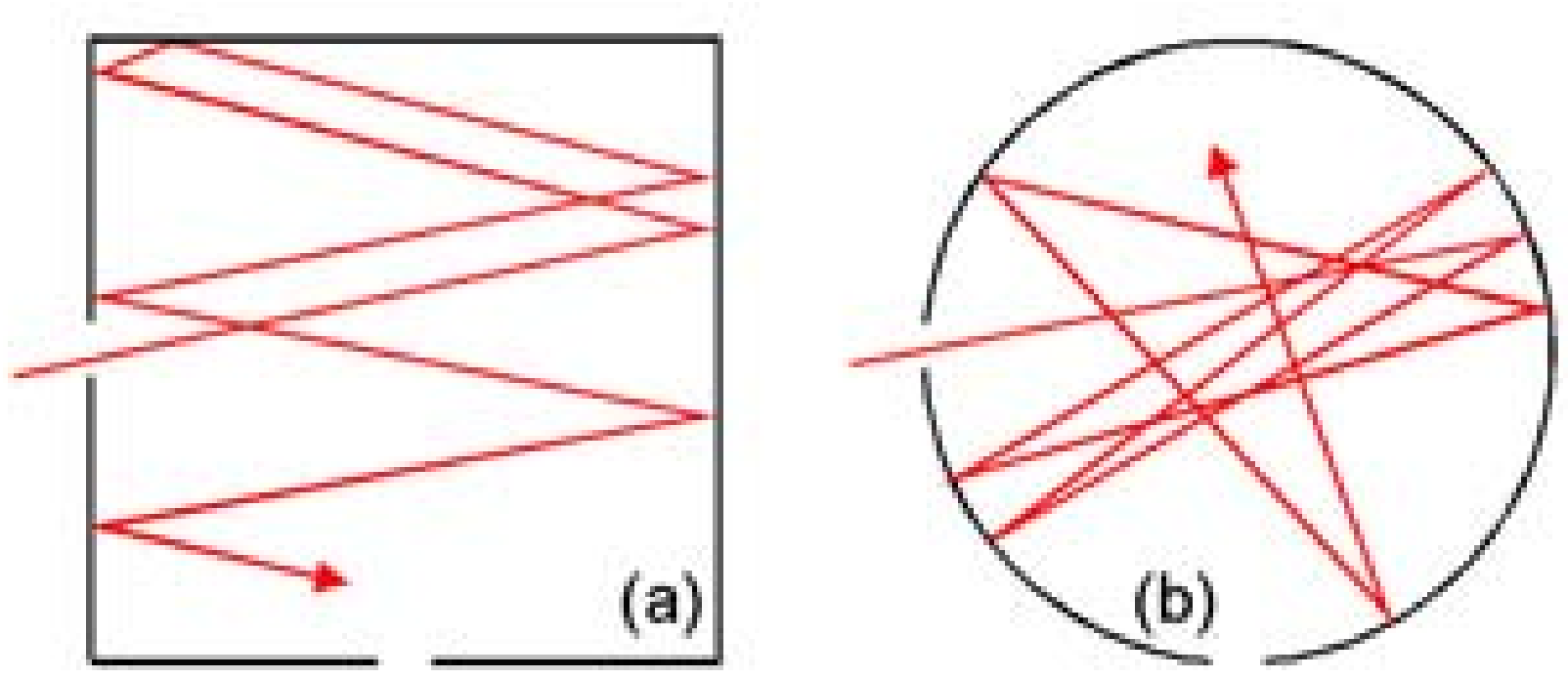

Figure 2.9: Schematic of typical electron trajectories in (a) a square and (b) a circular billiard. The spectral content of the MCF is directly dependent upon the trajectories through the billiard producing a 'magneto-fingerprint' of the billiard.

Jalabert, Baranger and Stone [53] first proposed the association between the MCF and the billiard geometry in 1990. They performed theoretical investigations of electron transport through a four-disk ballistic junction and a stadium-shaped billiard – both of which lead to classically chaotic electron dynamics. For such systems, the dwell-time $t_d$ (the time that a particular injected trajectory takes to exit the billiard) has an exponential probability distribution of the form:

$$P(t_d) \propto \exp(-\gamma t_d) \tag{2.3}$$

where $\gamma$ is the classical escape rate from the geometry [58,59]. Based on a semiclassical method following [58-60] it was determined that the probability distribution of trajectory loop areas $A$, also obeys an exponential distribution in the presence of classically chaotic electron dynamics:

$$P(A) \propto \exp(-\alpha|A|) \tag{2.4}$$

where $\alpha$ is the inverse of the typical area enclosed by an electron trajectory in the billiard. Jalabert, Baranger and Stone calculated $\gamma$ and $\alpha$ numerically and subsequently used their semiclassical approach to predict the magnetic-field correlation function of the MCF, $C(\Delta B) = \langle(G(B) - \langle G(B)\rangle_B)(G(B+\Delta B) - \langle G(B)\rangle_B)\rangle_B$, where $\langle\ \rangle_B$ indicates an average over magnetic field, obtaining a Lorentzian-squared form for $C(\Delta B)$:

$$C(\Delta B) = \frac{C(0)}{\left[1+(\Delta B/\alpha\phi_0)^2\right]^2} \tag{2.5}$$

In 1991, Jensen [61] examined further geometries in addition to those investigated in [53], deriving $\gamma$ and $\alpha$ as a function of the geometrical properties of the various geometries.

In 1992, Marcus *et al.* [9] reported the results of an experiment investigating two billiard geometries – a circle and a stadium. Reproducible MCF that were symmetric about $B = 0$ were obtained from both billiards. The symmetry about $B = 0$ is due to the Onsager relations [62,63] for four-terminal longitudinal and two-terminal magneto-resistance measurements (see §3.2.2 and §4.2.1 for details). There was a distinct difference in the MCF produced by the two billiards, with the circular billiard showing increased high-frequency spectral content compared to the stadium, confirming that the spectral content of the MCF is indeed geometry-dependent. Marcus *et al.* found that the correlation function of the MCF for the stadium obeyed the form (Eqn. 2.5) predicted in [53] and gave a value for $\alpha$ consistent with the numerical results presented in [53]. This form for the correlation function has also been observed in other experiments [64-67]. In contrast, the correlation function for the circular billiard, which is expected to generate classically regular (non-chaotic) electron dynamics, did not exhibit the form given in Eqn. 2.5. Finally, Marcus *et al.* noted the presence of a zero-field resistance peak in both billiard geometries. The width of these peaks was consistent with WL processes enclosing areas on the order of the device size.

Theoretical investigations of weak localisation in billiards were conducted by Baranger, Jalabert and Stone [48,49,54] following the experiment of Marcus *et al.* discussed above. Using Eqn. 2.5, they demonstrated that for a billiard with classically chaotic electron dynamics, the WL feature (peak) in the magneto-resistance should exhibit a Lorentzian dependence upon $B$:

$$\delta R(B) \propto \frac{1}{[1+(2B/\alpha\phi_0)^2]} \tag{2.6}$$

where $\delta R(B)$ is the WL correction to the magneto-resistance. In the case of geometries that produce classically non-chaotic dynamics, a precise universal form for the dwell-time, trajectory length and area distributions is not expected [49]. However these distributions usually exhibit a power-law dependence $P(A) \propto A^{-\gamma}$ [68-70]; this results in a linear form for the WL feature in the magneto-resistance:

$$\delta R(B) \propto -|B| \tag{2.7}$$

A number of experimental investigations of WL feature line-shape in billiards followed from the predictions of [48,49,54]. Berry *et al.* [10] investigated two stadium-shaped billiards, one of which was quasi-ballistic (width roughly $7l_{el}$) and the other ballistic (width roughly $\frac{1}{4}l_{el}$). Neither MCF, nor a WL feature was observed in the larger billiard because both $l_{el}$ and $l_\phi$ were both smaller than this billiard. However, the smaller billiard displayed reproducible MCF and a WL resistance peak with a Lorentzian line-shape consistent with the predictions of [48,49,54]. Berry *et al.* also noted that the MCF changed both as a function of the number of conducting modes in the QPCs, and following thermal cycling (to room temperature). The number of modes in the QPCs determine the injection properties of electrons into the billiard, which in turn determines the trajectories that contribute to the MCF spectral content, hence this observation was to be expected. The change in MCF following thermal cycling is due to redistribution of the random ionised donor potential in the heterostructure. This potential imposes random small-angle elastic scattering on the electron trajectories between large-angle scattering events from the billiard walls, also affecting the trajectories that contribute spectral content to the MCF. As a result of this thermal cycling effect, all measurements on a particular billiard should be performed in a single cool-down. This is a case for all of the results presented in this thesis.

The first observation of the predicted WL line-shape change was made by Chang *et al.* [11,71] in 1994. They examined the WL resistance peak in stadium-shaped and circular billiards, which are expected to produce chaotic and non-chaotic classical electron dynamics respectively. For each geometry, 48 billiards patterned on a single chip were measured at once to average out the MCF, providing a clearer WL feature. A Lorentzian line-shape was observed for the stadium and a linear line-shape for the circle. The linear line-shape was found to evolve to a Lorentzian line-shape as the temperature was increased, highlighting that phase coherence is essential in producing the linear line-shape. Upon increasing the temperature further, the WL feature was suppressed for both geometries. Berry *et al.* [72] and Chan *et al.* [73] subsequently conducted further investigations on the influence of geometry on transport finding similar results to those already discussed.

In 1995, Bird *et al.* [51,74] reported a similar change in WL resistance peak line-shape in a single square billiard. At lower gate biases, a Lorentzian WL line-shape was observed whilst at higher gate biases, corresponding to fewer conducting modes in the QPCs, and possibly a small reduction in billiard area or change in its shape, a linear WL line-shape was observed. This effect was confirmed by Micolich *et al.* in 1996 [75,76]. Numerous explanations have been offered for this effect to date. Initially it was thought that this transition was lead-induced, with rounding in the leads present at low gate biases but not at higher gate bias. It was proposed that this rounding led to the presence of classically chaotic electron dynamics, and in turn to the Lorentzian line-shape observed at low gate bias in a geometry that should otherwise exhibit non-chaotic electron dynamics [74]. Numerical simulations by Zozoulenko and Berggren [77] subsequently demonstrated that the line-shape of the WL feature is quite robust to rounding in the QPCs contrary to the suggestion in [74]. The work of Zozoulenko and Berggren suggested instead that rounding in the corners of the square may be responsible for the transition to chaotic electron dynamics indicated by the change in WL line-shape observed in [74]. Bird *et al.* [78] offered a slightly different explanation related to the phase-space filling of the billiard as a function of gate bias (and consequently QPC width). It was later observed in [51] that experiments performed by Taylor *et al.* [79,80] and Katine *et al.* [81] cast doubt on the reliability of the WL line-shape in indicating the presence of chaotic electron dynamics. Both of these experiments investigated geometries that cannot support chaotic dynamics, and Lorentzian WL line-shapes were observed in both cases. Recently, Akis *et al.* [82,83] have questioned whether the feature observed at $B = 0$ is due to weak localisation at all. At present this is still an issue of debate and since this thesis does not deal explicitly with weak localisation in billiards, this issue is not discussed further.

### 2.2.6 – Geometry-induced Transport in Billiards with Adjustable Geometry: The Sinai Billiard

Taylor *et al.* [84] reported the initial results of an experiment that investigated a single billiard where it was expected that a continuous evolution between a classically chaotic and a classically non-chaotic geometry could be achieved. Using a new fabrication technology known as a bridging interconnect (see §3.5) it was possible to independently bias a circular surface-gate located at the centre of a square surface-gate billiard. This capability allowed them to investigate the continuous transition between a square billiard (Fig. 2.10(a)) and a Sinai billiard (Fig. 2.10(b)). In the Sinai billiard, named after the Russian theorist Ya. G. Sinai, it is the combination of curved and straight walls that leads to the exponential sensitivity to initial trajectory conditions required for chaotic electron dynamics, as illustrated in the schematic in Fig. 2.10 [36]. Due to the action of the circle it is generally known as the Sinai diffuser. Taylor *et al.* noted that both the MCF and the WL feature evolved continuously as the Sinai diffuser was activated and then increased in radius. They also observed that structure was clustered on three distinct scales (coarse, fine and ultra-fine) with the Sinai diffuser activated. The finer scale was superimposed on the coarse scale, with features centred at $B = 0$ observed on both scales. In [84], this was attributed to two distinct WL processes with different trajectory areas. Whilst a transition in WL line-shape from linear to Lorentzian, as found in [11,74] was observed in this experiment, a far more significant and exciting observation was subsequently reported by Taylor *et al.* in 1997.

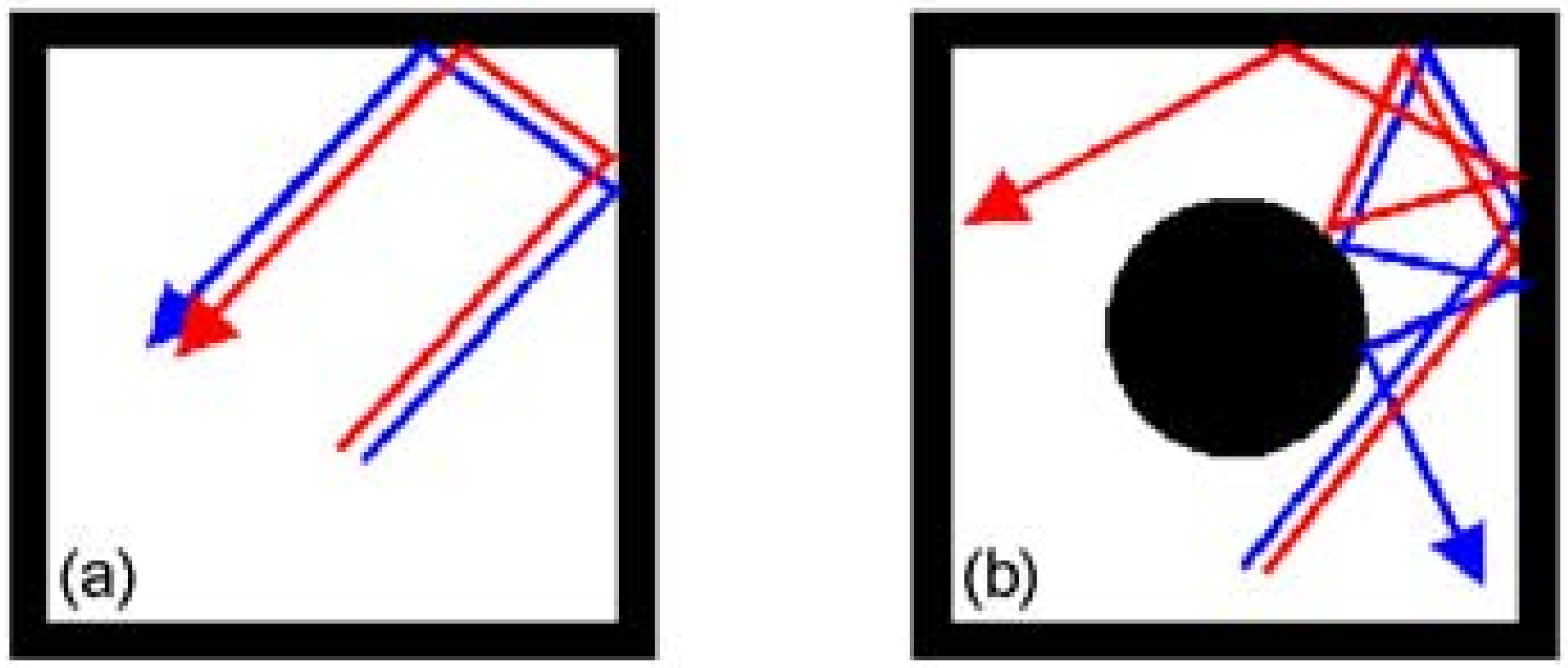

Figure 2.10: Schematic of two trajectories with similar initial conditions in (a) a square billiard and (b) a Sinai billiard. The introduction of the circle at the centre of the square leads to the exponential sensitivity to initial conditions that produces chaotic electron dynamics.

In their later paper [13], Taylor *et al.* reported exact self-similarity – a form of fractal behaviour – in the MCF with the Sinai diffuser activated. This behaviour was not present in the empty square however, suggesting that it was a product of classically chaotic electron dynamics in the billiard. Before proceeding with a discussion of fractal behaviour in billiards, an introduction to fractals and self-similarity is necessary.

# 2.3 – Fractals

## 2.3.1 – Exact Self-similarity

An object is said to display exact self-similarity if structural features of the object are exactly replicated at different scales[6]. This is demonstrated by the triadic von Koch curve shown in the top row of Fig. 2.11. At each subsequent magnification (Fig. 2.11(a) to (b) and (b) to (c)), the triangular structure is exactly repeated. Exact self-similarity is common in mathematically generated fractals and other examples include the Cantor set, Sierpinski gasket and Peano curve [2]. However, more important than the presence of repeated structure at different scales, is the scaling relationship between the structural levels that form the self-similar hierarchy.

[6] Note that there are numerous definitions of self-similarity ranging from the conceptual [34] to the mathematically rigorous [2,85]. A more conceptual definition, similar to that given by Theiler [86] is used in this thesis.

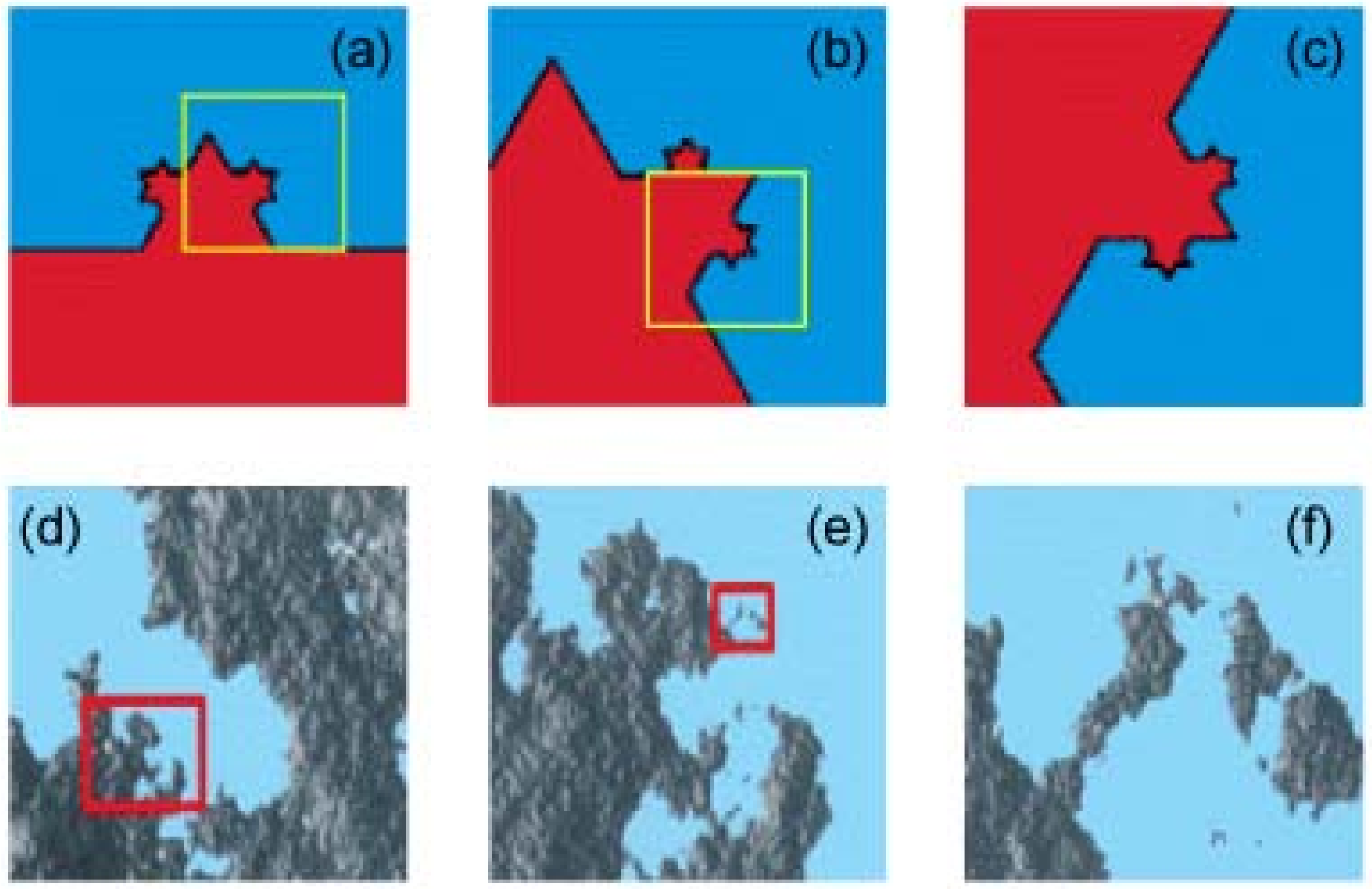

Figure 2.11: Subsequent magnifications of (a,b,c) the triadic von Koch curve and (d,e,f) an aerial view of the simulated coastline of an island (after [34]) demonstrating exact and statistical self-similarity respectively.

To examine this scaling relationship, consider the three trivial cases of exact self-similarity – a straight line, a square and a cube – as shown in Fig. 2.12. The straight line in Fig. 2.12(a) can be broken into $N$ identical line segments that are scaled by a ratio $r = 1/N$ compared to the original line (in this case 5 segments each $1/5^{th}$ the size of the original line). Treating the square and the cube similarly, the square can be broken into $N$ sub-squares, each scaled by $r = N^{-1/2}$ compared to the original square, and the cube into $N$ sub-cubes scaled by $r = N^{-1/3}$ compared to the original cube. Based on these observations it is clear that the scaling factor $r$ relating the sub-objects to the object is given by $r = (N^{-D})^{-1}$ where $D$ is the dimension of the object. This can be written in the form $N = r^{-D}$. By taking the logarithm of both sides it is then possible to extract the dimension $D$:

$$D = \frac{-\log N}{\log r} \tag{2.8}$$

In each of the three cases in Fig. 2.12, $D$ is an integer and is equal to the topological dimension $D_T$ of the object. For an object composed of line elements (Fig. 2.12(a)) $D_T = 1$, and object composed of area elements (Fig. 2.12(b)) $D_T = 2$, an object composed on volume elements (Fig. 2.12(c)) $D_T = 3$, and so on.

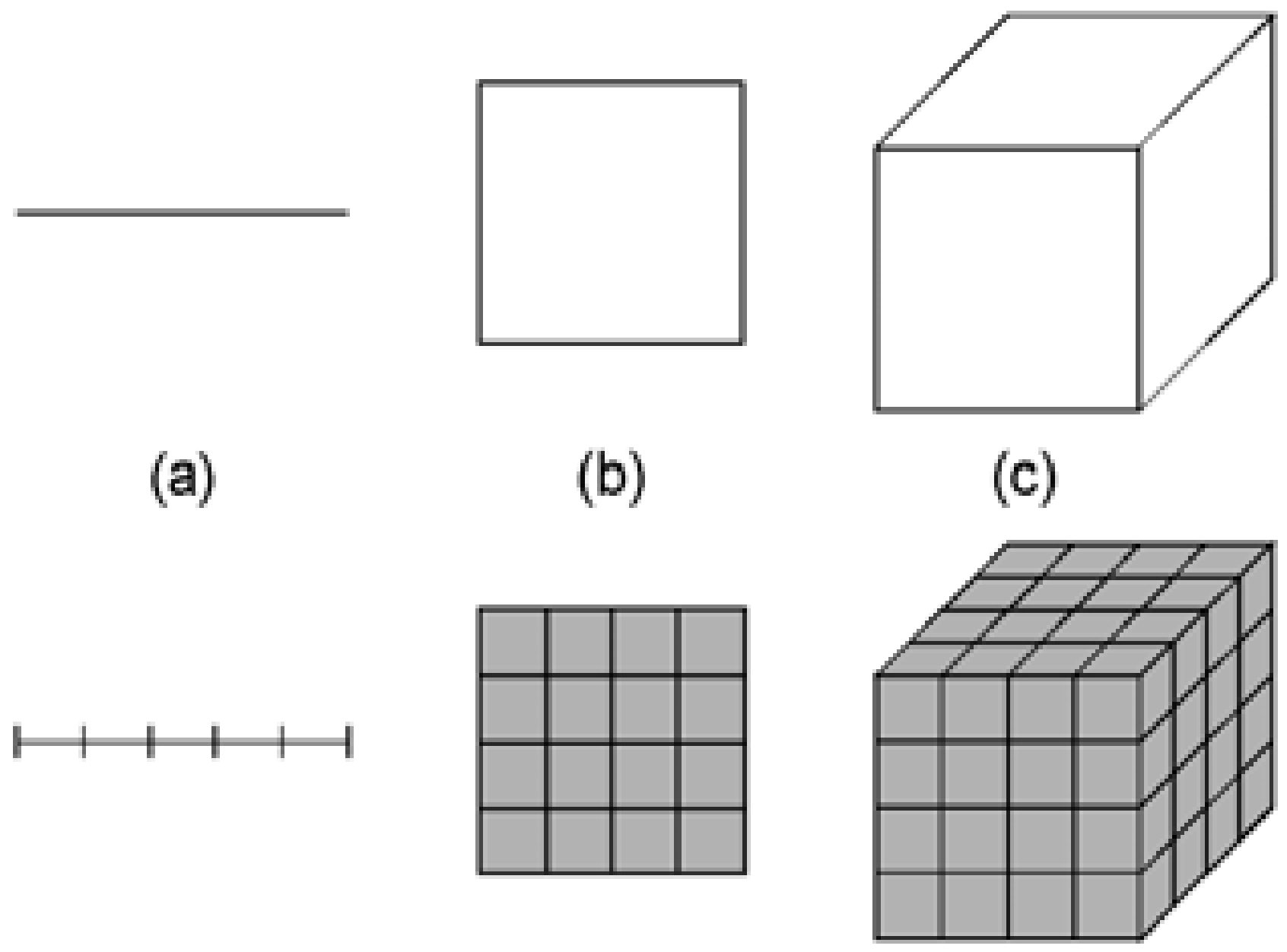

Figure 2.12: Three trivial cases of exact self-similarity: (a) a straight line, (b) a square and (c) a cube illustrating the concept of the dimension $D$.

Extending this method further, I will now consider the triadic von Koch curve (shown in Fig. 2.11) which was used as an example of exact self-similarity above. As shown in Fig. 2.13, the von Koch curve is generated by dividing a segment of length $L$ into 3 equal parts (Fig. 2.13(a)) and replacing the middle segment with two segments of length $L/3$ (Fig. 2.13(b)). In Fig. 2.13(b), there are $N = 4$ copies of the original segment in Fig. 2.13(a) each of which is scaled by $r = 1/3$. Using Eqn. 2.8, this gives a dimension $D = 1.26$. The most significant feature of this result is that $D$ is not only greater than $D_T$ (for the von Koch curve $D_T = 1$) but that it is *not* an integer.

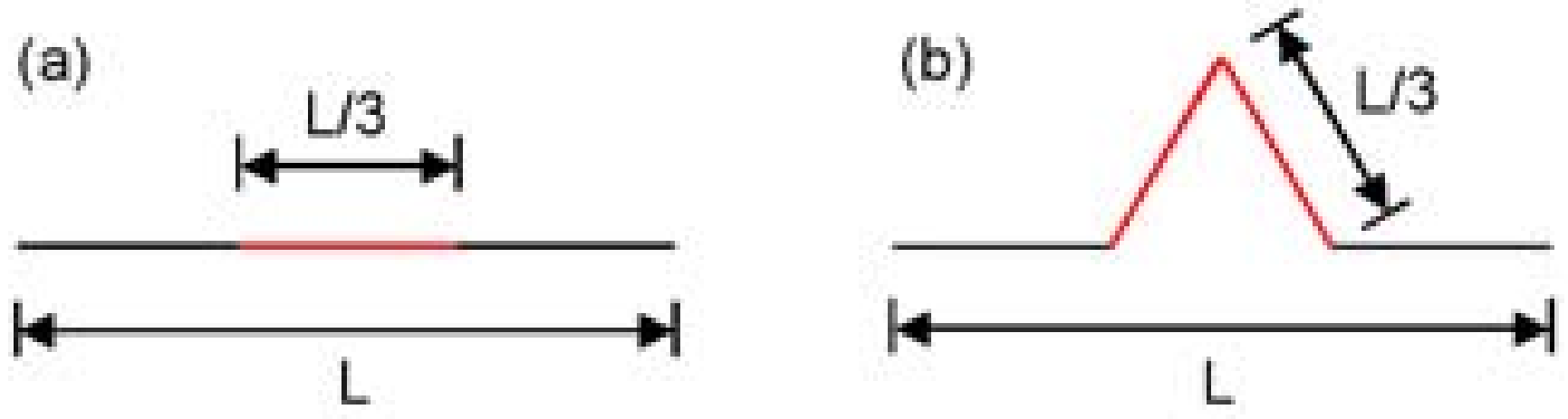

Figure 2.13: Schematic illustrating the process of generating a triadic von Koch curve.

Mathematicians such as Cantor, Peano, Lesbegue and Hausdorff have investigated objects with fractional dimension since the late 1800s. However, until the introduction of the concept of fractal geometry by Mandelbrot in 1975 [1,87], these objects largely remained mathematical curios – a "gallery of monsters" as described in [1]. Objects with a fractional dimension that exceeded their topological dimension were labelled as 'fractals' by Mandelbrot and hence the dimension $D$ is now commonly known as the fractal dimension $D_F$ when discussed in relation to fractals. Hence, in the case of exact self-similarity, the $D_F$ obtained by the method above is often known as the similarity fractal dimension. However, as the title of Mandelbrot's later essay *The Fractal Geometry of Nature* [2] suggests, his interest was more in describing the patterns observed in nature than the exactly self-similar mathematical objects. Natural patterns tend to display a second form of self-similarity known as statistical self-similarity.

## 2.3.2 – Statistical Self-similarity

In contrast to exact self-similarity, the only requirement for a statistically self-similar object is that the structure on different scales is linked by the same statistical relationship. This is demonstrated by the simulated coastline of an island shown in the bottom row of Fig. 2.11. From level to level ((d) to (e) and (e) to (f)), the precise structure is no longer identical. Instead, the levels are linked by the same statistical property, which in the case of the coastline, is the power-law scaling relationship between the coastline length and the measurement yardstick [2,88].

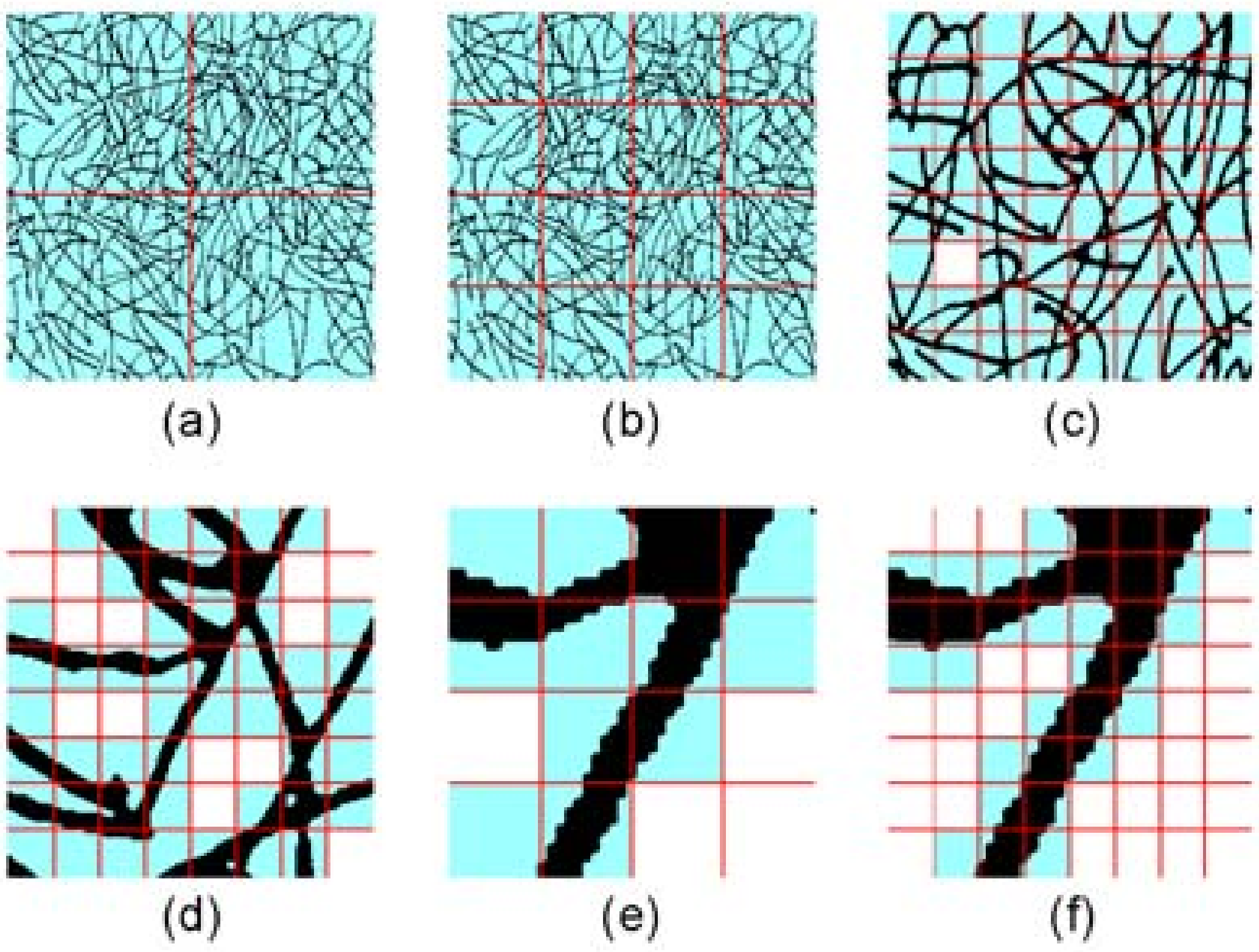

Figure 2.14: Schematic illustrating the box-counting method for progressively finer meshes (a) $\rightarrow$ (f). The blue squares contain some part of the object and hence contribute to the box-count $N(\varepsilon)$.

In the case where an object is statistically self-similar, a different approach to obtaining the fractal dimension must be taken. The most common of the possible approaches is the box-counting method, which assesses the minimum cover for the object as a function of cover element size. The box-counting method involves placing a mesh of $\varepsilon \times \varepsilon$ squares over the object and making a count of the number of squares $N(\varepsilon)$ that contain some part of the object as a function of $\varepsilon$. The counting procedure is demonstrated in Fig. 2.14 for progressively finer meshes (smaller $\varepsilon$). The object in Fig. 2.14 is an artwork by abstract painter Jackson Pollock, which serves as a good example due to the complexity of the pattern; the drip paintings of Jackson Pollock have been recently demonstrated to be fractal [89]. The blue squares contain some part of the object and hence contribute to the count $N(\varepsilon)$. For statistically self-similar fractal object, $N(\varepsilon)$ will obey a power-law as a function of $\varepsilon$ of the form $N(\varepsilon) \sim \varepsilon^{-D_F}$ where $D_F$ is the fractal dimension. Taking the logarithm of both sides allows $D_F$ to be extracted:

$$D_F \sim \frac{-\log N(\varepsilon)}{\log \varepsilon} \tag{2.9}$$

The box-counting fractal dimension $D_F$ is generally obtained by taking a linear fit to a graph of log $N(\varepsilon)$ versus log $\varepsilon$. Note that a common feature of fractals, both for exact and statistical self-similarity, is the presence of power-law scaling. Hence $D_F$ is an important parameter as it quantifies the power-law scaling relationship between self-similar levels of the structure. An extensive discussion of the application of the Box-counting method (and variants of this method) to MCF is presented in Chapter 6. Extended presentations of fractal geometry may be found in [2,90] and with more focus on the application of fractal geometry concepts in physics in [85,91,92].

## 2.3.3 – Fractals in Physical Systems

It is important to note that there is a key difference between mathematically generated fractals and fractals in physical systems. This difference is the range over which fractal behaviour is observed. For mathematically generated fractals such as the Mandelbrot and Julia sets [90], or the triadic von Koch curve in Fig. 2.11, the structure continues to infinitely large and infinitely small scales due to the recursive nature of the equation that defines these fractal objects. This is impressively demonstrated for the Mandelbrot set in [34]. In contrast, fractals in physical systems are observed over a finite range that is limited by upper and lower 'cut-offs'. These cut-offs can have one of two origins. The first origin is due to the physics that the particular system obeys and these are known as 'physical' cut-offs. The second origin is due to limitations in either the measurements or the fractal analysis techniques and these are known as 'analytical' cut-offs. The range over which fractal behaviour is observed is the overlap of the range between the physical cut-offs and the range between the analytical cut-offs. In other words, the upper cut-off on fractal behaviour is the minimum of the analytical and physical upper cut-offs and the lower cut-off on fractal behaviour is the maximum of the analytical and physical lower cut-offs. As an example, consider the coastline of an island presented in the bottom row of Fig. 2.11. The upper and lower physical cut-offs in this case are of the order of the circumference of a circle around the island and the size of the grains of sand that make up the coastline. Taking a typical example, these limits are on the order of km and mm respectively. Hence the physical cut-offs span a range of approximately 6 orders of magnitude in this example. Returning to the analysis of coastline lengths made by Richardson [88], the upper and lower analytical limits are provided by the maximum and minimum scales of the map of the coastline. In the case of Richardson's work, the analytical cut-offs span between 1 and 2 orders of magnitude. Whilst the coastline may still be fractal over the full 6 orders of magnitude, it is only valid to report the range of observation, which is the range common between the two types of cut-off and hence only 1-2 orders of magnitude. Suppose however, that the map can be sufficiently improved that it contains all of the details of the coastline and the analytical cut-offs lie outside the physical cut-offs. The range of fractal behaviour will then be limited by the physical cut-offs and observed over the 6 orders of magnitude, irrespective of any further improvement in the range between measurement and analysis cut-offs. Hence it is generally the aim in physical systems to expand the range between the analytical cut-offs, by improving the measurements and analysis, in order to observe the full range of fractal behaviour between the physical cut-offs and analyse the origin of these physical cut-offs.

There has been considerable debate recently regarding the validity of fractals observed over limited ranges [93-98]. Malcai *et al.* [93] and Avnir *et al.* [94] have investigated all reports of fractal behaviour in physical systems published in the Physical Review series of journals in the 1990s. They found that the reported ranges of observation of fractal behaviour lay between 0.5 and 3.8 orders of magnitude [93] with an average range of 1.3 orders of magnitude [94]. However, the few observations that occurred over more than 3 orders of magnitude in [93] seem to have been conveniently ignored in obtaining the average in [94]. In [94], Avnir *et al.* appear to claim that the limited range fractals may not be valid as fractals *per se*. However, Avnir *et al.* [97] later appear to retract from this view in light of criticism by Mandelbrot [95] and Pfeiffer [96]. In [97], Avnir *et al.* concur with the view of Mandelbrot and Pfeiffer [95,96] that limited range fractals are no less fractal over the range of observation than fractals observed over infinite ranges but that considerable care must be taken in assigning the label 'fractal' in the first place. It is not sufficient to simply fit a power-law to the data and take the fractal dimension as the power-law exponent. The cut-offs must be carefully determined and in order for a meaningful power-law fit and fractal dimension value to be assigned, these cut-offs should span at least one-order of magnitude since power-law fits over less than one order of magnitude are unreliable. Probably the most significant comment of this debate was made by Mandelbrot [95] "…many investigations [of fractal behaviour] in numerous fields started with few decades of experimental data and later moved to many." While fractals may be observed over limited ranges, providing the analysis is careful and the label of fractal is justified, with further investigation the range of measurement, and hence observation of fractal behaviour will increase and the confidence in the claim of fractal behaviour will increase with it. This is the case with the study of fractal behaviour in the MCF data of billiards presented in this thesis.

# 2.4 – Fractal MCF in Semiconductor Billiards

## 2.4.1 – The Sinai Billiard

Following the introduction to fractals and self-similarity in the preceding section, I will now return to continue the discussion of the study of fractal behaviour in semiconductor billiards prior to this thesis. As discussed in §2.2.6, Taylor *et al.* [13] reported the observation of exact self-similarity[7] in the magneto-resistance of a semiconductor Sinai billiard. The measurements obtained in this experiment are analysed as part of this thesis and are the subject of Chapter 5. The MCF for the Sinai billiard is found to exhibit exactly self-similar fractal behaviour over 3.7 orders of magnitude.

[7]Note that technically this is 'self-affinity' rather than self-similarity. Self-similarity relies on similarity under scaling by the same factor in the two mutually perpendicular directions. In contrast, for self-affinity the scaling in the two directions is by different amounts. Whilst all of the self-similarity observed in this thesis is technically self-affinity, the convention in this field thus far (and in many other fields) is to use the label of self-similarity for both cases. The term self-similarity is also used in this way throughout this thesis.

## 2.4.2 – The Billiard as a Two-Dimensional Potential Well

It is also possible to consider a billiard as a two-dimensional potential well with a geometry that closely matches the planar geometry of the confinement. All of the investigations presented in §2.2 assumed that the potential well was hard-walled, that is, with an infinite-walled square potential profile for a particular geometry, as shown schematically for the Sinai billiard in Fig. 2.15(a). However, this assumption is actually unrealistic because the potential well is defined electrostatically by the surface-gates, which lie some finite distance above the 2DEG. Instead the billiard has a soft-wall potential profile, the precise form of which is determined by the electrostatics of the device. An example of such a soft-wall potential profile for the Sinai billiard is shown in Fig. 2.15(b). The profile shown in Fig. 2.15(b) was obtained by self-consistently solving Poisson's equation throughout the 2DEG [99].

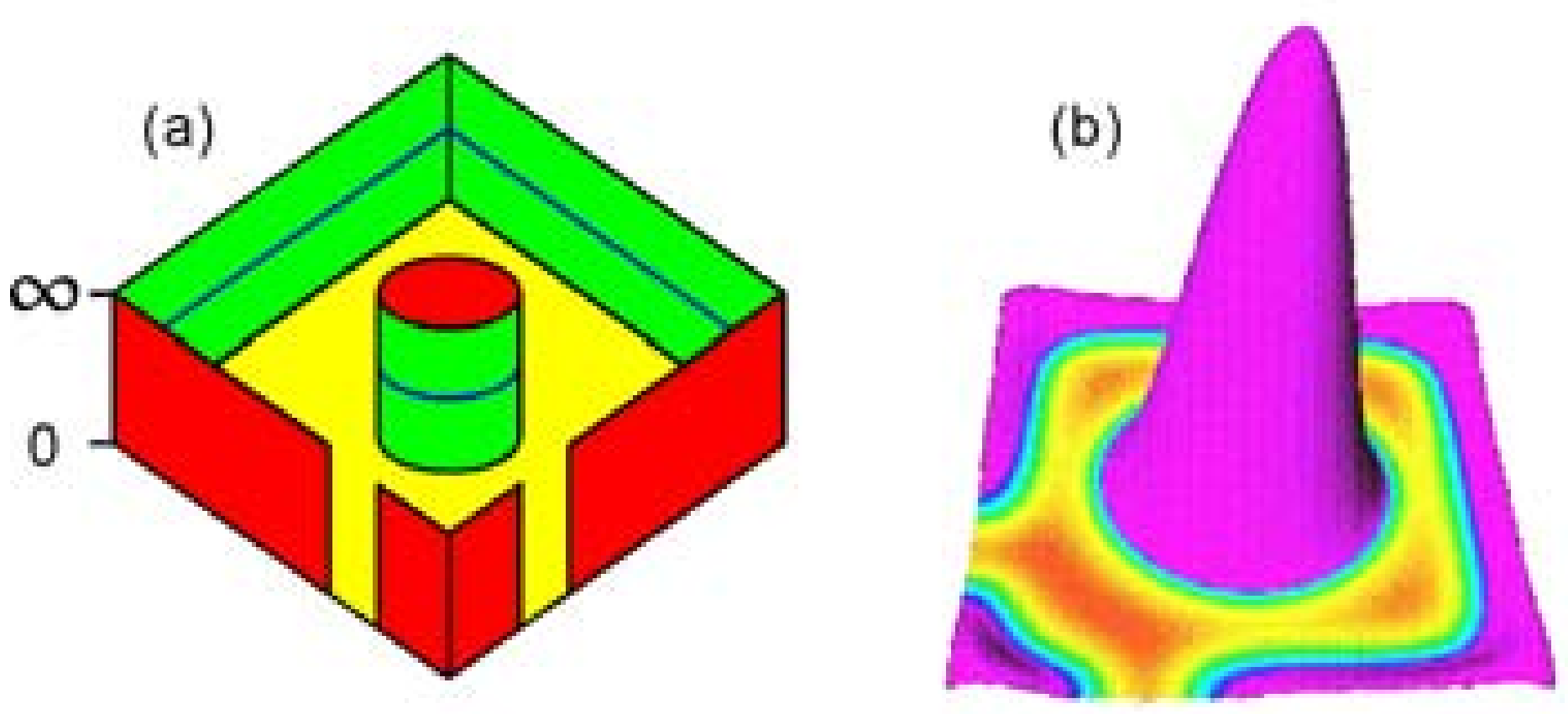

Figure 2.15: (a) Schematic of a hard-wall potential profile for a Sinai billiard. (b) Model of a soft-wall potential profile for a Sinai billiard. The model in (b) was obtained by self-consistently solving Poisson's equation for the device as discussed in [99] and was calculated by T.M. Fromhold (University of Nottingham, U.K.). The blue line in (a) and (b) indicates a typical Fermi energy in each of the potential wells.

With the rapid improvements in computer processing speeds over the last decade, fully self-consistent modelling of billiards has become viable. Such models [50,82,83,99-105] have become quite important in recent theoretical studies of electron transport through billiards. In spite of this however, it is still common for the potential profile to be assumed as parabolic for simplicity.

## 2.4.3 – Statistical Self-similarity: The Influence of a Soft-wall Potential Profile

In 1996, Ketzmerick [12] proposed that the presence of soft-wall potential profiles in mesoscopic systems in general would lead to the observation of fractal conductance fluctuations as a function of magnetic field. He demonstrated that the soft-wall potential profile leads to a mixed phase-space[8] containing both chaotic and non-chaotic regions. The non-chaotic regions form 'islands' in a chaotic 'sea' in this phase-space. Each of these islands are surrounded by a set of 'daughter' islands, each of which has its own set of daughter islands and so on [106]. These islands form an infinite self-similar hierarchy with the properties of the Cantor set and hence these islands are known as Cantori [107]. The infinite hierarchy of Cantori act as partial barriers to transport, increasing the dwell-time of trajectories that pass through these regions of phase-space [106,108]. The result of this is a power-law dwell-time distribution $P(t_d) \sim t_d^{-\beta}$ [12]. Using semiclassical theory similar to that used by Baranger, Jalabert and Stone [49,53,54] and Doron, Smilansky and Frenkel [41], Ketzmerick then obtained a trajectory area distribution that is also a power-law:

$$P(A) = A^{-\gamma} \tag{2.10}$$

This power-law area distribution led to a variance $\langle(\Delta G(\Delta B))^2\rangle_B = \langle[G(B)-G(B+\Delta B)]^2\rangle_B$ that scales as a power-law of the increment $\Delta B$ for $\gamma \leq 2$:

$$\left\langle(\Delta G(\Delta B))^2\right\rangle_B \sim \Delta B^{\gamma} \tag{2.11}$$

[8] For a dynamical system described by $N$ parameters that can vary as a function of time, a phase-space is an $N$-dimensional space where each possible state of the system corresponds to a single point in that space. As the system evolves in time, the state of the system follows a path in this space as a function of time that depends on the properties of the system. More extensive discussions of phase-space and dynamical systems may be found in [32,85,92].

where $\langle\ \rangle_B$ indicates that an average is taken over the magnetic field range $B_{min} < B < B_{max}$ of the MCF. For $\gamma \leq 2$, the MCF then has the same statistical properties as fractional Brownian motion. Hence the MCF is a statistically self-similar fractal curve with a fractal dimension given by:

$$D_F = 2 - \frac{\gamma}{2} \tag{2.12}$$

Since $0 \leq \gamma \leq 2$ then $1 \leq D_F \leq 2$, as expected for fractal behaviour exhibited by an object with a topological dimension of 1. However, Ketzmerick cites additional theory [109] in a later paper by Sachrajda *et al.* [15] to justify a further restriction of $1 \leq \gamma \leq 2$ which leads to $1 \leq D_F \leq 1.5$. The theory provided by Ketzmerick is expected to hold providing there is a mixed phase-space containing an infinite hierarchy of Cantori, and that the semiclassical theory remains valid. Surface-gate semiconductor billiards were proposed by Ketzmerick [12] as an ideal system for the observation of such fractal conductance fluctuations due to the presence of soft-wall potentials generated by the surface-gates. One final note regarding Ketzmerick's theory [12,15] must be made however. Given the large number of approximations and assumptions taken in deriving the final result, caution must be exercised in assuming the correctness or validity of the theory. This is particularly clear given the comment made by Sachrajda *et al.* in [15] that "… because the semiclassical calculations of [12] are based on many assumptions and approximations, one may even speculate if fractal [conductance] fluctuations exist at all". Hence, for the remainder of this thesis, the theory presented in [12,15] will be considered as a postulate rather than an established theory of fractal MCF in billiards.

## 2.4.4 – Observations of Statistically Self-similar Fractal MCF

The first observation of statistically self-similar fractal MCF was obtained by Hegger *et al.* [110] in quasi-ballistic Au nano-wires. Fractal MCF in this sample was observed over a range slightly exceeding one order of magnitude. The fractal dimension was obtained using both Ketzmerick's theory (i.e. calculating the variance of the MCF and using Eqns. 2.11 and 2.12) and a more conventional fractal analysis (i.e. the box-counting technique). Fractal dimension values ranging between 1.05 and 1.16 were observed with an apparent trend towards decreasing fractal dimension with increasing wire length. It is interesting to note that the observation of Hegger *et al.* initially presents as an anomaly to the theory provided by Ketzmerick [12]. The walls of the wire provide the confinement on electron transport in nano-wires. These walls however, are expected to provide a hard-wall confinement rather than the soft-wall confinement required for producing the mixed phase-space that generates the fractal MCF in [12]. Huckestein and Ketzmerick [111] later suggested that the mixed phase-space in this sample may be produced by a fortuitous 'bump' in one of the walls of the wire. However, as yet, this suggestion remains unpublished.

The first observation of statistically self-similar fractal MCF in a ballistic semiconductor billiard was achieved in 1996 as part of research for my Honours degree [75] and subsequently published [14,76,112]. In this initial study, which involved a square billiard, fractal behaviour was observed for between 1 and 1.2 orders of magnitude using the box-counting method for fractal analysis. The variance method proposed by Ketzmerick [12] and subsequently utilised by Hegger *et al.* [110] was used as confirmation of this observation of fractal behaviour. Linear behaviour in the variance analysis also extended over 1 order of magnitude. Fractal dimension values ranging from 1.18 up to 1.39 were observed. Note that in this sample, a soft-wall confinement was expected and hence the resulting fractal behaviour is consistent with the theory presented by Ketzmerick. Fractal MCF was also subsequently reported in billiards by Sachrajda *et al.* [15]. They claimed the observation of fractal behaviour over a range exceeding two orders of magnitude for a Sinai billiard and a stadium-shaped billiard. Note however, that the observation in the Sinai billiard was subsequently brought into question by Taylor *et al.* [113,114].

A comprehensive study of statistically self-similar fractal MCF is performed as part of this thesis and is discussed in Chapter 6. In particular, rigorously optimised fractal analysis techniques for use on statistically self-similar MCF are presented. Combined with improved experimental measurements (compared to [14,75,76,112]), statistically self-similar fractal MCF is clearly observed over a range exceeding 2.1 orders of magnitude in Chapter 6.

## 2.4.5 – Recent Developments in Billiards (1997-2000)

An extensive review of recent developments in the study of semiconductor billiards is presented by Bird in [115]. However, two developments relevant to later discussions will be covered here.

As mentioned in the preceding subsection, statistically self-similar fractal MCF were also observed by Sachrajda *et al.* [15]. Whilst the observation of statistically self-similar fractal MCF in the Sinai billiard is a subject of debate (see [113,114] and §6.5), their observation of fractal MCF in the stadium-shaped billiard is of interest. This device was intentionally designed to be large with wide entrance and exit ports in order to enhance the semiclassical nature of electron transport through the billiard. Furthermore, the wide entrance and exit ports allow all trajectories except for those affected by the presence of the infinite hierarchy of Cantori, to rapidly exit the billiard. The combination of these two features generates the optimum conditions for the observation of fractal MCF based on Ketzmerick's theory [12]. This device indicates that not only does the fractal behaviour appear to be robust to an increase in billiard size, it also appears to be robust for very wide leads that allow the easy escape of electrons from the billiard. One might suggest that this result is expected for two reasons. The first reason is that the influence of the soft-wall profile remains since this billiard is also a surface-gate device. The second reason is that the semiclassical approximation remains valid. Indeed, the semiclassical approximation becomes more valid given the increased billiard size and lead width. Overall, this is an interesting result as it is consistent with the suggestion [12,15] that the fractal behaviour is actually due to the trajectories trapped in the billiard by the infinite hierarchy of Cantori in the mixed phase-space.

Bird and coworkers at Arizona State University, USA and RIKEN Laboratories, Japan have investigated smaller billiards where the semiclassical approximation used by Baranger, Jalabert and Stone [49,53,54] is no longer valid. Instead, transport through the billiard must be treated using an entirely quantum mechanical approach [116]. As the billiard size was reduced, Bird *et al.* [117] observed a loss of spectral content in the MCF, which developed a more periodic character. This was attributed to the presence of wavefunction scarring [118] along the dominant periodic orbit in the billiard [116]. The dominant period of the MCF closely matches that expected from the dominant periodic orbit based on the Aharonov-Bohm relation $\Delta B = h/eA$ for different geometries with differing dominant periodic orbits and wavefunction scarring patterns [102,116].

Both larger and smaller billiards are investigated as part of a study of the physical dependencies of statistically self-similar fractal MCF. This investigation is presented in Chapter 7 and allows the investigation of fractal MCF in both the fully classical and fully quantum mechanical limits.

# Chapter 3 – Fabrication of Mesoscopic Semiconductor Billiards

The fabrication of a working semiconductor billiard requires a number of optimised semiconductor processing steps. Eight billiards, fabricated in laboratories in three different countries, were investigated for this thesis. Whilst the general process for fabricating these billiards is similar, the specific fabrication details differ between samples. §3.1 through §3.4 are devoted to the fabrication of simple (single layer gate architecture, single-2DEG) semiconductor billiards. Since a presentation of the specific, but differing details of each particular billiard would be prohibitively long, the general processing procedures presented are applicable to all samples, whilst any specific details discussed will be those used to fabricate billiards at the Cavendish Laboratory, University of Cambridge. It will be made clear where differing technical details have an important bearing on sample behaviour. §3.5 and §3.6 discuss the fabrication of the more advanced billiards presented in this thesis – the Sinai billiard and the double-2DEG billiard respectively. §3.7 outlines various important parameters for the full set of billiards investigated in this thesis.

## 3.1 – Production of a Semiconductor Heterostructure: Molecular Beam Epitaxy

The first requirement in the fabrication of a semiconductor billiard is the confinement of electrons to a 2DEG as discussed in §2.1.1. This confinement is achieved with semiconductor structures produced using Molecular Beam Epitaxy (MBE), whereby semiconductor layers with accurately defined compositions, dopant concentrations and thicknesses may be grown. MBE was developed by Cho and Arthur in 1970 [4] and is based upon the evaporation from elemental sources of component semiconductors onto a substrate in an ultra-high vacuum (UHV) ($<10^{-10}$mbar) environment. A schematic of the essential features of a MBE apparatus is shown in Fig. 3.1.

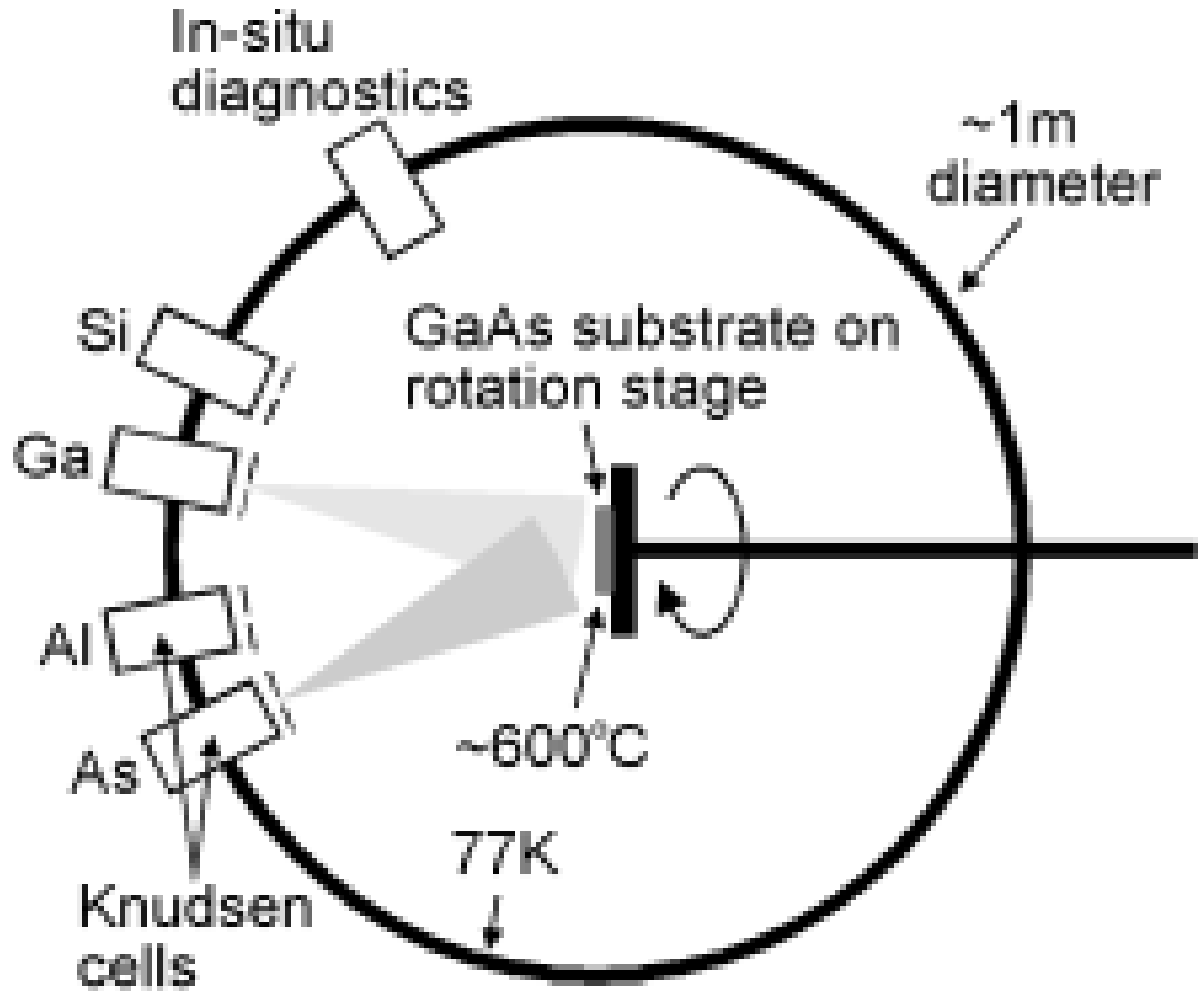

Figure 3.1: Schematic diagram demonstrating the essential features of a MBE apparatus.

A GaAs substrate is mounted on a rotation stage at the centre of a ~1m diameter stainless steel growth chamber maintained under UHV conditions. The substrate is typically a ~3mm thick wafer, sliced from a ~25mm diameter melt-grown GaAs cylindrical 'boule' where the boule axis coincides with the (100) GaAs crystallographic direction. Facing the substrate (see Fig. 3.1) is a set of Knudsen cells that contain the various constituent materials. Typically these are Al, Ga and As for producing layers of GaAs and $Al_xGa_{1-x}As$ semiconductor and Si for n-type doping of semiconductor layers. Each Knudsen cell is heated to provide a vapour of the particular material, which is collimated into a narrow beam directed at the substrate. Each Knudsen cell can be closed with a shutter to control the composition of the growing layers. The substrate is maintained at a temperature of approximately 600°C throughout the growth process to allow incident atoms that stick to the substrate to migrate along the surface to the correct lattice positions. The remainder of the growth chamber is maintained at 77K to capture any atoms that do not stick to the substrate. This prevents disruptions in crystal lattice growth due to constituent materials that are reflected from the chamber walls. The substrate is rotated at ~0.25Hz to average out non-uniformities in the incident flux of atoms due to some of the Knudsen cells being displaced off the normal to the substrate. Growth of the layers can be monitored in-situ by various techniques such as reflection high-energy electron diffraction (RHEED), X-ray diffraction (XRD) and secondary ion mass spectrometry (SIMS). In-situ growth monitoring techniques combined with the slow growth rate (~1μm/hr) and high shutter speed (~0.1s) allow precise control of individual layer depth and composition. The use of growth interrupts (where all shutters remain closed for ~1min) between layers allows the surface atoms to settle into place prior to growth of the next layer, minimising the presence of steps in the surface, and other growth defects such as dislocations and grain boundaries. Ultimately, growth of semiconductor heterostructures matching those discussed in Chapter 2 with control on doping concentration, semiconductor composition and layer thickness on the order of 1nm are possible. Growth details of the individual heterostructures used in each billiard are given in §3.7. More extensive discussions of MBE growth can be found in [119,120].

# 3.2 – Mesa Etching and Ohmic Contacts

The 2DEG confined at the AlGaAs-GaAs interface (see §2.1.1) lies parallel to, and typically ~100nm below, the heterostructure surface in the devices investigated in this thesis. The 2DEG extends laterally to within ~10nm of the edges of the substrate/heterostructure wafer as shown schematically in Fig. 3.2(a). The 2DEG does not extend completely to the edge due to electrostatic repulsion by negative surface states formed at the heterostructure edge [121]. The heterostructure often is cleaved (see §3.4) into smaller pieces following growth for ease of handling throughout the other processing stages. The second major step in fabrication involves further restricting the lateral extent of the 2DEG to a region known as a Hall-bar 'mesa' and providing electrical contact, enabling measurements of the electrical resistance of the 2DEG to be obtained.

## 3.2.1 – Mesa Formation

Restriction of the lateral extent of the 2DEG is obtained by etching the heterostructure to form a 'mesa' as shown in Fig. 3.2(e). The mesa etch process is outlined schematically in Fig. 3.2. The first step, labelled (a) in Fig. 3.2, involves depositing a layer of photo-resist (Shipley Microposit S1813 positive photo-resist) on the heterostructure surface. This is performed using a spin-and-bake technique. The wafer is mounted on a flat horizontal plate, coated in photo-resist, and spun at ~5000rpm for ~30s to form an even layer. The thickness of the resulting photo-resist layer is approximately 1μm and is inversely proportional to the rotation frequency.

Following the spinning process, the wafer is baked on a hotplate at 90°C for ~2min to dry the resist and remove any excess solvents. In step (b) a patterned chromium-on-glass mask is placed (with the chromium metallised surface closest to the wafer surface) in vacuum contact with the wafer. This mask defines the pattern within which the 2DEG is to be constrained. The wafer and mask are then illuminated with UV light.

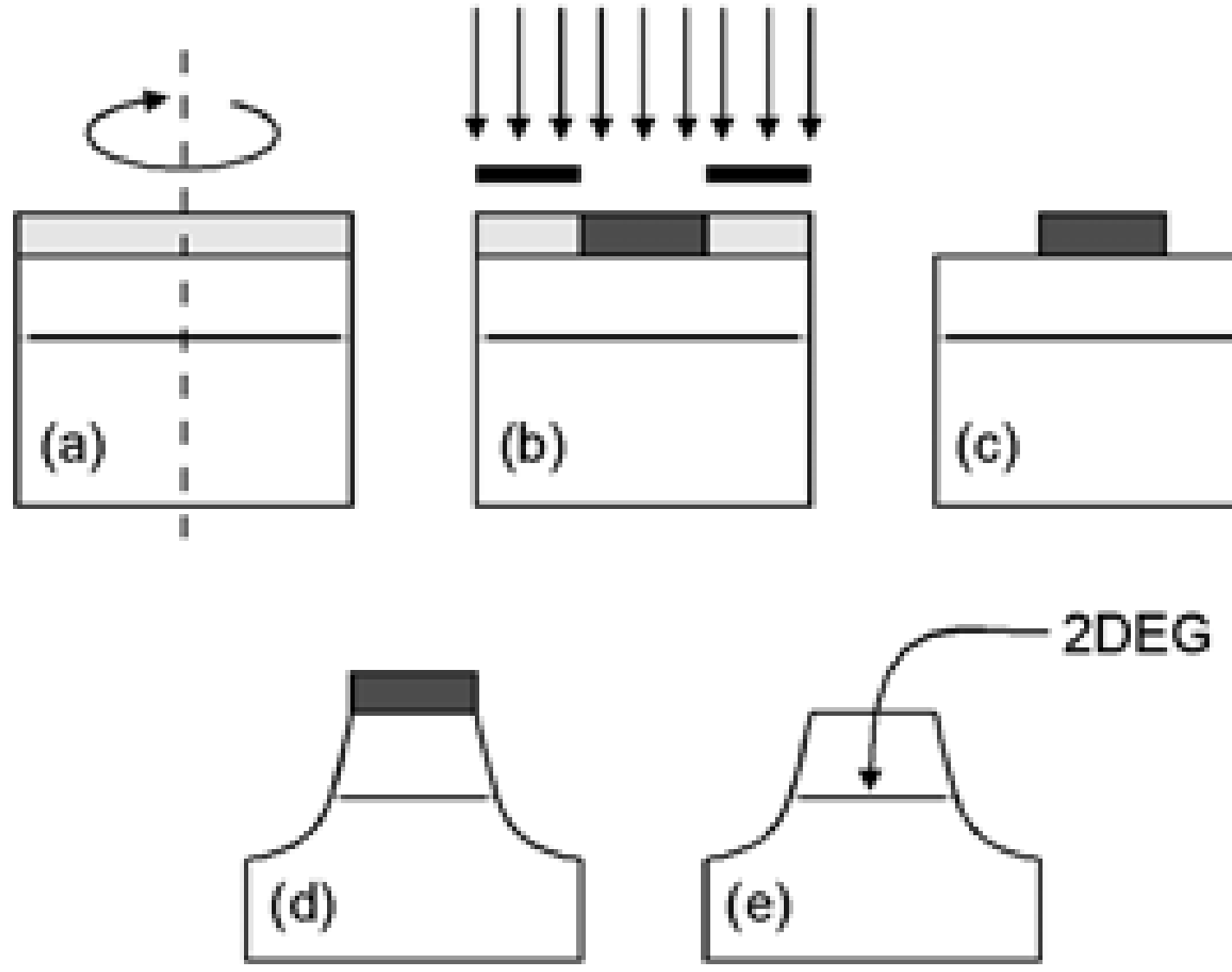

Figure 3.2: Schematic representation of the steps involved in mesa formation. (a) photo-resist deposition (shaded grey), (b) masked UV exposure, (c) development, (d) mesa etch, and (e) resist removal.

Regions to be etched are protected from UV exposure by the metallised regions of the mask whilst the remaining regions are exposed as shown in Fig. 3.2(b). The molecular structure of the photo-resist is modified by UV illumination, making the exposed regions more resistant to dissolution by the developer (Shipley MF19 developer). The wafer is soaked in the developer for ~30s (step (c)), dissolving the unexposed photo-resist and leaving the exposed photo-resist intact. The exposed photo-resist regions are not entirely robust to dissolution by the developer, so excess developer is removed by rinsing with $H_2O$ following the developing process. The next step is to perform a deep-etch through the 2DEG (step (d)). Note that in some cases a shallow-etch is used instead to achieve similar results. This step is performed by soaking the wafer in a buffered HF solution (3:4:60 HF:$H_2O_2$:$H_2O$). However, it is not uncommon for $H_2SO_4$ or $H_3PO_4$ based etchant solutions to be used [122,123]. The advantage of HF etching (Fig. 3.3(a)) over other etchants such as $H_2SO_4$ (Fig. 3.3(b)) is that it produces a gentler edge profile as shown in Fig. 3.3(a,c). This ensures that any metallisation running over the mesa edge remains continuous and is sufficiently thick to be robust to thermal expansion. A major disadvantage, however, are the safety considerations associated with HF use, demanding special facilities and considerable care in handling HF etching processes.

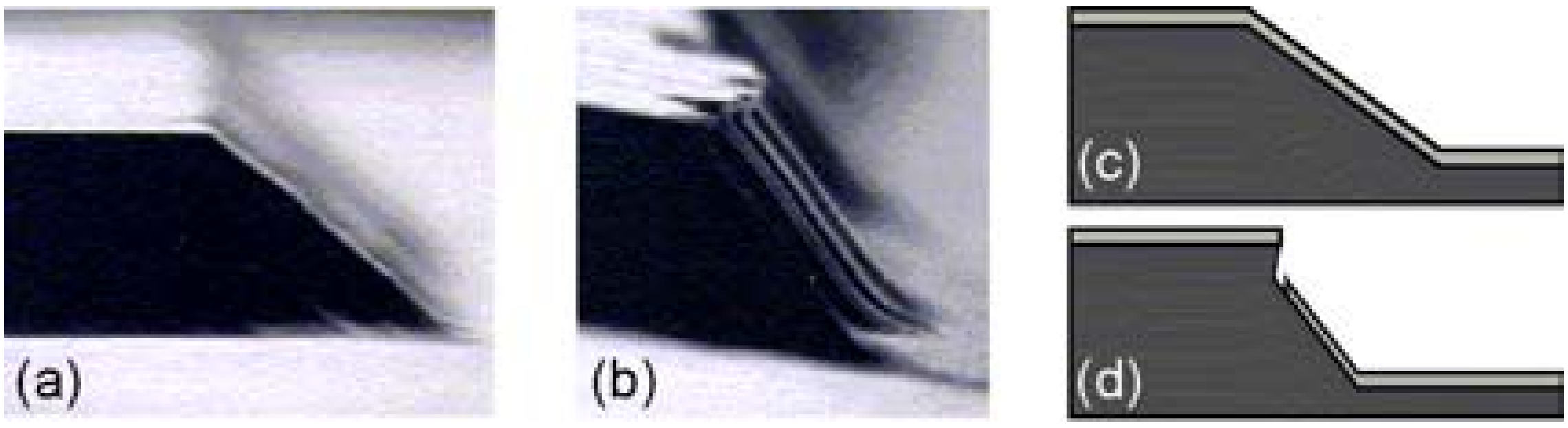

Figure 3.3: Etch profile produced by (a) HF and (b) $H_2SO_4$. (c) and (d) are schematics of metallisations deposited on the profiles shown in (a) and (b). Note that metallisation on the gentler slope (c) is thicker and hence more robust. (a) and (b) were provided by H.D. Clark and L.D. Macks (University of Cambridge, U.K.).

The depth of the etch is dependent upon a number of factors including the constituents of the etching solution and their relative concentrations, the time for which the wafer is exposed to the etching solution, temperature, degree of agitation, etc. As a result, etching is usually performed as a two stage process. Firstly the wafer is dried by placing it on a 110°C hot-plate for ~1min. The exact depth of the photo-resist is then measured using a Dektak profilometer. The first etching stage is performed by soaking the wafer in the etchant solution for 30s. The etch depth resulting from the first etching stage is found by making a second profilometer measurement, allowing the etch rate to be determined. Etch rates of 350-400nm/min are typical. The duration of the second etching stage is determined based on the measured etch rate and the remaining etch depth required for the total etch depth to extend below the 2DEG. The final step (e) involves the removal of the exposed photo-resist. This is usually performed by soaking the wafer in acetone for ~1min.

## 3.2.2 – The Hall-bar Geometry

Following the mesa etch process, the 2DEG extends to within ~10nm of the mesa edge and is confined to a geometry matching the lateral etch geometry defined by the chromium-on-glass mask pattern during the UV illumination step. Mesas are typically etched to form a Hall-bar geometry as shown in Fig. 3.4. This geometry has two current leads and four voltage leads. Typically the current is passed along the Hall-bar (from 1 to 4), with the voltage measured either from 1 to 4 (longitudinal two-terminal), 6 to 5 (longitudinal four-terminal) or 2 to 6 (transverse four-terminal or Hall). These measurement configurations will be discussed further in Chapter 4. The billiard is defined in the middle of the Hall-bar as discussed in §3.3.

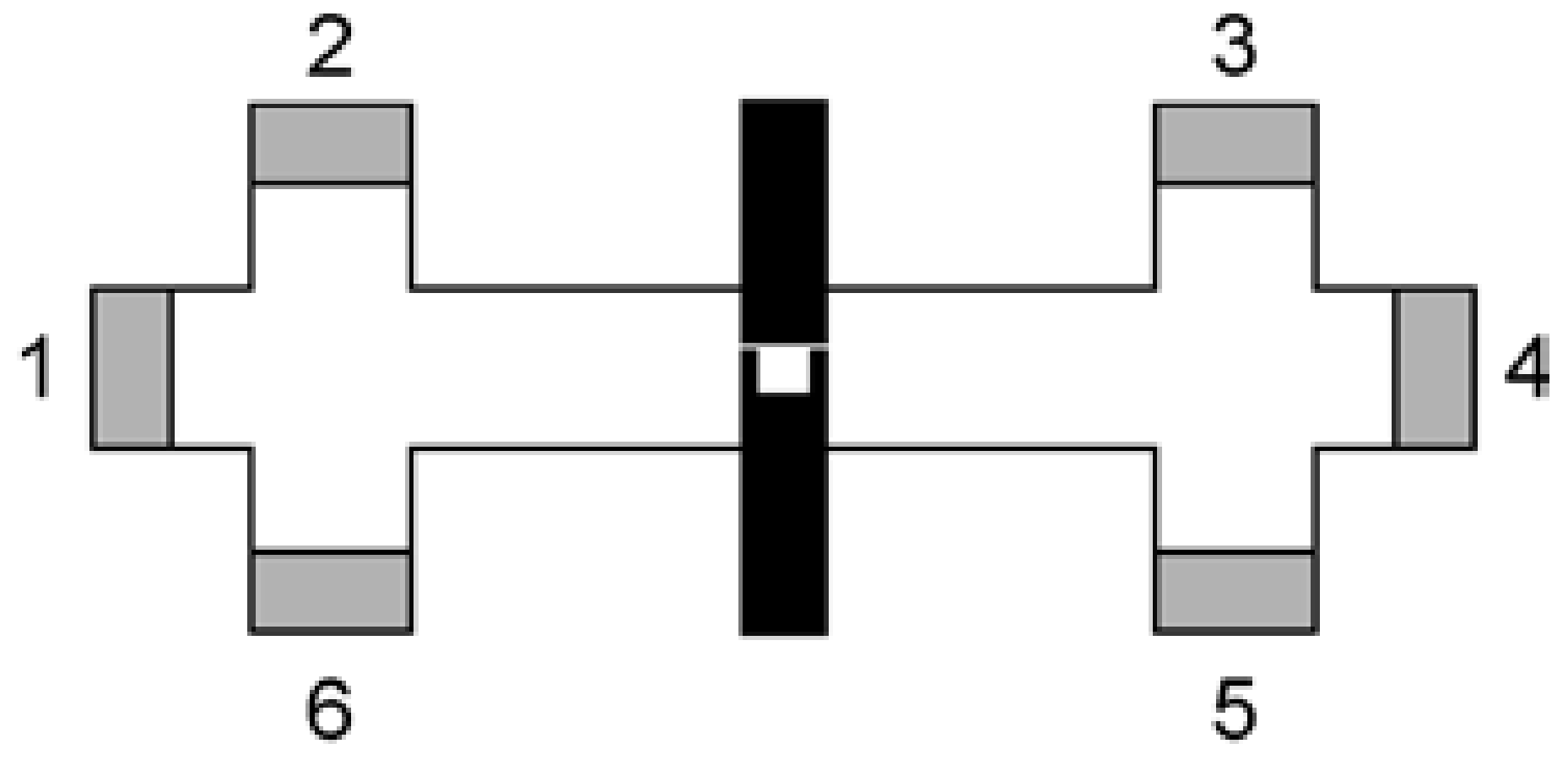

Figure 3.4: Schematic of the Hall-bar geometry. This geometry is the most common mesa geometry used in billiard devices. The locations of the ohmic contacts are shown in grey. The device (black) is shown much larger than actual size for clarity.

## 3.2.3 – The Schottky Barrier

Electrical contact to the 2DEG at the arms of the Hall-bar is made via ohmic contacts shown in grey in Fig. 3.4. Although the material between the surface and the 2DEG is a narrow layer of semiconductor, which should have a low electrical resistance, a metal placed in contact with a semiconductor surface actually produces a barrier (known as the Schottky barrier) to current flow between the metal and the semiconductor [124]. The mechanism for this is shown in Fig. 3.5. Prior to contact between a metal and a semiconductor (Fig. 3.5(a)), the energy reference level is the vacuum level. The work function $\phi$ of the metal and the electron affinity $\chi$ of the semiconductor determine the Fermi energy in the metal and the conduction band edge in the semiconductor respectively. The presence of a surface in the semiconductor leads to bending of the energy bands local to the surface in Fig. 3.5(a). Once the two materials come into contact, free charge redistributes across the interface to equalise the Fermi energies in the two materials. The height of the Schottky barrier is then determined by the presence of surface states in the semiconductor [124]. In the trivial case where surface states are not present, the barrier height is simply $\phi - \chi$. However, GaAs has a high density of surface states [124] and these act to fix the Fermi energy in the semiconductor leading to a Schottky barrier of height $E_c - E_F$ (Fig. 3.5(b)). Note that in theory this barrier height is independent of the work function of the metal and is determined entirely by a combination of the surface properties and doping of the semiconductor. In reality however, different metals can lead to slightly different barrier heights, largely due to the way they affect the surface of the semiconductor [124,125].

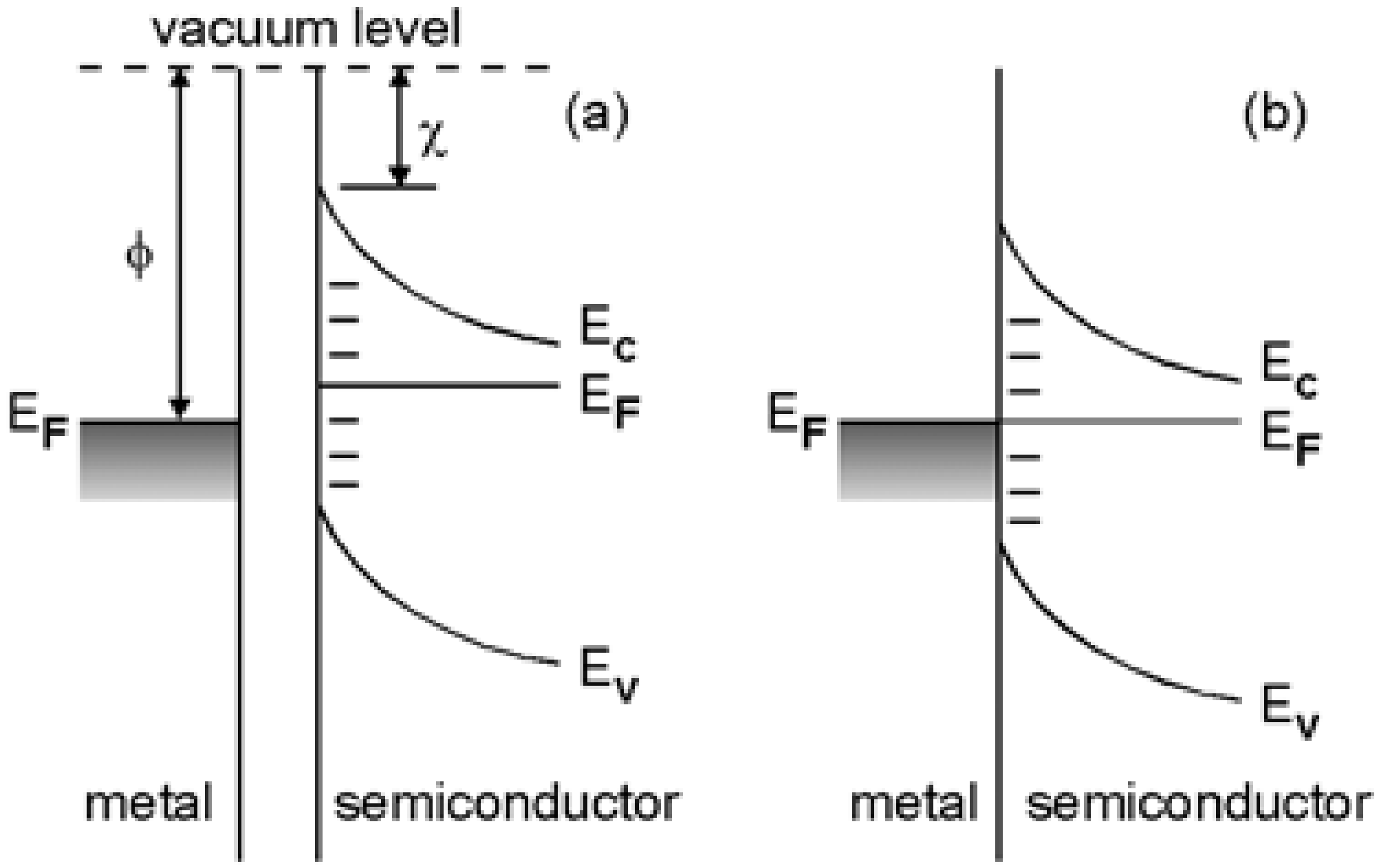

Figure 3.5: Schematic of the formation of a Schottky barrier at the interface of a metal and a semiconductor with no applied bias. (a) and (b) show the band structure before and after contact between the metal and semiconductor respectively.

Pinning of the Fermi energy by surface states explains why the Schottky barrier is 0.77eV in GaAs, irrespective of the metal placed in contact with the surface [123]. At low temperatures and typical GaAs doping concentrations (~$2\times10^{24}m^{-3}$), tunnelling and thermal processes are small [126] and a potential greater than +0.77V must be applied to the metal relative to the semiconductor before current will flow. Note, however, that current will only flow in one direction – reverse bias will only increase the barrier height. Hence simply putting metal on the heterostructure surface creates a Schottky diode, which is inappropriate for use in electrical measurements of the 2DEG.

## 3.2.4 – Ohmic Contacts

The solution to the problem at the end of §3.2.3 is twofold. Firstly, because the 2DEG is a 'buried' system, the metal needs to reach into the heterostructure to contact the 2DEG directly. Secondly, once the metal reaches the 2DEG, the Schottky barrier needs to be overcome in order for current to flow in both directions through the contact. These requirements are met by ohmic contacts [127], which are produced by the process outlined schematically in Fig. 3.6. Deposition of the photo-resist and its exposure with UV light through a patterned mask (steps (a) and (b)) follow the procedure previously outlined for the mesa etch (§3.2). Following development (step (c)) and immediately prior to metallisation (step (d)), the wafer is dipped in 1:1 HCl:$H_2O$ solution for 30s followed by a $H_2O$ rinse. The purpose of this is to remove surface oxides on the semiconductor, which can affect the behaviour of the ohmic contacts. Metallisation is performed by evaporation in a high vacuum environment. A Nickel-Gold-Germanium (NiAuGe) alloy typically is used for ohmic contacts. However, various other metals and alloys can be used instead [123]. The alloy ratio used in this case was 1:17:2 Ni:Au:Ge by weight. However, each laboratory typically has its own 'recipe' for ohmic contacts. The evaporated alloy covers the entire surface of the wafer, sticking either to the photo-resist or the semiconductor surface if it is exposed. It is important to evaporate the entire load of alloy to ensure the correct alloy ratio is preserved in the deposited metal. Sufficient alloy is loaded to generate a metallisation thickness of 150-200nm. 'Lift-off' (step (e)) is performed after metallisation by soaking in acetone.

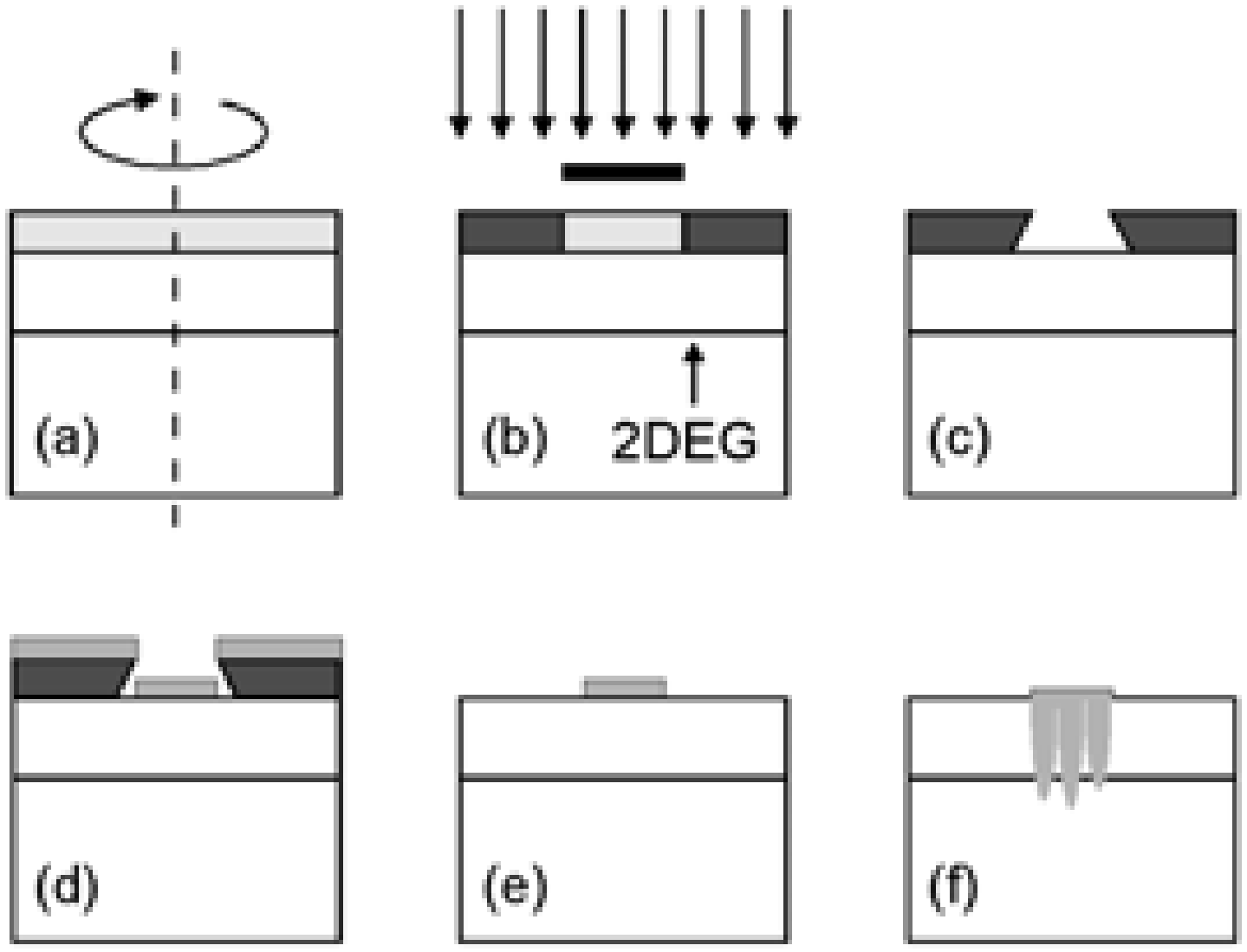

Figure 3.6: Schematic outline of the process involved in fabricating ohmic contacts: (a) photo-resist deposition, (b) exposure, (c) development, (d) metallisation, (e) 'lift-off', and (f) annealing.

The overhang profile and discontinuity of the metallisation shown in Fig. 3.6(c) and (d) is essential to ensure a successful lift-off as it allows the acetone to get underneath the metallisation, dissolve the photo-resist and lift the metal sitting on top of the photo-resist away from the semiconductor surface. The overhang profile is obtained by soaking the wafer in chlorobenzene for ~5min between the exposure and developing steps. This causes further hardening of the exposed photo-resist regions closest to the photo-resist surface compared to those deeper in the photo-resist, generating the overhang profile upon development [123]. Following lift-off, only metal evaporated onto the exposed semiconductor surface remains. The final step is rapid thermal annealing (step (f)), where the wafer is heated to 430°C for 80s in an atmosphere of 95% $N_2$, 5% $H_2$.

The annealing process is currently not well understood [123]. It is believed that once the annealing temperature is achieved, a complex diffusion process leads to the formation of the ohmic contact [123]. Firstly, Ga starts to diffuse out of the GaAs and accumulates in the Au. At the same time, Ge diffuses into the GaAs occupying the vacated Ga sites. Ge is an amphoteric donor in GaAs and increases the dopant concentration to well above $1\times10^{25}m^{-3}$. Dopant concentrations above $1\times10^{25}m^{-3}$ are crucial for the operation of ohmic contacts in GaAs [128] as discussed below. It is thought that the Ni serves two purposes. Firstly, Ni acts as a wetting agent to prevent molten AuGe forming balls on the surface of the semiconductor. Secondly, the Ni aids the diffusion of Ge into the GaAs [123]. However, TEM measurements by Kuan *et al.* [129] suggest that Ni compounds such as NiGe, NiAs and $Ni_2GeAs$ play an important role, possibly as important as Au compounds such as AuGa [130], in the chemistry that occurs during the annealing process. The purpose of the Au is threefold. Firstly, it acts to collect Ga, freeing up lattice sites for occupation by the Ge. Secondly, it acts as a capping layer, protecting the ohmic contact from oxidation. Thirdly, it minimises the contact resistance between bond wires (see §3.4) and the ohmic contact. An excess of Au in the alloy can free more Ga sites than the available Ge can fill, leading to vacancies that increase the resistance of the ohmic contact significantly [131]. For this reason, an extra Au capping layer is often deposited over the ohmic contact following annealing to achieve the ideal features – oxidation protection and lower contact resistance – without seriously degrading the optimum diffusion process. Ultimately, the diffusion process creates a set of spikes [132] that penetrate into the semiconductor as shown in Fig. 3.6(f). The duration of the annealing process is tuned so that these spikes extend through the 2DEG. The only remaining issue is the Schottky barrier between the ohmic spikes and the 2DEG. As mentioned earlier the height of the barrier is partially determined by the doping of the semiconductor. Furthermore, the width of this barrier is also determined by the doping and decreases as the dopant concentration is increased. The probability of electron tunnelling increases as the barrier width is decreased and leads to a tunnelling current that increases as the square root of the dopant concentration [124,125]. For dopant concentrations greater than $1\times10^{25}m^{-3}$ tunnelling processes are sufficient for low resistance ohmic contacts to be formed [123].

## 3.3 – Fabrication of Surface-gates

Following the mesa etch process and the formation of ohmic contacts, the surface-gates that define the billiard are fabricated. In contrast to ohmic contacts, surface-gates rely on the formation of a Schottky barrier for their operation. As discussed in §2.1.2, confinement of the 2DEG may be achieved by applying a negative bias to the surface-gates. This negative bias leads to electrostatic depletion of the 2DEG regions directly below the surface-gates. The presence of the Schottky barrier theoretically prevents current leakage from the surface-gates into the heterostructure for all surface-gate biases less than +0.77V. In reality however, current will flow for high negative biases and occasionally, in some devices, current may also leak from the surface-gates at lower biases. Leakage currents can cause major problems in electrical measurements of billiards, both directly by acting as an extra source of current, and indirectly by causing heating of the electrons. Typically, biases are restricted to between +0.7V and –3.0V to avoid leakage problems. Positive biases may also be applied to the surface-gates, typically only following illumination of the sample in order to increase the electron density via the persistent photo-conductivity effect [133]. Metallised surface-gates are generally opaque, shadowing the donors beneath them from illumination and leading to lower electron densities beneath the surface-gates compared to the rest of the 2DEG [79]. This leaves a partial depletion pattern of the surface-gate in the 2DEG at zero applied bias, which can be corrected by applying a positive bias to the surface-gate, drawing electrons into the 2DEG regions beneath the surface-gates and equalising the electron density throughout the 2DEG [79]. When this is performed, positive biases less than +0.7V are used to avoid crossing the Schottky barrier, which would allow current to flow between the surface-gate and the heterostructure.

Surface-gate fabrication requires the definition of features ranging from mm-scale wire-contact pads down to nm-scale features in the billiard itself. The resolution limit of optical lithography techniques is due to diffraction effects and is determined by the wavelength of the light used [122,123]. Conventional UV lithography techniques are currently capable of accurately defining features as small as 250nm [134], insufficient for semiconductor billiards which require features as small as 50nm. Whilst it would be possible simply to reduce the wavelength further, and achieve lower resolution limits, using X- or $\gamma$-ray lithography, these techniques induce large amounts of damage in the semiconductor. Use of these techniques is further inhibited by other difficulties such as expense and user safety. Lower resolution limits can also be achieved by resorting to particle beam techniques such as electron-beam and ion-beam lithography. Semiconductor damage with these techniques increases with the particle mass and hence electron-beam lithography (EBL) is the method of choice for defining sub-micron surface-gate features. Ion-beam lithography is used where patterned regions of semiconductor damage are desirable such as in the back-gates in double-2DEG billiards (see §3.6). EBL is capable of defining features down to 6nm [135,136], ideal for use in defining semiconductor billiards [137]. Whilst it is possible, in theory, to perform the entire surface-gate lithography process using EBL, in practical terms it is preferable to use optical techniques for defining large scale surface-gate features instead. This is because optical lithography illuminates the whole pattern simultaneously compared to EBL techniques where a small electron-spot needs to be raster-scanned over the whole pattern – the former takes only several seconds compared to many hours for the latter. Generally, the optical lithography is done prior to EBL because it is easier to align sub-micron features to sub-millimetre patterns, than vice versa, particularly since sub-micron features are at the resolution limit of even the best optical microscopes. Both lithography techniques are discussed in the following sub-sections. Detailed discussions of optical lithography may be found in [122] and [123], and electron-beam lithography in [138].

## 3.3.1 – Optical Lithography

The optical lithography process used to form surface-gates follows a set of steps similar to the formation of ohmic contacts. The optical lithography process is outlined schematically in Fig. 3.7. Steps (a)-(c) follow the same process used in producing the ohmic contacts (§3.2.4). A chlorobenzene soak is again used prior to developing to harden the surface and achieve an overhang resist profile. A short plasma etch (~25s in oxygen plasma) is performed following development to remove any remaining photo-resist from the exposed regions. Note however that this may lead to semiconductor damage, lowering the 2DEG electron mobility [139]. The metallisation process (step (d)) also follows from §3.2.4. For surface-gates, two metal layers are deposited. Their thicknesses depend on whether they are patterned using electron-beam or optical lithography. The first layer is typically NiCr and acts as a wetting layer to prevent the Au in the second layer from 'balling-up', aids adhesion of the surface-gate to the semiconductor surface, and prevents diffusion of the Au into the heterostructure. It is common for other surface-gate metal combinations such as Ti/Pt, Ti/Au, etc to be used [137]. For optically defined surface-gates the layer thicknesses are determined by the depth of the mesa etch. The NiCr layer is usually thin (~20nm) with the Au making up the remainder of the required surface-gate thickness. This is because Au prevents oxidation of the NiCr. The final step in the process is lift-off (step(e)) and also follows from §3.2.4.

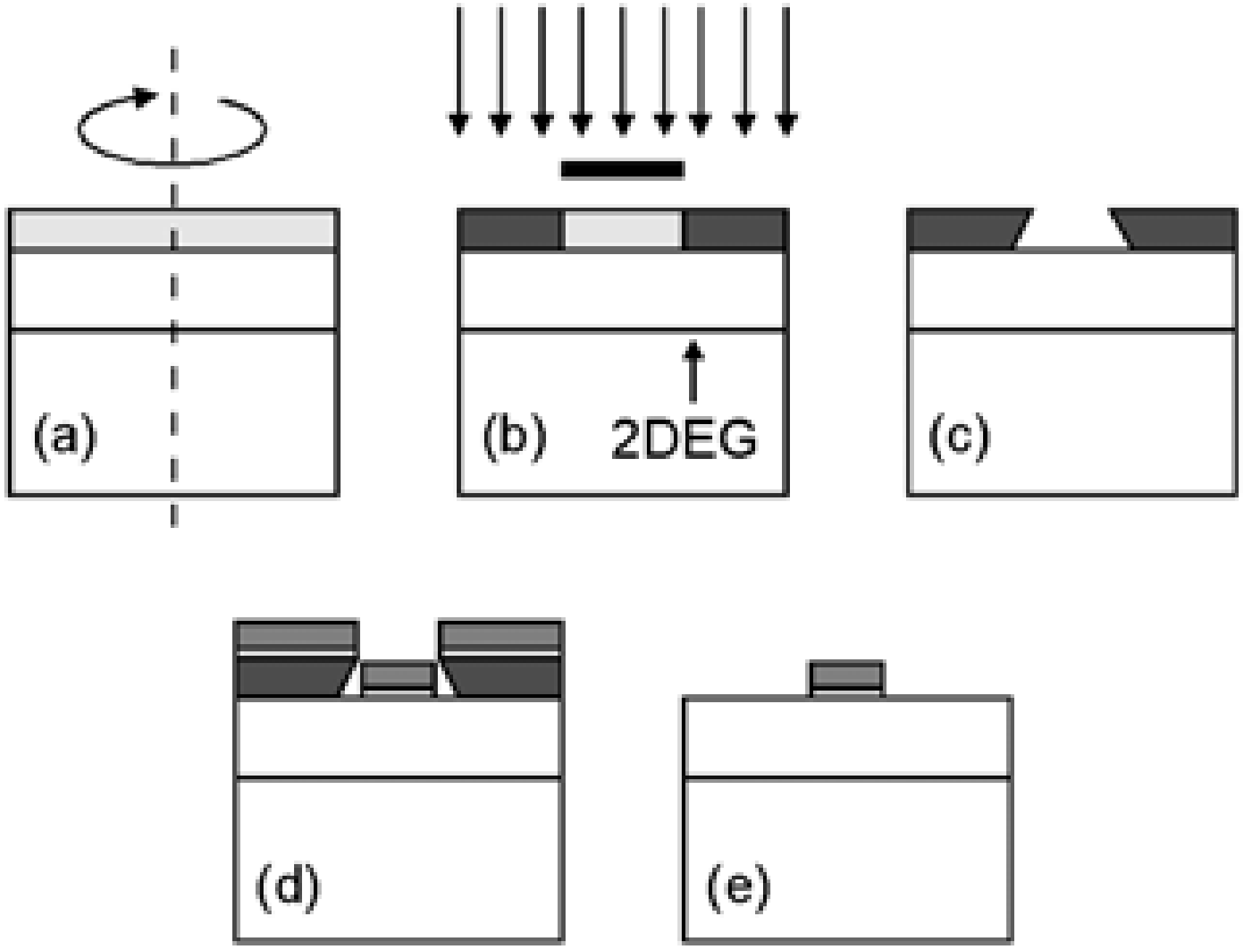

Figure 3.7: Schematic outline of the optical lithography process for large-scale surface gates: (a) photo- resist deposition, (b) UV exposure, (c) developing, (d) metallisation, and (e) lift-off.

## 3.3.2 – Electron-Beam Lithography

The EBL process is outlined schematically in Fig. 3.8 and while it follows a similar basic sequence of steps to optical lithography, there are a number of key differences. The first difference is the resist, which needs to be 'activated' by electron-irradiation. The standard EBL resist is poly-methyl-methacrylate (PMMA) which is also used as a photo-resist in deep-UV lithography [122]. Undiluted PMMA is a solid (plexiglass) and is diluted in either o-xylene or methyl-isobutyl-ketone (MIBK) to a viscosity suitable for forming sub-micron thickness layers by spinning prior to use in EBL[8]. Polymeric chains in PMMA are fragmented (chain scission) when exposed to electron irradiation doses between 2 and 8C/m$^2$ making them more susceptible to dissolution in the developer. PMMA has the added advantage that at high electron irradiation doses (greater than 50-70C/m$^2$) it becomes a highly robust cross-linked polymer [140]. This feature is important in §3.5.2. The spin-and-bake technique is used in step (a) to establish a thin layer of PMMA. The thickness of this resist is determined by the spin rate, with a resist thickness of ~150nm typically used. The resist is spun for 40s followed by a 20min bake at 150°C.

[8] The EBL resist used at Cambridge is known as P5. Whilst P5 is generally used without further dilution, MIBK dilutions of P5 are also used in order to achieve thinner resist layers.

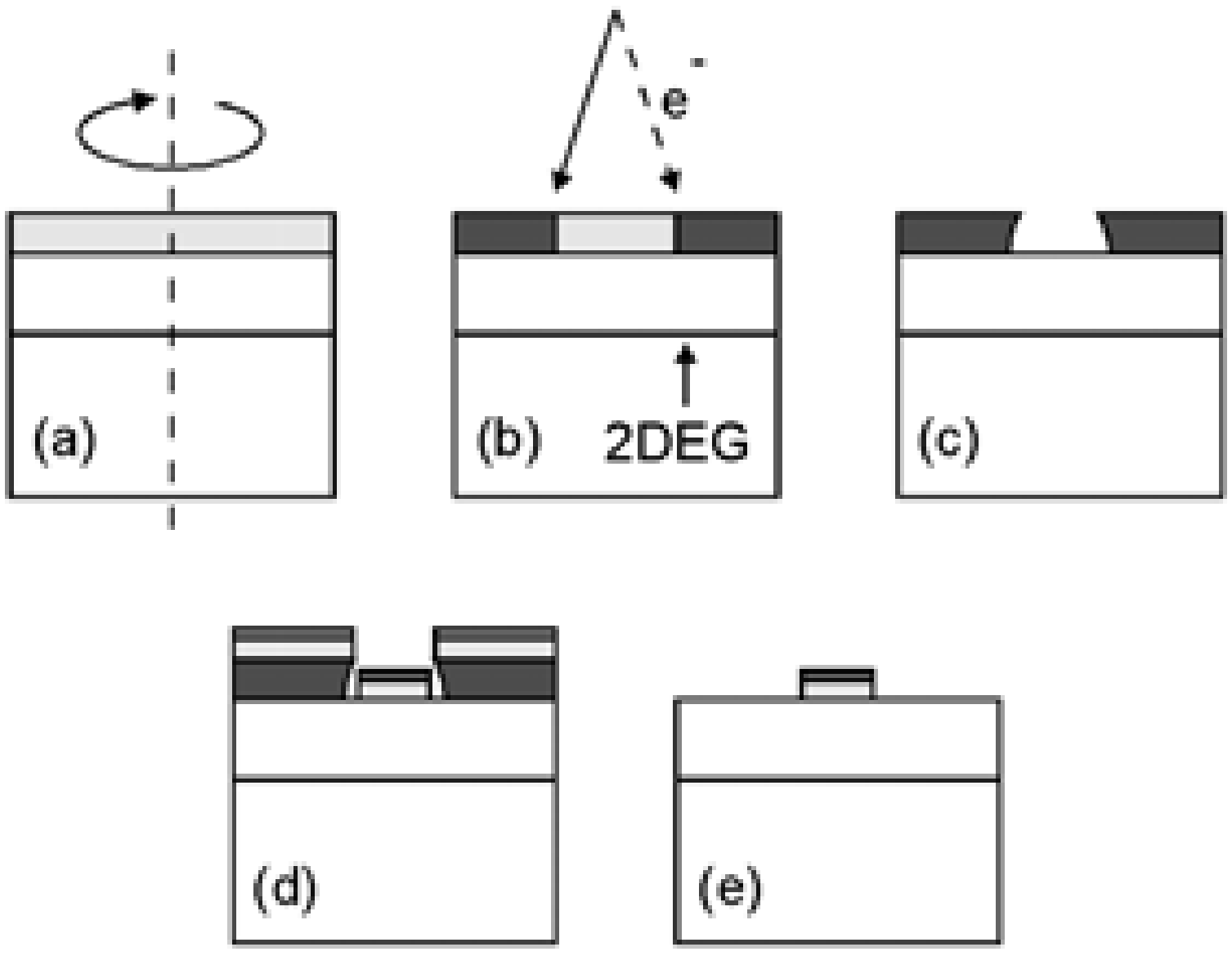

Figure 3.8: Schematic outline of the steps involved in EBL: (a) PMMA resist deposition, (b) electron-beam exposure, (c) development, (d) metallisation, and (e) lift-off.

Exposure (step (b)) is performed using a modified scanning electron microscope (SEM). In this step, a computer-controlled focussed electron-beam irradiates the regions of the resist where surface-gates are to be deposited. A schematic of a SEM modified for EBL is shown in Fig. 3.9. The various components of the SEM form a column above the sample. The electron-gun is located at the top of the column and consists of a heated metal filament, a collimating aperture and an adjustable accelerating potential difference of 20-50kV. Electrons are liberated from the heated filament by thermoelectric and field emission processes. Some of these electrons pass through the aperture into the accelerating potential to form a collimated electron-beam that travels down to the heterostructure surface. To form a small electron spot on the heterostructure surface however, the electron-beam needs to be focussed and this is performed in two stages using a pair of solenoids – the condenser lens and the objective lens. The focus of the condenser lens lies approximately half way down the column, at the centre of the electron-beam steering and blanking system. The electron-beam blanking system consists of a pair of plates, as shown in Fig. 3.9, and is used to prevent the electron-beam from reaching the sample. This is achieved by applying a large potential difference across the plates to deflect the electron-beam into a collector mounted in the column. Electrostatic beam-blanking is preferred to other methods of blanking the beam such as turning off the filament and masking due to its simplicity, speed and better stability. The electron-beam steering system consists of two pairs of plates. The electron-beam is deflected along the line joining the centres of opposite plates by applying a potential difference between the plates. Hence, by controlling the potential difference on the two opposing pairs of plates it is possible to steer the electron spot across the heterostructure surface.

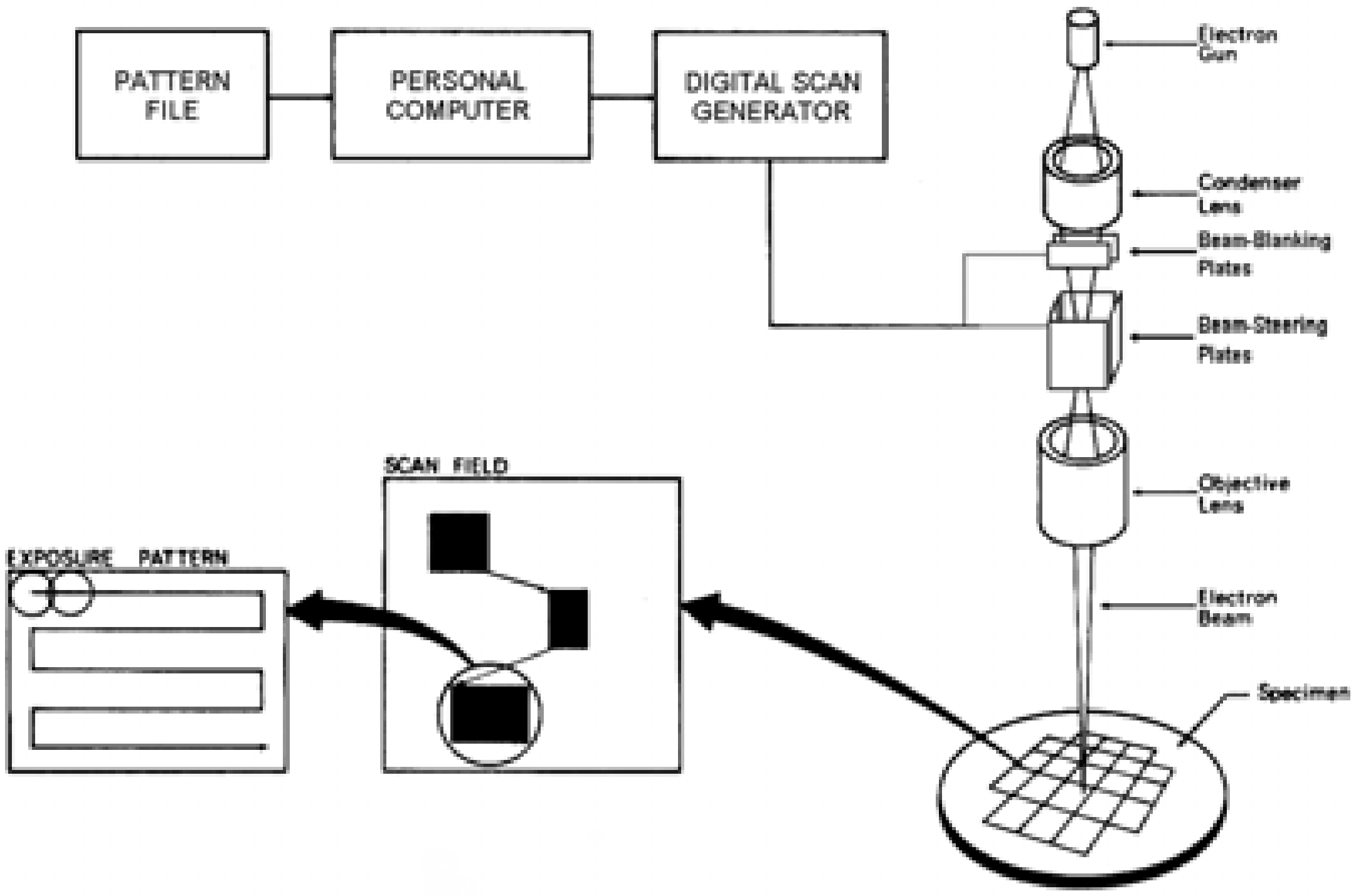

Figure 3.9: Schematic of a modified scanning electron microscope (SEM) used for electron-beam lithography in the fabrication of semiconductor billiards. Patterns are built up from rectangles that are defined by raster-scanning the electron-beam along a zig-zag path.

The objective lens lies further down the column and its purpose is to focus the electron-beam onto the heterostructure surface. Note that the focal length of the objective lens will depend on the deflection of the beam from column centre and hence it is often necessary to re-focus the beam when exposing separated regions of the sample. The potential differences applied to the two beam-steering plate pairs are computer controlled to define a particular pattern. Traditionally, the pattern is defined by raster-scanning a set of rectangles, however in the latest EBL-writers, more advanced techniques are employed to accurately pattern non-rectangular features without the presence of steps and other raster-scan artefacts. The beam current and the electron dosage required for exposure determine the raster-scan rate. The number of raster-scan passes required to expose an area is determined by the spot size of the electron-beam. Prior to exposure, the electron-beam is focussed and de-astigmatised to produce a circular spot approximately 10nm in diameter on the resist surface. Setting a small working distance (8mm) between the lens and the sample surface assists in minimising the spot size. A beam current of ~30pA is typically used during patterning. The intended pattern is aligned to the optical lithography gates at low magnifications, where the electron dose due to viewing is insufficient to expose the resist. For maximum patterning accuracy and resolution, exposure of the pattern is performed at the highest magnification at which the electron-beam can still access all parts of the pattern.

Development (step(c)) is performed by dipping the wafer in a 3:1 mixture of isopropyl alcohol (IPA) and MIBK. The duration of the development process is dependent on a number of factors including temperature and agitation, and is calibrated using a set of optimised, optically-visible dosage calibration boxes patterned in conjunction with the surface-gate pattern. In contrast to optical lithography, the electron-beam exposure process achieves a substantial overhang profile in the developed resist without extra processing. This profile is produced by enhanced exposure in the lower regions of the resist layer due to the combined effect of forward scattered, backscattered and secondary electrons [122,123,138]. The overhang profile plays an important part in the lift-off process, as discussed earlier in §3.2.4. Immediately prior to metallisation the sample is again dipped in 1:1 $HCl:H_2O$ for 30s followed by a $H_2O$ rinse to remove any surface oxides. The short plasma etch used in the optical lithography process is not used here because PMMA has a very low resistance to oxygen plasma etching [122]. Metallisation is performed as in the optical lithography process, typically with 30nm NiCr capped by 10nm Au. The Au layer again serves to protect against oxidation of the NiCr. The Au layer is kept as thin as possible to aid lift-off which is substantially more difficult in the EBL process, largely because the PMMA resist layer (~150nm) is much thinner than typical photo-resist layers (~1μm). Lift-off is achieved by soaking in acetone for 12-24 hours, followed by washing with a jet of acetone to remove any remaining metal that fails to fully lift off during soaking.

## 3.4 – Preparing the Device for Measurement: Cleaving, Packaging and Bonding

The final stage in the fabrication process involves taking the finished billiard device and packaging it so that it is safe from breakage, readily transportable and measurable using macroscopic electrical connections. Even in well optimised processing lines there is some loss of devices during fabrication due to poor lift-off, contamination, wafer damage, misalignment and other processing faults. The yield of useable devices generally decreases as the number of processing steps increases. Hence it is not uncommon for a large number of devices to be prepared in parallel on a single wafer to maximise the number of viable devices available at the end of the fabrication process. Viable and non-viable devices need to be separated at the end of fabrication and this is achieved by cleaving the wafer. Cleaving is performed by lightly dragging a diamond-tipped scribe across the GaAs surface. The scribed line serves as a stress concentration point for fracture when pressure is applied to the scribed surface on opposite sides of the scribed line, leading to a clean break along the scribed line. Individual devices are then packaged in either 20-pin Charntec flat-packs or 18-pin dual-in-line packages using silver epoxy adhesive or GE varnish. Electrical contact is made between the bonding pads and the pins of the package using 20μm Au wire attached using a ball-bonder, which utilises a combination of temperature, downward force and ultrasonic vibration to weld opposite ends of the Au wire to the device bonding pad and the package. For surface-gate devices, ball-bonding must be performed with care because these devices are static-sensitive and are easily destroyed by small electrical discharges between individual surface-gates. Hence at least one ohmic contact should be bonded first to allow dissipation of any potential difference between the sample and the package/bonder, and at least 1min should be left between bonding nearby surface-gates. Furthermore, automatic sparking used to form the next bond-wire ball should be disabled to avoid a destructive high-voltage discharge into the sample in the event of a bonder malfunction. Following bonding, surface-gate devices must be handled with care due to their extreme static sensitivity.

## 3.5 – Multi-level Gate Architectures: The Sinai Billiard

Multi-level architectures are well established in Si integrated circuit technologies [122]. However, they have only recently been applied to nano-scale surface-gate devices, in particular to enable the study of a continuous transition between a square and a Sinai geometry in a single semiconductor billiard [13]. Investigation of the Sinai billiard (Fig. 3.10(a)) has played a large part in this thesis. The results of this investigation are presented in Chapter 5.

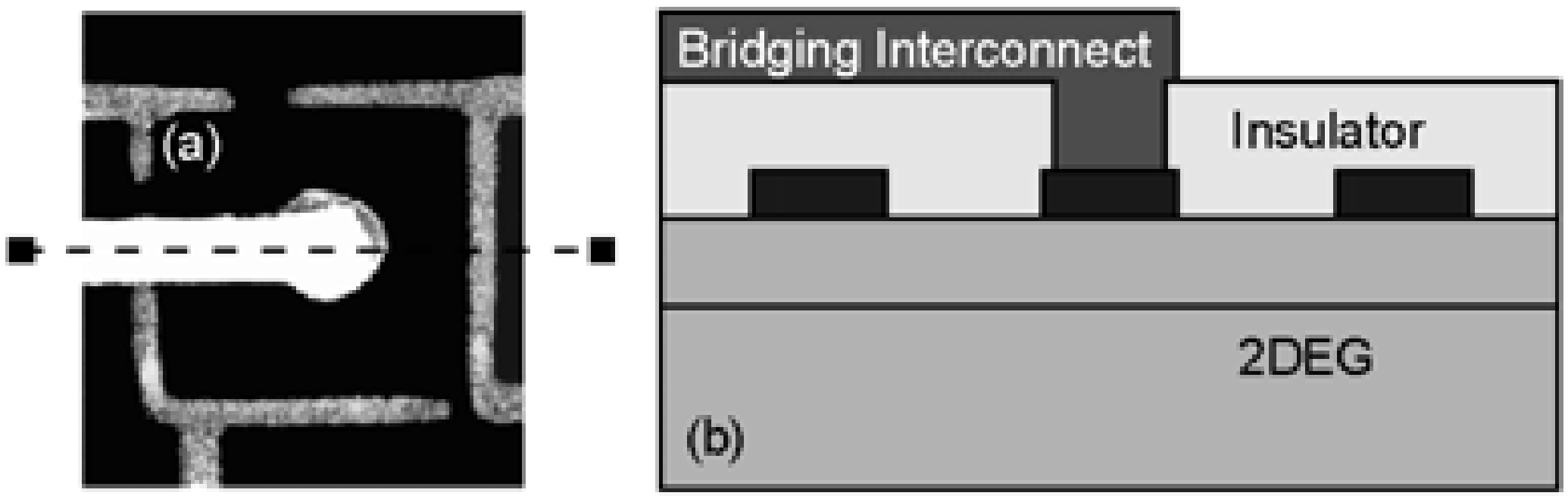

Figure 3.10: (a) Scanning electron micrograph of a Sinai billiard device. The first level of surface-gates, which are beneath the insulator, appear in grey. The white strip leading up to the central circle is the bridging interconnect. (b) Schematic cross-section through the dashed line in (a). The bridging interconnect crosses over the insulator layer to contact the central circle.

A multi-level gate architecture is required in order to bias the central circle independently of the square. The central circle is contacted using a bridging interconnect, which crosses a layer of insulator deposited on the heterostructure surface and connects to the central circular gate via a hole in the insulator, as shown in Fig. 3.10(b).

This section is divided into three parts to discuss the two distinct methods used to fabricate a bridging interconnect. §3.5.1 discusses the fabrication of the original Sinai billiard used for the investigations in Chapter 5. This billiard was fabricated by Y. Feng at the National Research Council, Canada. §3.5.2 and §3.5.3 discuss my work to fabricate a second Sinai billiard device at the Cavendish Laboratory, Cambridge. These billiards will be called the NRC Sinai billiard and the Cambridge Sinai billiard respectively from here onward.

## 3.5.1 – The NRC Sinai Billiard

Fabrication of the bridging interconnect follows the fabrication of surface-gates as discussed in §3.3. These surface-gates define a Sinai geometry (circle at the centre of a square) and appear in grey below the insulating layer in the scanning electron micrograph in Fig. 3.11(a). The dimensions of the surface-gates are shown in Fig. 3.11(b). Figure 3.12 outlines the steps involved in the fabrication of the bridging interconnect in the NRC Sinai billiard. Further details of this technique are presented in [141].

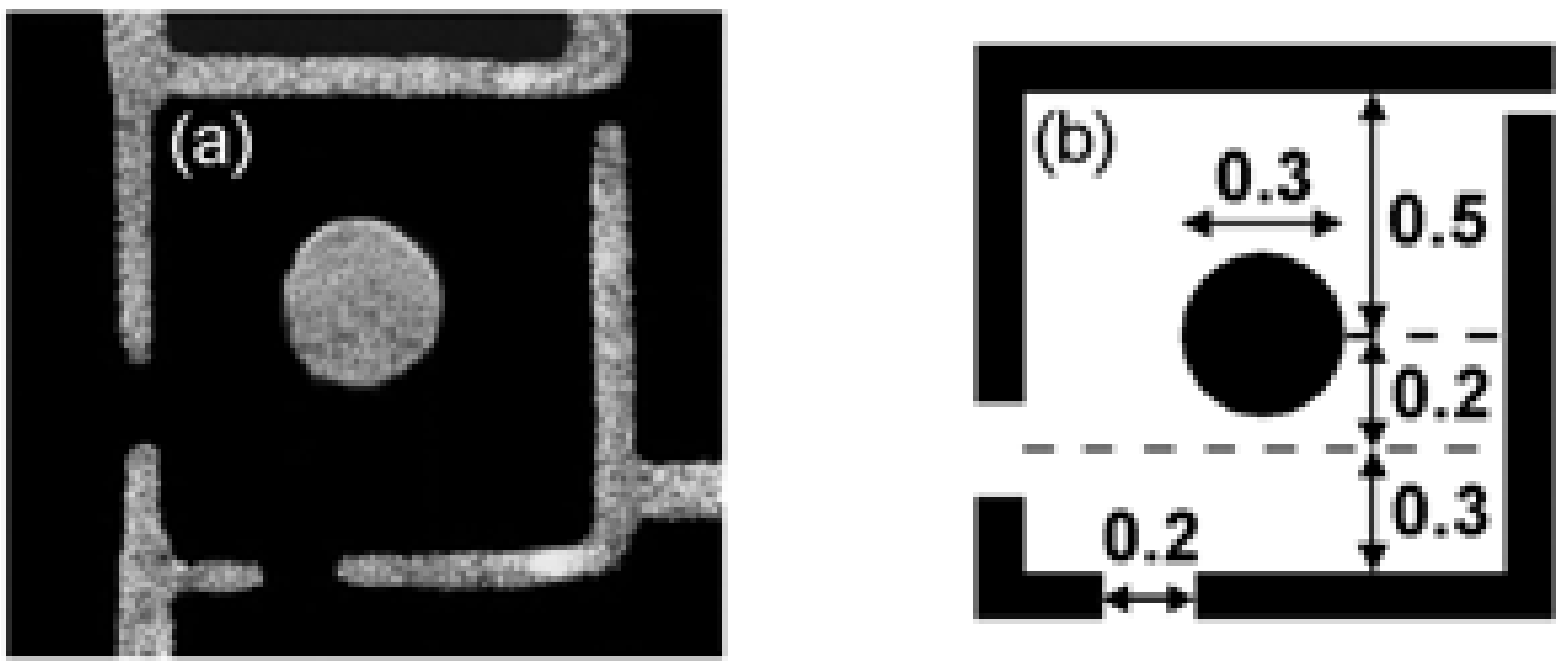

Figure 3.11: (a) Scanning electron micrograph of the NRC Sinai billiard prior to fabrication of the bridging interconnect. (b) Schematic of the NRC Sinai billiard. All dimensions are in μm.

Step (a) is the deposition of the insulating layer, which for the NRC devices is either polyimide or $Si_3N_4$, depending on the required thickness of insulator and the diameter of the hole through the insulator. These insulators are deposited using the spin-and-bake technique and plasma enhanced chemical vapour deposition (PECVD) respectively. The insulator in the particular Sinai billiard presented in this thesis was polyimide [13]. After insulator deposition, optical lithography and dry etching techniques are used to expose holes in the insulator layer directly over ohmic and surface-gate bonding pads in order to provide contact to the bond-wires. A further hole is exposed during this step to allow the bridging interconnect to contact a dedicated optical lithography gate which leads out to a bonding pad. Step (b) is the deposition of a layer of PMMA resist using the spin-and-bake technique. EBL is used to define a hole in this resist directly above the central circle (step (c)). Alignment is achieved using a set of coarse and fine scale alignment markers deposited during the optical and electron-beam surface-gate fabrication stages respectively. These markers allow alignment to the circle at successively higher magnifications. Following development (step (d)) the resist is used as an etch mask to open a hole in the insulator layer extending down to the central circular gate. The circular gate protects the heterostructure from damage during the etching process and acts as an etch stop. Etching was performed using a $CHF_3/O_2$ reactive ion etch, which is essential in order to achieve the insulator hole profile shown in Fig. 3.12(e). The PMMA resist was then removed by oxygen plasma etching and replaced with a new PMMA resist, which was thicker than the intended metal thickness of the interconnect. A second EBL step was then performed (step (f)) to re-expose the hole, and expose the path for the bridging interconnect, which contacts the dedicated optical lithography gate mentioned above. After development (step (g)), an oxygen plasma etch is performed to remove any remaining resist in the hole, which would prevent electrical contact between the bridge and the central circle. A two-stage metallisation process is used in step (h) to deposit Ti/Pt/Au followed by Ti/Au. Metallisation is performed with the metal source at an angle of approximately 20° with respect to the heterostructure surface normal. Between metallisation stages the sample is rotated by 180° about the normal. The purpose of this two-stage metallisation process is to ensure proper filling of the hole in the insulator layer. The final step in the process is lift-off (step (i)), which is achieved by soaking in acetone as in §3.3.

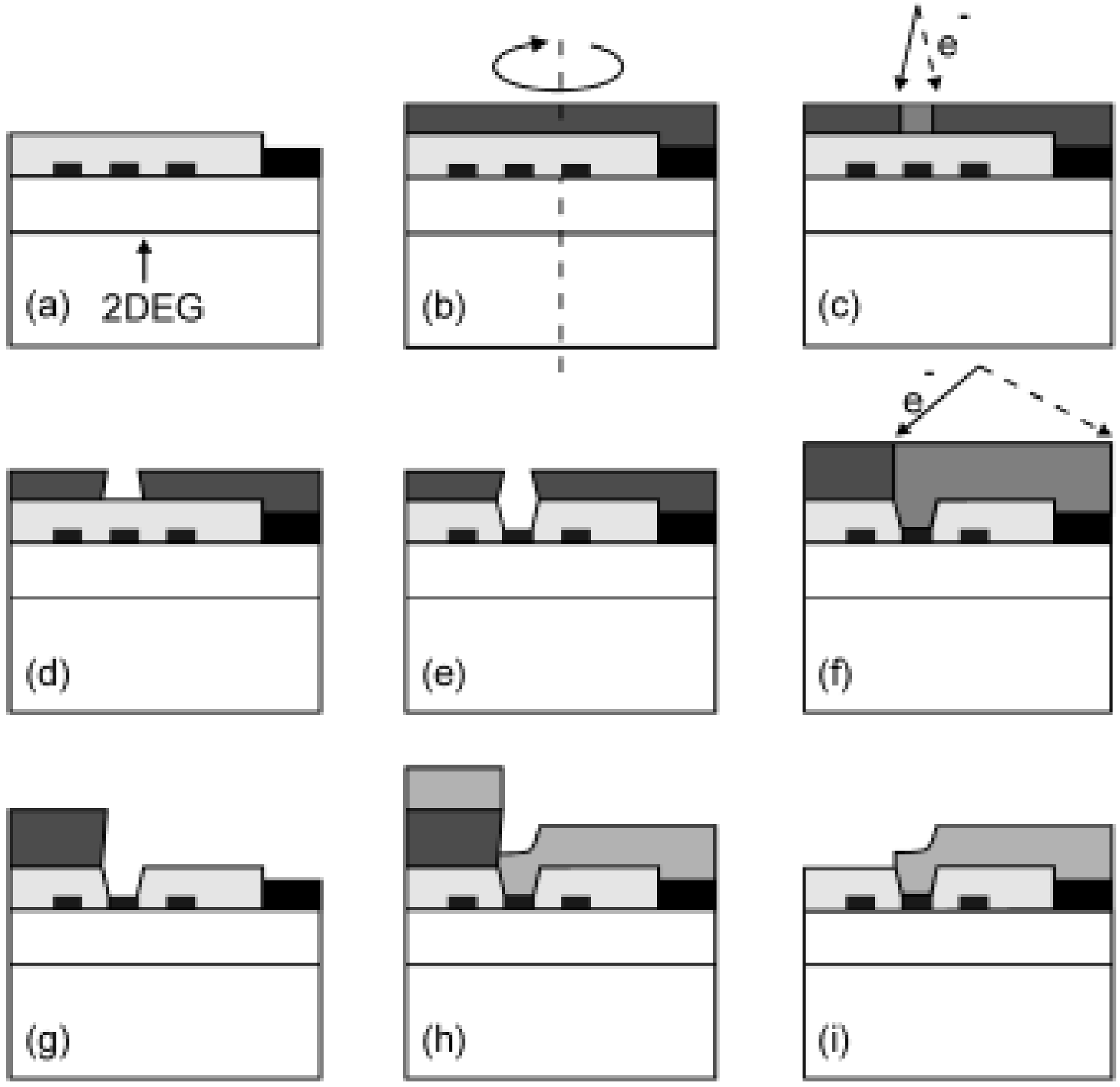

Figure 3.12: Schematic outline of steps involved in fabricating the bridging interconnect in the NRC Sinai device. (a) Insulator deposition, (b) PMMA deposition, (c) Hole exposure, (d) Development, (e) Reactive Ion Etching of hole, (f) Deposition of fresh PMMA and exposure of interconnect path and hole, (g) Development, (h) Metallisation, and (i) Lift-off.

## 3.5.2 – The Cambridge Sinai Billiard

Fabrication of the bridging interconnect for the Cambridge Sinai billiard also directly follows the fabrication of Sinai geometry surface-gates which are shown in Fig. 3.13(a). The dimensions of the surface-gates for this device are shown in Fig. 3.13(b). Figure 3.14 outlines the steps involved in the fabrication of the bridging interconnect for the Cambridge Sinai billiard.

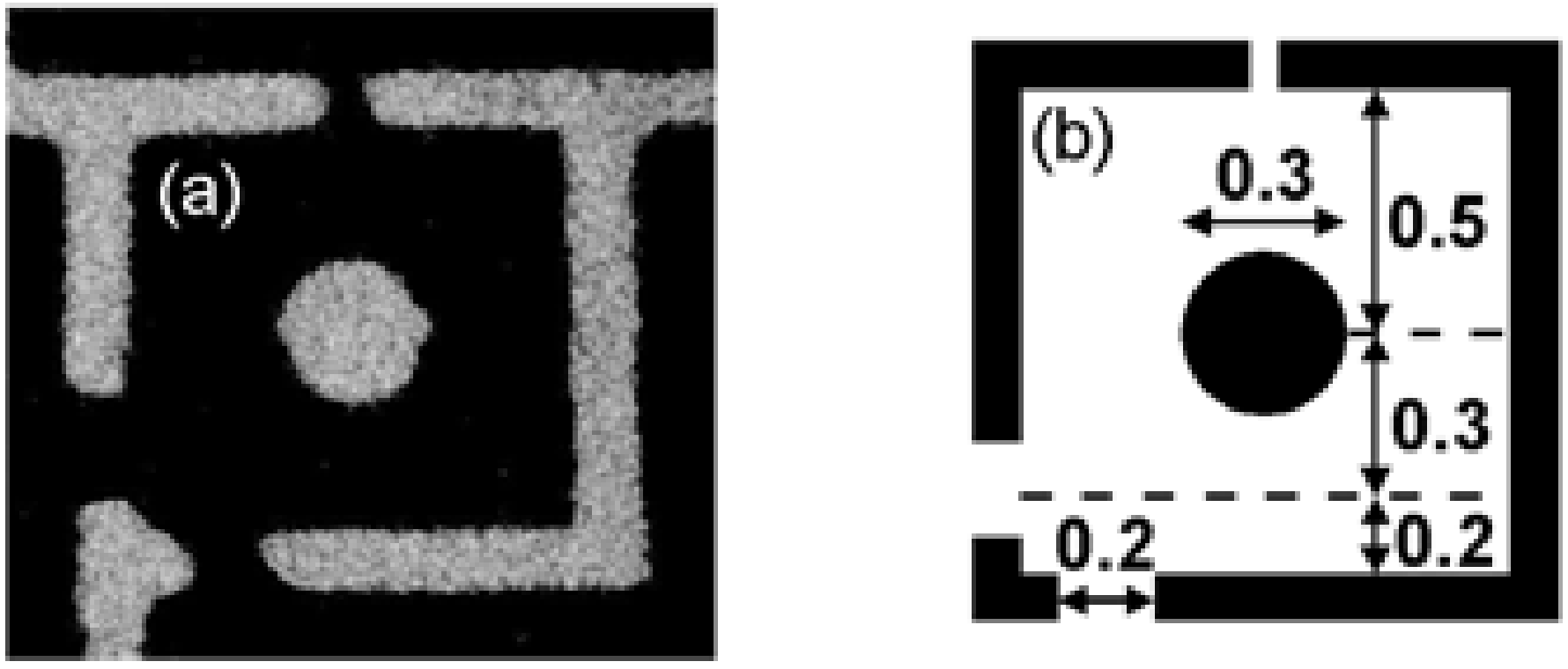

Figure 3.13: (a) Scanning electron micrograph of a Cambridge Sinai billiard prior to fabrication of the bridging interconnect. (b) Schematic of the Cambridge Sinai billiard. All dimensions are in μm.

Cross-linked PMMA was chosen as the support insulator for the bridging interconnect in the Cambridge Sinai billiard. In contrast to polyimide and $Si_3N_4$ insulators, reactive ion etching and PECVD (in the case of $Si_3N_4$) are not required to process cross-linked PMMA. All processing can be performed using only standard EBL equipment and techniques. An added advantage is the ability to correct any misalignment of the hole leading down to the circular gate. Step (a) is the deposition of an 80nm thick layer of PMMA using a spin-and-bake technique (2:1 PMMA (Undiluted[8] P5):MIBK spun for 40s at 6100 rpm followed by a 20min bake at 150°C). The sample is then mounted in a SEM for EBL exposure of the hole leading down to the circular gate.

Alignment is performed using the existing optical and electron-beam surface-gates rather than a set of alignment markers and occurs at four successive magnifications. Firstly, the device is located on the wafer, rotated to the correct orientation and centred in the field of view at 200× magnification. The pattern is then centred again at 800× magnification and the beam is focussed on one of the existing EBL surface-gates to ensure correct focussing (and removal of astigmatism) in the area local to the patterning region. Alignment at higher magnification requires a change in the way the sample is imaged using the SEM. Generally the SEM generates an image of the sample by continuously raster-scanning the electron-beam across a rectangular region of the sample surface. Based on the reflected electron intensity, a continuously updated image is produced on the SEM monitor. At higher magnifications a shorter time is required for continuous viewing to irradiate the resist with a sufficient electron dose to expose and eventually cross-link the resist.

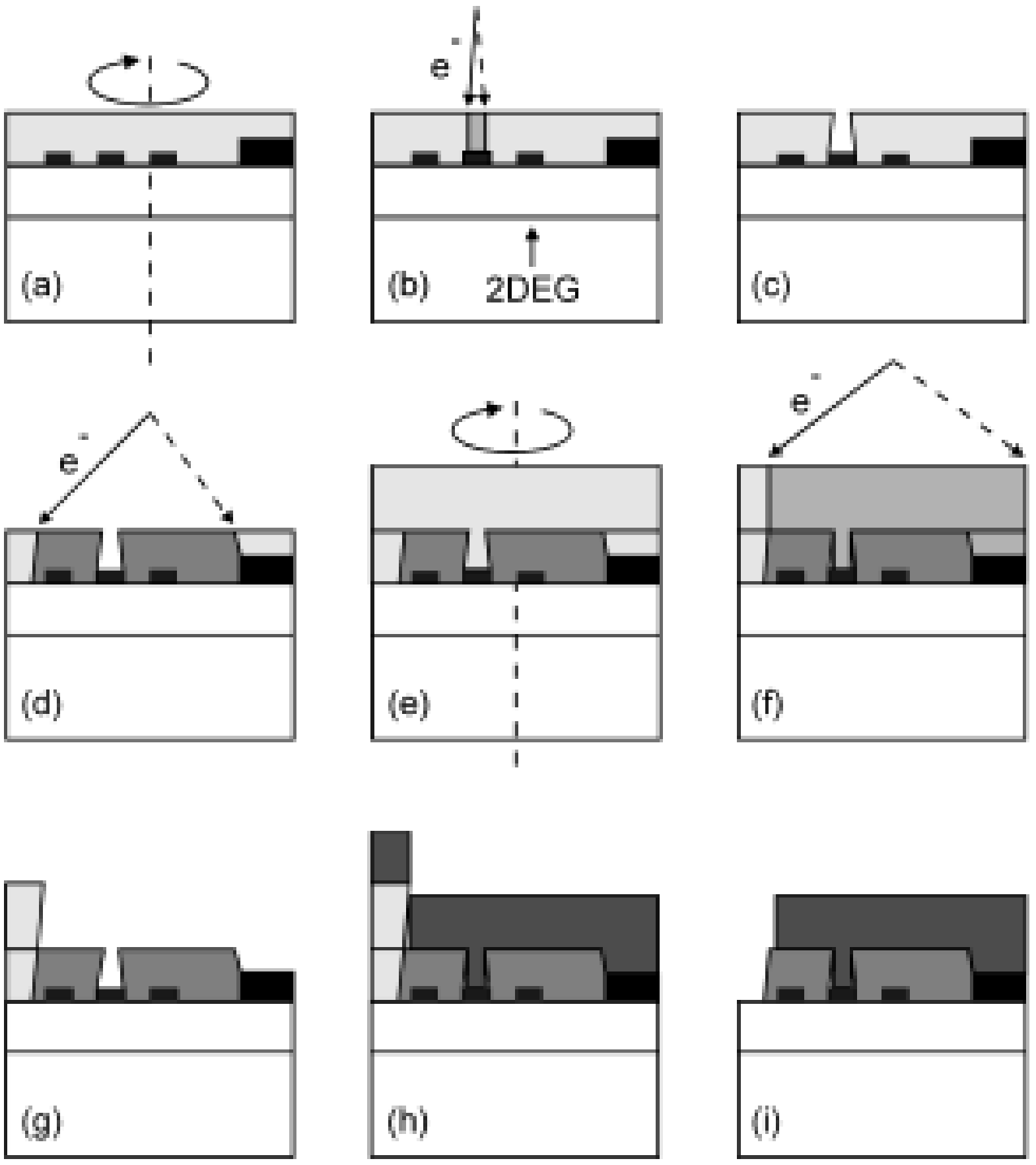

Figure 3.14: Schematic outline of steps involved in fabricating the bridging interconnect in the Cambridge Sinai device. (a) PMMA resist/insulator deposition, (b) Hole exposure, (c) Development, (d) Inspection for correct alignment and cross-linking of the PMMA in regions below the interconnect, (e) Deposition of a second PMMA resist, (f) Exposure of interconnect path, (g) Development, (h) Metallisation, and (i) Lift-off.

At magnifications higher than 800×, beam blanking is used to prevent continuous viewing, and instead a 'flash-view' technique is used where a single low-dose square occupying the full field of view is scanned in order to image the sample. A major disadvantage of this technique, particularly on the Cambridge SEM (which is not equipped with a 'screen-grabber') is that the 'flash-view' image is only visible on the monitor for a short time (2-3s). This makes it difficult to 'see' the sample properly. Furthermore, the use of low-doses in the 'flash-view' technique often leads to a lack of contrast making it difficult to differentiate between the surface-gates and the semiconductor background. Following alignment at 800× magnification, further alignment is performed at both 6000× and 15000× magnification; a maximum of 6 and 3 flash-views respectively can be used at the minimum possible 'flash-view' dose before significantly exposing the resist. If alignment cannot be achieved within the maximum number of flash-views, alignment must be abandoned and a new resist deposited prior to making another attempt. The alignment procedure is simplified by the fact that the pattern shown in Fig. 3.13(a) needs only to be within the field of view at 15000× magnification (corresponding to a field size of 4μm × 4μm). It does not need to be accurately aligned to any particular position within this field. The location of the pattern is marked on the SEM display with an overhead pen following a 'flash-view' at 15000× magnification. The pattern for the hole is then aligned to the pattern drawn on the screen and exposed.

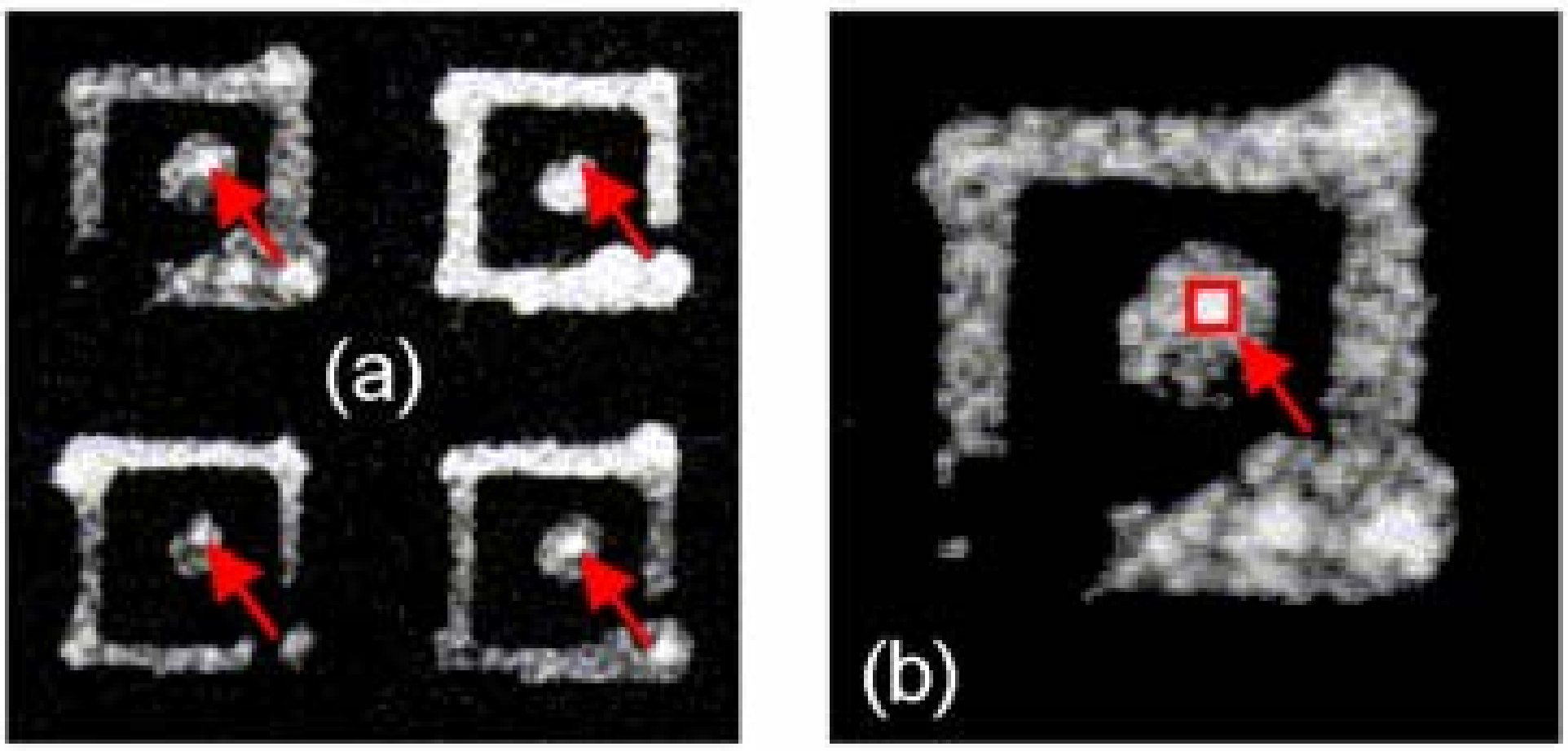

Figure 3.15: (a) Correctly aligned small holes in an array of test devices. (b) Magnified view of the billiard in the top left corner of (a). The red square indicates the patterned hole as an area of increased contrast on the circular gate.

The sample is developed (step (c)) in 3:1 IPA:MIBK for a duration determined by a set of dosage calibration boxes patterned at the edge of the sample. After development, the sample is placed back in the SEM for an inspection of the hole alignment. When the hole is over the circular gate, it appears as a region of increased contrast as indicated by the arrows in Fig. 3.15. If the hole is not correctly aligned the resist is dissolved in acetone and the process is repeated from step (a). Correction with this process is far easier than with the NRC techniques. In the case of the NRC devices the insulator is more difficult to remove. In addition, the optical lithography and dry etch steps prior to alignment, as well as potentially damaging reactive ion etching, need to be repeated at each attempt.

For a correctly aligned hole, cross-linking of the PMMA (step (d)) in regions under the bridging interconnect is performed using EBL with a very high dose (>50-70C/m$^2$) [139]. Cross-linked PMMA is insoluble in both MIBK and acetone. A second, 150nm thick layer of PMMA (step (e)) is then deposited on top of the original resist (PMMA (Undiluted[8] P5) spun for 40s at 8000rpm followed by a 20min bake at 150°C). A third EBL process (step (f)) is used to re-expose the hole, expose the path of the interconnect over the insulator, and expose a hole for the interconnect to contact its optical lithography surface-gate. Following development (step (g)) a short oxygen plasma etch is used to ensure there is no PMMA remaining on the circular gate, to prevent electrical contact between it and the interconnect. Metallisation (step (h)) is then performed by metal evaporation in a high vacuum environment. The bridging interconnect typically consists of 75nm NiCr under 15nm Au. The final step in the process is lift-off (step (i)). Excess metal and resist is removed by soaking in acetone for 24 hours, followed by rinsing with a jet of acetone as in §3.3. The final device would then be cleaved, packaged and bonded as described in §3.4.

## 3.5.3 – Fabrication Results: The Cambridge Sinai Billiard

I spent three months at the Cavendish Laboratory, Cambridge attempting to fabricate a Sinai billiard for follow-up experiments on the exact self-similarity observed in the magneto-conductance fluctuations of the NRC Sinai billiard [13]. Unfortunately, I was unable to fabricate a complete Sinai billiard during this time. Whilst it is possible that given sufficient time, a complete, working Sinai billiard could have been made, there were a number of problems and currently untested processing steps which may lead to further problems in this fabrication. The initial problem I experienced involved the surface-gate layer, in particular, the central circular gate.

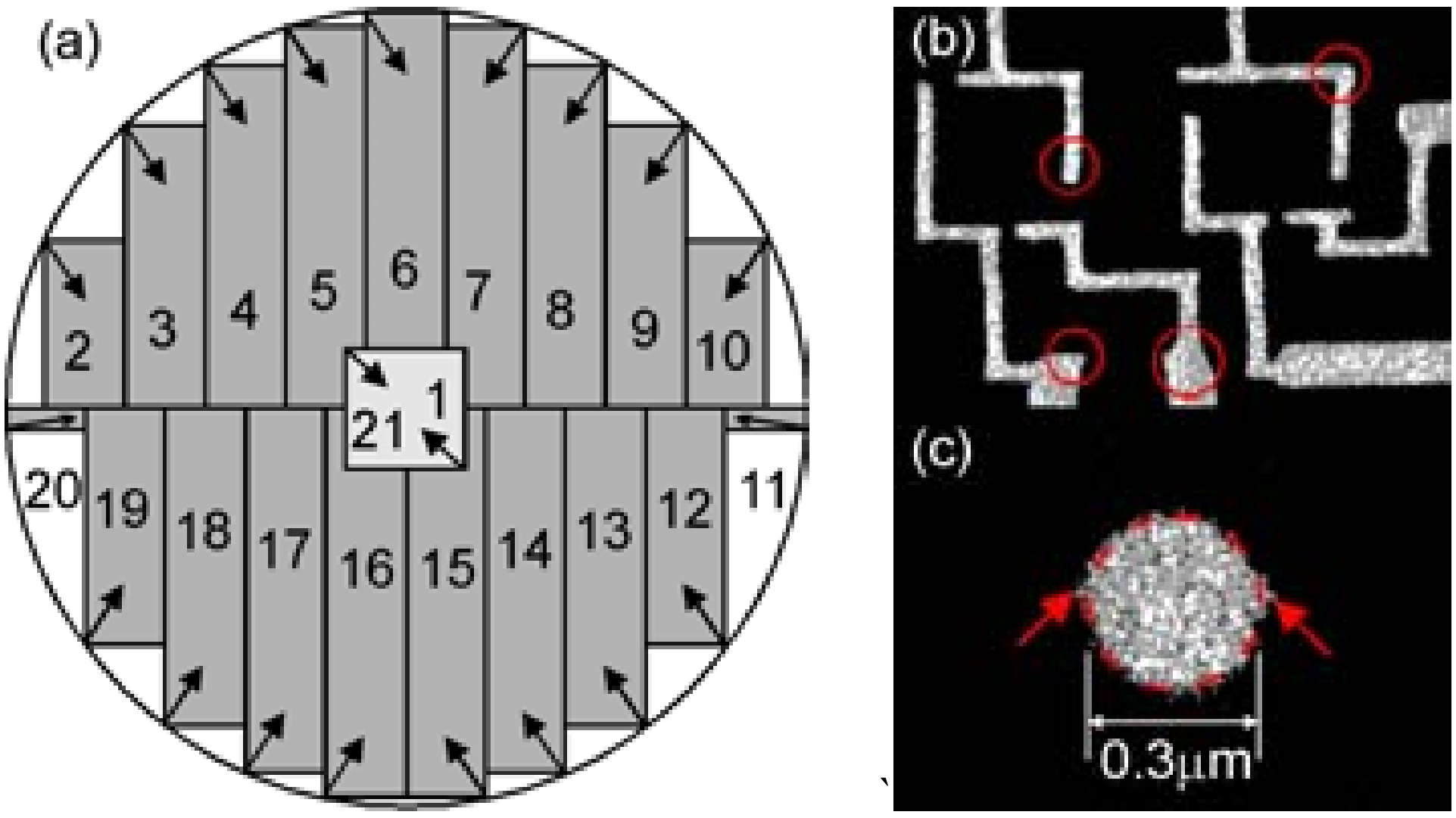

Figure 3.16: (a) The pattern used in defining the central circular gate. Four such patterns are drawn, each overlapping and rotated by 90° relative to the preceding one. The rectangle in the centre is patterned at the beginning and end of the pattern to reduce ticking. (b) Examples of ticking (circled in red) in surface-gates. (c) A circular gate patterned using four 90°-rotated copies of the pattern in (a). Note that there still remains a small amount of ticking in the circle as indicated by the red arrows.

Patterning a truly circular gate is difficult. Firstly, the pattern generator [142] on the SEM (Hitachi S-800) at Cambridge is only capable of drawing raster-scanned rectangles of a specified size and location. Complex patterns are built up from a set of these raster-scanned rectangles and this often leads to rectangular artefacts appearing in the edges of the circle. Furthermore, astigmatism and pattern stretching in one rectangle edge direction compared to the other can lead to an ellipse rather than a circle. I found that the optimum circle was drawn using sets of 19 rectangular strips (2-20 in Fig. 3.16(a)) that are drawn from where the strip touches the circle edge to the circle diameter, as shown in Fig. 3.16(a). Four of these sets of rectangular strips (each rotated by 90° relative to the one before it) are used to pattern the central gate. A ‘stretching factor' applied in the $x$-direction relative to the $y$-direction is used to eliminate errors in the relative scales in the $x$- and $y$-directions.

The second problem involves the beam blanking on the SEM and occurs when the beam is required to travel a long distance before patterning a rectangle. Various time delays, both in the pattern generator and the electron-optics, mean that the electron-beam is sometimes unblanked whilst it is still in transit, leading to an effect known as 'ticking', as shown in Fig. 3.16(b). Ticking effects can be minimised by employing a well-chosen patterning order and by patterning extra low-dose anti-tick rectangles to minimise the beam travel distance prior to patterning higher dosage rectangles. This strategy is employed in the strip pattern in Fig. 3.16(a) where an anti-tick rectangle is drawn at the beginning and end of the pattern (1,21 in Fig. 3.16(a)), and each strip is drawn from the corner lying on the circle edge, as indicated by arrows within each rectangle.

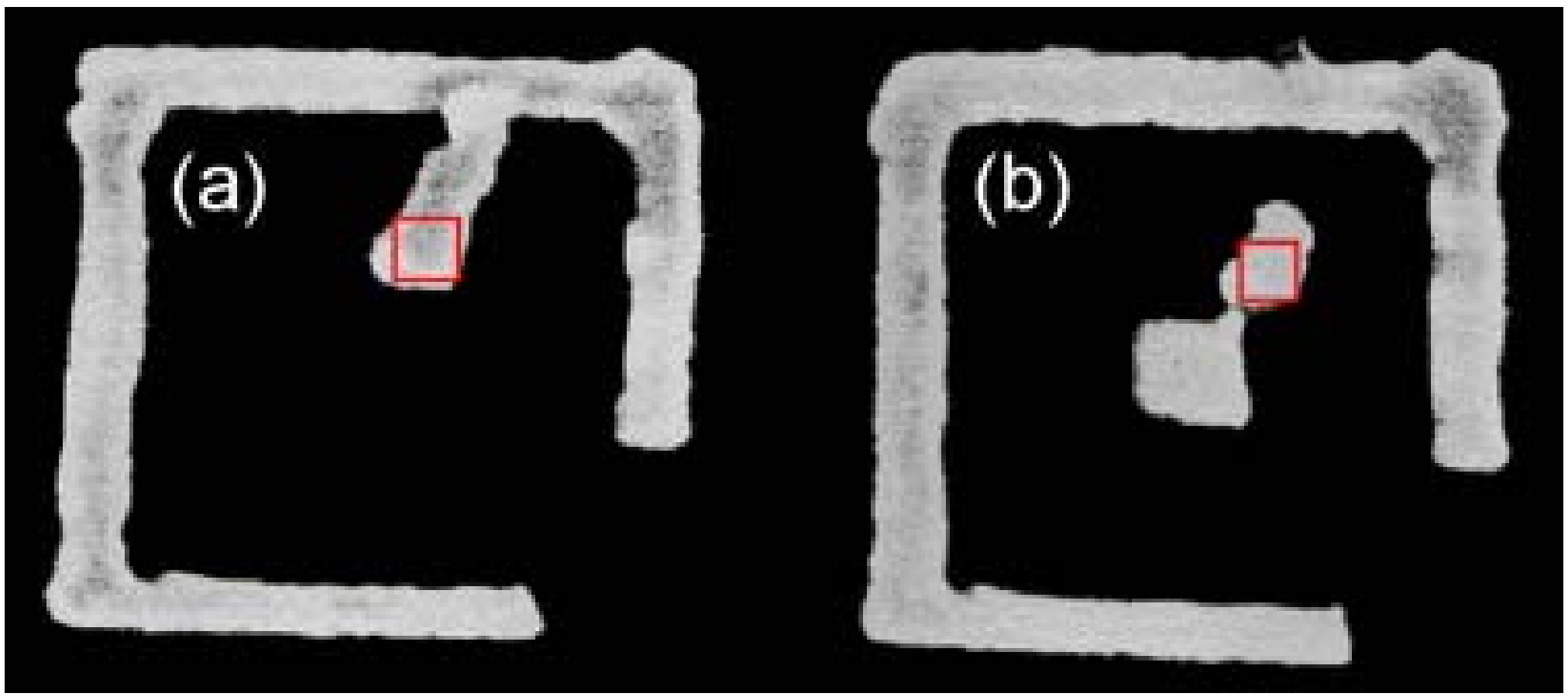

Figure 3.17: (a) Ticking on the insulator hole before employing strategies to minimise ticking. (b) Insulator hole patterned using the optimum technique described in the text. Ticking is still present but not as significantly as in (a). The red squares in both patterns indicate the approximate size and location of the intended hole. The hole in (b) is deliberately misaligned to illustrate the effect.

Whilst this strategy is largely successful in the patterning of the surface-gates, ticking was a more substantial problem in patterning the insulator hole. Whilst Fig. 3.15 shows square holes aligned to the central gate in the test patterns, and these holes appear square on viewing the resist, upon metallisation there is actually a substantial tick on the corner of the hole, as shown in Fig. 3.17(a). Elimination of this tick is essential in producing a truly circular gate. However, this is difficult since the small hole is the only pattern in the field. A relatively high dose (~15C/m$^2$) is required for proper exposure due to its size. A number of strategies were employed in an attempt to eliminate this problem, including switching from automatic to manual beam-blanking, anti-tick rectangles, programmed waiting times, lowering the current to extend the patterning time and drawing the hole using numerous low-dose rectangles rather than a single high-dose rectangle. I found that a combination of these is required to minimise ticking whilst maintaining good alignment of the hole. It is preferable to maintain a high beam current (~30pA) despite the fact that this increases ticking, because the alignment process is then considerably easier and more accurate. I also found automatic beam-blanking led to significantly reduced ticking (Fig. 3.17(b)) compared to manual beam-blanking (Fig. 3.17(a)). Anti-tick rectangles appeared to be more effective than beam-waiting, particularly in the case where large anti-tick rectangles are used. The hole in Fig. 3.17(b) is defined by first patterning a large, low-dose, anti-tick rectangle that originates[9] in the corner of the patterning field furthest from the hole and ends at the corner of the hole furthest from the origin of the anti-tick rectangle. Thus, any tick caused by travel between the end of the anti-tick rectangle (centre of the hole) and the pattern origin of the hole (one of the corners) is entirely contained by the hole.

The hole itself is patterned in two stages, firstly by a low-dose rectangle (~2C/m$^2$) followed by a rectangle with the remaining dose required to fully expose the hole. Further improvements on Fig. 3.17(b) could be made by patterning another anti-tick rectangle after the hole, and possibly by using the pattern in Fig. 3.16(a) rather than a square, since this slows down the patterning time, further minimising the beam travel delays that cause ticking. There was also a significant amount of random alignment error in this process, most likely caused by the pattern generator used on the Cambridge SEM. Whilst patterning an accurately aligned, well-shaped hole proved difficult on the Cambridge SEM, modern SEM/EBL systems with high-speed electrostatic beam-blanking and advanced pattern generation/alignment features should easily overcome these problems. Later steps in this process remain either untested, or not thoroughly tested, and may lead to further problems.

[9] Note that rectangles are patterned by raster scanning where the electron follows a zig-zag path (see Fig. 3.9). Hence it is meaningful to speak of an origin and an end point for a rectangle.

As a final note, it may be possible to use ‘air-bridge’ technology [122,123,126,143] instead to overcome some of the problems inherent in the bridging interconnect methods discussed in this section (i.e. §3.5).

# 3.6 – Double-2DEG Billiards

Recent advances in semiconductor growth and fabrication technologies have led to the development of heterostructures with two parallel, independently contactable 2DEGs [144,145]. Using conventional surface-gate methods it is then possible to define a pair of billiards – one directly below the other – with a nominally identical geometry since they are defined by a common set of surface-gates. These double-2DEG billiards provide the opportunity for a number of interesting experiments, one of which has been investigated as part of this thesis. A complete discussion of this experiment and the results obtained is presented in Chapter 6. This section is devoted to a discussion of the modifications required to processes described in §3.1 to §3.4 in order to fabricate a double-2DEG billiard.

The most notable difference between single- and double-2DEG billiards is the structure and growth of the heterostructure. Figure 3.18(a) shows a schematic of a double-2DEG heterostructure. The active region of the device involves a set of five layers indicated by the bracket at the right of Fig. 3.18(a). These five layers establish a pair of narrow potential wells as shown in Fig. 3.18(b). Whilst the confinement is induced on both sides of each well by AlGaAs layers, as opposed to only one side in the single-2DEG heterostructure, the mechanism leading to the formation of a 2DEG is the same as that discussed in §2.1.1. That is, electrons are constrained to travel in only two-directions by the confinement imposed by the narrow potential well. The depths of the two 2DEGs beneath the heterostructure surface are determined by the thicknesses of the various layers during the growth process. The composition and thickness of the barrier layer separating the two 2DEGs is also important in determining the strength of any interaction effects between the 2DEGs. Hence it is not uncommon for materials such as AlAs, which provides a higher barrier, to be used instead of AlGaAs in the barrier layer [146]. Each of the two 2DEGs has its own modulation-doping layer as shown in Fig. 3.18(a).

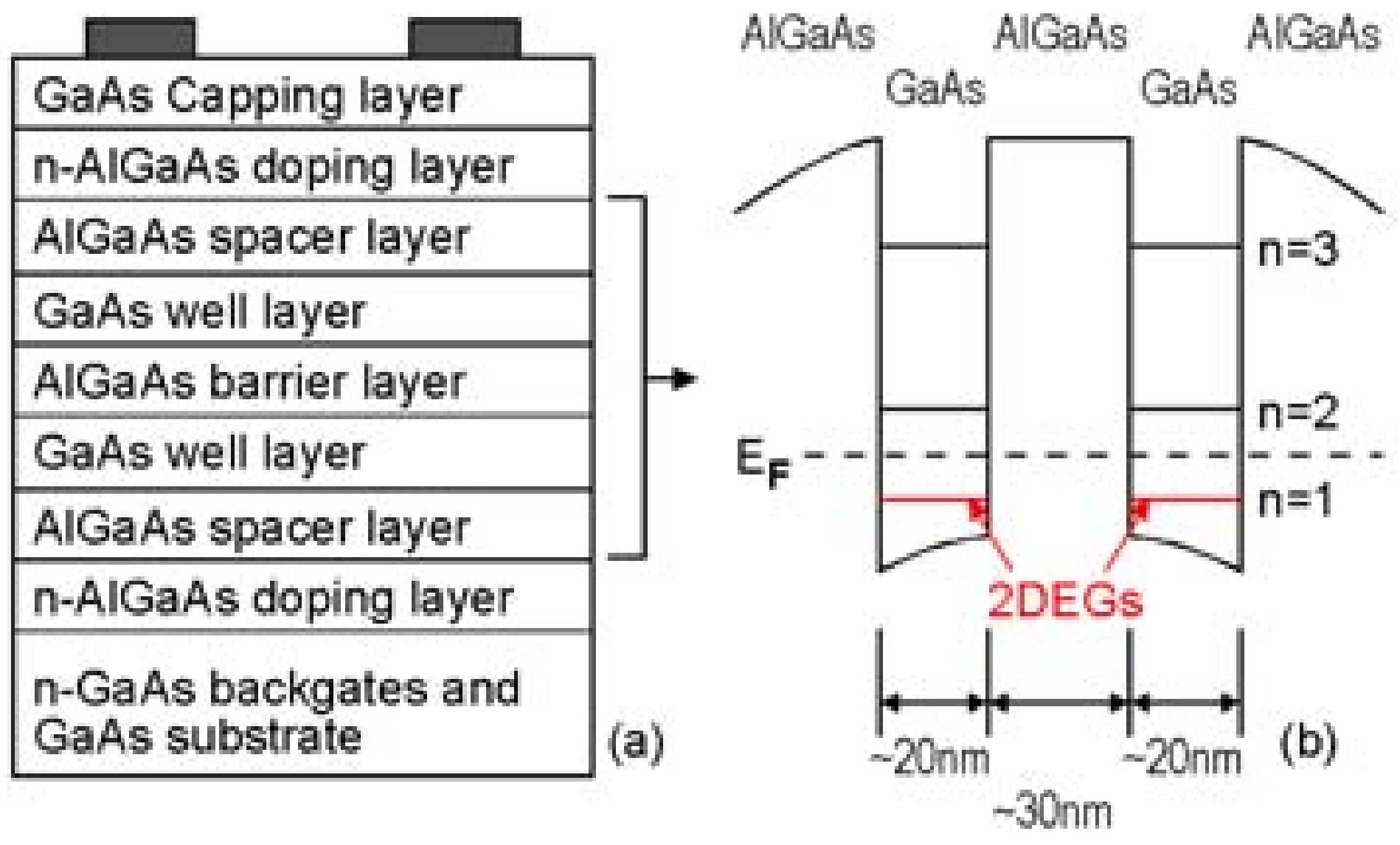

Figure 3.18: (a) Schematic of a double-2DEG semiconductor heterostructure. (b) The pair of narrow potential wells formed in the active region of the device. 2DEGs are confined in each of the potential wells.

The main difference in the growth process itself is due to the requirement for independent electrical contactability of the two 2DEGs. Both 2DEGs share a single set of ohmic contacts that penetrate from the surface down to below the lower 2DEG (see later). Independent contact is achieved using sets of surface- and back-gates positioned at the ends and sides of the Hall-bar as shown in Fig. 3.19. These isolation gates, when sufficient negative bias is applied, deplete the nearest 2DEG in the regions beneath/above them severing the electrical path between the particular 2DEG and the ohmic contact. The isolation gates for the top 2DEG are surface-gates defined using optical lithography. Isolation gates for the bottom 2DEG (back-gates) are defined in a conductive $n^+$-GaAs layer located below the lower 2DEG using focussed ion-beam lithography (FIBL) [145]. An extensive discussion of ion-beam lithography techniques is presented in [138]. FIBL is used to damage selected regions of the $n^+$-GaAs layer, rendering them non-conductive. Hence, patterned conducting regions are formed and these may be negatively biased to allow depletion of the 2DEG analogously to surface-gates. In addition to the isolation gates, a back-gate patterned to match the dimensions of the Hall-bar is also provided for fully depleting the lower 2DEG where necessary. Since FIBL causes large amounts of damage to the semiconductor, it is necessary to pattern the back-gates prior to the growth of the 2DEG layers shown in Fig. 3.18(a). As a result, double-2DEG heterostructures undergo a two-stage growth process in a combined MBE/FIBL system with FIBL used to define the back-gates between the MBE growth stages. MBE and FIBL are performed in separate chambers of the combined MBE/FIBL system allowing the wafer to remain under UHV conditions for the entire MBE/FIBL process [145]. This is essential in preventing oxidation between the growth stages, which would result in a significant loss in electron mobility.

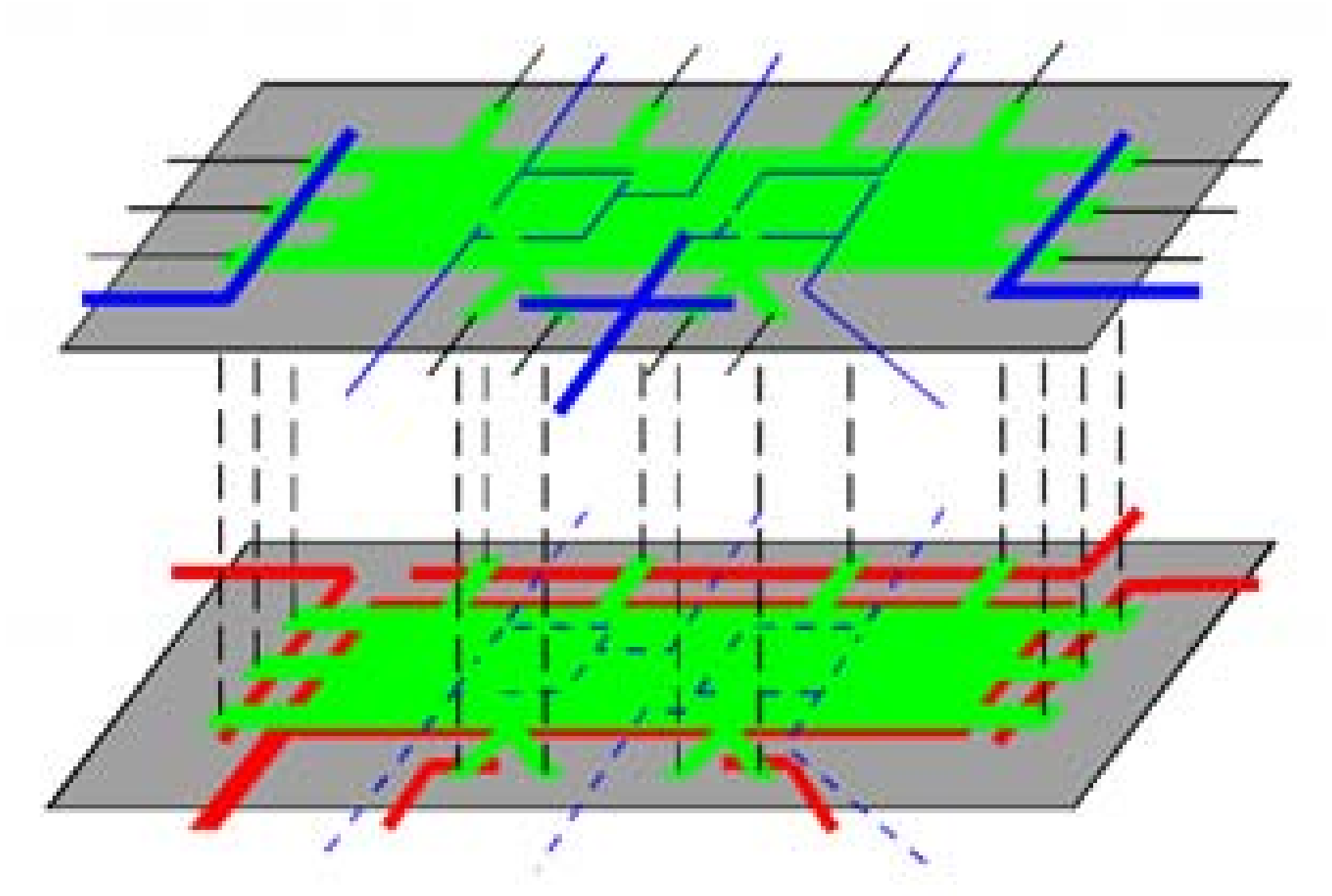

Figure 3.19: Schematic of a double-2DEG billiard. The Hall-bar is shown in green with surface-gates in blue and back-gates (conductive $n^+$-GaAs) in red. The black dashed lines between layers are the ohmic contacts. The blue dashed lines in the bottom layer indicate that the lower billiard is defined using the same surface-gates that define the upper billiard.

In the first growth stage, a ~1μm GaAs layer is grown on a semi-insulating GaAs substrate. The purpose of this first layer is to grow over any surface roughness in the substrate, providing an atomically flat surface for further layers to be grown on [119,120]. The second layer in the first growth stage is a 50nm layer of $3\times10^{24}m^{-3}$ Si-doped GaAs. Initially this layer is conducting with a resistivity of $\sim 500\Omega Y^{-1}$ at room temperature [145]. FIBL is performed using a focussed 30keV beam of Ga ions which is raster-scanned over the $n^+$-GaAs layer to damage selected regions according to a computer-generated pattern. An ion-beam spot size of ~50nm and beam current ~100pA is used in this process. Damaged regions are exposed to an ion-beam dose of $\sim 3\times10^{17}$ions/m$^2$ leading to an increase in resistivity by a factor of $\sim 10^7$ over that of the unexposed regions [145].

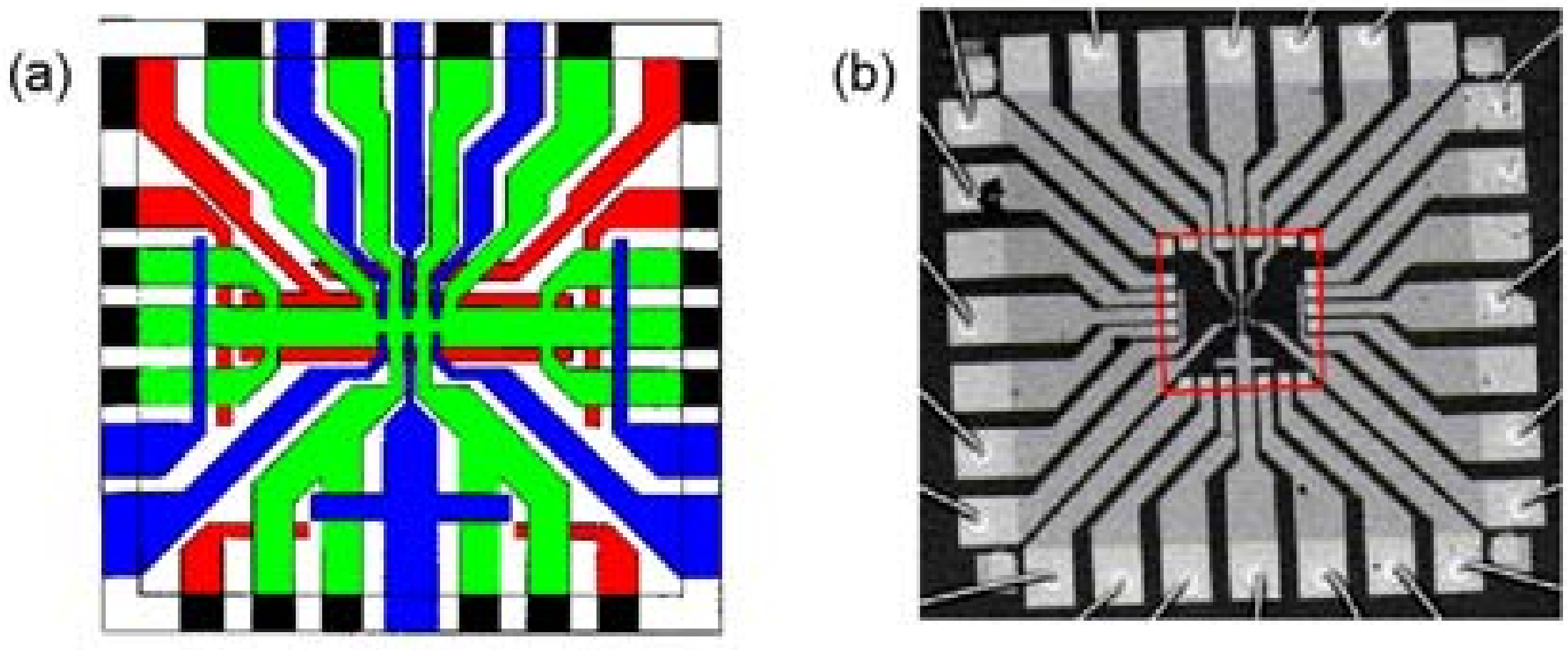

Figure 3.20: (a) Schematic of the various masks/patterns used in processing the double-2DEG billiards: mesa etch in green, optical lithography surface-gates in blue and ion-beam lithography back-gates in red. Ohmic contacts are indicated in black. (b) Scanning electron micrograph of a complete double-2DEG billiard. The outer edge of the schematic in (a) corresponds to the superimposed red square. The ohmic contacts are visible as squares of slightly higher contrast just inside the red square.

Following patterning of the back-gates, a set of alignment markers are defined by ion-beam milling at the corners of the back-gate field using a dose of $\sim 1\times10^{20}$ions/m$^2$ of 30keV Ga ions. Due to the layer-by-layer nature of MBE growth, the alignment marker depressions generated by the ion-beam milling process are transferred to the surface of the wafer, albeit slightly larger in area, allowing alignment to the back-gate field upon completion of the growth process. The second growth stage is then performed, beginning with 20nm GaAs and 250nm AlGaAs which act as a spacer layer between the back-gates and the active region of the device, and even out any surface roughness induced by the FIBL process in the undamaged regions. The active region of the device is then grown as outlined in Fig. 3.18(a) with two 40nm $2\times10^{24}m^{-3}$ Si-doped AlGaAs doping layers, 30nm AlGaAs spacer layers, 20nm GaAs well layers and a 30nm AlGaAs barrier layer. The final layer is a 10nm thick GaAs capping layer to prevent oxidation of the upper n-AlGaAs doping layer.

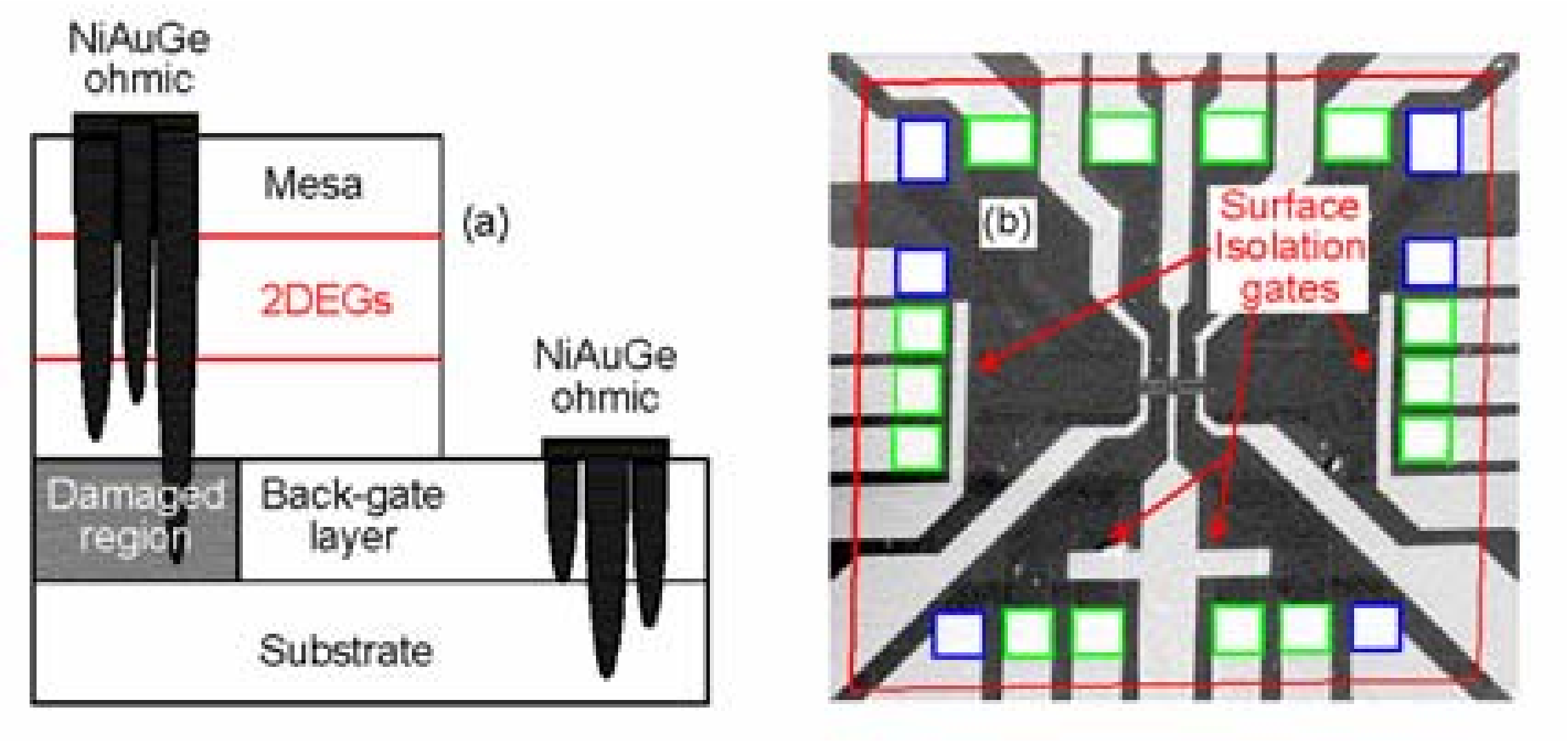

Figure 3.21: (a) Schematic illustrating the use of ohmic contacts in making electrical connections to the 2DEGs and the back-gates (After [145]). (b) Scanning electron micrograph of the central region of a double-2DEG billiard, highlighting the ohmic contacts for the 2DEGs (green) and back-gates (blue). The red square marks the inner perimeter of the isolation etch region and is aligned to the FIBL markers. The red arrows indicate isolation surface-gates. The remaining optical lithography surface-gates connect to the surface-gates that define the billiard.

Processing commences with the mesa etch and ohmic contact formation processes following the procedures outlined in §3.2.1. and §3.2.4. The mesa etch pattern is shown in green in Fig. 3.20(a), aligned to the FIBL back-gate pattern (red). In double-2DEG billiards, ohmic contacts (black squares in Fig. 3.20(a)) are used to provide electrical contact to the FIBL back-gates in addition to both 2DEGs. This requires special precautions to be taken to avoid the back-gate ohmics connecting to the 2DEGs and vice versa. A deep mesa etch to below the lower doping layer is used to physically constrain both 2DEGs to the Hall-bar region as shown in Fig. 3.21(a). The back-gate pattern is carefully designed so that highly resistive damaged regions lie below all of the 2DEG ohmic contact locations. This is clear in Fig. 3.20(a) where none of the back-gates (red) cross under any of the 2DEG ohmics. Furthermore, each of the back-gates lead out beyond the Hall-bar to regions off the mesa where they can be contacted without connecting to the 2DEGs (see Fig. 3.21(a)). As a result there are actually two sets of ohmics (as highlighted in Fig. 3.21(b)) despite the fact that they are fabricated in a single processing stage. Since the substrate is semi-insulating, penetration of the ohmic spikes beyond their intended depth is not a major problem. However, care must be taken to ensure that the NiAuGe deposition only occurs in the intended ohmic contact locations prior to annealing. Ohmic metallisation in other regions can lead to leakage paths between the 2DEGs, the back-gates and the surface-gates.

The isolation surface-gates shown in Fig. 3.21(b), which are used to independently contact the bottom 2DEG by cutting off the top 2DEG, are fabricated using optical lithography as discussed in §3.3.1. The mask pattern for these gates is shown in blue in Fig. 3.20(a). Optical lithography surface-gates are used to connect all ohmics (back-gate and 2DEG) to bonding pads at the edges of the sample, and support EBL surface-gates, as shown in the micrographs of Figs. 3.20(b) and 3.21(b). Standard EBL techniques (§3.3.2) are used to pattern the billiard gates shown in the micrograph of Fig. 3.22.

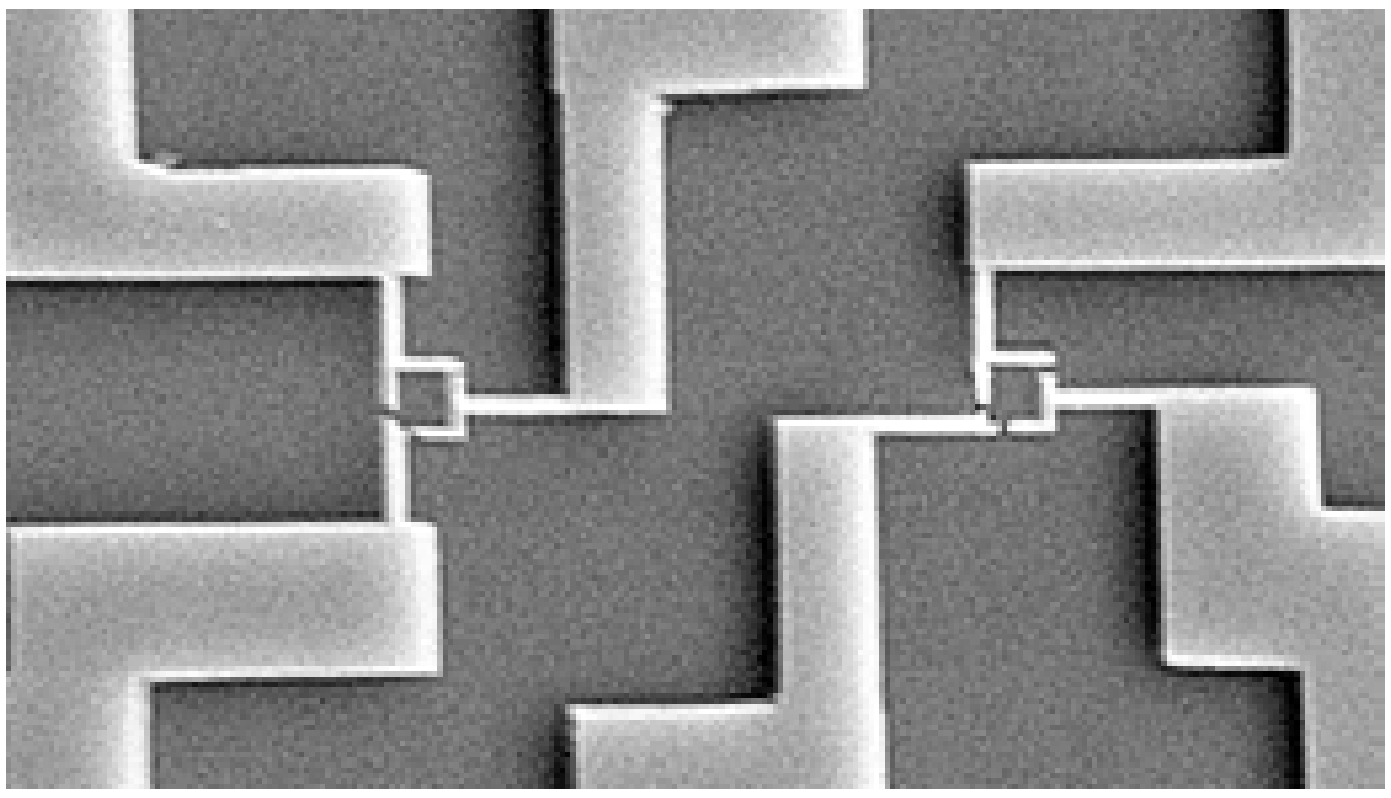

Figure 3.22: Scanning electron micrograph of the central regions of the double-2DEG device showing the two 1μm square billiards patterned using EBL techniques.

Whilst there are two billiards patterned on the Hall-bar, typically only one billiard is bonded upon completion of the device. Before cleaving, packaging and bonding, a final isolation etch is performed. Prior to this etch all of the back-gates are electrically connected by the undamaged regions which lie outside the FIBL patterning field shown by the red square in Fig. 3.21(b). This is particularly important since it is common to deposit ohmic contacts at the bonding pads to enhance bond-wire adhesion, and these ohmics will lead to shorting between the bond pads via the back-gate layer. The isolation etch process commences with an optical lithography stage, which defines a square covering the FIBL patterning field (i.e. the red square in Fig. 3.20(b)) to protect the FIBL regions during the isolation etch. A 1:8:8 solution of $H_2SO_4$:$H_2O_2$:$H_2O$ is used to etch through the back-gate layer (typically a 15s soak is required) to isolate the individual back-gates and bond pads. The wafer is then soaked in acetone for 30s to remove the photo-resist etch mask.

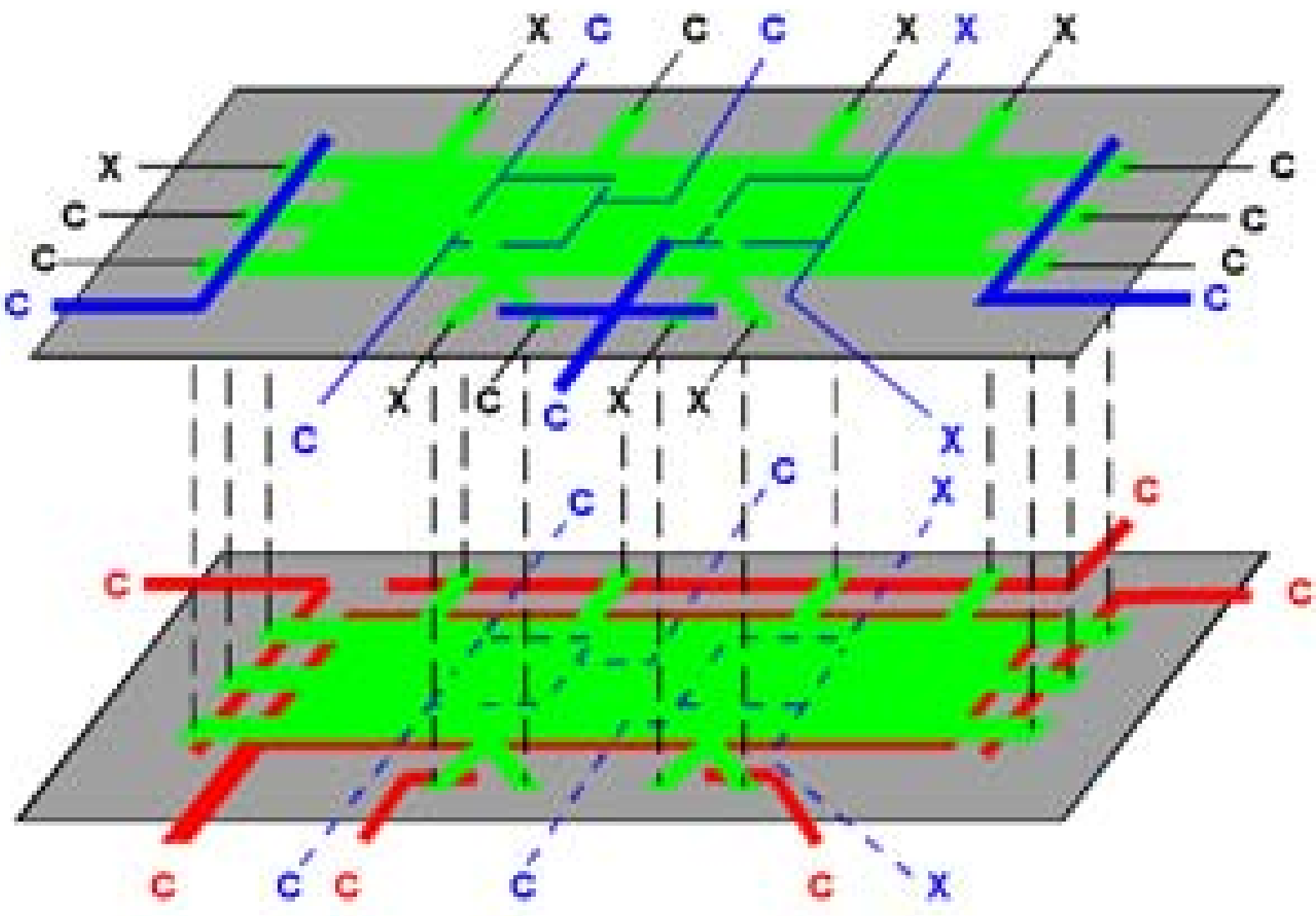

Figure 3.23: Schematic of a double-2DEG billiard with ohmics (black), back-gates (red) isolation surface-gates (thick blue) and EBL billiard surface-gates (thin blue). Bonded connections are marked with a C whilst those remaining unbonded are marked X.

Cleaving, packaging and bonding is then carried out as discussed in §3.4. Due to the limited number of pins on the package (and wires running down the dilution unit) it is not possible to connect all of the ohmics, back-gates, isolation surface-gates, and billiard surface-gates. Instead only the left-hand billiard is bonded, mainly since it is the only one which can be measured with current and voltage probes at the ends of the Hall-bar. The bonded ohmics and gates are indicated by a 'C' in Fig. 3.23 whilst those remaining unconnected are marked by an 'X'.

# 3.7 - Comparison of Devices used in this Study

A number of devices fabricated in different laboratories are discussed in this thesis. The important parameters and details regarding these devices are compared below:

## 3.7.1 - NRC[10] 1.0μm Sinai billiard

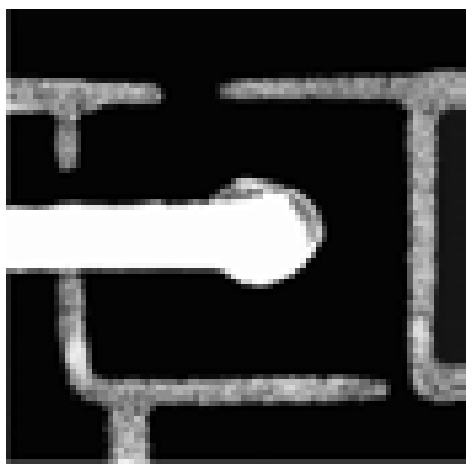

Electron Density: $2.3\times10^{15}m^{-2}$
Electron Mobility: $316m^2/Vs$
2DEG Depth: 183nm
Minimum donor-2DEG separation: 60nm
Donor concentration/type: Dual layer doping. $1\times10^{22}m^{-3}$ Si at 60nm from 2DEG and $2\times10^{22}m^{-3}$ Si at 22nm below surface.
Fabrication: NRC Canada by Y. Feng on wafer grown at NRC
Experiment: NRC Canada, Sept-Oct 1996, by R.P. Taylor and R. Newbury

## 3.7.2 - RIKEN[11] Square Billiards: 0.4, 0.6, 1.0 and 2.0μm

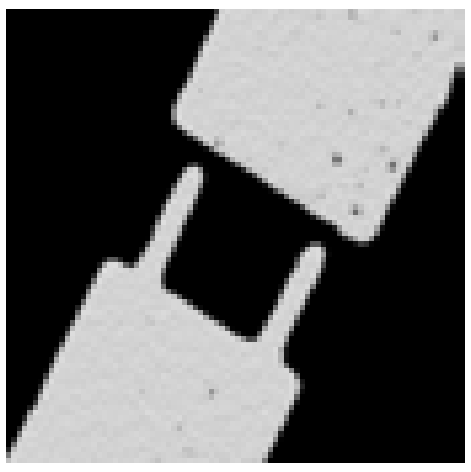

Electron Density: $4.4\times10^{15}m^{-2}$
Electron Mobility: $40m^2/Vs$
2DEG Depth: 71nm
Minimum donor-2DEG separation: 16nm
Donor concentration: $1\times10^{24}m^{-3}$ Si
Fabrication: RIKEN, Japan by K. Ishibashi, J.P. Bird on wafer from Sumitomo Electric Co. Ltd., Japan.
Experiment: RIKEN Japan and UNSW at various times between 1994 and 1996 by J.P. Bird, J. Cooper, A.P. Micolich, R. Newbury, R.P. Taylor and R. Wirtz.

---

[10] Institute for Microstructural Sciences, National Research Council of Canada, Ottawa, K1A 0R6 Canada.

[11] Nanoelectronics Materials Laboratory, RIKEN, 2-1 Hirosawa, Wako-shi, Saitama 351-01, Japan.

## 3.7.3 - RIKEN[11] Side-wall Sinai Billiard

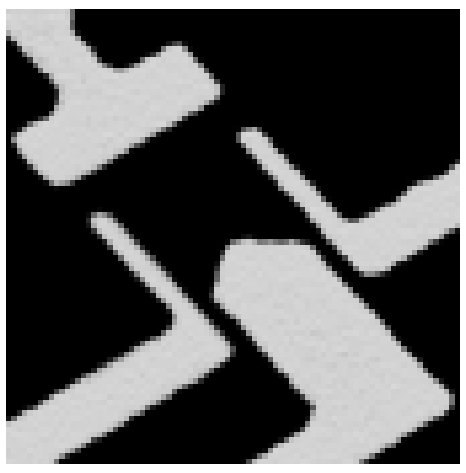

Electron Density: $4.5\times10^{15}m^{-2}$

Electron Mobility: $51m^2/Vs$

2DEG Depth: 60nm

Minimum donor-2DEG separation: 15nm

Donor concentration: $2\times10^{24}m^{-3}$ Si

Fabrication: RIKEN, Japan by J. Cooper, J.P. Bird on wafer from Sumitomo Electric Co. Ltd., Japan.

Experiment: UNSW, May 1997 and October 1998 by A.P. Micolich, R.P. Taylor, R. Newbury, and M. Hallett.

## 3.7.4 - NRC[10] AXA Device

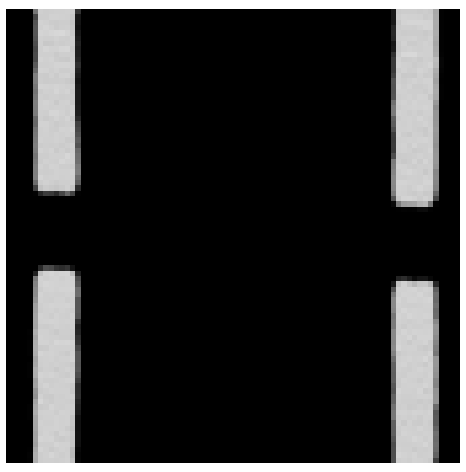

Electron Density: $3.6\times10^{15}m^{-2}$

Electron Mobility: $300m^2/Vs$

2DEG Depth: 90nm

Fabrication: NRC, Canada by P.T. Coleridge on wafer grown at NRC

Experiment: UNSW, July 1994 by R.P. Taylor and R. Newbury.

## 3.7.5 - Cambridge[12] 1.0μm Double-2DEG Square Billiard

Electron Density: Upper 2DEG: $2.9\times10^{15}m^{-2}$ Lower 2DEG: $2.8\times10^{15}m^{-2}$

Electron Mobility: Upper 2DEG: $130m^2/Vs$ Lower 2DEG: $110m^2/Vs$

2DEG Depth: 90nm and 140nm

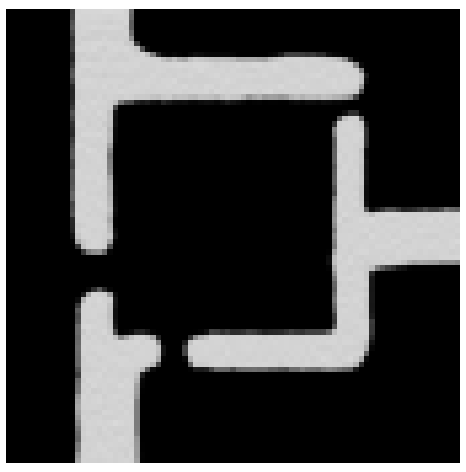

Minimum donor-2DEG separation: 30nm

Donor concentration/type: $2\times10^{24}m^{-3}$ Si

Fabrication: Cambridge, UK by H.D. Clark, W.R. Tribe on wafer grown at Cambridge by E.H. Linfield and P.D. Rose.

Experiment: UNSW, October 1999 by A.P. Micolich, R.P. Taylor, R. Newbury, A. Ehlert and M. Neilson.

---

[12] Semiconductor Physics Research Group, Cavendish Laboratory, Madingley Road, Cambridge, CB3 0HE, U.K.

# Chapter 4 – Experimental Methods and Characterisation Techniques

This chapter is divided into two sections. §4.1 discusses the various cryogenic and electrical methods used in experiments on semiconductor billiards. §4.2 discusses the techniques used to obtain the value of several important parameters of the billiard.

## 4.1 – Experimental Methods

The observation of electron quantum interference effects is an important feature of billiard experiments. Minimisation of inelastic electron scattering events, such as electron-phonon and electron-electron scattering, is essential in maintaining a sufficient electron phase coherence length for observable quantum interference effects to occur [147]. This is achieved by cooling the sample to millikelvin temperatures using a $^3He/^4He$ dilution refrigerator. However, it is important to bear in mind that a material can have a number of temperatures relating to the entropies of the various parts of the material (i.e. the crystal lattice, electrons, nuclear spins, etc) [148,149]. Above 1K the various parts are thermally well coupled and it makes sense to talk about a single temperature for the material. However, at temperatures below 1K the various temperatures can be significantly different in a practical situation. Whilst the cryogenic methods used in billiard experiments cool the heterostructure lattice to millikelvin temperatures, the most important temperature in these experiments is the electron temperature. Providing any electrical measurements are sufficiently non-invasive (cause minimal Joule heating of the electrons) these temperatures are expected to be quite close since the lattice and electrons can remain thermally coupled at millikelvin temperatures provided care is taken. This means that the measurement electronics needs to be carefully designed so that it uses sufficiently low currents and biases (nA/μV) that electron heating is minimal without serious loss in measurement accuracy. At such low currents/biases, separation of the measurement signals from the background noise is difficult, requiring sophisticated measuring techniques and well-designed shielding and filtering systems. Shielding and filtering is doubly important since electrical noise picked up from the environment may appear across the sample and can lead to significant electron heating. §4.1.1 discusses the various techniques and equipment required to obtain millikelvin temperatures in billiard experiments, with particular focus on the $^3He/^4He$ dilution refrigerator. §4.1.2 deals with electrical measurements of billiards including brief discussions of shielding, filtering and grounding in the measurement circuit, as well as computer control of the experiment and data acquisition. More specific electrical and cryogenic details are discussed in Appendix A.

## 4.1.1 – Cryogenics: Obtaining Millikelvin Temperatures

A number of methods for obtaining temperatures well below 273K were developed during the last century, each with its own minimum achievable temperature [148,149,150]. Billiard experiments are generally performed on one of four types of cryostat: helium dip-probe, pumped-$^4He$ cryostat, pumped-$^3He$ cryostat, or $^3He/^4He$ dilution refrigerator – with base temperatures of 4.2K, 1.4K, 300mK and ~20mK respectively. Whilst the dilution refrigerator is more expensive to set-up and more complicated to operate than the other cryostats mentioned above, it allows experiments over the widest temperature range (4.2K – 20mK) with the lowest base temperature. As a result, the dilution refrigerator is the cryostat of choice for the experiments performed in this thesis. This section is divided into three subsections. §4.1.1.1 will briefly discuss the basic operating principles of the other three cryostats mentioned above since these principles play an important part in the operation of the dilution refrigerator. §4.1.1.2 will discuss the properties of $^3He/^4He$ liquid mixtures that allow the dilution refrigerator to achieve temperatures far below those of the other three cryostat types. §4.1.1.3 will discuss the actual operation of the dilution refrigerator. Finally, §4.1.1.4 will discuss the specific equipment used in experiments at UNSW.

### 4.1.1.1 Liquid Helium and Helium Evaporation Cryostats

With the widespread use of superconducting magnet systems in medical applications, liquid $^4He$ is now commonly commercially available. The $^4He$ dip-probe is by far the simplest of the four cryostats commonly used in the study of mesoscopic semiconductor systems. The sample is attached to the end of a probe and immersed in liquid $^4$He to bring the sample to a temperature of 4.2K. Whilst this temperature is insufficient for investigations of quantum interference phenomena, it is commonly used in the low temperature assessment of properties such as electron density and mobility (see §4.2.1), and to identify working samples prior to experiments. The pumped helium cryostats achieve temperatures below 4.2K using the principle of evaporative cooling [148,149]. In order for an atom or molecule of liquid to evaporate, some amount of energy is required to enable it to break free from the interatomic/intermolecular forces holding it as part of the liquid. This amount of energy is known as the latent heat of vaporisation and is usually made available at the expense of the heat in the liquid (and other objects in thermal equilibrium with the liquid), resulting in cooling. The vapour pressure determines the rate of evaporation of a liquid and is defined as the partial pressure of evaporated gas required for the rate of condensation to be equal to the rate of evaporation. Hence, once the partial pressure of evaporated gas above the liquid becomes equal to the vapour pressure there is no longer any net evaporation, and hence no further cooling. This is the case for liquid helium at atmospheric pressure, which has a temperature of 4.2K, as mentioned above. Using a vacuum pump, the partial pressure of evaporated gas can be maintained at a low value, leading to a high net rate of evaporation and increased cooling. Indeed at sufficiently low partial pressures, the liquid helium will boil, further enhancing the evaporation rate and the resultant cooling. The latent heat of vaporisation and the net rate of vaporisation determine the minimum temperature achievable using this technique. The net rate of vaporisation is dependent on the vapour pressure and the partial pressure of evaporated gas, which depends on the pumping rate of the vacuum pump used. Typical minimum temperatures achieved using this technique are ~1.2K for $^4$He and ~300mK for $^3$He [148]. A significantly lower temperature is achieved for pumped $^3$He, largely due to its higher zero-point energy. The higher zero point energy in $^3$He compared to $^4$He means that $^3$He has a higher vapour pressure (by a factor of 74, 610, and 9800 at 1K, 0.7K and 0.5K respectively [149]), leading to a higher net rate of evaporation and increased cooling. Another important difference between the isotopes is statistics – $^4$He is a boson and $^3$He is a fermion. $^4$He undergoes Bose condensation at 2.18K to form a superfluid, in contrast to $^3$He which also becomes a superfluid but only at temperatures below 3mK [148]. The superfluid behaviour of $^4$He requires special design of the container/pumping connection to prevent superfluid film flow out of the container. Such precautions are unneccessary with $^3$He, which is a normal fluid throughout the operation of both

pumped-$^3$He cryostats and dilution refrigerators. This further enhances the minimum temperature of $^3$He cryostats since the heat leaks present in pumped-$^4$He cryostats due to superfluid film flow are eliminated. The major complication in the use of $^3$He is expense. $^3$He is far less abundant than $^4$He and is largely produced as a byproduct in the nuclear decay of tritium ($^3$H $\rightarrow$ $^3$He + $e^-$ $\tau_{½}$ = 12.5yr) [149]. This route of production leads to a $^3$He cost of ~\$200/litre, roughly 20 times that of $^4$He. For this reason, loss of $^3$He is to be avoided. Closed $^3$He systems need to be used with special attention paid to leak-proofing to prevent the use of $^3$He pumping systems becoming prohibitively expensive.

#### 4.1.1.2 Properties of $^3$He/$^4$He Liquid Mixtures

At temperatures below 3.19K both $^4$He and $^3$He are liquids and may form a combined $^3$He/$^4$He liquid mixture. The $^3$He concentration (or mole-fraction) in this mixture is given by $x = n_3/(n_3+n_4)$ where $n_3$ and $n_4$ are the number of moles of $^3$He and $^4$He respectively. Figure 4.1 shows the $T$-$x$ phase diagram of a liquid $^3$He/$^4$He mixture at saturated vapour pressure. For temperatures above the tricritical point ($T_3$ = 0.86K) the mixture can be in one of two states: a superfluid where the $^3$He concentration is sufficiently low for the $^4$He bosonic behaviour to dominate, and a normal fluid at higher $^3$He concentrations. At temperatures below 0.86K, three distinct phases are possible with a dual-phase liquid mixture present at intermediate $^3$He concentrations. The dual-phase region is bounded by the 'coexistence curve' (which includes the tricritical point) and on crossing this curve from either the superfluid or normal regions the mixture spontaneously separates into two phases, one with a higher concentration of $^3$He and the other with a higher concentration of $^4$He. The $^3$He-rich (concentrated) phase floats on top of the $^4$He-rich (dilute) phase due to its lower density.

The mixture in a dilution refrigerator typically has a $^3$He concentration of $x \sim 0.15$ which remains constant throughout operation. After the mixture is liquified, it is cooled, first crossing the λ-curve and entering the superfluid phase, and then meeting the co-existence curve at a lower temperature. This path is vertical in the $T$-$x$ phase diagram because $x$ is fixed, as shown in Fig. 4.1. Once the co-existence curve is reached the mixture spontaneously separates into dilute and concentrated phases.

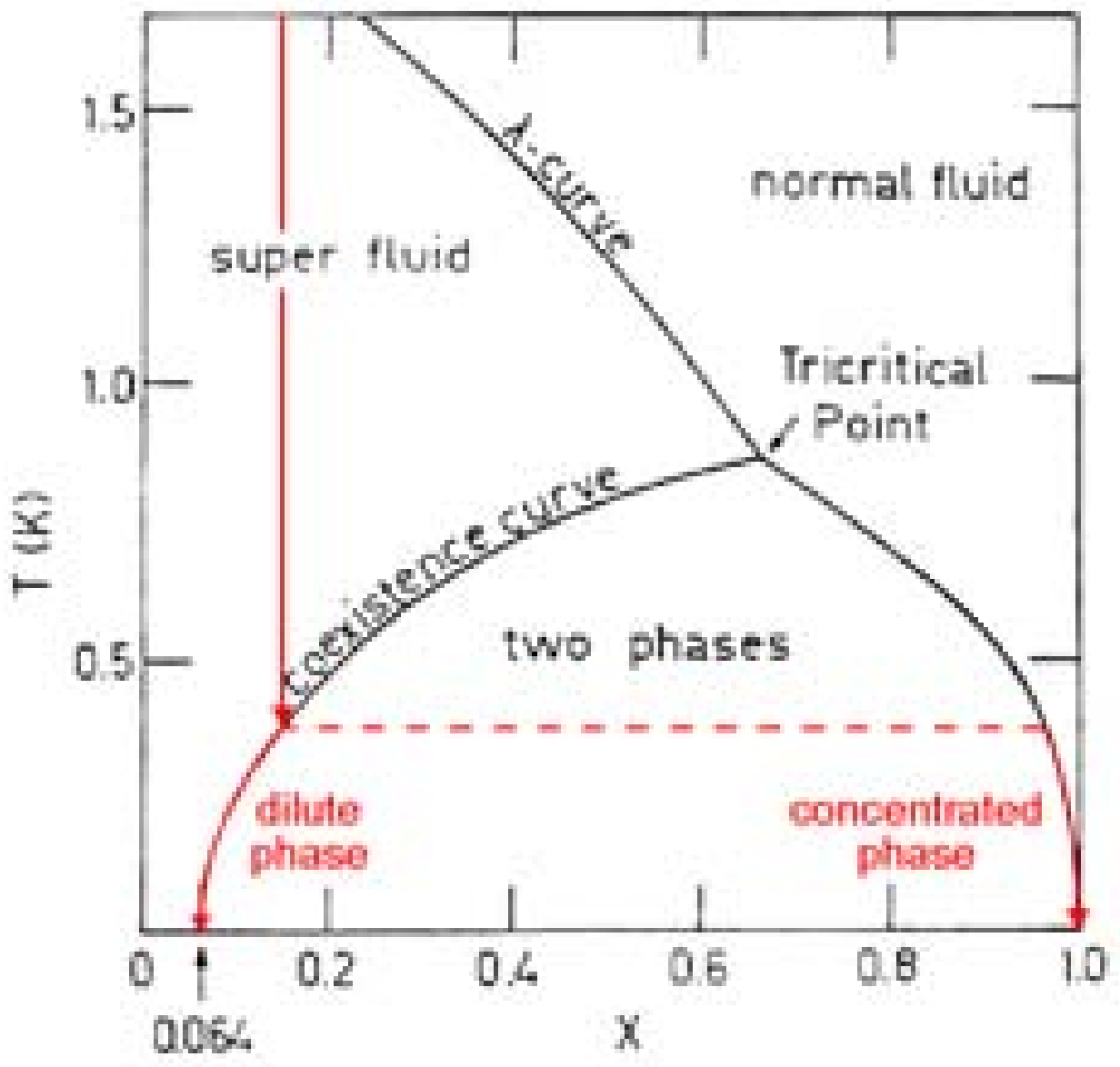

Figure 4.1: Temperature versus $^3$He concentration (*T*-*x* phase diagram) for a liquid $^3$He/$^4$He mixture (after [151]).

The $^3$He concentration of the two phases is determined by the *x*-values of the coexistence curve branches at that particular temperature – the concentrated phase ($x_C$) on the right branch, and the dilute phase ($x_D$) on the left branch. As the temperature is reduced further the concentrations of the phases continue to be determined by the coexistence curve. Two important features of the *T*-*x* phase diagram with respect to operation of a dilution refrigerator are that $x_C \sim 1$ at temperatures below ~100mK and more importantly, $x_D$ has a minimum concentration of 0.064 at 0K. If $x_D$ went to zero as *T* approached absolute zero, operation of a dilution refrigerator would not be possible. The physical reasons behind this minimum $^3$He concentration in the dilute phase are discussed in [148,149].

#### 4.1.1.3 Operation of a $^3$He/$^4$He Dilution Refrigerator

A schematic of a dilution refrigerator, cryostat and external gas-handling system is shown in Fig. 4.2. The cryostat consists of three concentric metal cans – the outer vacuum can (OVC), liquid nitrogen ($LN_2$) jacket and the main bath – working from the outside inwards. The main purpose of the cryostat is to shield the dilution unit ($T$ < 4K) from the environment ($T$ ~ 300K). However, the cryostat also serves as a liquid $^4$He reservoir for the $^4$He pot and, in the case of the experiments discussed in this thesis, contains the superconducting solenoid which needs to be immersed in liquid $^4$He whilst in use (see §4.1.1.4). The dilution unit is contained inside the inner vacuum can (IVC), with the sections below the still often shielded by an extra copper radiation shield that is concentric and internal to the IVC. The dilution unit consists of five main parts – the $^4$He pot, condenser, still, heat exchangers and mixing chamber – the latter four linked by pipes that form a circulation path for the $^3$He/$^4$He mixture.

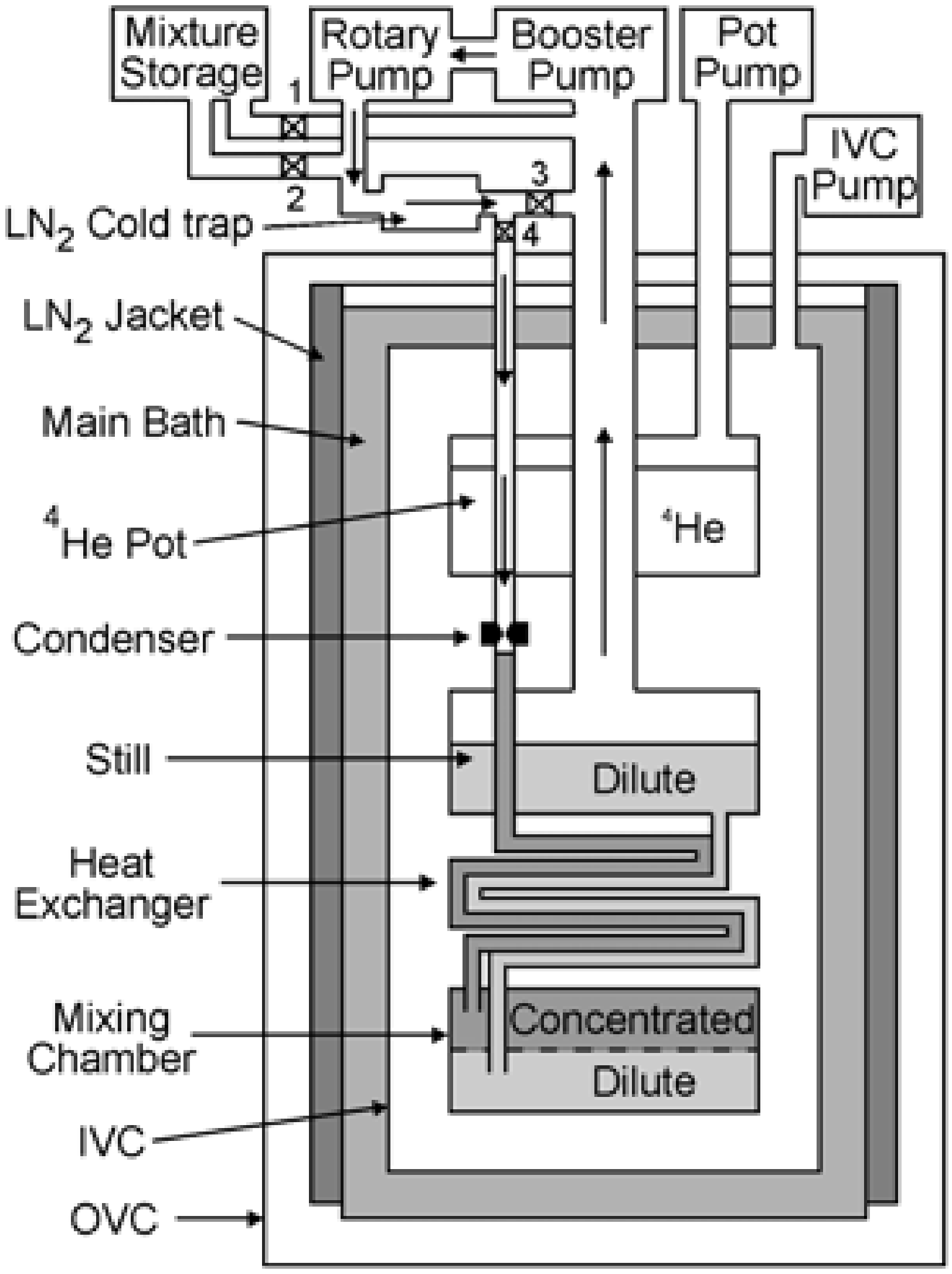

Figure 4.2: Schematic of the essential features of a $^3$He/$^4$He dilution refrigerator, cryostat and external pumping system.

In normal operation, temperatures range from 4.2K at the top of the IVC to 1.2K at the $^4$He pot (pumped-$^4$He cryostat), ~600mK at the still and ~20mK at the mixing chamber. The external gas-handling system consists of two pumps to circulate the mixture, a pump for the $^4$He pot, and a pump for evacuating the IVC, as well as a vessel for storage of the mixture whilst the dilution unit is not in use.

Operation commences by precooling the cryostat and dilution unit to 4.2K. This is performed in two stages – first to 77K and then to 4.2K by filling the main bath with liquid $N_2$ and liquid $^4$He respectively. During the precooling process the IVC is filled with a small amount of exchange gas ($N_2$ and then $^4$He) to allow heat transfer between the dilution unit and the main bath ensuring the whole dilution unit comes to 4.2K. Once the main bath is full of liquid $^4$He and the dilution unit has reached 4.2K, the IVC is pumped to high vacuum and kept in this state for the remainder of the experiment, thermally isolating the dilution unit from the main bath. As mentioned in the previous section, $^3$He has a boiling temperature of 3.19K at 1atm and hence the mixture will not condense completely at 4.2K. Further cooling of the mixture is achieved by pumping on the $^4$He pot. The tube carrying the mixture to the condenser passes through the $^4$He pot, precooling the mixture to 1.2K prior to liquification by Joule-Thomson expansion in the condenser aperture. The $^4$He pot is filled continuously via a tube passing through the IVC and into the main bath. The fill rate is controlled using a needle valve to prevent the $^4$He pot from overfilling or emptying.

The dilution unit can be divided into two halves either side of the mixing chamber – the condenser side and the still side. The condenser side contains a number of narrow tubes as well as the condenser aperture, and is aimed at ensuring the mixture is completely liquified and precooled to base temperature before entering the mixing chamber. The pipe-work on the still side has a larger diameter to aid pumping of the mixture that evaporates from the still. This difference in pipe-work diameter is important in the initial condensation of the mixture, which is performed via the still side rather than the condenser side. This is to prevent any contaminants that have leaked into the mixture during storage, and have not been captured by the 77K $LN_2$ cold trap, from solidifying in the narrow pipes of the condenser side and blocking the mixture path through the dilution unit. Instead, any contaminants are safely frozen out onto the walls of the wider pipes on the still side. Once all of the mixture has condensed into the dilution unit, the storage vessel is closed (spring-loaded valves allow one-way retreat of the mixture back into the storage vessel in the event of overpressure in the dilution unit) and circulation of the mixture is commenced. The external rotary and booster pumps pump evaporated helium from the still and supply it to the condenser line for re-liquification, driving the circulation process. The heat exchangers allow the mixture in the still side to cool the re-liquified mixture, ensuring that it is close to base temperature prior to re-entering the mixing chamber. Despite the fact that the mixture was precooled to 1.2K during the condensation process, its temperature on the commencement of circulation is closer to 2.5 – 3K. This is because of heating by the remainder of the dilution unit, which is at ~4.2K during condensation of the mixture. As discussed in §4.1.1.2, for a $^3$He concentration of $x$ = 0.15, a temperature below 375mK is required for phase separation of the mixture [151]. This temperature is achieved by pumped evaporative cooling of the mixture as part of the mixture circulation process. Cooling from ~2.5K immediately after condensation to base temperature (~20mK) actually occurs in three stages as described in Fig. 4.3.

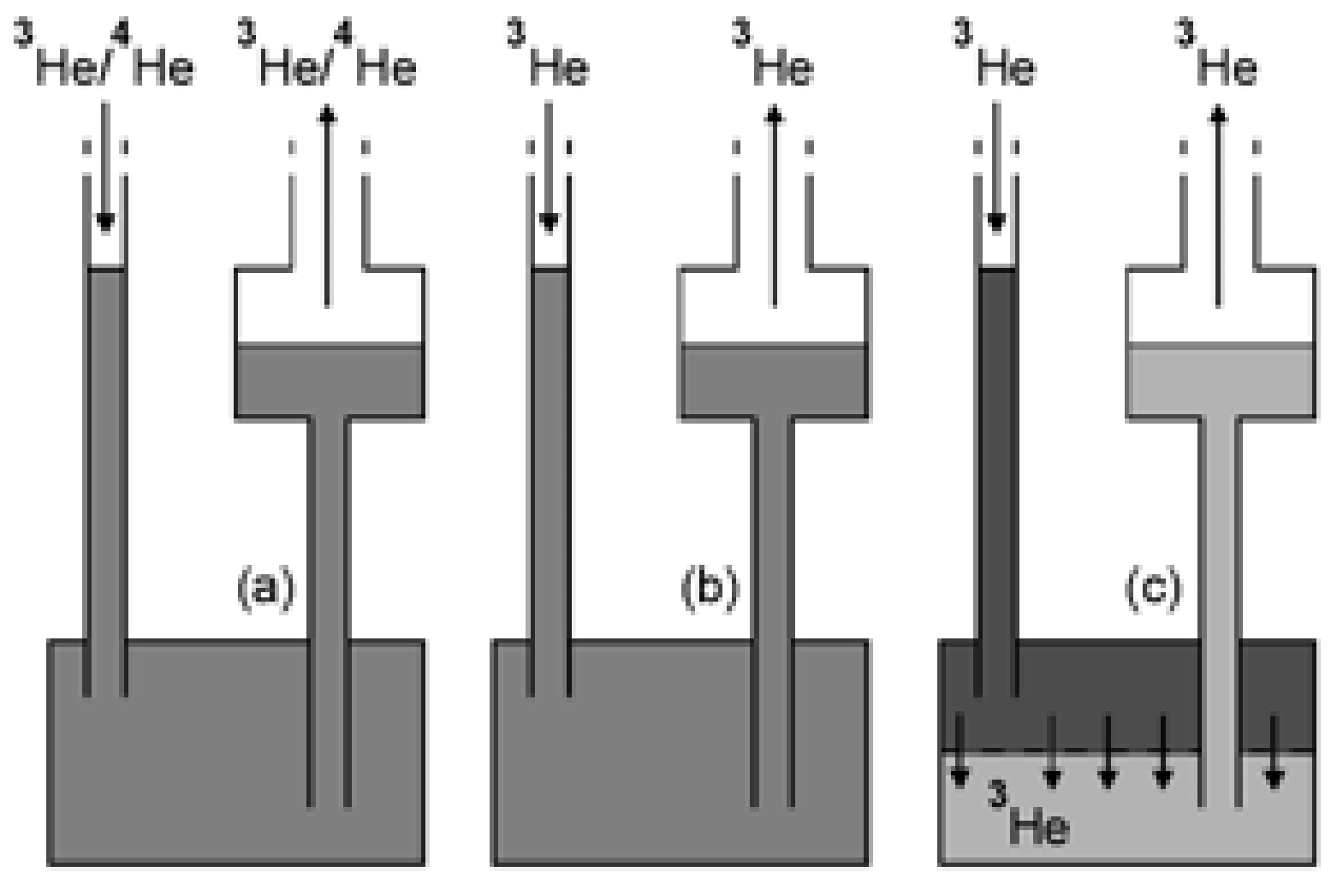

Figure 4.3: Schematic of the three operation regimes of a dilution refrigerator. (a) $T$ > 1K: $^4$He vapour pressure comparable to $^3$He vapour pressure and cooling is by evaporation of both $^4$He and $^3$He. (b) $T$ ~ 1K to phase separation temperature: $^4$He vapour pressure much smaller than $^3$He vapour pressure with cooling largely by $^3$He evaporation only. (c) below phase-separation temperature: Cooling occurs by $^3$He evaporation at the still and $^3$He crossing the phase boundary at the mixing chamber.

Initially both $^{3}He$ and $^{4}He$ are evaporated off the still (Fig. 4.3(a)) leading to evaporative cooling as discussed in §4.1.1.1. As the mixture temperature begins to fall below ~1.5K the vapour pressure of $^{4}He$ begins to decrease rapidly compared to that of the $^{3}He$. Hence evaporation of $^{3}He$ begins to dominate the cooling of the still with very little evaporation of $^{4}He$ occurring (Fig. 4.3(b)). Throughout these stages cooling of the mixing chamber only occurs by heat conduction through the mixture itself.

Once 375mK is reached, the mixture phase-separates into dilute and concentrated phases. The specifications of the dilution unit and quantity of mixture used are set so that the phase separation boundary lies inside the mixing chamber. Upon phase-separation, the cooling mechanism changes significantly to that shown in Fig. 4.3(c) where there are two centres of cooling – the mixing chamber and the still. Cooling of the still continues via $^{3}He$ evaporative cooling as it did prior to phase separation. Cooling in the mixing chamber, however, occurs entirely due to the minimum $^{3}He$ concentration of 6.4% in the dilute phase. As mentioned in §4.1.1.2, the concentrated ($^{3}He$-rich) phase floats above the dilute ($^{3}He$-rare) phase, with the condenser line entering the concentrated phase and the still line entering the dilute phase. The still also contains the dilute phase of the mixture as shown in Fig. 4.2. As the rotary and booster pumps continue to draw $^{3}He$ off the still, the $^{3}He$ concentration in the dilute phase in the still begins to decrease. This establishes a $^{3}He$ concentration gradient in the dilute phase, resulting in an osmotic pressure gradient between the mixing chamber and the still. This osmotic pressure acts to eliminate the concentration gradient by drawing $^{3}He$ from the dilute phase in the mixing chamber and transferring it to the still, causing a reduced $^{3}He$ concentration in the mixing chamber. In order to maintain the minimum 6.4% $^{3}He$ concentration in the dilute phase, $^{3}He$ is drawn across the phase boundary from the concentrated phase. This process is known as mixing [149]. Since the $^{4}He$ in both phases is well below the λ-point temperature, it is a superfluid, in its ground state, and effectively a mechanical vacuum for the $^{3}He$ dissolved in it. As a result, the transfer of $^{3}He$ across the phase boundary can be seen as an evaporation of $^{3}He$ atoms from the concentrated phase ($^{3}He$ quasi-liquid) into the dilute phase ($^{3}He$ quasi-gas). As discussed in §4.1.1.1, an atom/molecule evaporating from a liquid to a gas requires some energy, the latent heat of vaporisation, which it generally obtains at the expense of the heat contained in the liquid, resulting in cooling of the liquid. Analogously, the mixing process also requires energy, the latent heat of mixing, and leads to cooling of the mixture and the sample, which is maintained in thermal contact with the mixture. The drop in $^3$He concentration in the concentrated phase is compensated by the $^3$He pumped off the still, via the external pumping circuit and condenser line, resulting in continuous operation of the dilution refrigerator. Once cooling due to mixing commences, the mixing chamber temperature rapidly drops below that of the still, which will not drop below ~300mK (i.e. the minimum temperature of a pumped $^3$He cryostat). Heat is usually applied to the still (~3mW) to accelerate the $^3$He evaporation process and hence increase the mixing rate to further lower the mixing chamber temperature. As a result, the normal operating temperature of the still is ~600mK. The mixing chamber cooling power is typically on the order of 5μW, depending on a large number of factors including the size of the mixing chamber, $^3$He cycling rate, etc. The minimum mixing chamber temperature typically achievable is ~20mK, depending on external heat leaks, both from the sample and the cryostat/dilution unit.

#### 4.1.1.4 UNSW Dilution Refrigerator

Billiard experiments conducted at UNSW were performed using the equipment shown in Fig. 4.4. The system consists of a top-loading dilution unit, cryostat, and external gas handling systems as shown in Fig. 4.4(a), as well as a superconducting solenoid and measurement/diagnostics/control electronics. A close-up photograph of the dilution unit is shown in Fig. 4.4(b) with the various parts discussed in preceding sections indicated. The dilution unit consists of an Oxford Instruments Kelvinox 300 dilution unit (still, heat exchangers and mixing chamber) built onto a pre-existing insert with the original $^4$He pot and most of the initial support structures intact. The dilution refrigerator is specified capable of a base temperature below 10mK with no added heat load (sample wiring, cold finger, etc). Following a complete rewiring of the dilution unit, which I conducted in early 1999, the base temperature at the mixing chamber with low current (<5nA) measurements in progress is lower than 20mK. Details of the wiring refit are discussed in Appendix A. A second $RuO_2$ resistor mounted next to the sample indicated a sample lattice temperature of approximately 35mK, concurrent with the above mixing chamber temperature measurement.

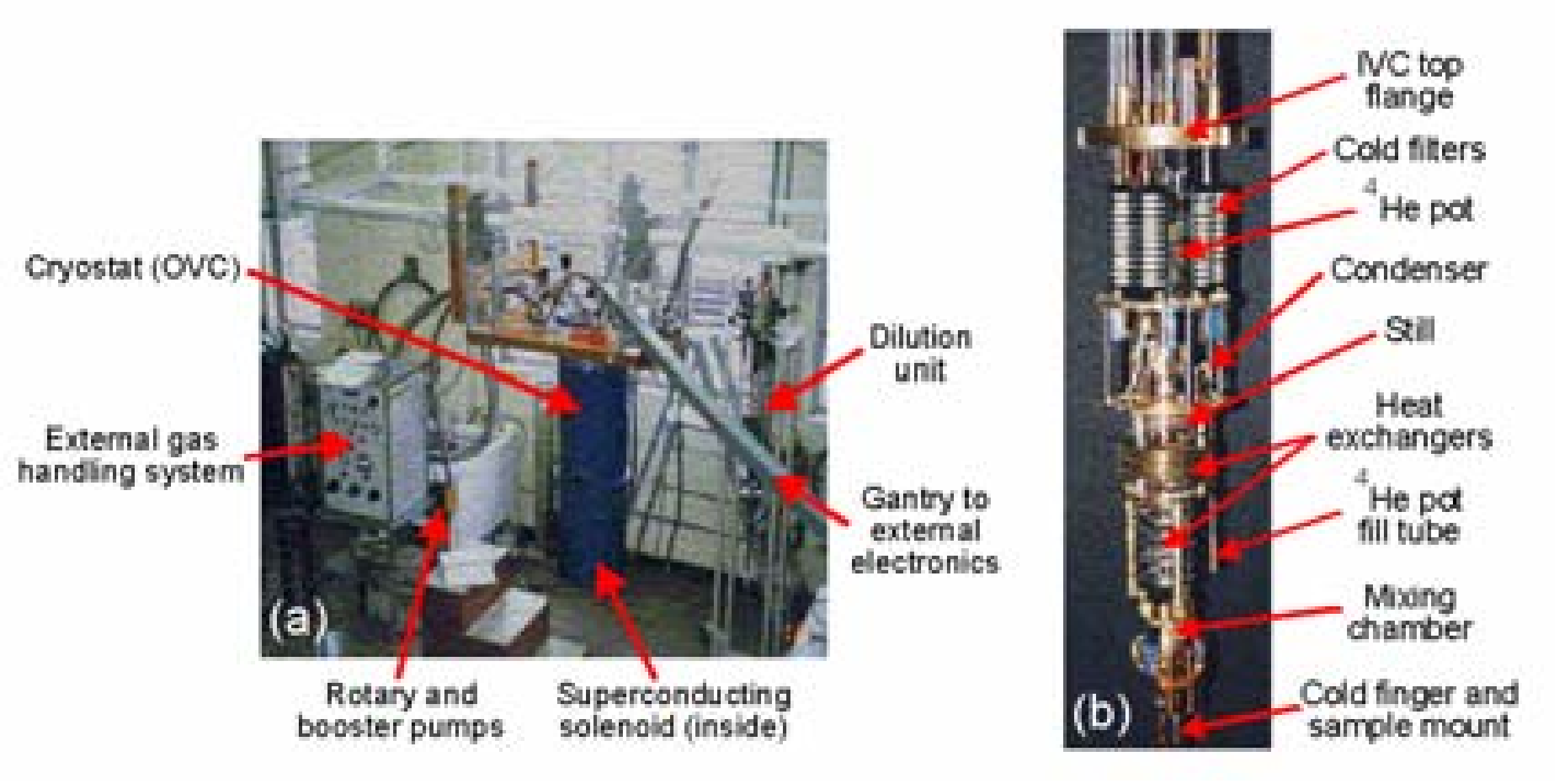

Figure 4.4: (a) Photograph of the cryostat, dilution unit and external gas handling systems used for billiard experiments conducted at UNSW. (b) Close-up photograph of the dilution unit indicating the various items discussed in preceding sections.

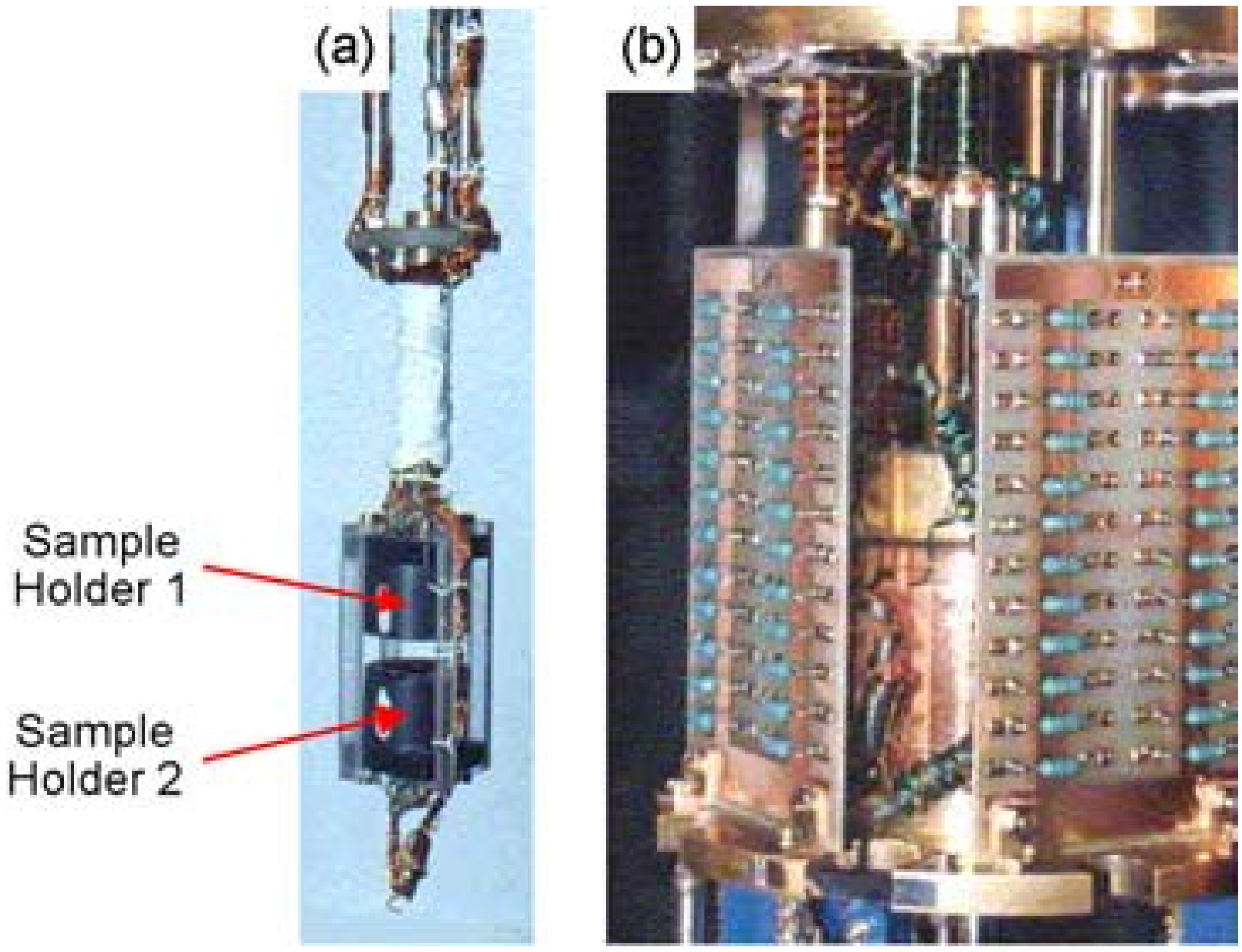

Figure 4.5: (a) Photograph of the cold finger and sample mounts. (b) Photograph of the LC cold-filters mounted on the $^4$He pot plate.

The cryostat broadly matches that described in preceding sections with the addition of a NiTi/$Ni_3Sn$ superconducting solenoid mounted at the bottom of the main bath. The superconducting solenoid is capable of producing magnetic fields up to 15T at 4.2K. Magnetic fields in the bore of the solenoid are homogenous to 1 in 1000 within a 5mm radius about the centre of the solenoid, and are controllable to 0.1mT resolution via computer interface to the solenoid power supply. The mixing chamber is located well above the superconducting solenoid. The sample is located at the centre of the solenoid by a cold finger mounted on the bottom of the mixing chamber. Thermal contact between the sample and the mixing chamber is provided by the sample wires rather than the cold finger itself. A tri-strut cold-finger (see Appendix A) is used to minimise eddy current heating during magnetic field sweeps and allows easy alignment of the sample holders within the radiation can to prevent touching. Facilities are available for up to two samples, each with up to 18 electrical contacts, to be mounted at the end of the cold finger during any single cool-down and are measurable simultaneously. As shown in Fig. 4.5(a), the samples are mounted facing each other, with the solenoid centre lying directly between them. Both samples lie well within the magnetic field homogeneity region specified above.

### 4.1.2 – Electronics: Measurement, Diagnostics, Control and Data Acquisition

A block diagram of the electronics set-up used during billiard experiments performed at UNSW is shown in Fig. 4.6. Experiments discussed in this thesis that were conducted outside UNSW were performed using systems similar to the one discussed in this section. Dilution unit diagnostics and control electronics are standard and will not be discussed further in this thesis. The remaining electronics serve to perform four main tasks during the experiment. These include making accurate measurements of the electrical resistance across the sample, maintaining stable DC biases on surface- and back-gates, control of the applied magnetic field and monitoring and control of the sample temperature. Electrical resistance is measured using a four-terminal technique as shown in the circuit diagram in Fig. 4.7.

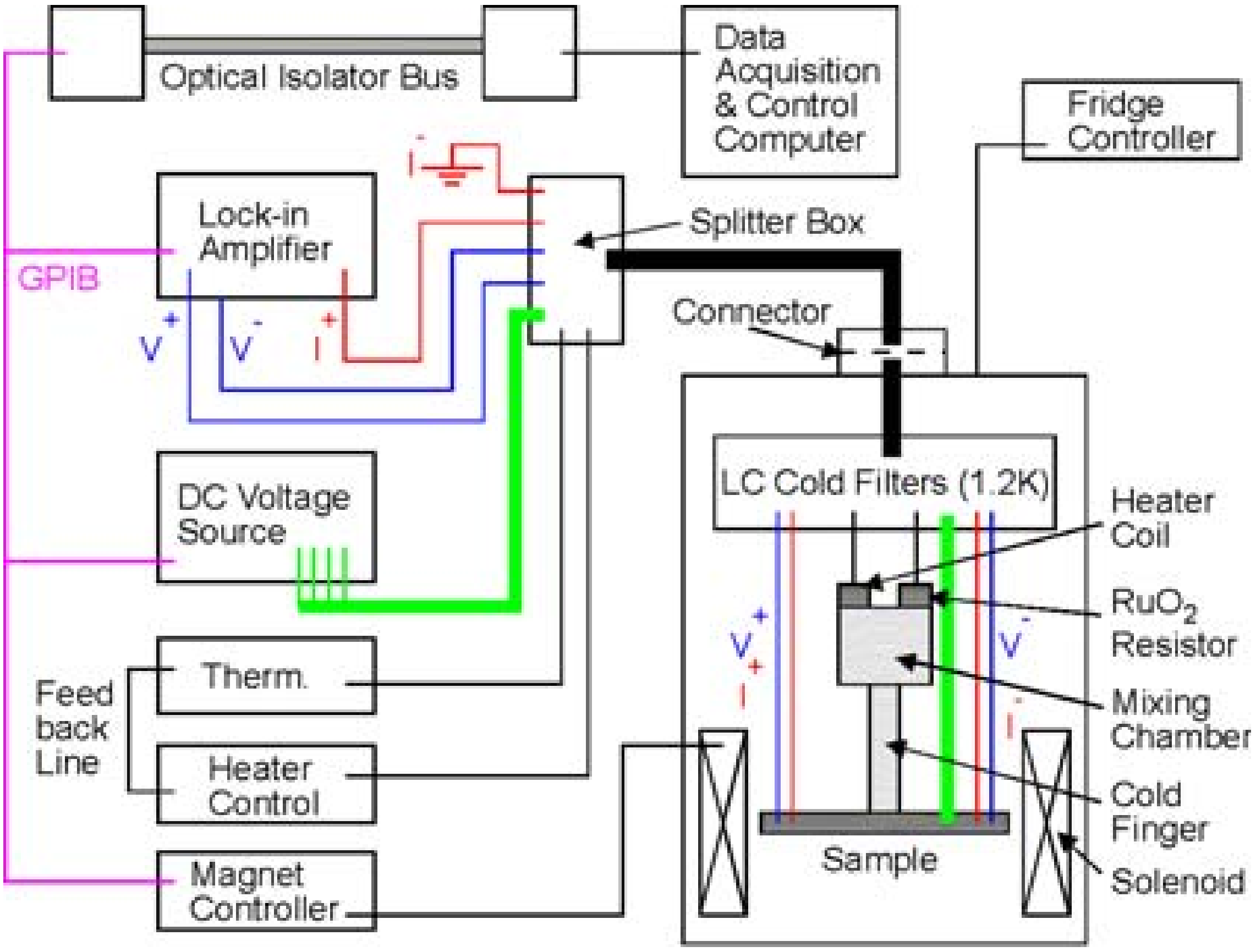

Figure 4.6: Block diagram of the electronics set-up used during billiard experiments conducted at UNSW.

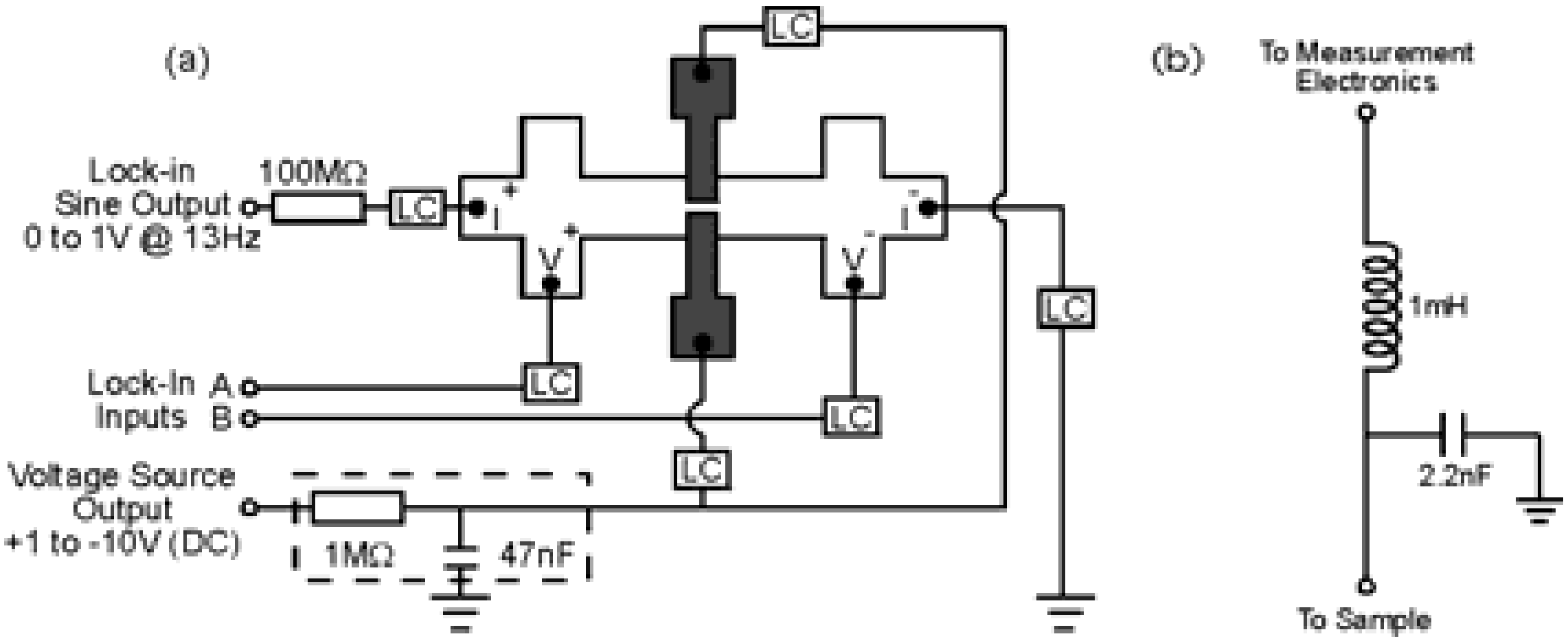

Figure 4.7:(a) Measuring circuit used for four-terminal longitudinal resistance measurements across the billiard. (b) shows the circuit diagram for the LC cold filters present on each line immediately adjacent to the sample contacts in (a).

In this technique, a constant current is passed along the sample while the potential difference along/across[13] the sample is measured using a separate pair of contacts. The resistance of the current path between the voltage contacts is then just the measured potential difference divided by the constant current. The four-terminal technique is used, as opposed to the two-terminal technique common in multimeters, to eliminate contributions to the resistance by the leads, which run several meters across the lab, and contact resistances between the wires, ohmics and the 2DEG. Note that the two-terminal technique is actually the four-terminal technique in the limit where the potential difference is measured across the entire current path. The lead and contact contributions can add up to as much as 500Ω (compared to the sample resistance, which can typically lie anywhere between 50Ω (unbiased billiard) and 40kΩ) and can vary during the course of the experiment, obscuring the real resistance behaviour of the sample. Whilst the intention is to make a measurement of the DC resistance across the billiard, this is not possible using DC techniques because the requirement for low currents/biases to avoid heating means that noise becomes a serious problem. Instead, a low frequency (<100Hz) AC technique involving phase sensitive detection is required. Providing that the period of the AC excitation (>0.01s) is much greater than any of the characteristic times for electron transport in the billiard (~μs or below) the AC technique effectively behaves as a DC measurement. The potential difference between the voltage probes is measured using a digital signal processing lock-in amplifier (Stanford Research Systems SRS530). Phase reference for the measurement is provided via the current, which is established by placing the lock-in internal oscillator phase reference output (0-1V sine wave at the reference frequency) across a MΩ-range ballast resistor. The resulting current is in the nA-range and is tuned to order of magnitude by the resistor value (0.1,1,10,100MΩ) and exact value by adjusting the phase reference output voltage. Reference frequencies of 13, 17 and 78Hz have been used during these experiments and are chosen to avoid the noisy 50Hz harmonics. Current calibration is performed by substituting the sample with a 1000Ω standard resistor, hence providing calibration of the final resistance measurement. The current is passed into the sample via the source contact ($I^+$) and exits via the current drain ($I^-$), which acts as the earth for the measurement circuit and is connected to the lock-in ground line, which in turn serves as earth for the entire electronics set-up. All leads connecting the splitter box to the various instruments are coaxial cables twisted into pairs with their outer shields earthed [152]. The wiring gantry extending from the splitter box, which is mounted near the measurement/control instruments, across the lab to the top of the cryostat, is composed of 18 wire double shielded cables. Wires internal to the cryostat form twisted pairs and are shielded by the cryostat itself, which is grounded to the lock-in earth via the gantry/coaxial cable shields [152]. Care is taken to ensure that the cryostat is attached to no other ground (e.g. via the pumps, gas handling system or cryostat support). The electronics set-up used in these experiments provides a sufficiently good signal-to-noise ratio that the use of screened rooms or current/voltage pre-amplifiers is not required. The electrical supply to the entire measurement set-up is passed though a power filter/surge protection, followed by an isolation transformer and a second stage of power filtering, before being supplied to any of the instruments. In high field operations ($B$ > 2T) the magnet supply must be directly connected to mains power to avoid overloading the power-filtering equipment. This has little effect on the observed signal-to-noise ratio. Lastly, all lines aside from the dilution unit control/diagnostics lines pass through a set of LC filters (Fig. 4.5(b)) that are mounted on the $^4$He pot. The circuit diagram for these LC filters is shown in Fig. 4.7(b) and their construction is discussed further in Appendix A. Electrically, these filters are immediately adjacent to the respective components/sample contacts as shown in Fig. 4.7(a). The purpose of these LC filters is to minimise radio-frequency (RF) noise, picked up by the external wiring, from entering the cryostat and leading to heating of electrons in the sample. The values of the components are chosen (1mH/2.2nF) to provide zero attenuation in the measurement frequency range (<100Hz) but high attenuation of frequencies in the RF range and above (>500kHz).

[13] For a discussion of longitudinal and transverse resistance measurements see §3.2.2 and §4.2.1.

Gate biases are produced by a computer controlled four-port DC voltage source (IOtech DAC488HR/4) capable of voltages between +1 and −10V. The voltage source shares a common ground with the lock-in ensuring biases are applied with respect to $I^-$. Each voltage source line passes through a RC (1MΩ/47nF) filter before entering the cryostat to minimise the presence of electrical noise on the gates. Surface- and back-gates need to be handled carefully to protect them from possible damage caused by sparking between gates. All gates are connected to $I^-$ when not in use and throughout the cool-down. When a change in the bias applied to a gate is required, this is done in a continuous fashion at no faster than ±0.2V/min. Surface-gates are taken no further than +0.7V in the positive bias direction to avoid crossing the Schottky barrier, which allows current leakage from the gate into the heterostructure. In the negative bias direction, surface-gates are taken no further than the bias required to pinch off the 2DEG for isolation surface-gates, and no further than 40kΩ resistance in the case of gates forming QPCs. 'Back-of-the-wafer' Au back-gates were not used in any of the experiments discussed in this thesis. The FIBL back-gates used in the double-2DEG billiard samples were not taken further than –1.8V to avoid them becoming 'stuck' due to the charge-trapping behaviour of the FIBL-damaged regions at the edges of the back-gates at high (> –2.0V) negative biases [153]. FIBL back gates were taken no further than 0.5V in the positive bias direction to avoid noise caused by excessive current leakage from the back-gates.

Sample temperature control is provided separate to the dilution unit thermometry and control systems. A pair of $RuO_2$ thick film resistors are used for sample thermometry. The first is mounted on the sample wires just below the mixing chamber while the second is mounted on the sample wires at the end of the cold finger just before the wires connect to the sample holder. Note that the sample wiring is the only thermal path between the mixing chamber and the sample. The upper $RuO_2$ resistor is measured in a four-terminal configuration using an AC resistance bridge (RV-Electronikka AVS-47). The resistance-bridge is connected to a heater controller (RV-Electronikka TS-530) that drives a 100Ω heater coil mounted in thermal contact with the sample wires. This system is driven in a feed-back loop, with the heater output adjusting itself to maintain a constant temperature as measured by the $RuO_2$ resistor. The reason for driving the heater via the upper $RuO_2$ resistor rather than the lower one, which is likely to be more representative of the sample temperature, is to provide temperature stability during magnetic field sweeps. This is because the upper $RuO_2$ resistor is mounted well outside the solenoid bore and not only does it experience significantly lower magnetic fields, but $dB/dt$ is also lower, minimising the effect of eddy current heating on the heater output. The lower $RuO_2$ resistor is provided to give some indication of the sample lattice temperature compared to that at the mixing chamber since these can be significantly different (up to 15%). Control of the magnet is provided by an Oxford Instruments IPS-120 power supply under computer control. A pair of red LEDs are mounted on optical fibres that provide optical access to each sample independently. Samples are cooled to base temperature in the dark and these LEDs are used to provide short bursts of light for increasing the electron density of the 2DEG in single-2DEG samples via the persistent photo-conductivity effect [133]. Illumination is not used on double-2DEG samples as it only increases the electron density of the upper 2DEG, leading to an undesirable electron density mismatch between 2DEGs.

The lock-in amplifier, DC voltage source and the magnet power supply are computer interfaced via general purpose interface bus (GPIB). The sample thermometry/heating is operated under manual control because the feed-back loop can break into large oscillations if not carefully adjusted. To prevent noise generated by the computer appearing in the measuring circuit, the computer is electrically isolated from the GPIB by an optical fibre isolation system and the computer operates on a separate power supply and earth line to the measurement instruments. Data acquisition and control is performed using a set of Labview programs I wrote specifically for use in billiard experiments. This set includes a program for measuring resistance as a function of gate bias using up to four bias sources in a single sweep, another for measuring resistance as a function of applied magnetic field with preset gate biases and a control program which allows multiple magnetic field sweeps to be programmed for overnight operation.

## 4.2 – Characterisation Techniques

The study of low-dimensional electron systems has been pursued since the 1960s and has led to a number of significant discoveries along the way. However, effects that were once cutting-edge research are now commonly observed in the research devices of today. These effects have been thoroughly investigated and now serve as tools for understanding the new effects that are being observed in today's research. This is particularly the case with investigations of semiconductor billiards, which are complicated systems with a number of variable parameters that need to be known prior to any attempt to understand the effects observed in these devices. This section discusses techniques for measuring five important parameters in billiard experiments.

These are the Fermi energy $E_F$, the elastic mean free path $l_{el}$, the billiard area $A_B$, the number of conducting modes in the QPCs $n$, and the quantum lifetime $\tau_Q$. Both the Fermi energy and the elastic mean free path are properties of the 2DEG rather than the billiard *per se*, and are directly related to the electron density $n_s$ and mobility $\mu$ of the 2DEG respectively.

## 4.2.1 – Fermi Energy and Elastic Mean Free Path

These parameters are generally measured at the beginning of the experiment, prior to biasing any surface-gates. The relationship between the Fermi energy and the 2DEG electron density follows from the 2DEG density of states $\rho_{2D}(k) = (2\pi^2)^{-1}$ [154] and the fact that all $k$ states up to the Fermi wavevector $k_F$ are occupied. The number of electrons per unit area (i.e. 2DEG electron density $n_s$) is then given by the density of possible electron states $\rho_{2D}(k)$ multiplied by the **k**-space area of the Fermi circle $\pi k_F^2$. Rearranging this gives $k_F = (2\pi n_s)^{1/2}$ which, via the free electron dispersion relation $E = \hbar^2 k^2/2m^*$, results in the Fermi energy:

$$E_F = \frac{\pi\hbar^2 n_s}{m^*} \tag{4.1}$$

It is important to note at this point that the 2DEG is not necessarily homogeneous. The 2DEG electron density $n_s$ can vary at different locations of the 2DEG, and this is particularly the case in the billiard, where stray potentials from the surface-gates can lead to a slight depletion of the 2DEG within the billiard. The precise electron density within the billiard $n_{billiard}$ is difficult to measure however, and generally this is assumed to be equal to the electron density of the bulk 2DEG. The electron density of the bulk 2DEG is generally measured using the Shubnikov-de Haas effect discussed below.

The application of a magnetic field $B$ perpendicular to the plane of the 2DEG causes the electrons to follow circular paths with radius $r_{cyc} = \hbar k_F/eB$. For magnetic fields sufficiently strong that $\hbar\omega_c >> kT$ and $\omega_c\tau_Q >> 1$, harmonic oscillator quantisation of the cyclotron orbit occurs. The density of states then condenses into a discrete set of energy levels called Landau levels given in the ideal case by [55]:

$$\rho_{2D-LL}(E) = \frac{2eB}{h}\sum_n \delta(E - E_n) \quad \text{where} \quad E_n = \left(n + \frac{1}{2}\right)\hbar\omega_c \tag{4.2}$$

and $n$ is an integer called the Landau level index. As the magnetic field $B$ is increased, successive Landau levels rise above the Fermi energy and depopulate, leading to oscillatory structure that is periodic in $1/B$ in the longitudinal resistance $R_{xx}$ (see Fig. 4.8(inset)). This is known as the Shubnikov-de Haas effect [154] and is shown for a typical sample in Fig. 4.8. The period of these oscillations is given by:

$$\Delta\left(\frac{1}{B}\right) = \frac{g_{LL}e}{hn_s} \tag{4.3}$$

where $g_{LL}$ is the Landau level degeneracy which equals either 1 or 2 depending on whether or not $B$ is sufficient for Zeeman splitting to occur.

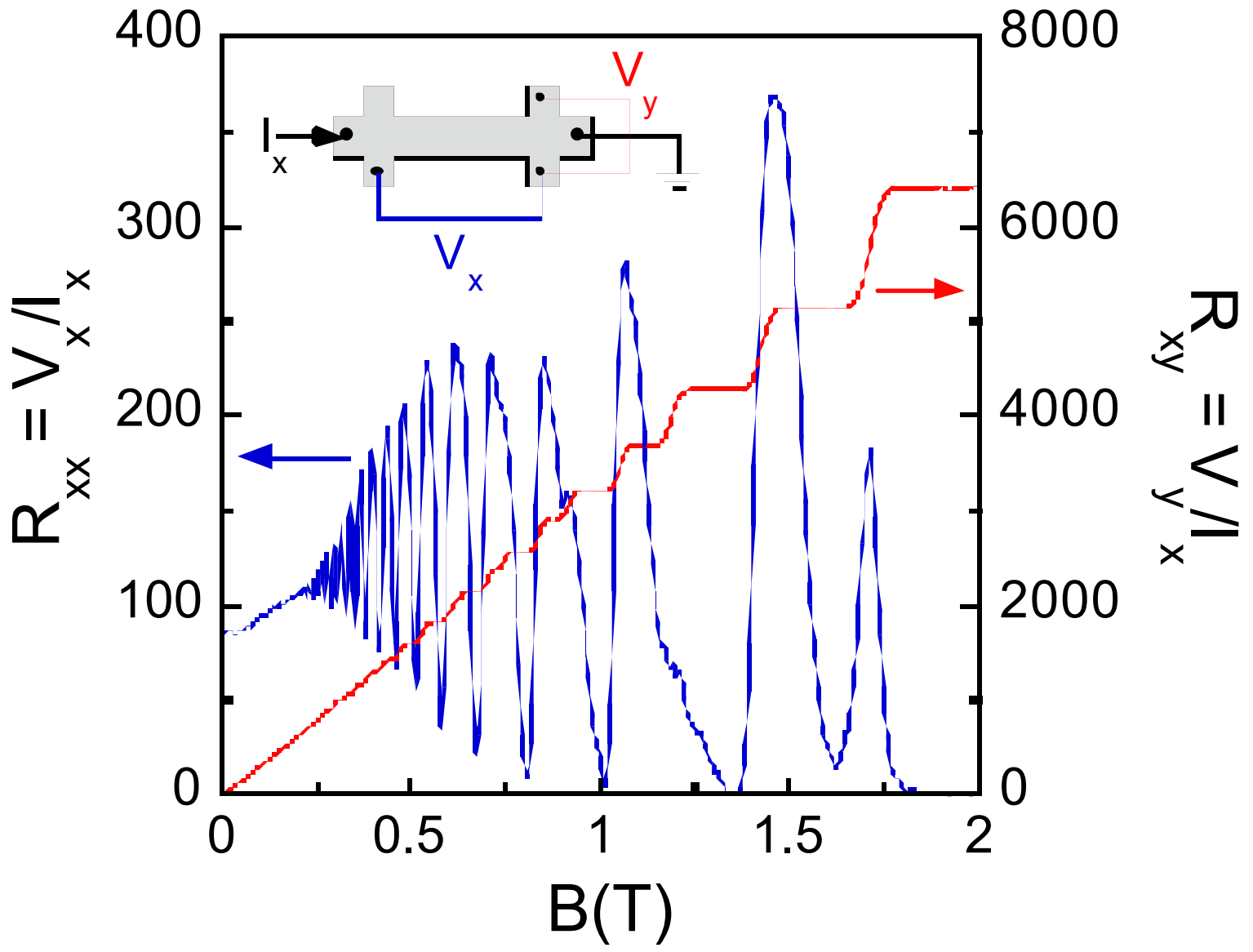

Figure 4.8: Shubnikov-de Haas effect (blue) and quantum Hall effect (red) measured as part of the characterisation of one of the samples investigated in this thesis. Inset is a schematic of a Hall-bar (see §3.2.2 for details) illustrating, for a current along the Hall-bar, the longitudinal resistance $R_{xx} = V_x/I_x$ and the transverse resistance $R_{xy} = V_y/I_x$ with the potential difference measured along and across the Hall-bar respectively.

It is of course also possible to measure $n_s$ using the classical Hall effect [55] where $n_s = B/eR_{xy}$ and $R_{xy}$ is the transverse resistance (see Fig. 4.8(inset)). Note also that Landau level quantisation is also responsible for the quantum Hall effect, which produces transverse resistance plateaux at resistances of $h/(\nu e^2)$ where $\nu$ is the Landau level filling factor, as shown in Fig. 4.8. An in-depth discussion of the quantum Hall effect may be found in [155].

The elastic mean free path can be obtained from the mobility via the Drude model for electrical conduction [55]. In this model, the mobility is related to the average time between impurity scattering events (i.e. the scattering relaxation time $\tau$) by $\mu = e\tau/m^*$. The elastic mean free path is directly related to the scattering relaxation time via $l_{el} = v_F\tau$, giving:

$$l_{el} = \frac{m^* v_F \mu}{e} \tag{4.4}$$

where $v_F$ is the Fermi velocity, which can be determined from $E_F$ via $v_F = (2E_F/m^*)^{1/2}$. The electron mobility of a 2DEG is given by [156]:

$$\mu = \frac{1}{en_s R_{xx}(B=0)}\left(\frac{L}{W}\right) \tag{4.5}$$

where $(L/W)$ is the aspect (length:width) ratio of the 2DEG, and $R_{xx}(B = 0)$ is the zero field longitudinal resistance.

## 4.2.2 – Characterisation of the Entrance and Exit QPCs

Figure 4.9 shows a schematic diagram of the surface-gates used to define a QPC. Application of a negative bias $V_g$ to the surface-gates leads to depletion in the 2DEG regions below the surface-gates. Due to the vertical separation between the surface-gates and the 2DEG however, the depletion regions in the 2DEG are larger than the surface-gates, as indicated by the lines around the surface-gates in Fig. 4.9. Areas of the depletion region that extend beyond the vertical projection of the surface-gate pattern onto the 2DEG form the depletion edge, which has a width $W_d$ that is dependent upon $V_g$. Hence the surface-gates, which lithographically define an opening of width $W_l$, form an opening in the 2DEG of variable width $W = W_l - 2W_d$, which depends upon $V_g$. Note however, that $W$ does not necessarily behave linearly with $V_g$. Providing that both $W$ and $L$ are comparable to $\lambda_F$ and much smaller than $l_{el}$, the channel thus formed is a variable width QPC.

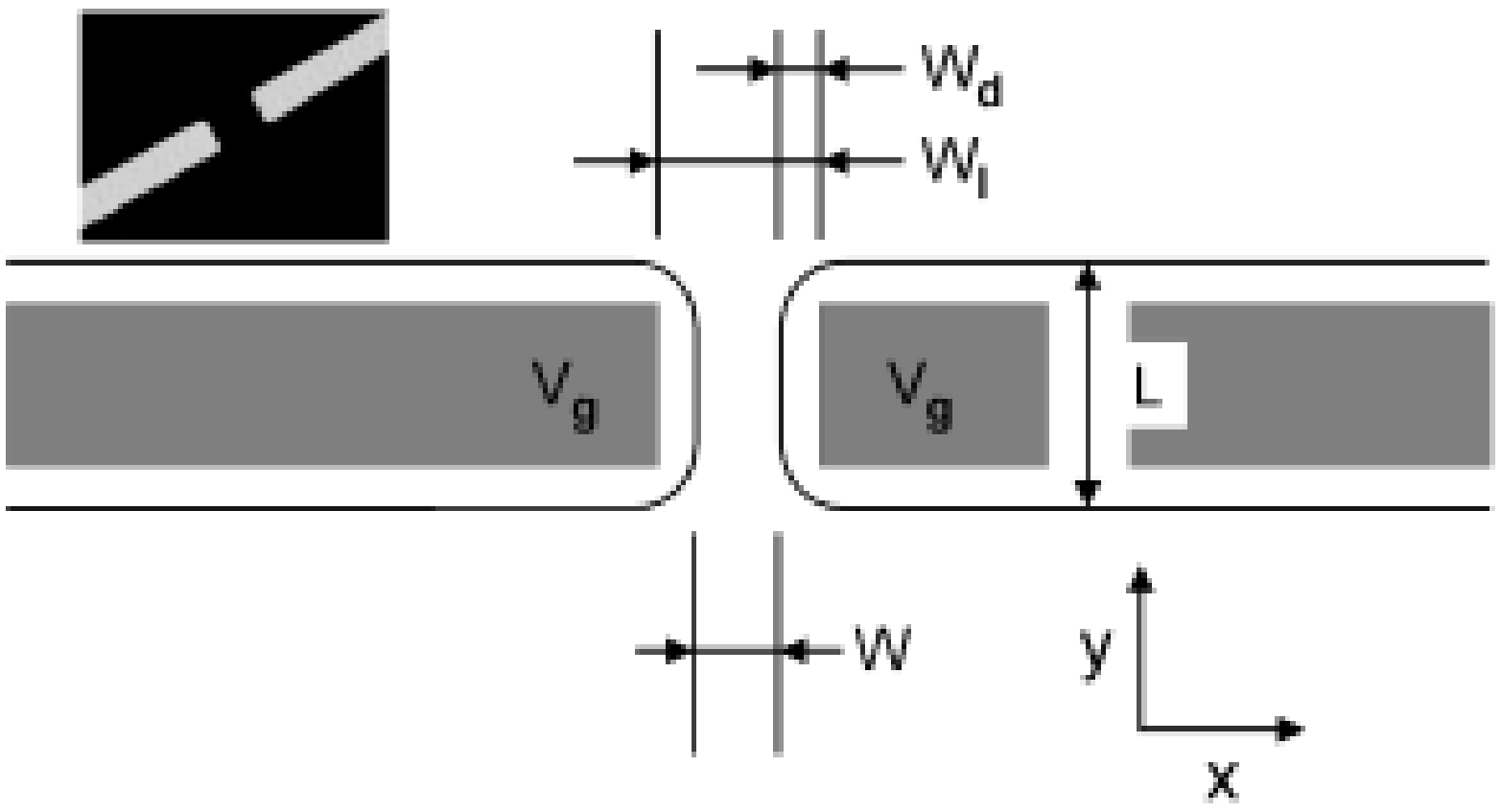

Figure 4.9: Schematic of surface-gates used to define a QPC. With a negative bias $V_g$ applied to the surface-gates, a QPC of width $W$ and length $L$ is formed in the 2DEG. The inset shows a scanning electron micrograph of surface-gates used to define a QPC.

Along the $x$-direction in Fig. 4.9, the QPC may be approximated as a 1D infinite square potential well of width $W$. Note that the electrons remain free (non-quantised) in the $y$-direction. In such a potential well, the energy is quantised into discrete energy levels. These energy levels satisfy the condition that an integer number $n$ of half wavelengths $\lambda = h/(2m^*E)^{½}$ is equal to $W$. At low temperatures, thermal smearing is minimal and electrons flowing through the QPC will have an energy very close to $E_F$. Hence the condition for the discrete energy levels becomes:

$$\frac{\lambda_F}{2} n = W \tag{4.6}$$

where $n$ is known as the mode number. Using $k = 2\pi/\lambda$ and $E = \hbar^2k^2/2m^*$, the discrete QPC energy levels (called 1D-subbands) are then:

$$E_n = \frac{\hbar^2 \pi^2 n^2}{2m^* W^2} = \frac{\hbar^2 k_x^2}{2m^*} \tag{4.7}$$

Assuming spin degeneracy, the one-dimensional density of states $\rho_{1D}$ and group velocity $v_g$ are dependent upon $k_x$ and hence $n$:

$$\rho_{1D} = \frac{1}{\pi} \frac{m^*}{\hbar^2 k_x} \qquad v_g = \frac{\hbar k_x}{m^*} \tag{4.8}$$

The current carried by a particular mode $n$ for a given difference in chemical potential $\delta\mu$ (corresponding to an electrical potential difference $V = \delta\mu/e$) is then $I_n = ev_g\rho_{1D}\delta\mu = (2e/h)\delta\mu$. Of particular note is the fact that $I_n$ is independent of mode-number. In other words, the current is divided equally between the modes. This means that the total current $I$ is given by the number of modes $N$ with energy less than $E_F$. The conductance through the QPC is then obtained as:

$$G = \frac{I}{V} = \frac{NI_n e}{\delta\mu} = \frac{2e^2}{h} N \tag{4.9}$$

Note that this means that the conductance through the QPC is quantised in integer multiples of $2e^2/h$. As $V_g$ is increased, the width of the depletion regions $W_d$ increases, narrowing the QPC. Since $E_n$ is proportional to $W^2$, the sub-band spacing increases as $W$ is reduced. Hence as $V_g$ is increased, each mode in turn rises above the Fermi energy, decreasing $N$ in integer steps and hence producing the $2e^2/h$ steps in conductance as a function of $V_g$, as shown in Fig. 4.10. This effect was first observed experimentally in 1988 by Wharam *et al.* [7] and independently by van Wees *et al.* [8]. Note that the step locations as a function of $V_g$ are non-trivial. This is due to the complicated relationship between $V_g$ and $W$ and the fact that $E_F$ within the QPC is dependent upon the electron density $n_{qpc}$ inside the QPC. The QPC electron density is dependent upon $V_g$ and usually does not match the bulk electron density in the 2DEG.

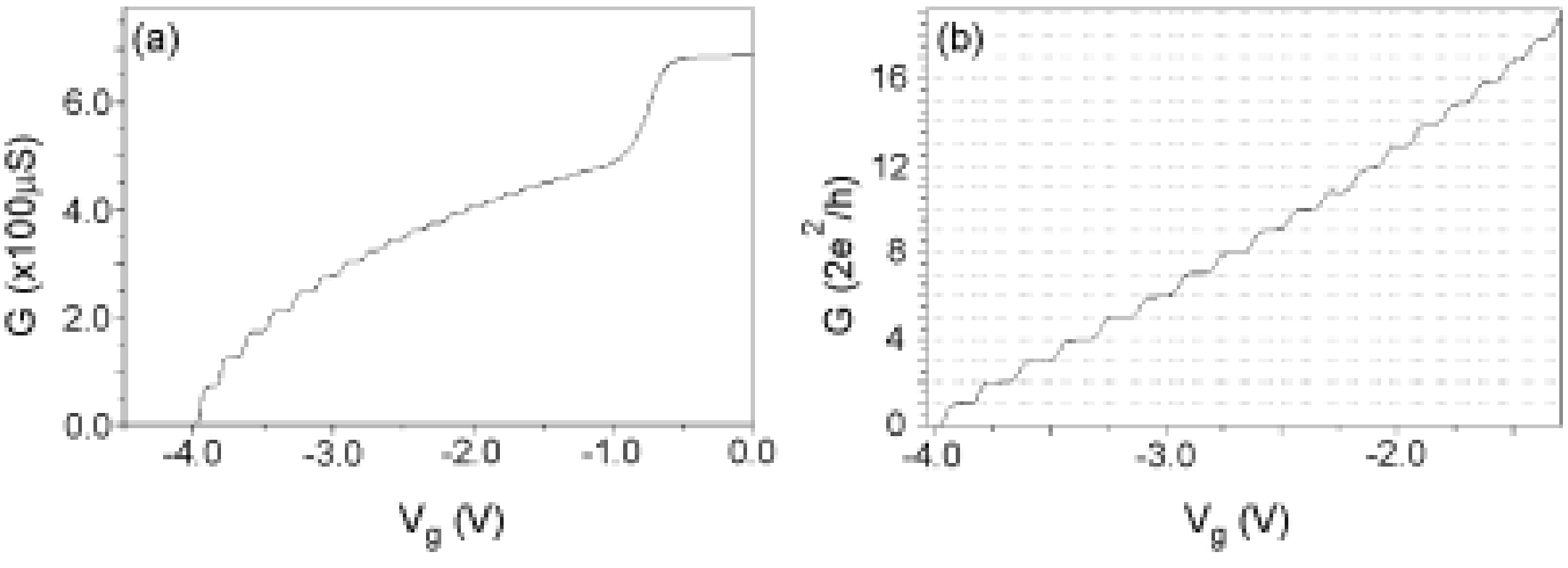

Figure 4.10: (a) Conductance $G$ through the QPC as a function of $V_g$ for the QPC device shown in Fig. 4.9 (inset) showing the presence of quantised steps. (b) Close-up of (a) for $-4.0V < V_g < -1.25$ demonstrating the quantisation of conductance in steps of $2e^2/h$. Data provided by R. Wirtz (University of Cambridge, U.K.) from [157].

The majority of devices investigated in this thesis are designed so that the QPCs can be biased independently. In this case, traces similar to those in Fig. 4.10 are obtained for both the entrance and exit QPCs at the beginning of the experiment to allow a calibration of the gate bias $V_g$ required to achieve a particular number of conducting modes in the QPCs. Whilst it is possible that fringing fields from other surface-gates forming the billiard will affect this calibration slightly, any such deviations are expected to be small. In the case where the QPCs are unable to be biased independently, $V_g$ is either set based on plateaus observed for the QPCs in series or based on the total resistance of the billiard.

## 4.2.3 – The Quantum Lifetime

An important parameter in Chapter 7 is the quantum lifetime $\tau_Q$, which determines the total broadening of the billiard energy levels discussed in §7.4. The quantum lifetime is determined using a correlation field analysis of MCF obtained in the skipping orbit regime developed by Bird *et al.* [158,159]. In this regime, electron transport through the billiard occurs via skipping orbits, where the electron follows a circular path of radius $r_{cyc} = \hbar k_F/eB$ between specular reflections from the billiard walls, as discussed in §2.2.4. The derivation of $\tau_Q$ in [158,159] commences by considering an 'average' skipping orbit along the billiard wall as shown in Fig. 4.11.

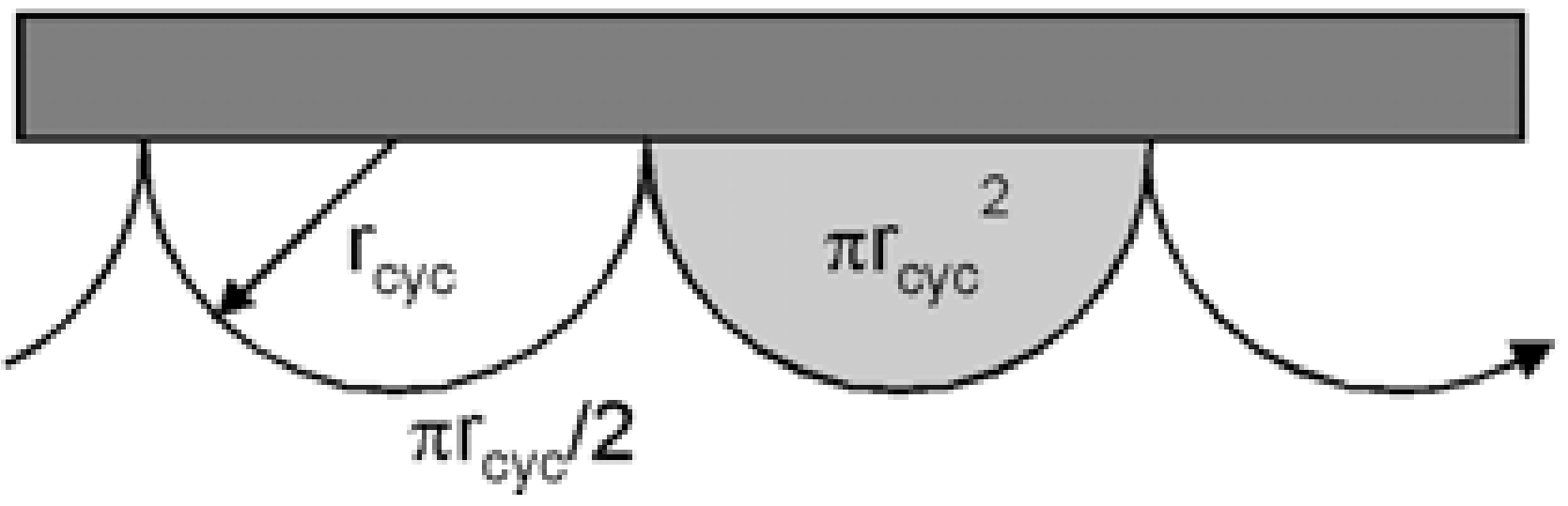

Figure 4.11: Schematic illustrating the 'average' skipping orbit discussed in the derivation of the correlation field analysis method for obtaining $\tau_Q$. Each skip has a radius of $r_{cyc}$, a length of ½$\pi r_{cyc}$ and encloses and area of $\pi r_{cyc}^2$ as shown.

The area enclosed between different orbits $A(B)$ is directly related to the area enclosed between the 'average' orbit and the billiard wall $A_c(B)$ [159]. It is expected that $A(B) = \kappa A_c(B)$, where $\kappa$ is a constant close to 1 which is related to the overlap of the different orbits [158]. Note that both of these areas shrink as the magnetic field is increased because of the inverse relationship between $r_{cyc}$ and $B$. The latter area $A_c(B)$ is given by $A_c(B) = ½(N\pi r_{cyc}^2)$, where $N$ is the average number of reflections from the billiard wall before the electron no longer contributes to quantum interference processes. It is important to note that at this point I have diverged slightly from the original derivation of Bird *et al.* [158,159] wherein $N$ is the average number of reflections before the electron loses phase coherence. To avoid interrupting the derivation, the significance of this difference in the definition of $N$ is discussed further in the next paragraph. The length that the electron travels along this 'average' skipping orbit prior to ceasing to contribute to quantum interference is just $l_Q = N\pi r_{cyc} = v_F\tau_Q$. Combining $A_c$ and $l_Q$ and using $v_F = \hbar k_F/m^*$ yields:

$$A_c = \frac{1}{2}\left(v_F \tau_Q r_{cyc}\right) = \frac{\hbar^2 k_F^2}{2m^* eB}\tau_Q \tag{4.10}$$

Bird *et al.* [159] define the correlation field $B_c(B)$ of the MCF using the correlation function $F(B,\Delta B) = \langle (G(B) - \langle G(B)\rangle_B)(G(B+\Delta B) - \langle G(B)\rangle_B)\rangle_B$, where $\langle\ \rangle_B$ indicates an average over some suitable range centred on $B$. The correlation field is then defined as the half-width of the correlation function $F(B_c) = F(0)/2$. The correlation field is essentially an 'average' period of the MCF and in the skipping orbit regime can be related to $A_c$ via the Aharonov-Bohm relation $B_c(B) \cong h/(eA_c(B))$. Bird *et al.* [158] note that the Aharonov-Bohm relation in this case is expected to be equal to within a small numerical constant. Using Eqn. 4.10 gives:

$$B_c(B) \cong \frac{4\pi m^* B}{\hbar k_F^2 \tau_Q} \tag{4.11}$$

$\tau_Q$ is then obtained by a linear fit of $B_c(B)$ versus $B$ in the skipping orbit regime as described in [158,159].

Returning to the discussion of the previous paragraph, Bird *et al.* assume that the electron continues to accumulate area (that contributes to quantum interference effects) until it loses phase coherence via an inelastic scattering event. In doing this, Bird *et al.* must assume either that the characteristic dwell-time $\tau_D$ of the electron is longer than the phase coherence time, or that the electron can continue to contribute to quantum interference effects once it has left the billiard. In contrast, the definition that I have taken requires neither assumption, as it takes into account the fact that the electron can cease to contribute to quantum interference effects by exiting the billiard. However, instead of giving a phase coherence time as in [158,159] it instead gives a combination of the phase coherence time and the dwell-time via the addition of rates:

$$\frac{1}{\tau_Q} = \frac{1}{\tau_\phi} + \frac{1}{\tau_d} \tag{4.12}$$

Hence in terms of the level broadening discussed in §7.4, the quantum lifetime broadening $\Delta E_Q = \hbar/\tau_Q$ takes into account both the broadening due to finite phase coherence and finite dwell-time of electrons within the billiard.

## 4.2.4 – The Billiard Area

The area of the billiard in the 2DEG may be determined using the edge-state Aharonov-Bohm effect [160,161]. As discussed in §2.2.4, as the magnetic field is increased electron transport begins to take place via skipping orbits along the walls of the billiard. Once the magnetic field becomes large enough for $\omega_c \tau_Q >> 1$ and $\hbar\omega_c >> kT$ to hold, Landau level quantisation imposes a restriction upon the possible skipping orbits [162] to form a discrete set of edge-states [22,56]. Each edge-state is associated with a different skipping orbit along the wall although for each skipping orbit the radius of curvature $r_{cyc} = \hbar k_F/eB$ is the same, as shown in Fig. 4.12. Note that each skipping orbit is actually a cycloid and has an associated guiding centre path, which is the path of the centre of a rolling circle such that a point on the circumference of the circle traces out the cycloid as the circle is rolled. The distance between the guiding centre and the billiard wall is different for the various edge-states, as shown in Fig. 4.12. As the magnetic field is increased, successive Landau levels rise above the Fermi energy and depopulate, reducing the number of edge-states travelling along the walls of the billiard. Note that edge-states on opposite walls of the billiard travel in opposite directions around the billiard due to the action of the Lorentz force, which causes them to circulate around the curved path in a single direction (either clockwise or anticlockwise).

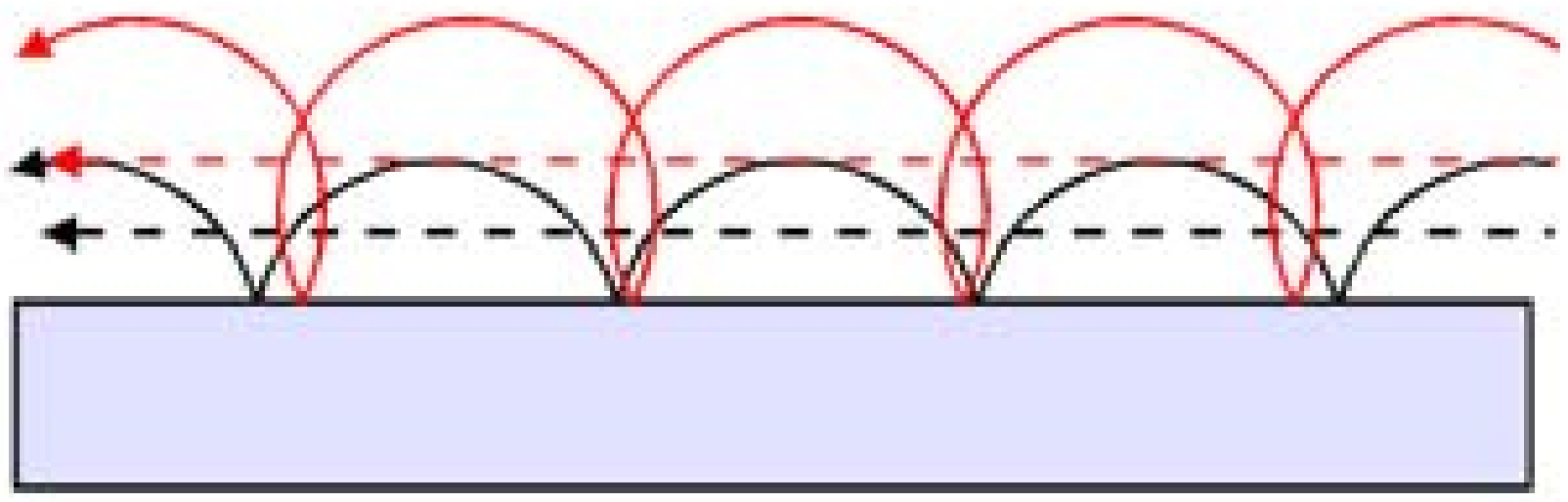

Figure 4.12: Schematic illustrating a pair of edge-states propagating along a wall. Both edge-states have the same radius of curvature $r_{cyc}$. The dashed lines are the guiding centre paths for the two edge-states.

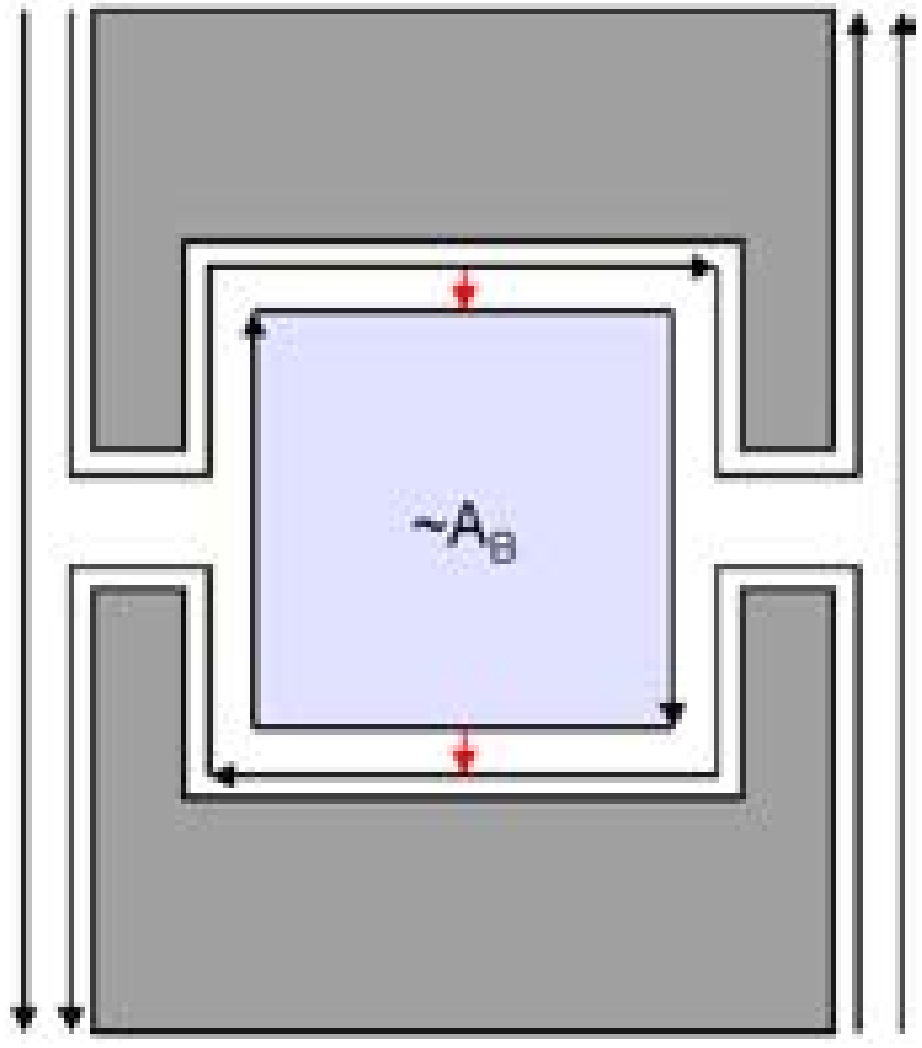

Figure 4.13: Schematic illustrating the edge-state Aharanov-Bohm effect in a billiard. The red arrows indicate tunnelling between the edge-states. Note that in reality, the edge-states are much closer to the billiard walls.

It is possible for electrons to tunnel between edge-states on opposite walls of the billiard via edge-states that are trapped inside the billiard, as shown in Fig. 4.13. In this way, a loop is formed that generates magneto-conductance oscillations via the Aharonov-Bohm effect discussed in §2.1.4. This loop encloses an area very close to that of the 2DEG since the edge-states travel close to the billiard walls (closer than it appears in the schematic of Fig. 4.13). Hence the period $\Delta B$ of these magneto-conductance oscillations can be used to determine the billiard area $A_B$ via the Aharonov-Bohm relation $\Delta B = h/(eA_B)$. Note that unlike at low fields where fluctuations due to a large number of magneto-conductance oscillations are observed (see §2.2.3), in the edge-state regime, there are very few Aharonov-Bohm loops formed and the magneto-conductance oscillation is usually quite clear as a single periodic oscillation in the magneto-conductance. Further discussion of the edge-state Aharonov-Bohm effect may be found in [160,161].

# Chapter 5 – Exact Self-similarity in Billiard MCF

The subject of this Chapter is the experiment performed on the NRC Sinai billiard device by R.P. Taylor and R. Newbury at the National Research Council in Ottawa, Canada in October 1995. Their initial report of this experiment [13] presented the first observation of self-similarity and fractal behaviour in the magneto-conductance fluctuations (MCF) of a semiconductor billiard, and the only observation of exact self-similarity in the magneto-conductance of a semiconductor billiard to date. This Chapter discusses my analysis of the data produced in this experiment.

This Chapter commences with a discussion of the NRC Sinai billiard experiment in §5.1. §5.2 presents the experimental data obtained in this experiment and an initial analysis of the data, which led to the discovery of exact self-similarity in the magneto-conductance discussed in §5.3. A correlation function approach for quantifying the presence of exact self-similarity and identifying the scaling factors that relate the self-similar levels is also presented in §5.3. The observation of exact self-similarity is examined further in §5.4 with a detailed discussion of the properties, observation limits and physical dependencies of this behaviour. §5.5 presents a model of the exact self-similarity based on the Weierstrass function – a well-known generator of self-similar structure. This model generates structure remarkably similar to that observed in the experiment. The effect of the transition between the Sinai and square billiard geometry on the exact self-similarity is discussed in §5.6. The conclusions are presented in §5.7.

## 5.1 – The NRC Sinai Billiard Experiment

Fabrication of the NRC Sinai billiard used in the experiment presented in this Chapter is discussed in Chapter 3. Figure 5.1(a) shows EBL-defined surface-gates, which form a 1μm × 1μm square with a 0.3μm diameter circular gate located at the centre of the square. Entrance and exit leads that are 0.2μm wide are situated in the lower left-hand corner of the billiard as shown in the schematic in Fig. 5.1(b). As discussed in §2.1.2, application of a negative bias to the surface-gates leads to electrostatic depletion of the 2DEG in regions directly below the surface-gates, generating 2DEG depletion regions closely matching the surface-gate pattern. Note that while there are actually three openings in the square, the opening in the upper right corner of Fig. 5.1(b) is sufficiently narrow (50nm) that it remains depleted throughout the experiment. The purpose of this opening is to allow gates 1 and 2 to be biased independently. In this particular device, however, this opening did not form correctly, shorting gates 1 and 2 together. Throughout the experiment, gates 1, 2 and 3 (referred to as the 'outer' gates from here onwards) were maintained at a common bias of $V_O$. In the 2DEG, the width of the QPC entrance and exit ports, and the area inside the square is dependent on $V_O$ due to the depletion-edge surrounding the surface-gate pattern (see Fig. 5.2(a) and §4.2.2). The 'inner' circular gate is connected via the bridging interconnect (shown in white in Fig. 5.1(a)) as discussed in §3.5. The bridging interconnect allows the inner gate to be biased independently of the outer gates at a voltage $V_I$. Characterisation of other devices demonstrates that the bridging interconnect itself causes no perturbation in the billiard potential for the range of $V_I$ used in this experiment [163].

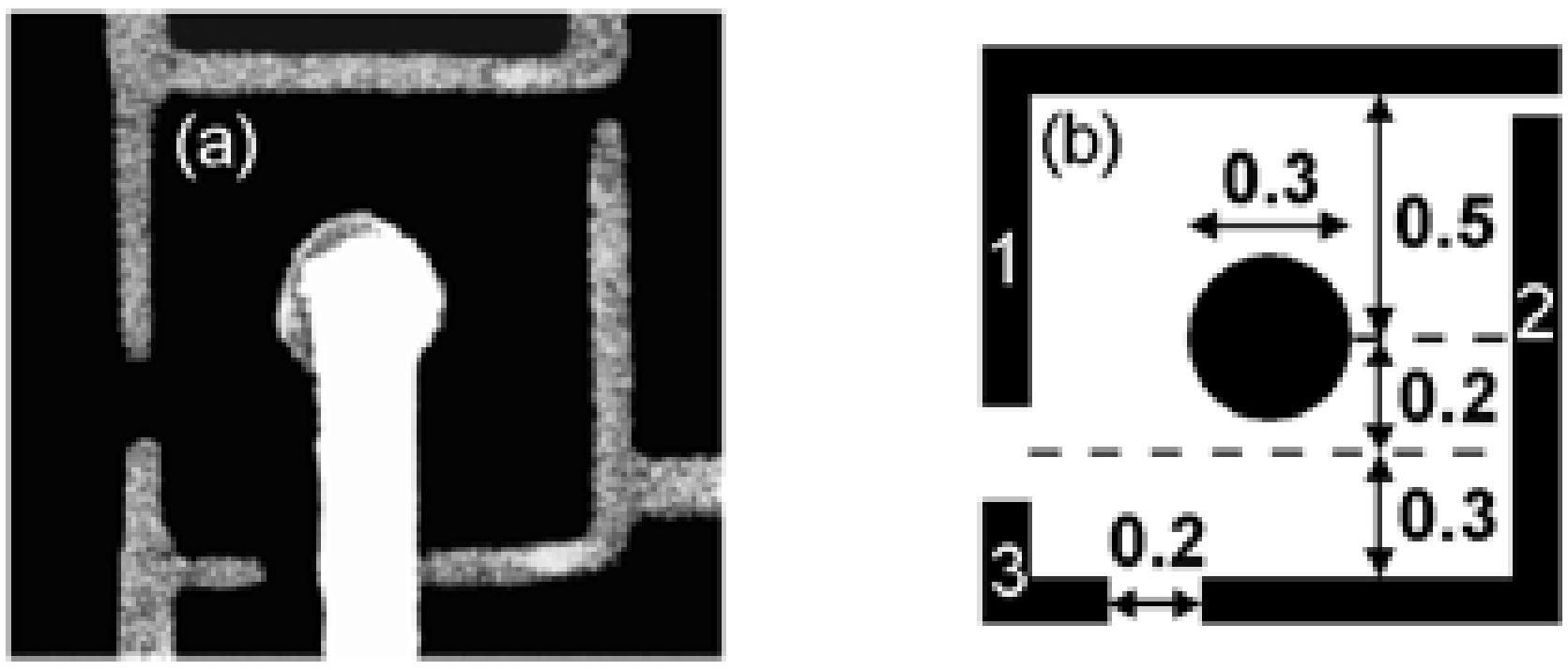

Figure 5.1: (a) Scanning electron micrograph and (b) schematic of the NRC Sinai billiard discussed in this Chapter. Numbers on the gates in (b) are referred to in the text.

The device was cooled to millikelvin temperatures in the dark with all surface-gates grounded. Prior to the electrical measurements, the sample (with the surface-gates still grounded) was illuminated in short bursts using a red LED, to increase the 2DEG electron density via the persistent photo-conductivity effect [133]. Post-illumination characterisation of the 2DEG gave an electron density of $2.3\times10^{15}m^{-2}$ and an electron mobility of $316m^2/Vs$, corresponding to an elastic mean free path of $l_{el} \sim 25\mu m$ and a Fermi wavelength of $\lambda_F \sim 50nm$. Electrical measurements of this billiard were performed using techniques similar to those described in §4.1.2. Unless otherwise specified, billiard magneto-resistance measurements were performed using a four-terminal longitudinal measurement configuration in order to remove contributions from the wires leading to the sample and the ohmic contacts.

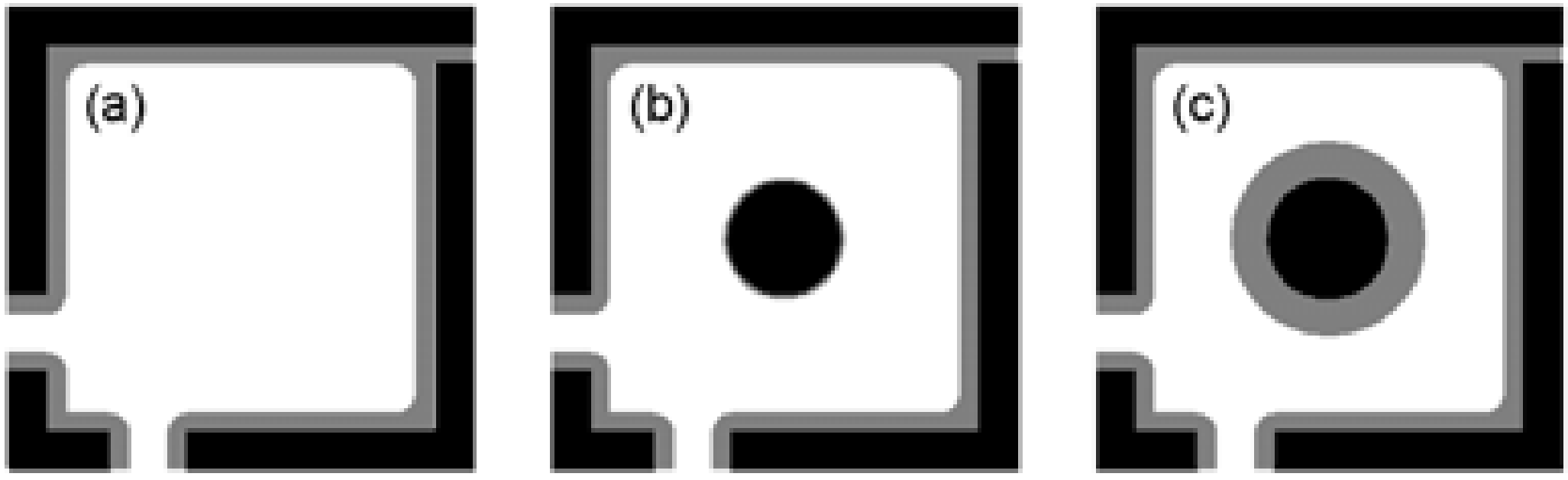

Figure 5.2: Effect of the applied inner gate bias. (a) At $V_I = +0.7V$ diffuser presence is minimised giving a square geometry. (b) At $V_I = 0V$ the region under the inner gate is fully depleted due to shadowing during illumination, giving a Sinai geometry. (c) For $-3.0V < V_I < 0V$ the depletion edge expands around the inner gate increasing the size of the Sinai diffuser.

The surface-gates in this device are opaque (at the LED wavelength), leading to shadowing of the heterostructure regions under the surface-gates during the illumination process. Shadowing during illumination causes the 2DEG under the surface-gates to be depleted even though there is no negative bias applied to the surface-gates [79]. This illumination-induced depletion pattern can be minimised by applying a positive bias (+0.7V) to the surface-gates [79]. This is important for the inner gate in particular, since for $V_I = +0.7V$ and $V_O < 0V$ the resulting billiard is a square (i.e. the circle is undefined) as shown in Fig. 5.2(a). As the positive bias on the inner gate is reduced the region under this gate becomes partially depleted, followed by full depletion at $V_I = 0V$ [13], changing the device geometry from a square to a Sinai billiard as shown in Fig. 5.2(b).

The radius of the Sinai diffuser increases as $V_I$ is made more negative, as shown in Fig. 5.2(c). For $V_I$ more negative than –3.3V, the amplitude of the MCF is observed to decrease. This is due to electron heating by current leakage from the surface-gates into the heterostructure. As a result, a maximum negative bias $V_I$ of –3.3V is used.

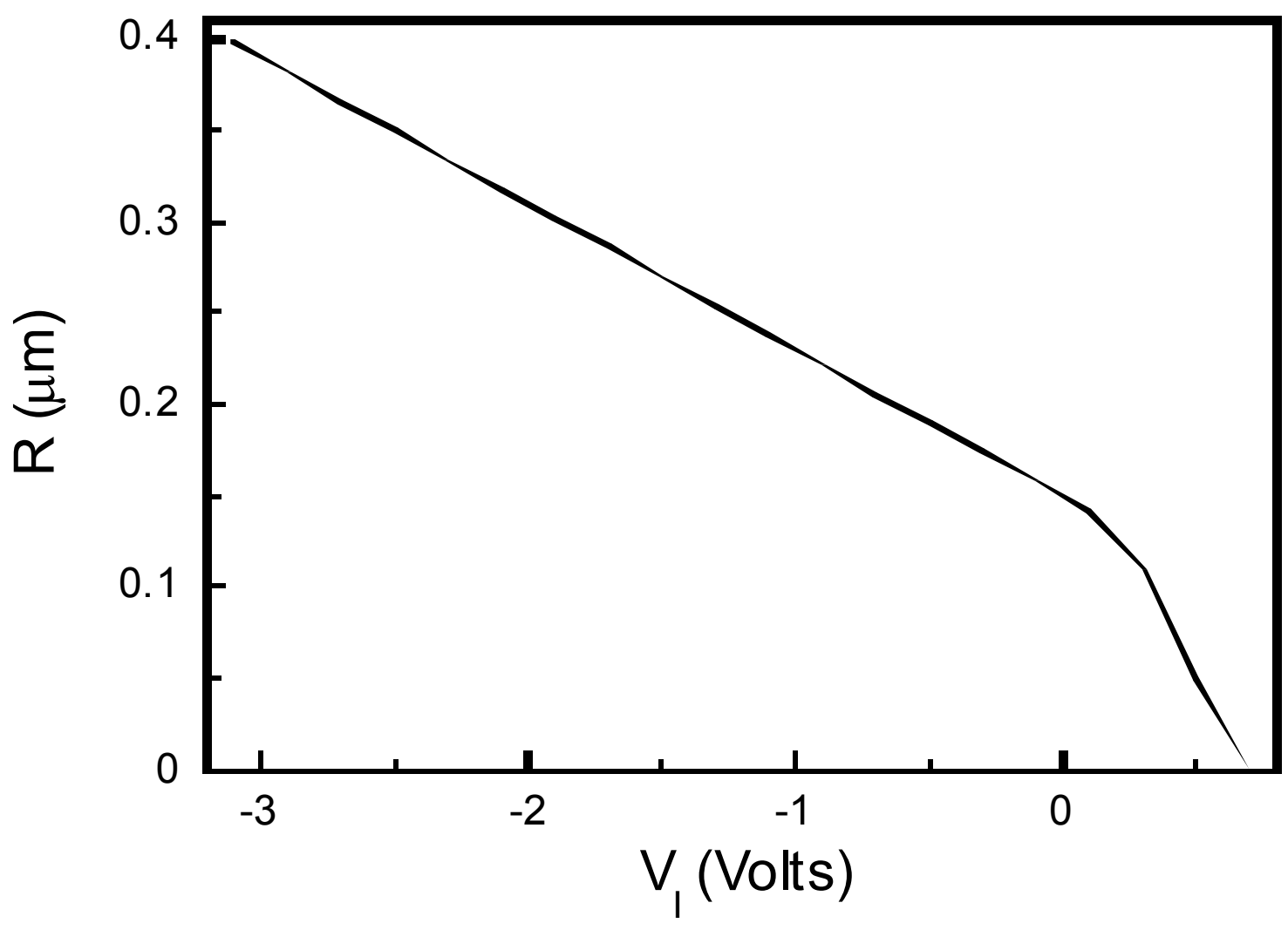

Figure 5.3: Expected values of Sinai diffuser radius $R$ versus $V_I$ for the Sinai billiard based on characterisation of previous devices in [163].

Characterisations of previous devices suggest that for $V_I < 0$V the radius of the Sinai diffuser changes linearly in increments of $\Delta R \sim +40$nm for $\Delta V_I \sim -0.5$V with $R = 0.15\mu$m at $V_I = 0$V as shown in Fig. 5.3 [163]. For $0\text{V} < V_I < +0.7$V the Sinai diffuser radius is expected to follow approximately linear behaviour, reaching $R = 0\mu$m at $V_I = +0.7$V. Outer gate biases of $V_O$ = –0.5, –0.51, –0.52 and –0.55V were used in this experiment. These biases correspond to 9,7,5 and 3 modes transmitted in the QPCs. As discussed in §2.2.3, the number of modes transmitted in the entrance QPC determines the injection properties of electrons entering the billiard, which in turn strongly affects electron transport through the billiard.

# 5.2 – Initial Investigation of the Experimental Data

In order to make a comprehensive investigation of the behaviour of the Sinai billiard and its transition to a square billiard, measurements of magneto-resistance across the billiard were obtained as a function of $V_I$ between +0.7V and –3.3V for each of the four $V_O$ values listed in §5.1. Measurements also were performed as a function of temperature $T$ between ~30mK and 4.2K. Classical and quantum contributions to the magneto-resistance can be distinguished using the temperature dependencies of the square ($V_I$ = +0.7V) and Sinai ($V_I$ = −3.0V) billiards shown in Figs. 5.4(a) and (b) respectively. At the highest temperature ($T$ = 3.3K, top traces), electron phase-breaking mechanisms are sufficiently strong that contributions to the magneto-resistance from quantum interference effects are negligible. The remaining features are of a purely classical origin and may be used to monitor the evolution of the electron trajectories as the Sinai diffuser is activated. The general characteristics of the 3.3K traces have been confirmed by a classical trajectory analysis used previously to successfully model circular billiards [164,165].

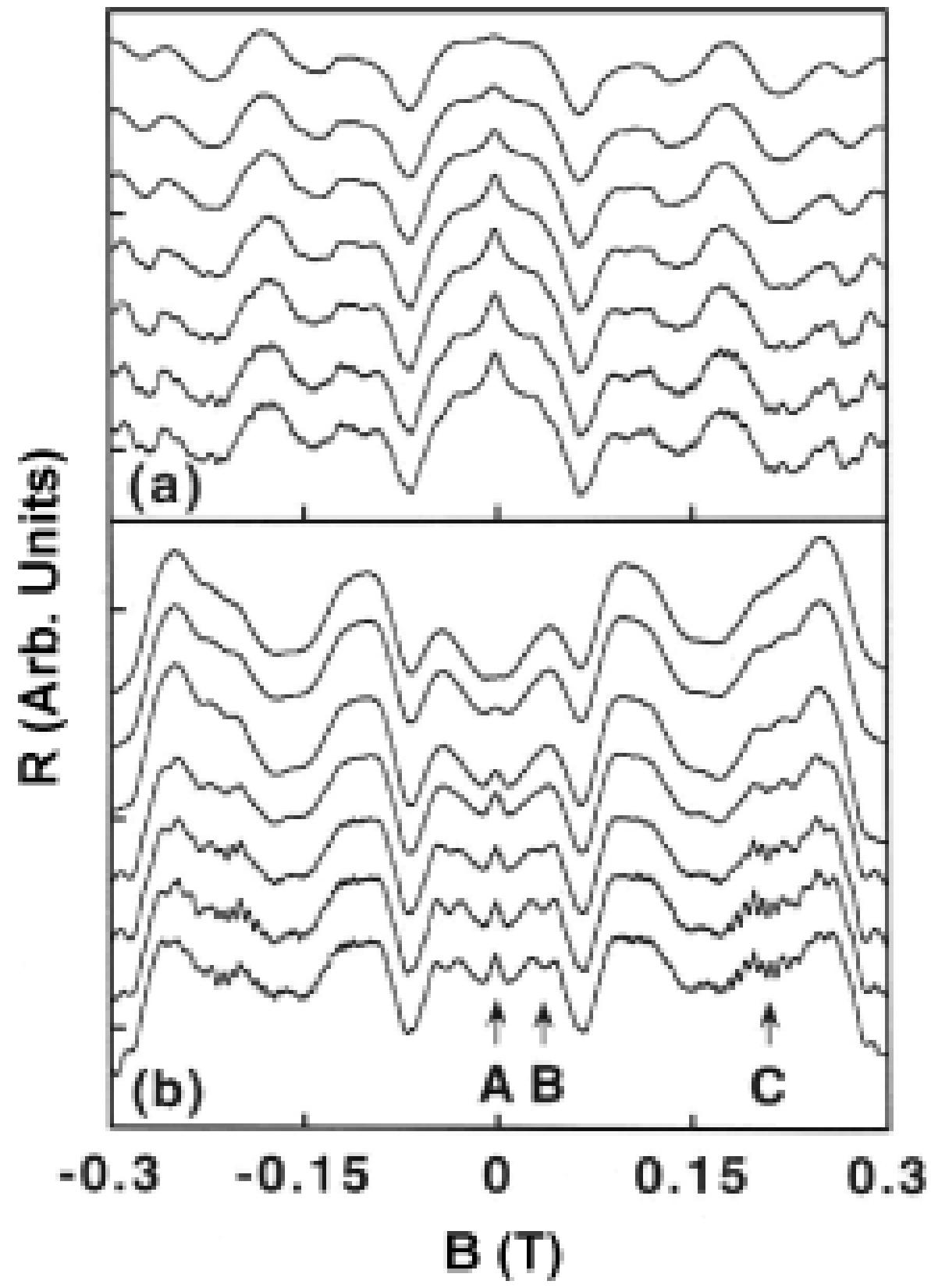

Figure 5.4: Magneto-resistance of the square (a) and the Sinai billiard (b) for temperatures of 30mK (bottom), 0.1, 0.4, 0.8, 1.6, 2.5 and 3.3K (top). $V_O$ = −0.52V. See text for an explanation of labels A, B and C.

In particular, for the square billiard there is a resistance maximum at zero magnetic field due to reflections from the far wall back into the entrance lead. In contrast, the Sinai billiard has a minimum at zero magnetic field due to electrons that are focussed into the exit lead by the Sinai diffuser. Features marked A, B and C in Fig. 5.4 are not observed in the classical analysis and display a temperature dependence sharper than that expected for classical features, indicating that they originate from quantum interference processes. The structure marked C matches Shubnikov-de Haas oscillations observed in the bulk 2DEG [154] and hence is not investigated further. Instead discussion will be restricted to features A and B, which are the resistance peak at zero magnetic field and the resistance fluctuations on the shoulders of this central peak, referred to as the 'shoulder features' hereafter, respectively. Figure 5.5 shows the evolution of the magneto-resistance structure as the device undergoes the transition from the square ($V_I$ = +0.7V - bottom) to the Sinai ($V_I$ = −2.9V - top) geometry in steps of $\Delta V_I$ = 0.4V for fixed $V_O$ at a lattice temperature of 30mK. In Fig. 5.5, features A and B are both observed to evolve as $V_I$ is made more negative. A rise in the background resistance of the magneto-resistance traces with decreasing $V_I$ is also apparent in Fig. 5.5 indicating the activation of the Sinai diffuser followed by a gradual increase in its radius. This rise in background resistance is due either to increased reflection of electrons back into the entrance port by the Sinai diffuser or narrowing of the circulating (roughly annular) channel around the Sinai diffuser by the expanding Sinai diffuser. However, the background resistance rise is not expected to be due to the QPCs pinching off due to stray potential from the Sinai diffuser [163].

To investigate features A and B in detail, it is necessary to examine magneto-conductance instead of magneto-resistance. Electron quantum interference effects lead to fluctuations in the transmission of electrons through the billiard as the magnetic field is changed. Following from the discussion of the Landauer formula in §2.1.3, these fluctuations in electron transmission are directly related to fluctuations in conductance $\Delta G$. The magneto-resistance fluctuations $\Delta R$ however, are strongly influenced by the background resistance since $\Delta R = -R^2 \Delta G$ , and it is clear from Fig. 5.5 that this changes quite markedly over the set of magneto-resistance traces obtained in this experiment. Hence, hereafter all discussion will consider conductance instead of resistance to eliminate the contribution of the background resistance to the fluctuations. As a result, the resistance maxima in the previous discussion will subsequently correspond to conductance minima.

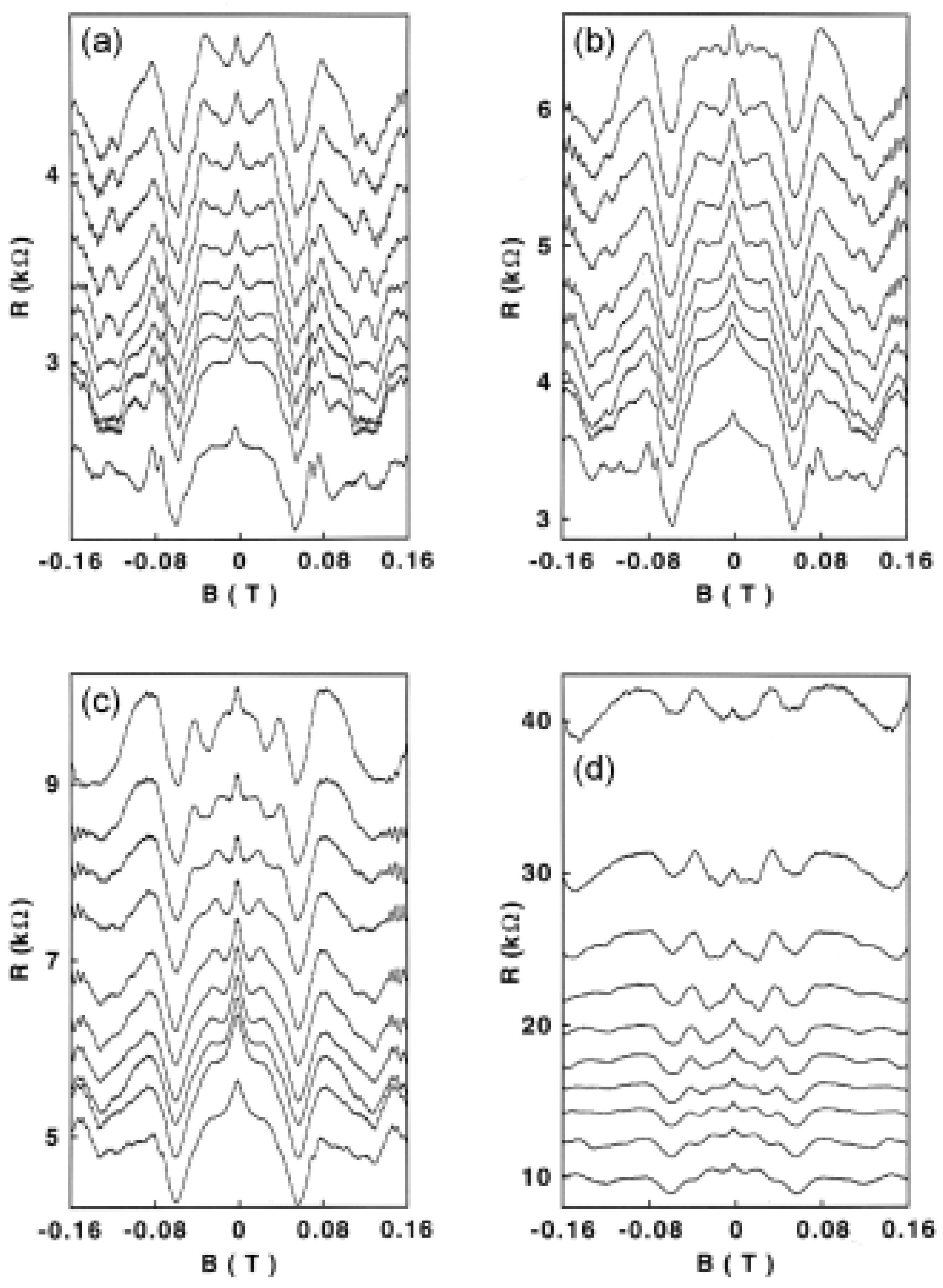

Figure 5.5: Magneto-resistance traces taken at $V_O$ values of (a) –0.5V, (b) –0.51V, (c) −0.52V and (d) –0.55V. Each set shows the transition from $V_I$ = +0.7V (bottom) to −2.9V (top) in 0.4V increments. Each trace was obtained at $T$ = 30mK.

Discussion of the Sinai geometry and its transition to the square geometry will largely focus on the data presented in Fig. 5.5(b) for the remainder of this chapter. A closer inspection of feature A in Fig. 5.4 is presented in Fig. 5.6(a). Structure in Fig. 5.6(a) is observed to be clustered on two distinct scales – the 'coarse' scale corresponding to feature A in Fig. 5.4 and 'fine' structure superimposed upon the coarse scale structure. Note that the fine scale structure also consists of a central trough with shoulder features. These two levels of structure ('coarse' and 'fine') are assigned characteristic magnetic field scales (defined as the full width at half-maximum of the central trough at each level) of $\Delta B_C \approx$ 20mT and $\Delta B_F \approx$ 1mT respectively, as indicated in Fig. 5.6(a).

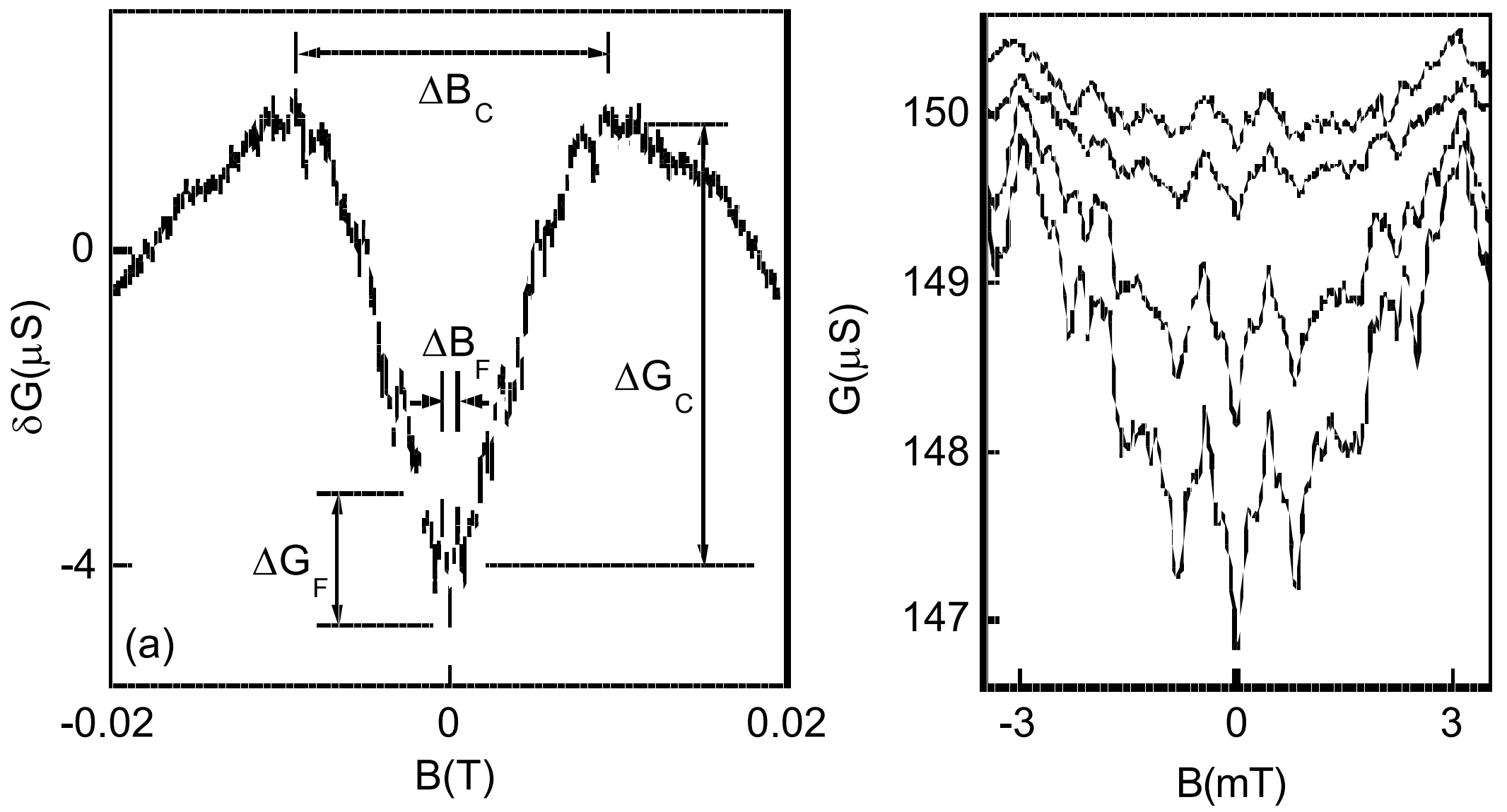

Figure 5.6: (a) Fluctuation from mean conductance $\delta G(B) = G(B) - \langle G(B)\rangle_B$ versus $B$ for the Sinai billiard indicating the amplitude and magnetic field periods of the coarse and fine structure. The average $\langle\ \rangle_B$ is performed over the range –0.02T < $B$ < 0.02T (b) Fine magneto-resistance structure for the Sinai billiard as a function of temperature. The temperatures and zero-field conductances are (0.03K, 146.8μS, bottom), (0.8K, 134.8μS), (1.6K, 116.1μS), (1.9K, 115.7μS, top).

Conductance amplitudes (defined as the full height of the central trough at each level) for the fine and coarse scales ($\Delta G_F$ and $\Delta G_C$) are also indicated in Fig. 5.6(a). Note that hereafter, the coarse and fine structure have been obtained as separate data-traces at different resolutions. This is because the time required to obtain both levels in a single data-trace at a resolution sufficient to adequately resolve features in the fine structure is prohibitively large. As discussed earlier, the coarse structure was found to have a quantum-mechanical origin based on the coarse structure temperature dependence shown in Fig. 5.4. The temperature dependence of the fine structure shown in Fig. 5.6(b) indicates that the fine structure also has a quantum-mechanical origin. Based on the quantum-mechanical origin of the coarse and fine levels, and assuming an Aharonov-Bohm flux relationship (see §2.1.4 and §2.2.3) then $\Delta B_C/\Delta B_F = A_F/A_C$, where $A_C$ and $A_F$ are the characteristic areas enclosed by the trajectory loops that determine the coarse and fine quantum interference processes.

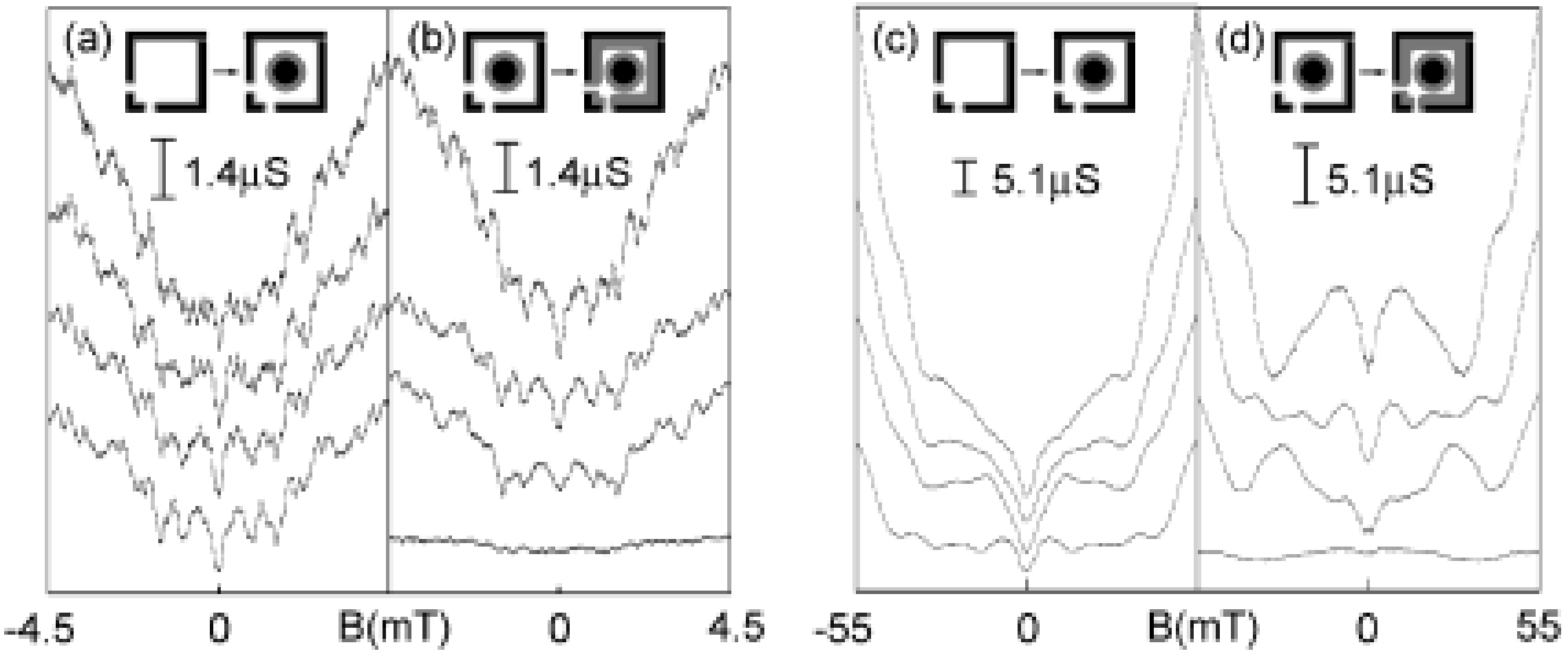

Figure 5.7: (a) Fine structure for $V_O = -0.51$V. $V_I$ and $G(B = 0)$ values are (+0.7V, 265μS, top), (−0.5V, 209μS), (−1.7V, 173μS) and (−3.1V, 151μS, bottom). (b) Fine structure for $V_I = -3.1$V. $V_O$ and $G(B = 0)$ values are (−0.5V, 231μS, top), (−0.51V, 151μS), (−0.52V, 108μS) and (−0.55V, 26μS, bottom). (c) Coarse structure for $V_O = -0.51$V. $V_I$ and $G(B = 0)$ values are (+0.7V, 265μS, top), (−0.5V, 209μS), (−1.7V, 173μS) and (−3.1V, 151μS, bottom). (d) Coarse structure for $V_I = -3.1$V. $V_O$ and $G(B = 0)$ values are (−0.5V, 231μS, top), (−0.51V, 151μS), (−0.52V, 108μS) and (−0.6V, 26μS, bottom).

Hence $A_F$ is ~20 times greater than $A_C$. This means that the fine structure is generated by trajectories considerably longer than the trajectories that generate the coarse structure. An initial suggestion was that the action of the Sinai diffuser is to form two sub-billiards – one where electrons can orbit around the Sinai diffuser and another where trajectories are restricted to the corner where the entrance and exit ports are located. In this case the fine structure would be due to longer trajectories winding around the Sinai diffuser whilst the coarse structure would be due to the shorter trajectories in the corner sub-geometry. Whilst the ratio $A_F/A_C$ seems consistent with this suggestion, three pieces of experimental evidence demonstrate this not to be the case.

The first piece of evidence is that pinching off the conducting channel around the Sinai diffuser destroys both the coarse and the fine structure. Figures 5.7(a) and (c) show the behaviour of the fine and coarse structure respectively as $V_I$ is increased from +0.7V (top) to –3.1V(bottom) at constant $V_O$. In both the fine and coarse structure a continuous evolution in the central feature and the shoulder feature is apparent. Note also that in both the coarse and fine structure the shoulder features evolve after the Sinai diffuser is activated (i.e. $V_I < 0$V), suggesting that both features are formed by the Sinai geometry. Due to gate-leakage problems $V_I$ is restricted to negative biases less than −3.3V. This means that it is not possible to pinch off the conducting channel surrounding the Sinai diffuser by adjusting $V_I$ alone. Pinch off is instead achieved by increasing the negative bias $V_O$, squeezing the conducting channel from the outside as shown in the inset of Fig. 5.7(b). Figures 5.7(b) and (d) show the effect of increasing $V_O$ from –0.5V (top) to –0.55V (bottom) with $V_I = -3.1$V for the fine and coarse structure respectively. Both $\Delta G_F$ and $\Delta G_C$ decrease to zero as the conducting channel is pinched off indicating that both the fine and the coarse structure are generated by trajectories that circulate around the Sinai diffuser. It should be noted that this decrease in the amplitude of the coarse and fine structure is not due to changing QPC width as $V_O$ is made more negative. Indeed, this should have the opposite effect, since increased amplitude in the central feature is expected with narrowing QPC width under a semiclassical approach [166].

The second piece of evidence is that as the temperature is raised, reducing the electron phase coherence, the conductance amplitudes of the coarse $\Delta G_C$ and fine structure $\Delta G_F$ both decrease exponentially. This is apparent both in the data shown in Figs. 5.4(b) and 5.6(b) and quantitatively as seen later in Fig. 5.13. Finally, the third piece of evidence is the remarkable similarity in the appearance of the two levels of structure, as discussed in detail in the following section.

# 5.3 – Observation of Exact Self-similarity in the Sinai Billiard

Further investigation of the fine and coarse structure suggests that there are more than just similarities in the dependencies of the two structural levels on billiard parameters. Figure 5.8(b) shows an overlay of the coarse (blue) and fine (red) structure shown in Fig. 5.8(a) for $V_O = -0.51$V and $V_I = -2.7$V. Note that in Fig. 5.8 the coarse structure appears smoother than the fine structure. This is because the two structural levels are obtained as separate traces at different resolutions. The relative resolution (number of data points across the data-trace) is considerably smaller for the coarse (105 data points) compared to the fine (788 data points) leading to a removal of finer scale structure in the coarse data-trace. However, aside from this reduced relative resolution of the coarse data-trace, the two data-traces appear remarkably similar, suggesting the presence of exact self-similarity (see §2.3.1). In order to confirm that the similarity observed in Fig. 5.8(b) is not a coincidence, a parameter of the billiard was changed ($V_O$ increased from –0.51V to –0.55V). Whilst the appearance of the coarse and fine structure obtained after changing $V_O$ is different, the similarity between the two levels is retained, as demonstrated in Fig. 5.9(a). However, note that it is possible to destroy the similarity between the coarse and fine structure in a number of ways. The first is by decreasing $V_O$, which allows the electrons to escape the billiard more rapidly, and have fewer interactions with the Sinai diffuser. The effect of decreasing $V_O$ is demonstrated in the coarse and fine data presented in Fig. 5.9(b). The second way is to substantially increase $V_O$, which pinches off both the QPCs and the conduction channel around the Sinai diffuser, as discussed earlier, and demonstrated in Fig. 5.7(b) and (d). Finally, the third way is by reducing the radius of the Sinai diffuser. This is discussed further in §5.6. Note that each of these ways of suppressing the self-similarity involves reducing the interaction of the electrons with the Sinai diffuser, strongly suggesting further that trajectories interacting with (or circulating) the Sinai diffuser are responsible for the exact self-similarity observed in the experimental data.

The self-similarity observed in the coarse and fine levels in Fig. 5.8 may be verified and investigated quantitatively by introducing a correlation function approach [167] to the analysis of the data. The correlation function is a more rigorous way of dealing with self-similarity because rather than relying on a single feature (e.g. the central feature height and full-width at half maximum in Fig. 5.6(a)) the correlation function assesses all of the features in the trace.

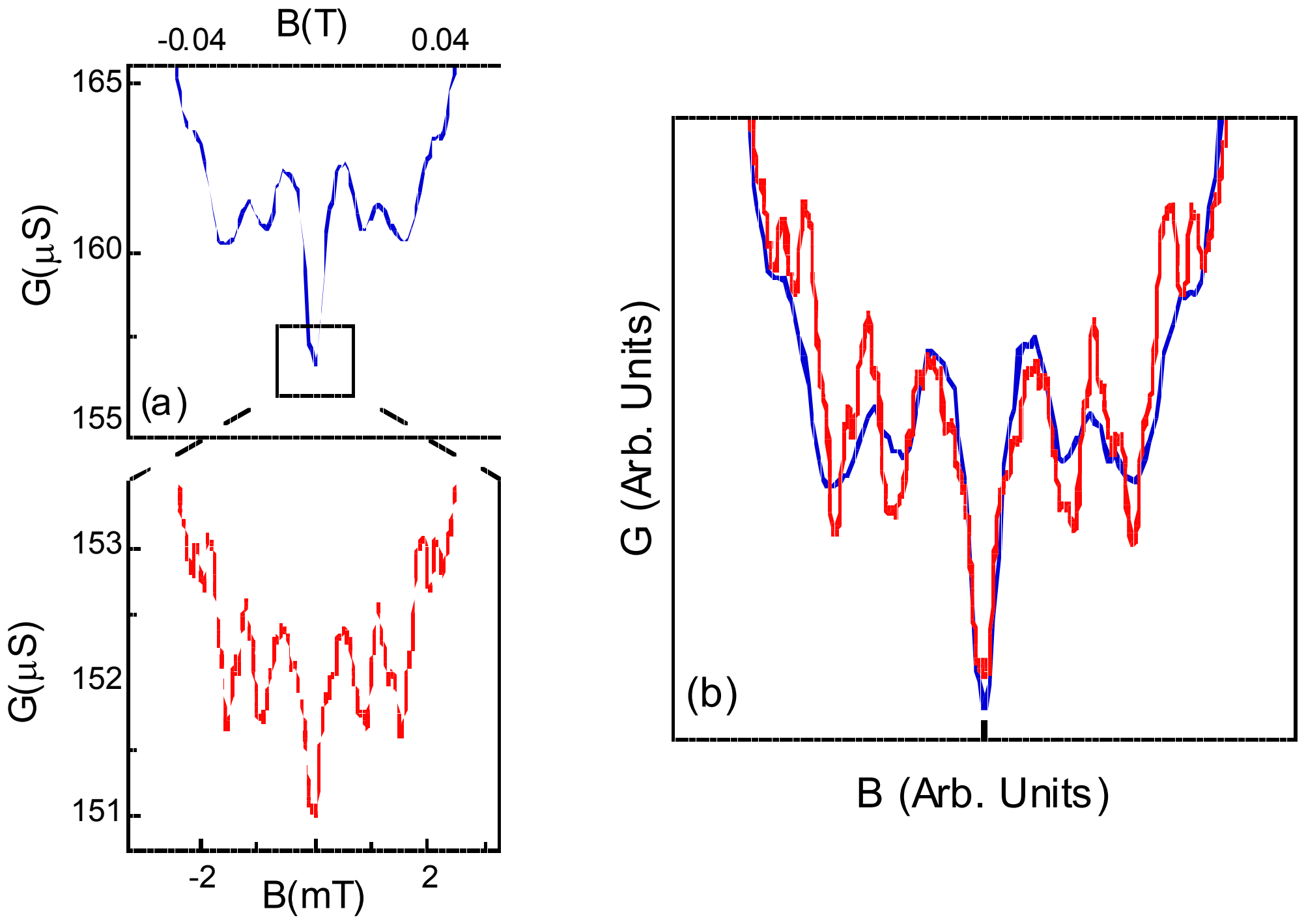

Figure 5.8: (a) Coarse structure (top) with fine structure (bottom) superimposed on the central feature for $V_O = -0.51$V and $V_I = -2.7$V. (b) Overlay of the coarse structure and the fine structure (after scaling by a factor of approximately 20 in magnetic field and 4 in conductance) demonstrating the remarkable similarity between these two structural scales.

In order to do this however, the conductance amplitudes of the structural levels must first be defined in a slightly different way. Instead of using the amplitude of a single feature in each level (e.g. the height of the central feature $\Delta G_0$), the coarse and fine scale amplitudes are defined as functions of magnetic field as follows: $\delta G_C = G_C(B) - \langle G_C(B)\rangle_B$ and $\delta G_F = G_F(B) - \langle G_F(B)\rangle_B$. The brackets $\langle\ \rangle_B$ represent an average performed over the entire magnetic field range of the particular level as defined in Fig. 5.8(a). Note that $\langle G_C(B)\rangle_B$ and $\langle G_F(B)\rangle_B$ are not necessarily equal since the averages are taken over different ranges. Bearing in mind the overlay plot shown in Fig. 5.8(b), it should then be possible to select conductance and field scaling factors $\lambda_G$ and $\lambda_B$ such that the coarse $\delta G_C(B)$ and scaled-fine $\lambda_G \delta G_F(\lambda_B B)$ are similar, if not identical, data-traces. The correlation function takes the coarse and scaled-fine data-traces, overlays them as shown in Fig. 5.8(b) and calculates the difference between the two data-traces for each magnetic field point.

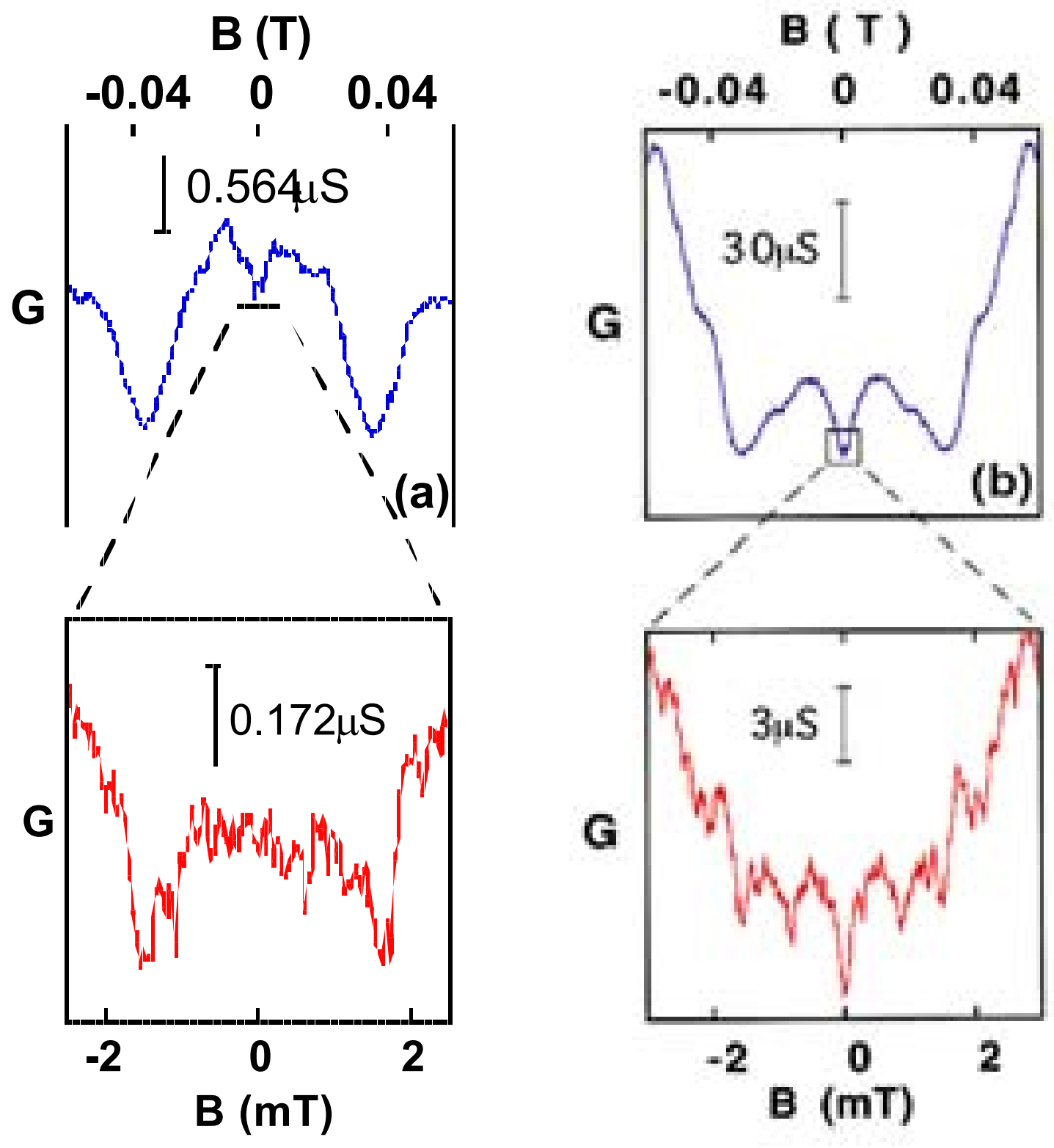

Figure 5.9: Coarse and fine structure for (a) $V_O$ = −0.55V demonstrating that the remarkable similarity observed in Fig. 5.8 is not coincidental and (b) $V_O$ = −0.50V indicating that the similarity is not preserved for all parameter values.

The root-mean-square difference between the coarse and scaled-fine data-traces is then normalised to produce the correlation function $F$ defined in Eqn. 5.1:

$$F = 1 - \frac{\sqrt{\left\langle \{\delta G_C(B) - \lambda_G \delta G_F(\lambda_B B)\}^2 \right\rangle_B}}{N} \tag{5.1}$$

The averaging $\langle\ \rangle_B$ is performed over the entire range of $B$ common to the coarse and scaled-fine data-traces. Note that the coarse and scaled-fine data-traces do not necessarily have the same data resolution either. Hence the data-trace with the higher resolution has its resolution reduced using an interpolation process so that both data-traces have a common number of data points (105 points) to allow comparison. The interpolation process is performed so that the data points in each data-trace have common $B$ values. This resolution-reduction process has little effect on the overall appearance of the scaled-fine data-trace. This can be seen by comparing the original fine structure data-trace in Fig. 5.8(a) with the fine structure in Fig. 5.14(g) where the data has been interpolated so that it has a resolution matching that of the coarse structure in Fig. 5.14(f). The normalisation constant $N$ is calculated by averaging 1000 values of the expression in Eqn. 5.2 where $X(B)$ and $Y(B)$ are functions that generate random traces over the $B$ range common to the coarse and scaled-fine data-traces with data resolution matching that used in comparing the coarse and scaled-fine data-traces.

$$N = \left\langle \sqrt{\left\langle \{X(B) - Y(B)\}^2 \right\rangle_B} \right\rangle_{1000} \tag{5.2}$$

Due to the Onsager relationships [62,63], $\delta G_C(B)$ and $\delta G_F(B)$ are symmetric about $B$ = 0T, and $X(B)$ and $Y(B)$ are therefore reflected about $B$ = 0T to ensure the same basic symmetry as the data. The amplitude ranges of $X(B)$ and $Y(B)$ are equated to that of $\delta G_C(B)$ for $V_I$ = +0.7V. The role of $N$ is to set $F$ = 0 when $\delta G_C(B)$ and $\lambda_G \delta G_F(\lambda_B B)$ are randomly related data-traces and $F$ = 1 if the two data-traces are mathematically identical. In this way, $F$ identifies similarities in patterns seen at two different field scales and is fundamentally different from equations that assess a fractal dimension $D_F$, which simply identifies the scaling relationship between structure at different scales. Note, however, that values between $F$ = 0 and $F$ = −1 are in fact possible and correspond to anti-correlation (i.e. correlation that is less than random). However, negative $F$ values are not expected in this analysis and will not be discussed further. Within the range $0 \leq F \leq 1$, it is informative to gauge when self-similarity becomes 'visually absent'. As an example of 'visibly dissimilar' data-traces, a correlation of the coarse structure at $V_I$ = +0.7V with that of a straight line with value $\delta G_C(B = 0)$ produces an $F$ value of 0.24 which is significantly above the randomly related case of $F$ = 0. This indicates that $F$ does not need to reach zero for self-similarity to be absent, only that it should be significantly below 1.

The correlation function was introduced to make a quantitative assessment of self-similarity between coarse and fine data-traces for a given set of scaling factors $\lambda_G$ and $\lambda_B$. However, it is also possible to use the correlation function to assess the precise values of the scaling factors that relate the self-similar coarse and fine structure in Fig. 5.8(b) by calculating $F$ as a function of $\lambda_G$ and $\lambda_B$. The scaling factor map presented in Fig. 5.10(a) shows $F$ as a function of $\lambda_G$ and $\lambda_B$ for the case of $V_I = -2.7$V and $V_O = -0.51$V, which corresponds to the data presented in Fig. 5.8.

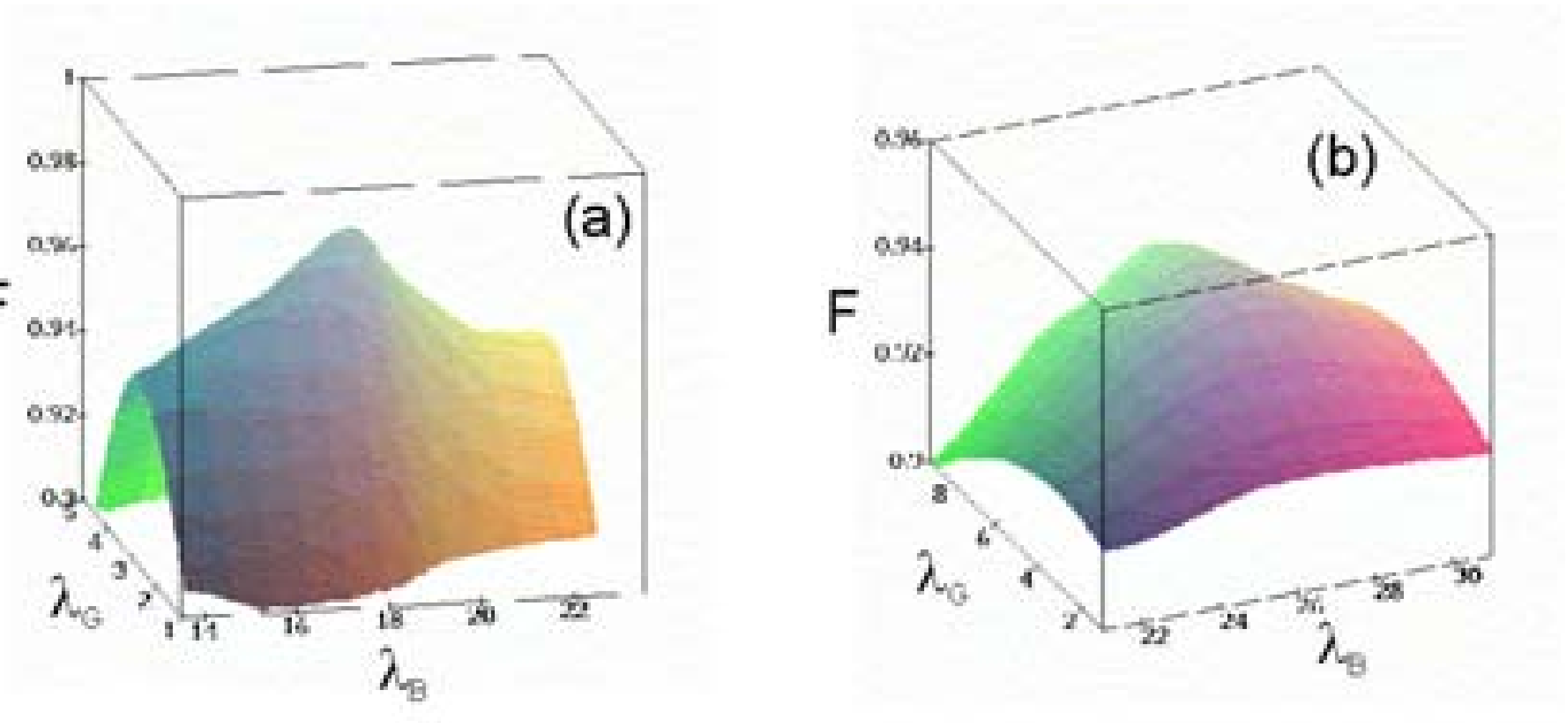

Figure 5.10: Scaling factor maps for (a) $V_O = -0.51$V and $V_I = -2.7$V and (b) $V_O = -0.55$V and $V_I = -2.7$V corresponding to the self-similar coarse and fine structure shown in Figs. 5.8(b) and 5.9(a) respectively.

A clear maximum of $F = 0.97$ is obtained for $\lambda_G = 3.7$ and $\lambda_B = 18.6$. The presence of a single peak indicates that there is a single pair of scaling factors describing the system, resulting in structure clustered around the two distinct structural levels (coarse and fine). The high value of $F$ at the maximum ($F \sim 0.97$) confirms that the coarse and fine structures are self-similar. The scaling factor map for the data presented in Fig. 5.9(a) is shown in Fig. 5.10(b). This scaling factor map also has a peak which lies at $\lambda_G = 6.1$ and $\lambda_B = 26.4$ with $F \sim 0.94$. Whilst the peak in Fig. 5.10(b) has a slightly lower $F$ value and is not as sharp as the peak observed in Fig. 5.10(a) it confirms the retention of self-similarity with the change in $V_O$ discussed earlier.

Exact self-similarity is a signature of fractal behaviour indicating that the MCF for the Sinai billiard may be fractal. The presence of fractal behaviour is determined by calculating the fractal dimension $D_F$. The two scaling factors relating the self-similar coarse and fine structure are related by a power-law relationship $\lambda_G = (\lambda_B)^{\beta}$. Such a scaling behaviour is defined as fractal if $D_F = 2 - \beta$ lies in the range $1 < D_F < 2$ [2]. Using the scaling factors obtained from the correlation function for the data presented in Fig. 5.8, gives $D_F = 1.55$, demonstrating the presence of fractal behaviour in the Sinai billiard. The fractal dimension $D_F$ is observed to change as a function of $V_O$, which in turn determines $n$, the number of modes transmitted in the QPCs. Note that $D_F$ is determined by the scaling factors $\lambda_B$ and $\lambda_G$ via $\lambda_G = (\lambda_B)^{\beta}$ where $D_F = 2 - \beta$. Values of $n$, $\lambda_B$, $\lambda_G$ and $D_F$ are plotted as a function of $V_O$ in Fig. 5.11.

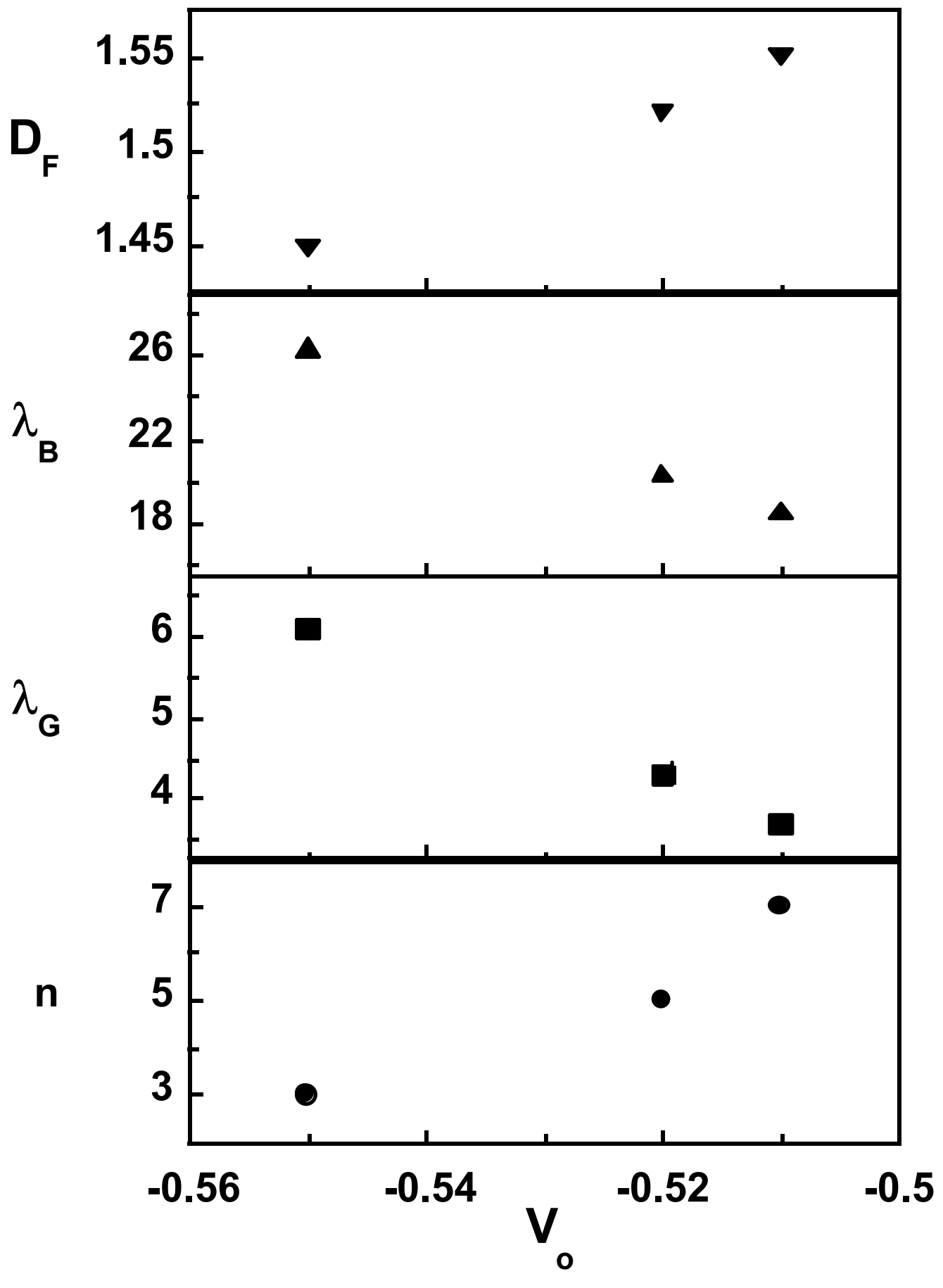

Figure 5.11: Evolution of the parameters $n$, $\lambda_B$, $\lambda_G$ and $D_F$ as a function of $V_O$ for the Sinai billiard ($V_I = -2.7$V).

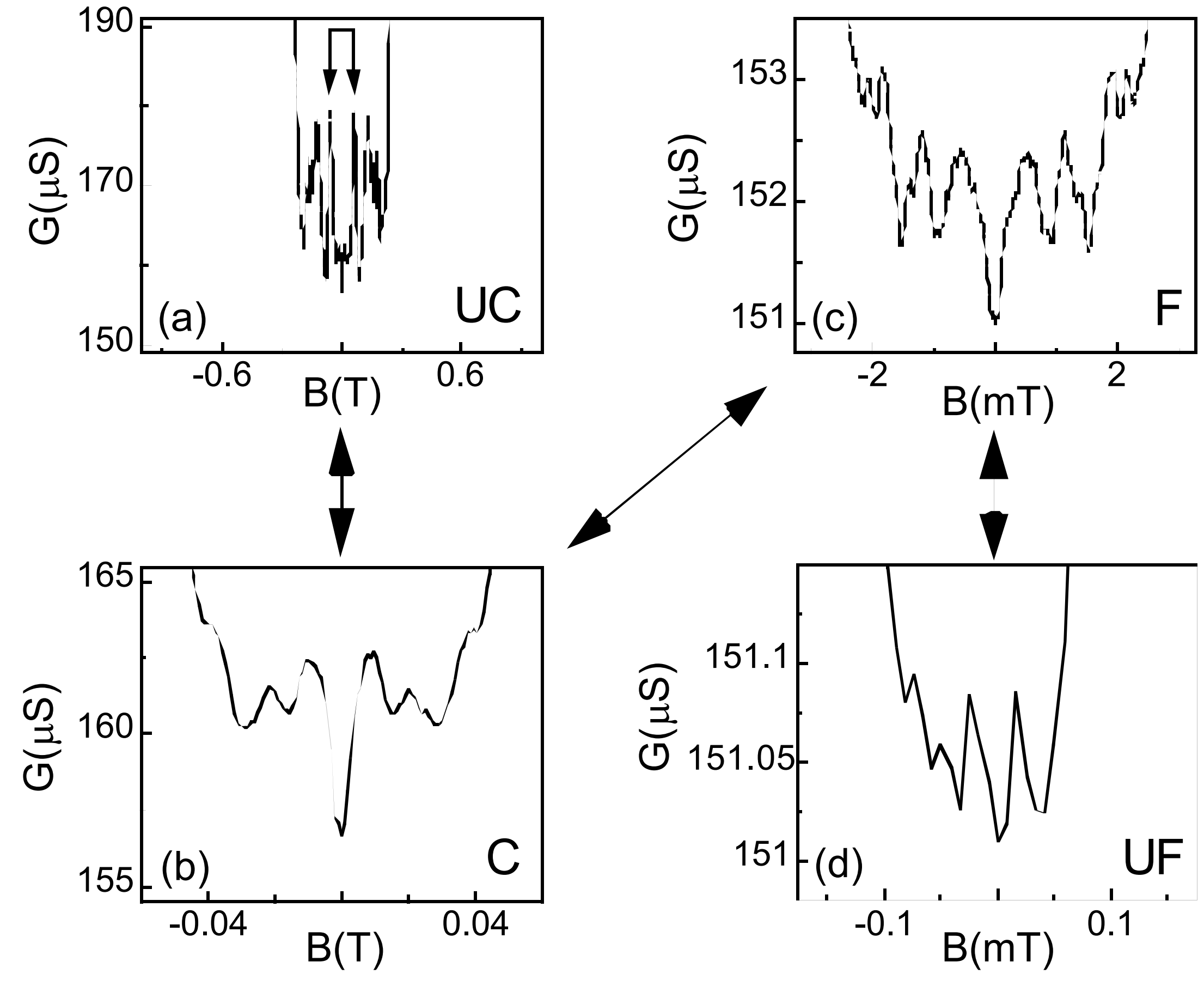

Figure 5.12: The four structural levels observed in the data for $V_O = -0.51$V and $V_I = -2.7$V: (a) ultra-coarse, (b) coarse, (c) fine and (d) ultra-fine. Arrows in the ultra-coarse trace (a) indicate the upper cut-off as discussed in §5.5.

Based on the relation between self-similarity and fractal behaviour it would be expected that further structural levels would be found on field scales ~20 times smaller than $\Delta B_F$ and ~20 times larger than $\Delta B_C$. Closer inspection of the fine structure (Fig. 5.12(c)) reveals the presence of 'ultra-fine' structure superimposed on the fine-scale central feature as shown in Fig. 5.12(d) at a field scale ~20 times smaller than the fine structure. Furthermore, expanding the field scale for the coarse structure (Fig. 5.12(b)) by a factor of ~20 reveals that the coarse structure is superimposed on the central trough of the 'ultra-coarse' structure shown in Fig. 5.12(a). A detailed analysis of the scaling properties of these four structural levels is presented in the following section.

# 5.4 – Scaling Properties of Self-similar Structure in the Sinai Billiard

A summary of the scaling properties of the four structural levels is shown in Fig. 5.13(a). The four levels – ultra-coarse (UC), coarse (C), fine (F) and ultra-fine (UF) – are assigned the indices $i = 1 - 4$ respectively. For simplicity, this analysis considers only one selected feature of the magneto-conductance structure – the central feature. In Fig. 5.13(a), $\Delta B_{FWHM}$ corresponds to the full width at half-maximum of this feature and $\Delta G_0$ corresponds to its full height. Figure 5.13(b) shows that $\Delta G_0$ follows an exponential increase for each of the four levels as the temperature is lowered and the electron phase coherence length is increased. By extrapolating these lines to zero temperature the four filled circles shown in Fig. 5.13(a) are obtained. With the exception of the ultra-fine data point (discussed later), the data condenses onto the zero temperature power-law line shown. This power-law line therefore describes the case of maximal phase coherence length, where the dependence on the billiard geometry of the distribution of trajectory areas for loops contributing to the quantum interference effect is least affected by decoherence processes. Discussion will be restricted to these zero temperature points for the remainder of this section.

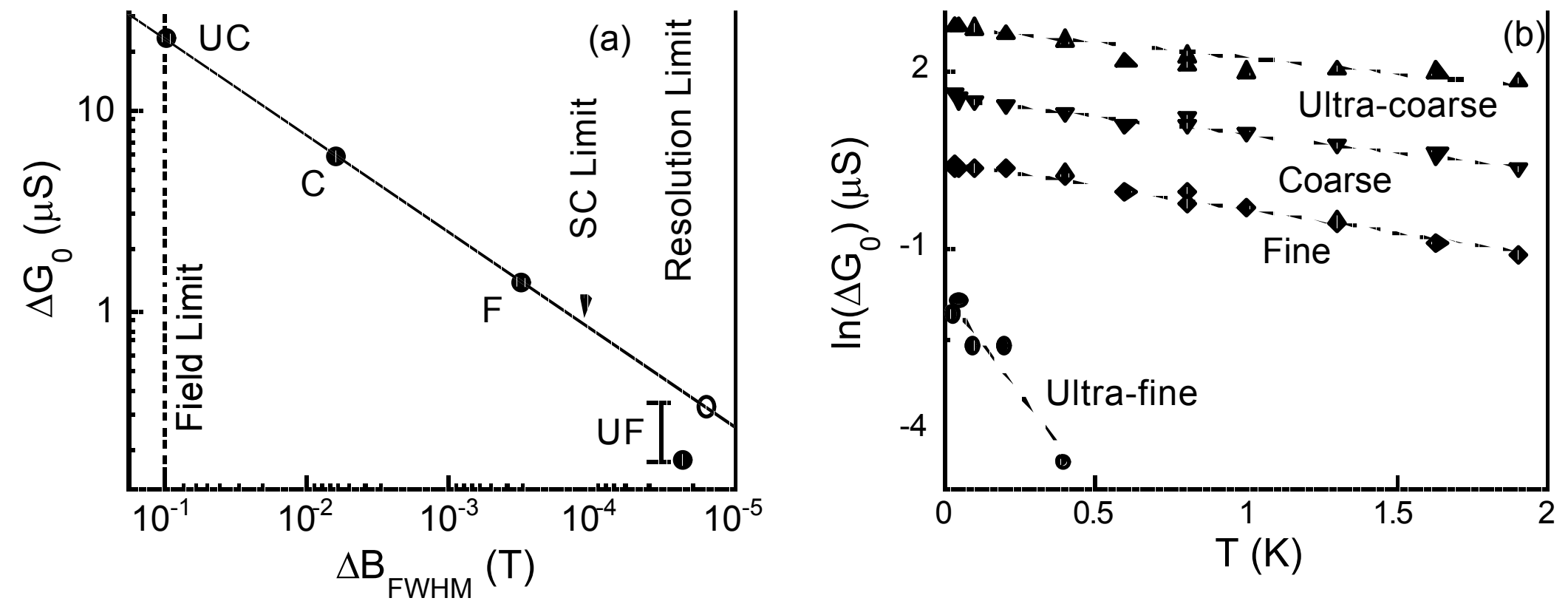

Figure 5.13: (a) Scaling properties of the levels observed in the experimental data based on the central feature amplitude $\Delta G_0$ and the full width at half-maximum $\Delta B_{FWHM}$. (b) Central feature amplitude $\Delta G_0$ as a function of temperature shows an exponential increase with decreasing temperature and is extrapolated to obtain zero temperature values of $\Delta G_0$.

To commence the investigation of the scaling properties, it is important to note that the four levels lie at equal increments along the power-law line. This confirms that the amplitudes and periods of consecutive levels are related by common field and conductance scaling factors $\lambda_B = \Delta B_i/\Delta B_{i+1}$ and $\lambda_G = \Delta G_i/\Delta G_{i+1}$, respectively. Returning to the discussion in the preceding section, it is possible to determine the fractal dimension based on the power-law relationship between the scaling factors. Allowing for uncertainties in the ultra-fine point (see later), the data points in Fig. 5.13(a) all lie on a line whose gradient $\beta$ gives $D_F = 1.55$ (i.e. all four points follow the fractal scaling relationship). The fractal dimension value obtained here matches the value obtained based on the scaling of the coarse and fine levels alone in §5.3.

The dashed vertical lines in Fig. 5.13(a) indicate the *experimental* limits of this fractal behaviour. As discussed in the preceding section, the coarse and fine levels, which lie well within these limits, exhibit exact self-similarity as shown in Fig. 5.8. In contrast, the ultra-coarse and ultra-fine levels are measured close to the respective experimental limits. A discussion of how this produces the observed reduction in self-similarity follows. The high-field cut-off occurs when the cyclotron radius of the electron becomes smaller than the billiard width. The calculated upper field limit is marked by arrows in Fig. 5.12(a) and Fig. 5.14(e). Thus parts of the ultra-coarse pattern lie outside this limit as indicated by the dashed data line in Fig. 5.14(e). For magnetic fields higher than the arrows, the skipping orbit regime (see §2.2.4) becomes established and the similarity between this pattern (the dashed data line) and the equivalent section of the coarse pattern deteriorates. As expected, for magnetic field values smaller than the arrows, the ultra-coarse pattern still bears a similarity to that observed in the coarse level, and the height and width of the central feature are used to calculate the ultra-coarse coordinate in Fig. 5.12(a). Unlike the ultra-fine level (see below), because the central feature of the ultra-coarse level lies within the observation limit, its coordinate lies on the fractal power-law line with no correction required. The low-field cut-off is determined by the magnetic field resolution limit for the experiment. In order to observe a MCF feature, a minimum of three data points is required, corresponding to an interval of 0.016mT. The ultra-fine structure shown in Fig. 5.12(d) consists of only 46 data points compared to the 788 data points within the equivalent fine structure shown in Fig. 5.12(c). This reduction in resolution by 94% distorts the ultra-fine structure, resulting in a loss of similarity between the fine and ultra-fine structures. To quantify this distortion, a comparison is made between the fine structure before and after removing data points to artificially reduce its resolution by 94%. The introduction of reduced resolution causes $\Delta G_0$ to fall by 49%. This percentage distortion corresponds to 80nS for the ultra-fine level. Whereas a visual inspection confirms the resulting loss in self-similarity for the ultra-fine structure in Fig. 5.12(d), the distortion can be compensated for when calculating the fractal scaling behaviour of this level in Fig. 5.13(a). It is the corrected $\Delta G_0$ value that is indicated by the filled circle in Fig. 5.13(a), as well as the points in the ultra-fine temperature dependence presented in Fig. 5.13(b). The difference between the position of this point and its anticipated position on the power-law line (open circle), as indicated by the 2Ω adjacent bar, is within the noise limit of the experiment (0.05% of the signal, corresponding to ~3Ω).

Improvements to the resolution limit of the data-trace alone (i.e. by improved magnetic field resolution in the measurements) will not allow the observation of a fifth level (super-fine (SF)) below the ultra-fine structure. Extrapolating the scaling plot in Fig. 5.13(a) gives $\Delta B_{SF} \approx 1\mu T$ and $\Delta G_{SF} \approx 47nS$ for the super-fine level. The amplitude of the super-fine structure (~2Ω or 0.03% of signal) is smaller than the typical noise level in the experiment and hence would be unobservable. A solution is to move the levels up the power-law line (to the left in Fig. 5.13(a)) by reducing the billiard area $A_B$. This in turn reduces the characteristic loop areas and increases the $\Delta B$ values of each level of MCF structure. This has the added benefit of allowing the features to become better resolved. A study of the area dependence of fractal behaviour in the square billiard is discussed in Chapter 7. However, it is not entirely clear how a decrease in $A_B$ will affect the nature of the quantum interference processes that produce exact self-similarity in the Sinai billiard. For a 1μm billiard, the energy level spacing (see §7.4 for details) in the billiard $\Delta = 2\pi\hbar^2/m^*A_B$ is 9μeV. The corresponding Heisenberg time $\tau_H = \hbar/\Delta$ can be converted into an approximate $\Delta B$ assuming $L = v_F\tau_H$, $A \sim L^2/4\pi$ and $\Delta B = h/eA$. This is indicated in Fig. 5.13(a) as the semiclassical limit (SC limit). To the right of this value, the trajectory traversal times exceed $\tau_H$ and the semiclassical picture of quantum interference processes would not be expected to hold [43]. For a 1μm billiard, the ultra-coarse, coarse and fine levels satisfy the semiclassical condition but the ultra-fine data lie slightly beyond. If the size of the billiard is reduced, the semiclassical limit will shift to the left and may start to affect the exact self-similarity of the levels. This potential problem with scaling the billiard size also restricts improvements to the upper cut-off. One possible way to move the field limit to the left is to reduce the size of the billiard, but as discussed above, this might affect the self-similarity of the existing levels. An alternative solution is to lower the electron density $n_s$ which reduces $k_F$ through $k_F = (2\pi n_s)^{1/2}$, however, this may also affect the exact self-similarity. Billiards fabricated with a back-gate could fine-tune $n_s$ in order to shift the field limit to the left. This means that the effect of changing $n_s$ can be investigated in a single sample, unlike changing the billiard area, which usually requires a number of samples, eliminating variations between samples (e.g. impurity distribution, lithographic variations, etc) as a potential cause of any observed effects.

Finally, in addition to increasing the magnetic field range over which the power-law behaviour is observed, it is also desirable to increase the number of levels observed within a given magnetic field range. The results indicate that an increase in the number of modes in the entrance and exit ports reduces $\lambda_B$ and hence the spacing between the different levels. The range of observation of fractal behaviour, defined by the upper and lower cut-offs shown in Fig. 5.13(a), is 3.7 orders of magnitude in magnetic field. There has been considerable debate recently regarding the validity of observations of fractal behaviour in various physical systems [93-98]. A recent survey of fractal measurements in physical systems revealed that the average range over which fractals are observed is only 1.3 orders of magnitude [94]. Whereas fractals that are observed over limited ranges are, in theory, no less fractal than those observed over a larger range [95], the reliability of fitting to power-law behaviour is intrinsically linked to the range over which the fit can be applied. A large range therefore lends confidence to the observation of fractal behaviour. The extended range of observation (3.7 orders of magnitude) makes the Sinai billiard an ideal system in which to model the scaling relationships of this fractal phenomenon. Having established the precise scaling properties of the fractal MCF observed in the Sinai billiard, the next section presents a model for the self-similar MCF based on the scaling properties introduced in this section.

# 5.5 – The Weierstrass Model

As a starting point for the introduction of the Weierstrass model, I will consider a Sinai billiard defined by a hard-wall potential profile. Recent theoretical predictions for the hard-wall Sinai billiard express the magneto-conductance $G(B)$ as a sum of damped cosines [168,169,170]. These cosine oscillations arise from an effect analogous to the Aharonov-Bohm effect where the magnetic flux enclosed by electron trajectory loops alters the electron phase and hence the quantum interference which determines the conductance through the billiard [168]. Accordingly, the period of the $i$th cosine in this summation is given by $b_i = h/2eA_i$. The summation of cosine contributions to $G(B)$ is pictured as commencing with a loop enclosing the fundamental area $A_1$ for $i = 1$, followed by harmonics with enclosed areas $A_i = iA_1$. The amplitudes of the cosines contain a field-independent component $a_i$ and also a field-dependent term which decreases with increasing magnetic field according to a field scale set by $b_i$ [168,170]. This results in a set of damped cosines analogous to the Al'tshuler-Aronov-Spivak (AAS) effect [28]. In order to model a soft-wall Sinai billiard, these basic characteristics of the hard-wall billiard are retained by constructing $G(B)$ using a summation of damped cosines:

$$G(B) = \frac{ce^2}{h}\sum_{i=1}^{\infty}\frac{a_i}{(\alpha B/b_i)^2+1}\cos\left\{\frac{2\pi B}{b_i}\right\} \tag{5.3}$$

where $c$ and $\alpha$ are dimensionless constants. Adopting a power-law relationship for the coefficients $a_i$ and $b_i$ based on the scaling properties shown in Fig. 5.13(a), it is possible to construct a model that describes the experimental data. The summation of cosines in Eqn. 5.3 then becomes a member of a class of functions known as Weierstrass functions [32,171] that are well-known generators of self-similar structure. The power-law scaling is introduced as follows. The enclosed areas $A_i$ are related to $A_1$ through the expression $A_i = \lambda_B^{i-1}A_1$ rather than through $A_i = iA_1$. Thus for successive $i$ values, the period becomes a factor of $\lambda_B$ smaller than for the preceding level. The coefficient $a_i$ is made proportional to $\lambda_G^{1-i}$ so that for successive $i$ values the amplitude at zero field becomes a factor of $\lambda_G$ smaller and the damping occurs over a field scale $\lambda_B$ smaller than the preceding level. In order to model the experimental data accurately, the required fitting parameters are $A_1 = 7.1\times10^{-15}\text{m}^2$, $\alpha = 0.08$, $\lambda_B = 18.6$, $\lambda_G = 3.7$ and $c = 0.514$. Figures 5.14(a), (b), (c) and (d) show four levels of a hierarchy of self-similar structure generated using this model and the listed parameter values. At this stage the model curve bears a strong resemblance to the experimental data, replicating not only the self-similarity of the coarse and fine levels, but also the general features of the MCF structure.

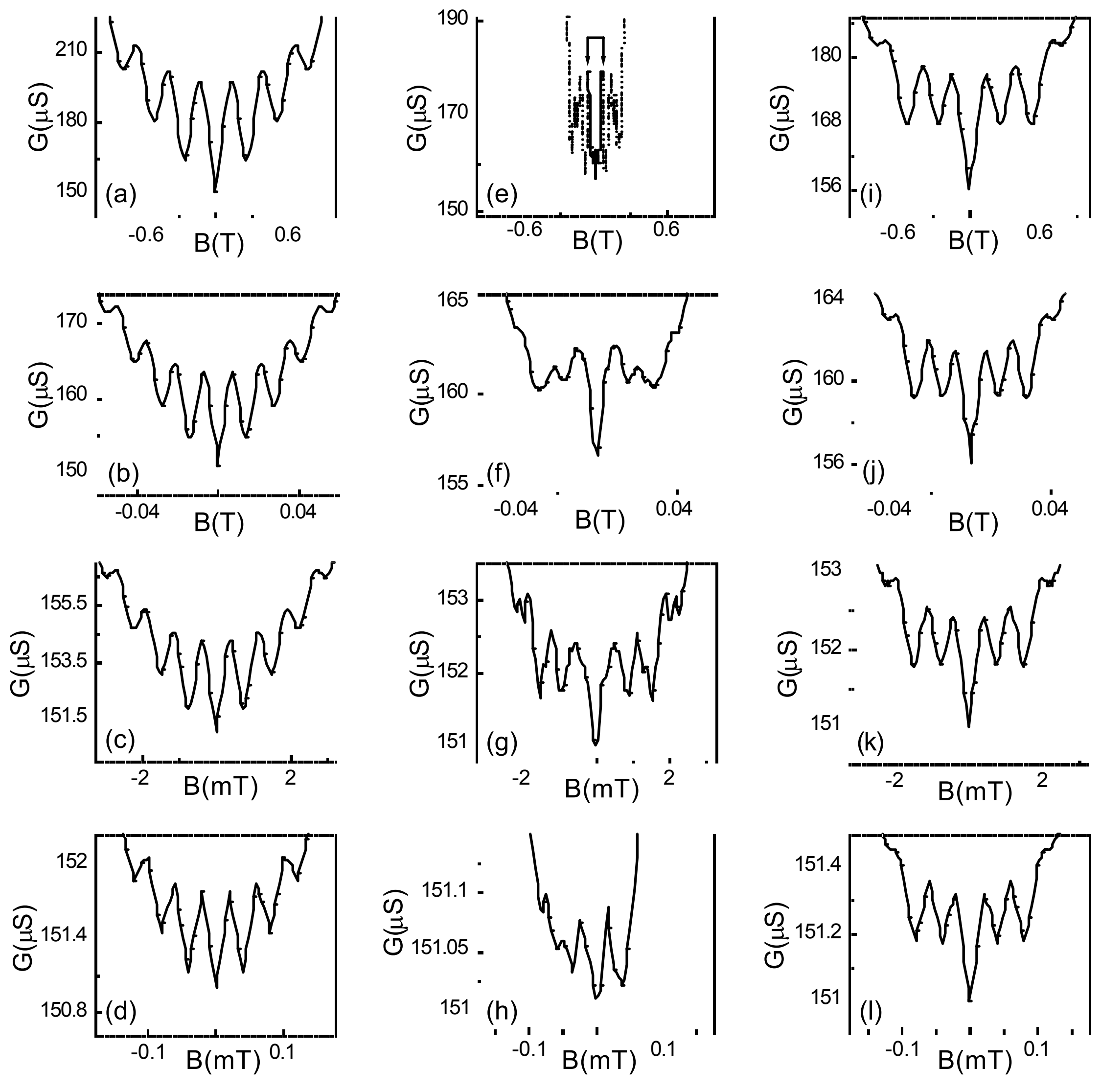

Figure 5.14: From top to bottom – ultra-coarse, coarse, fine and ultra-fine levels for: (a,b,c,d) the initial Weierstrass model, (e,f,g,h) the experimental data and (i,j,k,l) the refined Weierstrass model. Arrows in the ultra-coarse data (e) indicate the upper cut-off.

The Weierstrass model presented thus far ignores two key physical considerations. Firstly, the circulating channel in the semiconductor billiard has a finite width and therefore the unique areas $A_i$ that give rise to the single points in Fig. 5.15(a) should be replaced by a narrow distribution of areas $\Delta A_i$ centred on $A_i$, as indicated by the bars in Fig. 5.15(b). For each level, the ratio $\Delta A_i/A_i$ is fixed at 0.23 with $\Delta A_1 = 1.6\times10^{-15}\text{m}^2$ centred about $A_1 = 7.1\times10^{-15}\text{m}^2$ for the $i = 1$ level. Secondly, the theoretical predictions for the hard-wall Sinai billiard suggest that the $h/2e$ oscillations should coexist with $h/e$ oscillations [168,170]. The initial Weierstrass model is constructed purely from a summation of scaled damped $h/2e$ oscillations. Therefore the second refinement to the initial model is to introduce a second hierarchy with $A_i$ and $a_i$ identical to the first, but generating $h/e$ oscillations rather than $h/2e$ oscillations by using $b_i = h/eA_i$. Based on fitting this refined model to the data, I found that the contribution of the $h/e$ oscillations, set by the constant $c$, is half that for the $h/2e$ oscillations. The second hierarchy then produces the bottom line in Fig. 5.15.

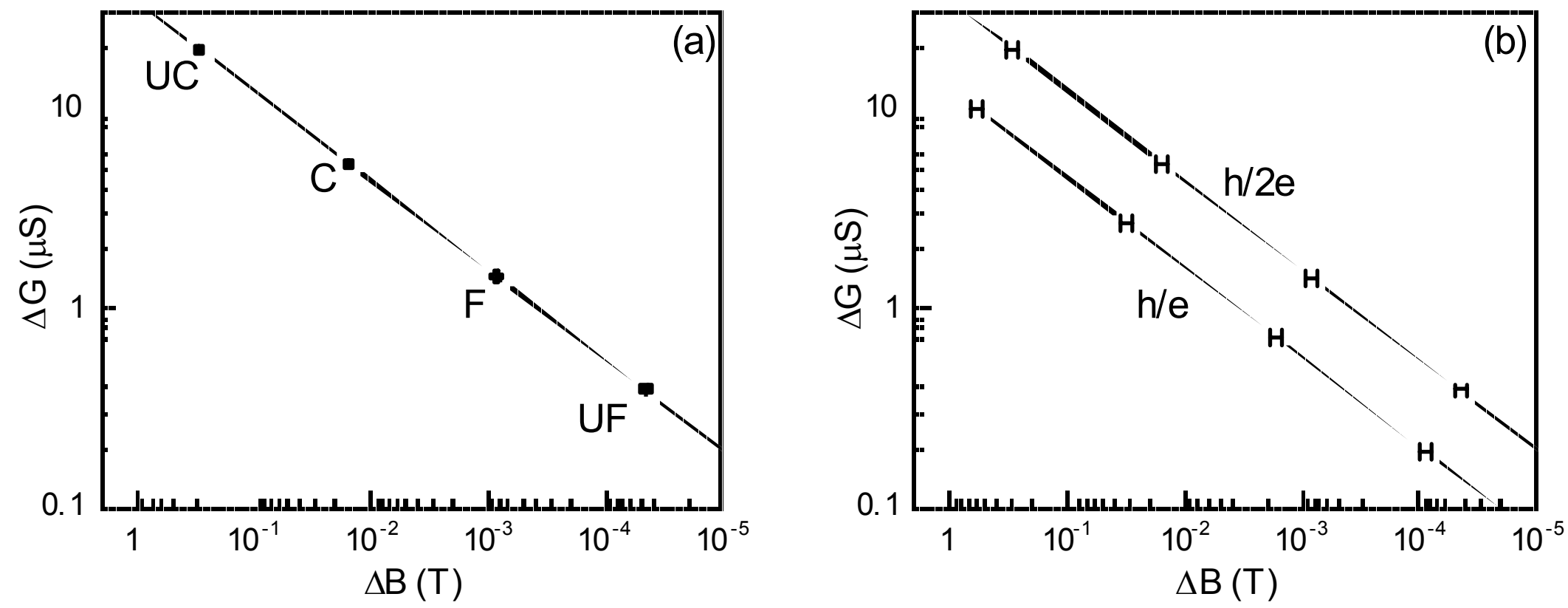

Figure 5.15: Scaling properties of levels of (a) the initial Weierstrass model and (b) the refined Weierstrass model. $\Delta G$ is the amplitude and $\Delta B$ is the period of the cosines used to produce the levels in both cases. The upper line in (b) corresponds to the single line presented in (a).

In Fig. 5.13(a), $\Delta B_{FWHM}$ and $\Delta G_0$ correspond to the central feature width and height, whilst in Fig. 5.15, $\Delta B$ and $\Delta G$ relate to the period and amplitude of the cosine used to generate a particular level. However the hierarchies in Figs. 5.13(a) and 5.15 still have the same scaling properties, the same point separations $\lambda_B$ and $\lambda_G$, and lie on power-law lines with the same gradient. Hence each hierarchy has the same fractal dimension $D_F$. The results of these two refinements to the Weierstrass model are presented in Figs. 5.14(i,j,k,l). The similarity between the refined Weierstrass model and the experimental data is more convincing than that observed with the initial Weierstrass model.

The striking similarities between the Weierstrass model and the experimental data for the coarse and fine levels suggests that the Weierstrass model has identified a possible set of trajectories that generate the observed exact self-similarity. However, the Weierstrass model does not directly investigate electron transport through the billiard, and hence does not allow precise trajectories to be identified – only possible characteristics such as enclosed area and periodicity (*h*/*e* or *h*/2*e*). Fromhold *et al.* [99] and Akis and Ferry [172] have performed quantum-mechanical calculations of electron transport through the Sinai billiard taking realistic account of the potential profile of the billiard produced by the negatively biased surface-gates. These models have demonstrated the presence of a complicated mixed phase-space due to the soft-wall nature of the billiard potential, as suggested in [12]. To date the exact self-similarity observed in the experimental data has not yet been explained fully in these models, although recent results appear promising [52].

# 5.6 – Removing the Sinai Diffuser: Loss of Exact Self-similarity

This section discusses the effect of reducing the radius of the Sinai diffuser on the exact self-similarity observed in Fig. 5.8. Figure 5.16 provides a comparison between the coarse (left column) and fine (right column) structure as the Sinai diffuser is gradually reduced in size and finally removed from the billiard (−2.9V at the bottom to +0.7V at the top in steps of 0.2V). The self-similarity observed in the Sinai billiard (bottom traces) is gradually suppressed as the diffuser radius is reduced. Ultimately, for the case where the Sinai diffuser is absent, exact self-similarity is no longer observed, however the self-similarity appears to have been removed well before this point. This suggests that the self-similar behaviour is actually due to fine-tuning of the diffuser radius rather than merely the presence of the diffuser. For a classical hard-wall billiard, removal of the Sinai diffuser would lead to a removal of any chaotic regions in the phase-space, producing a fully non-chaotic system. In a soft-wall billiard however, the phase-space is mixed, irrespective of the presence (and radius) of the Sinai diffuser, which means that any effects observed in the data can only be due to adjustments in this mixed phase-space, rather than the removal of chaos.

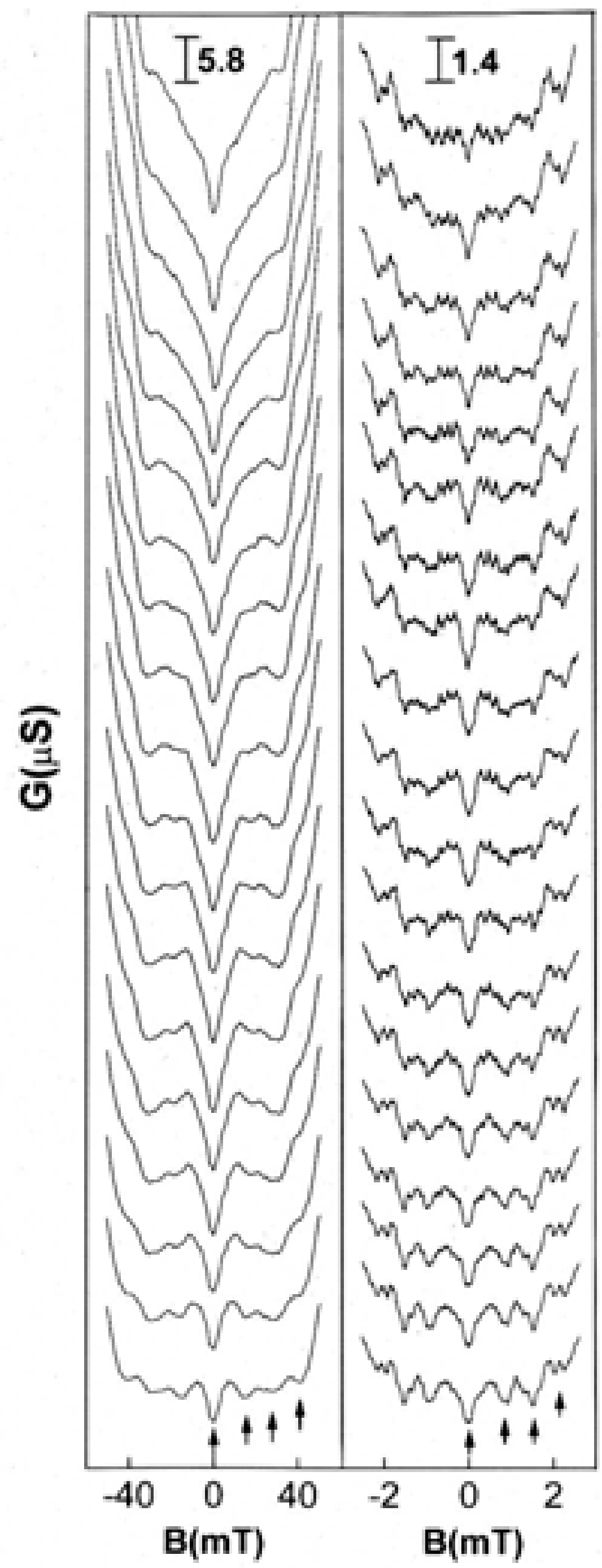

Figure 5.16: Cascade of coarse structure (left column) and corresponding fine structure (right column) for $V_O = -0.51$V. The $V_I$ and $G(B = 0)$ values are from top to bottom: (+0.7V, 264.5μS), (+0.5V, 231.1μS), (+0.3V, 225.7μS), (+0.1V, 221.2μS), (−0.1V, 217.8μS), (−0.3V, 214.2μS), (−0.5V, 208.7μS), (−0.7V, 203.5μS), (−0.9V, 198.9μS), (−1.1V, 192.7μS), (−1.3V, 187.5μS), (−1.5V, 183.2μS), (−1.7V, 177.9μS), (−1.9V, 172.7μS), (−2.1V, 168.8μS), (−2.3V, 164μS), (−2.5V, 160.4μS), (−2.7V, 156.6μS), (−2.9V, 151.0μS).

The correlation function is now used to confirm this loss in exact self-similarity quantitatively. Figure 5.17(a) shows the scaling factor map presented in §5.3 to establish the presence of self-similarity in the Sinai billiard. This scaling factor map corresponds to the bottom data-trace in the left and right columns of Fig. 5.16. In contrast to the scaling factor map for the Sinai billiard, the equivalent scaling factor map for $V_I$ = +0.7V, shown in Fig. 5.17(b), remains below 0.7 for all value of $\lambda_G$ and $\lambda_B$. Furthermore, there are no peaks in the $V_I$ = +0.7V scaling factor map, indicating that for the square geometry, the magneto-conductance no longer exhibits exact self-similarity. Note that although 0.7 is substantially greater than zero, it is a considerable decorrelation. This is clear from a visual comparison of the top two data-traces (coarse and fine) in Fig. 5.16. It is also important to bear in mind that a comparison of the coarse data-trace to a straight line (a more radical decorrelation than that observed in the data from the square) led to $F$ = 0.24 as discussed in §5.3.

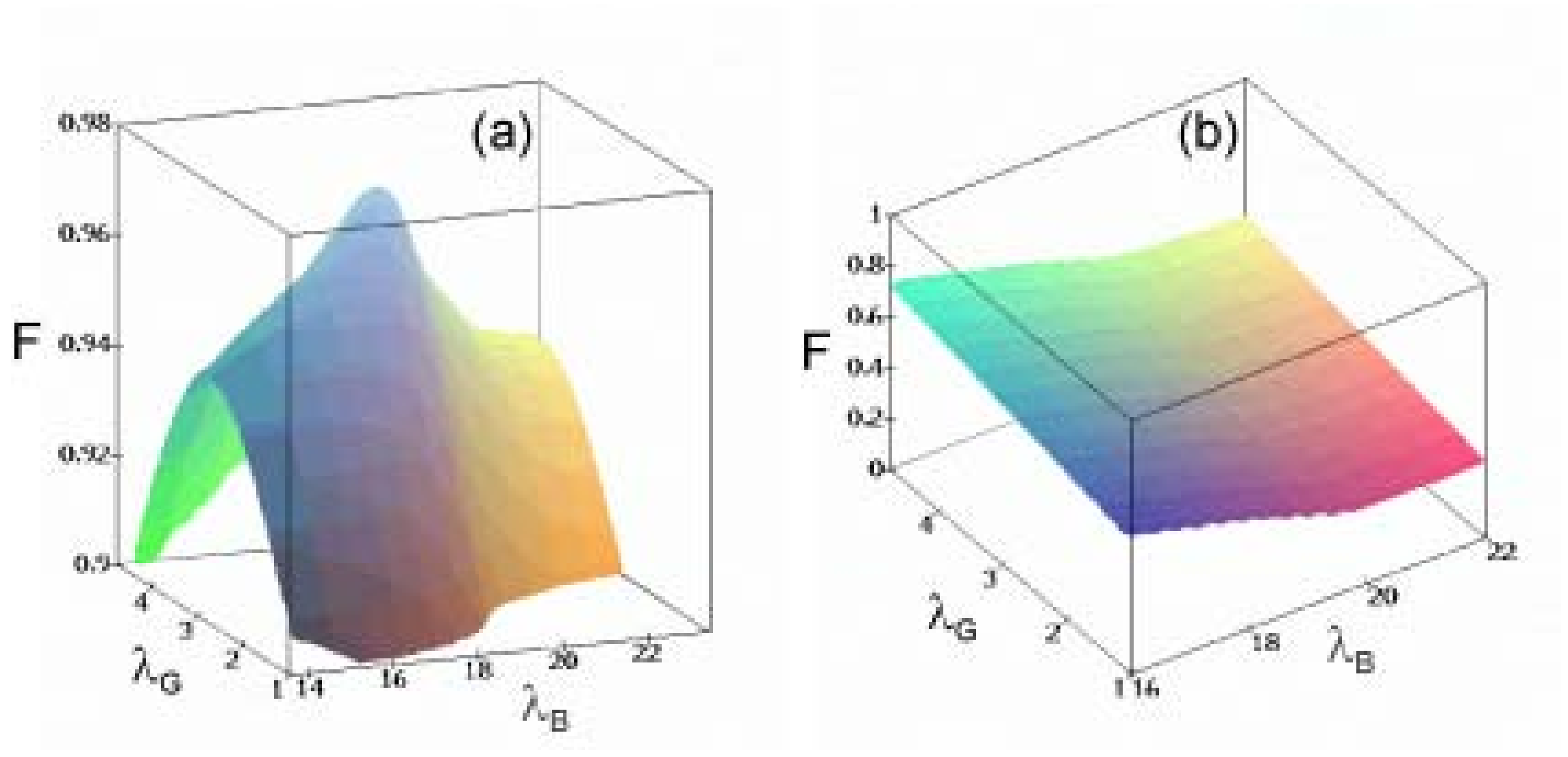

Figure 5.17: Scaling factor maps for $V_O$ = −0.51V with (a) $V_I$ = −2.7V and (b) $V_I$ = +0.7V.

The correlation function is now used to assess how the exact self-similarity is suppressed as the Sinai diffuser is removed. Figure 5.18 shows plots of $F$ versus $V_I$ for $V_O = -0.51$V (thick line) and $V_O = -0.55$V (thin line). Using the scaling factor maps presented in Fig. 5.10, the appropriate scaling factors ($\lambda_G$, $\lambda_B$) were calculated to be (3.7, 18.6) and (6.1, 26.4) respectively. The error bars in Fig. 5.18 represent the difference between the $F$ value obtained by correlating two nominally identical $\delta G_C(B)$ traces taken at $V_I = -3.0$V in opposite field directions, and the expected value of $F = 1$. In both cases there is a clear decrease in correlation from $F > 0.95$ at $V_I = -3.0$V down to $F < 0.6$ at $V_I = +0.7$V. Indeed, for $V_O = -0.55$V the correlation parameter falls to below 0.1 as the Sinai diffuser is removed. This demonstrates that the exact self-similarity is gradually suppressed as the radius of the Sinai diffuser is decreased rather than being suddenly removed at some critical Sinai diffuser radius. This is consistent with the trends observed in the data in Fig. 5.16. It is interesting to note that in both traces in Fig. 5.18, the maximum $F$ value is achieved at $V_I = -2.7$V with a trend towards decreasing $F$ value occurring for more negative $V_I$ values. The presence of a maximum $F$ value for some distinct Sinai diffuser radius confirms the notion that exact self-similarity can be fine-tuned by the Sinai diffuser.

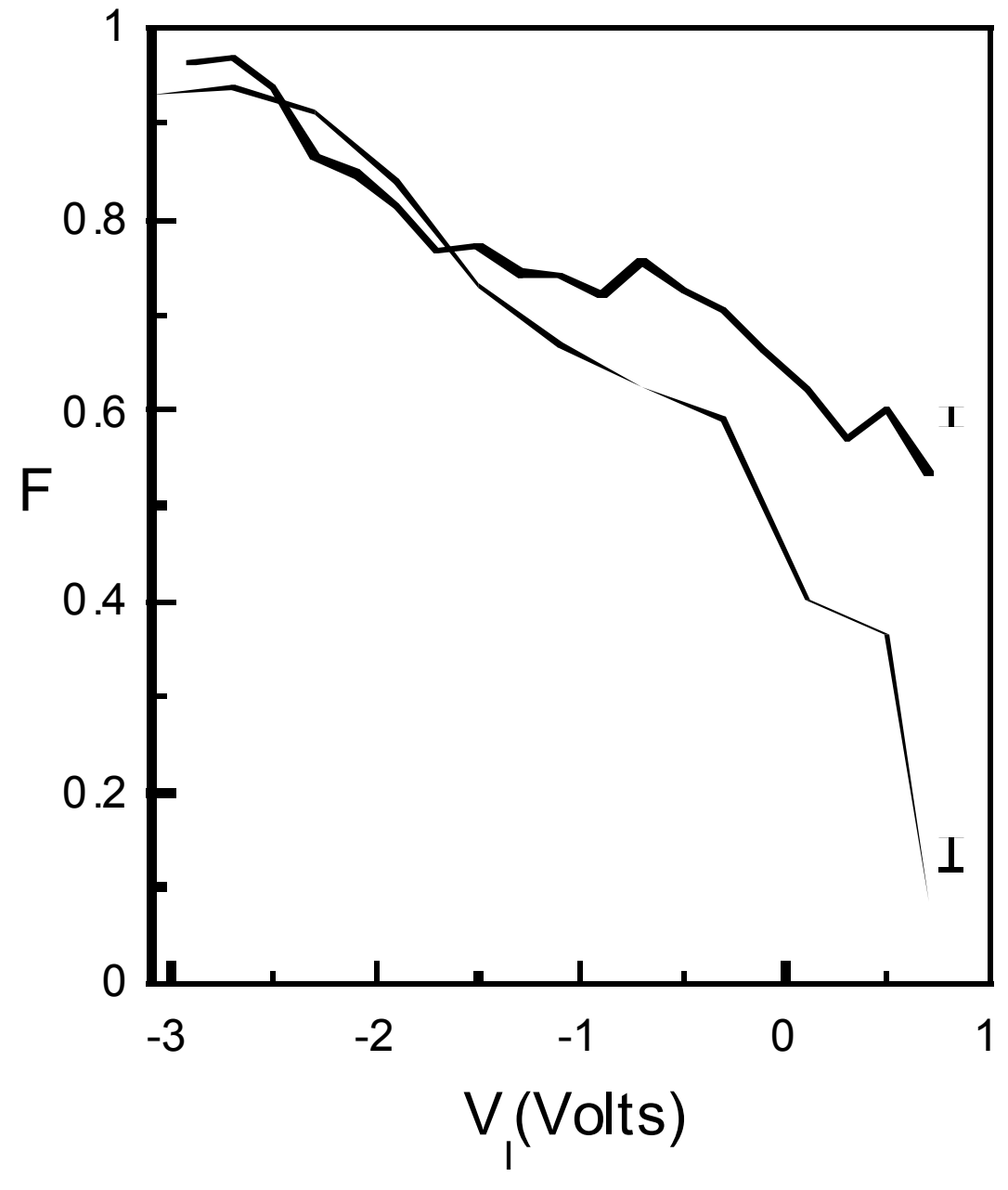

Figure 5.18: Plots of $F$ versus $V_I$ for $V_O = -0.51$V (thick line) and $V_O = -0.55$V (thin line).

Unfortunately it was not possible to achieve more negative $V_I$ values to investigate how $F$ decreases for $V_I < -2.7$V because $V_I$ is limited to ~ −3V in order to prevent gate-leakage. Furthermore, it is also interesting to note that the case of fully decorrelated ($F = 0$) coarse and scaled-fine data-traces can be approached in the empty square by adjusting $V_O$. Increasing $V_O$ from −0.51V to –0.55V reduces the number of conducting modes in the entrance and exit ports from 7 to 3, reduces the port width $W$, and hence increases the billiard characteristic dwell length $L_D = v_F \tau_D = A_B \pi / 2W$ ~ 13μm by 20% [64]. By increasing $L_D$ (i.e. increasing $V_O$), the probability of trajectories interacting with the diffuser before exiting the billiard increases. This is manifested as a sharper gradient in Fig. 5.18, and ultimately as a lower final $F$ value at $V_I = +0.7$ for higher $V_O$ values. Whilst at this point I have simply used the correlation method to demonstrate a loss in exact self-similarity as the Sinai diffuser is removed, I will be revisiting the correlation method in §6.6 to present a model for how this transition occurs.

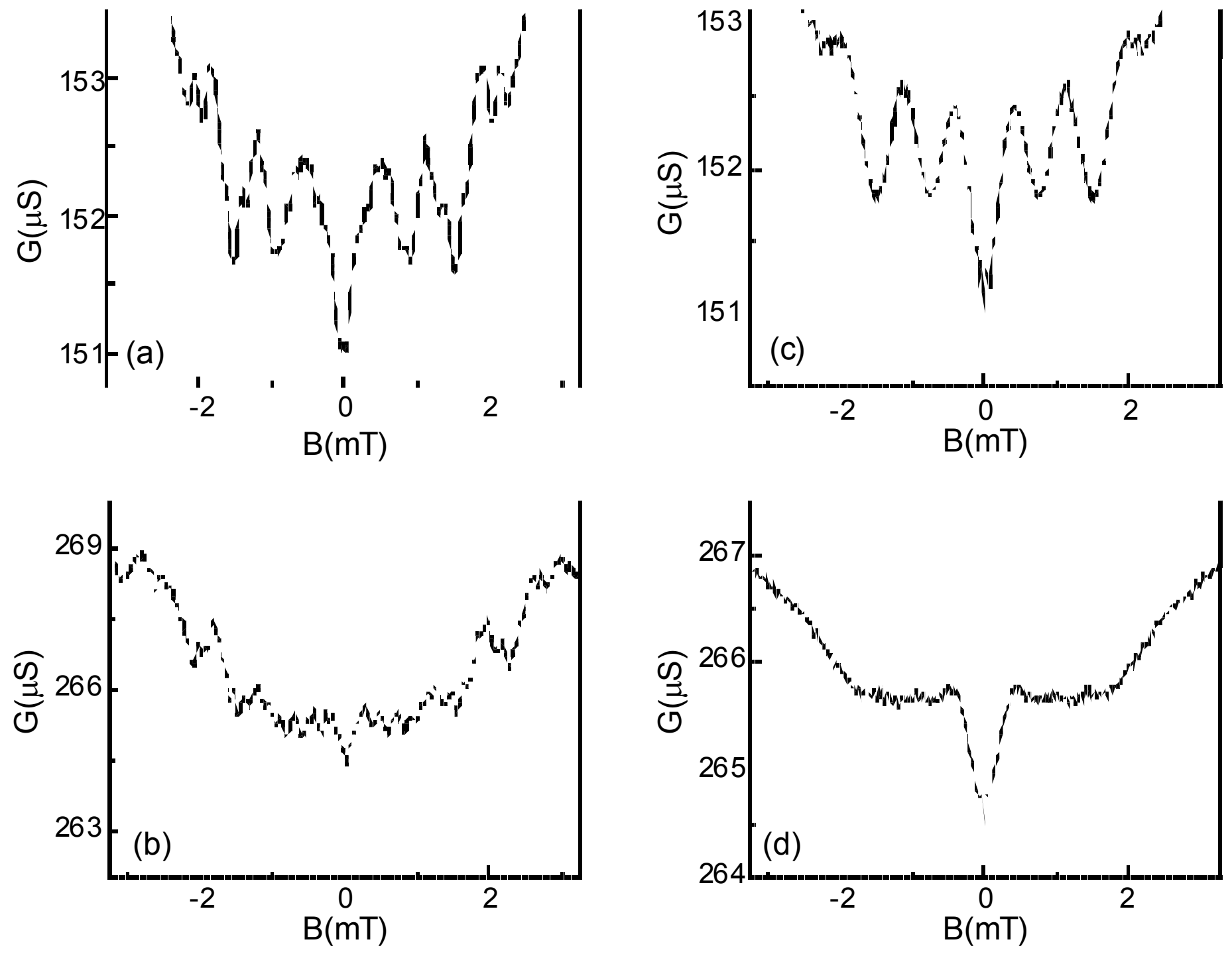

Figure 5.19: Fine structure data for (a) $V_I = -2.7$V and (b) $V_I = +0.7$V corresponding to the Sinai and square geometries respectively. (c) shows the fine structure level produced by the refined Weierstrass model for $\Delta A_I = 1.6\times10^{-15}\text{m}^2$. (d) shows the fine structure level produced by the refined Weierstrass model when $\Delta A_I$ is increased to $3.0\times10^{-16}\text{m}^2$.

Given the success of the Weierstrass model in reproducing the experimental data, further investigations were performed to see if the Weierstrass model could also reproduce the Sinai-square transition without having to break from the basis of the model. That is, by retaining the summation of narrow distributions $\Delta A$ of damped cosines ($h/eA$ and $h/2eA$) with a power-law relationship between the area and amplitude of subsequent structural levels. By strictly adhering to the basis of the Weierstrass model, the removal of exact self-similarity can not be achieved. This is because it is this basis that generates the exact self-similarity in the model. However, the appearance of the data in the transition was found to match that of the experimental data by increasing the finite area distribution $\Delta A_1$ in the refined Weierstrass model. Figure 5.19(a) and (b) show the fine structure for $V_I = -2.7$V and +0.7V respectively. Figure 5.19(c) shows fine structure obtained from the refined Weierstrass model which is remarkably similar to the experimental data shown in Fig. 5.19(a). By increasing the area distribution $\Delta A_1$ from $1.6\times10^{-15}$m$^2$ to $3.0\times10^{-16}$m$^2$ the fine structure shown in Fig. 5.19(d) is obtained. Note that aside from the considerably larger central peak in Fig. 5.19(d) compared to Fig. 5.19(b), the resemblance between the Weierstrass model and the data is striking. These results suggest that the action of the Sinai diffuser in the transition to exact self-similarity may be to eliminate all but a few possible electron trajectories leading to the clusters of self-similar structure observed in the experiment. This proposed explanation for the suppression of exact self-similarity is discussed in further detail in §6.6.

Prior to concluding, it is important to note that the suppression of exact self-similarity observed as the Sinai diffuser radius is decreased raises an important question. Is the MCF for the square billiard still fractal despite the fact that it no longer exhibits exact self-similarity? This issue is the subject of the next Chapter.

# 5.7 – Summary and Conclusions

In summary, an experimental investigation of electron transport through a semiconductor Sinai billiard was performed. The aim of this experiment was to investigate the continuous transition between a classically non-chaotic (square) and a classically chaotic (Sinai) geometry in the regime where quantum interference effects are possible. Magneto-conductance fluctuations (MCF) were used to investigate electron transport through the billiard. Comparisons between high and low temperature traces (4.2K and ~30mK respectively) identified the presence of two features – a central conductance minimum and a set of minima on the shoulders of this central feature – generated by quantum interference effects. As the shape was continuously evolved from the square to the Sinai geometry a continuous evolution in these features was observed. Further investigation of the magneto-conductance revealed structure clustered on two distinct scales – coarse and fine – differing by a factor of approximately 20 in magnetic field. For the Sinai billiard, the two features that evolved through the transition were present on both scales. Visual comparison between structure on the two scales was strongly suggestive of the presence of exact self-similarity in the MCF. A correlation analysis approach was introduced to quantify the exact self-similarity observed in the data. The correlation parameter $F$, gives a value of 1 for structural levels that are exactly identical and a value of zero for data-traces that are randomly related. For the Sinai geometry a comparison between the coarse and fine levels gave $F$ values exceeding 0.95 indicating the presence of exact self-similarity between these levels.

Further investigations revealed the presence of ultra-fine structure approximately a factor of 20 smaller than the fine structure and superimposed on it, as well as ultra-coarse structure approximately 20 times larger than, and supporting the coarse structure. Whilst the visual similarities between the ultra-coarse and coarse, and ultra-fine and fine structures are not as striking as for the coarse and fine structures, a reasonable match is observed given that the ultra-coarse and coarse levels lie at the upper and lower experimental limits. The upper and lower limits of the exact self-similar behaviour were found to be due to a changeover to skipping orbit transport once the cyclotron radius of the electron becomes smaller than half the width of the billiard, and limited resolution and noise in the data, respectively. The hierarchy of self-similar levels is observed to lie on a power-law line in a plot of conductance scale versus magnetic field scale. The gradient of the power-law line is directly related to the fractal dimension $D_F$ of the MCF. The fractal dimension was observed to vary as a function of $V_O$ changing from 1.55 at $V_O = -0.51$ to 1.45 at $V_O = -0.55$V. In the Sinai billiard, fractal behaviour, observed between the upper and lower cut-offs, extends over 3.7 orders of magnitude, comfortably exceeding the average range of fractal behaviour reported in other physical systems.

A simple model based on theoretical studies of semiclassical transport in hard-wall Sinai billiards, the self-similar properties of the Weierstrass function and the scaling behaviour observed in the MCF was formulated. Following a fit of this model to the experimental data to establish the values of various parameters in the model, an initial resemblance to the data was apparent. Further modifications to the initial model to account for two key physical factors in the Sinai billiard – the presence of both $h/e$ and $h/2e$ oscillations, and the finite distribution of areas contributing to each scale – generated a remarkable match to the observed data.

The effect of a continuous transition between the Sinai and square geometries – by varying the bias applied to the central surface-gate – demonstrated that the exact self-similarity observed in the Sinai billiard is removed as the Sinai diffuser radius is decreased. This was confirmed using a correlation analysis where the correlation factor $F$ was observed to decrease to as low as 0.1 during the transition to the square geometry. The observed transition raises an important question. Does the MCF in the square geometry remain fractal despite the suppression of exact-self similarity? This question is addressed in the following Chapter.

# Chapter 6 – Statistical Self-similarity in Billiard MCF

In Chapter 5 the presence of exact self-similarity in the Sinai billiard was established. It was also observed that by removing the Sinai diffuser to transform the Sinai billiard into an empty square billiard, this exact self-similarity was gradually suppressed. This observation raised an important final question in Chapter 5 of whether the MCF observed in the square billiard remains fractal, despite the loss of exact self-similarity. In this Chapter I pursue this issue further.

This Chapter commences with a discussion of methods for assessing fractal behaviour in MCF where exact self-similarity is not observed. §6.1 discusses a common method for assessing the fractal dimension of data-traces – the box counting method – and its application to the analysis of MCF data. The box-counting method is the foundation for two of the three fractal analysis methods used in this chapter – the variation method and the horizontal structured elements (HSE) method. The third method is the variance method, which is based on the theory presented in [12]. These methods and their application to the MCF data are presented in §6.2. §6.3 presents an analysis of MCF data obtained from a square billiard, establishing that the MCF does indeed remain fractal, despite the loss of exact self-similarity. Further, the scaling properties revealed in the fractal analysis indicate the presence of statistical self-similarity (see §2.3.2) in the MCF of the square billiard. A discussion of the properties of this statistical self-similarity is presented in §6.4. §6.5 discusses work presented by Sachrajda *et al.* [15] that claimed the presence of statistical self-similarity in the MCF of the Sinai billiard. I provide a re-analysis of the data presented in [15] demonstrating that this claim is actually incorrect and that their analysis instead confirms the presence of exact self-similarity in the Sinai billiard. Following from these results it is clear that fractal behaviour is present in the MCF of both the Sinai and square geometries and that a transition between exact and statistical self-similarity occurs as the Sinai diffuser is removed. §6.6 presents a unified picture of exact and statistical self-similarity in billiards that offers a possible mechanism for the Sinai-square transition. The conclusions reached in this chapter are presented in §6.7.

# 6.1 – Assessing the Fractal Dimension of Data-traces: The Box-counting Method

There are a large number of methods available for assessing the fractal dimension of an object, such as a data-trace, that has a topological dimension of one [2,3]. One of the commonly used methods is the box-counting method, largely due to the simplicity of implementing it compared to other methods. A brief initial discussion of the box-counting method for assessing the covering fractal dimension was presented in §2.3.2. I will now go on to discuss issues relating to the implementation of the box-counting method to assess the fractal dimension of data-traces rather than geometrical objects.

The most significant difference between geometrical patterns, such as the triadic von Koch curve and the Pollock artwork discussed in §2.3, and data-traces, such as the magneto-conductance traces obtained from semiconductor billiards, are the units of the two axes. For the geometrical objects, both axes have dimensions of length and are hence directly comparable, allowing the easy definition of a square of size $\varepsilon \times \varepsilon$. In contrast, data-traces have different dimensions on the two axes (e.g. $B$ and $G$ for magneto-conductance traces). The relative scaling of the two axes is arbitrary, causing the $\varepsilon \times \varepsilon$ squares, and consequently the fractal dimension, to depend upon the aspect ratio of the trace (i.e. relative scaling of the two axes) [15]. This issue is most easily resolved by a linear mapping of the data-trace onto a two-dimensional space that ranges from 0 to $R$ in both axes, where $R = B_{max} - B_{min}$, prior to analysis. The advantage in this choice of $R$ (as opposed to say $R = 1$) is that the width of the squares in the mesh bear a direct relationship to the field scale $\Delta B$ of the magneto-conductance trace. Indeed, hereafter the squares will be described as $\Delta B \times \Delta B$ rather than $\varepsilon \times \varepsilon$ to emphasise this. The added advantage of this solution to the aspect ratio problem as a whole, is that dividing the two axes by integers $i$ (i.e. $\Delta B = R/i$) guarantees a mesh of non-overlapping squares that exactly covers the entire $[0 \rightarrow R, 0 \rightarrow R]$ space. Note however, there are other solutions to the aspect ratio problem; dividing the $G$ axis into boxes of width $\Delta B$ as proposed in [15] is amongst them.

After mapping the data-trace onto the $[0 \rightarrow R, 0 \rightarrow R]$ space, the box-counting method follows that discussed in §2.3.2. A mesh of squares $\Delta B \times \Delta B$ covering the $[0 \rightarrow$ $R$, $0 \rightarrow R$] space is generated where $\Delta B = R/i$ is obtained by dividing the range $0 \rightarrow R$ by a positive integer $i$. The box-count $N(\Delta B)$ is then obtained by counting the number of $\Delta B \times \Delta B$ squares that contain some part of the data-trace, as shown in Fig. 6.1.

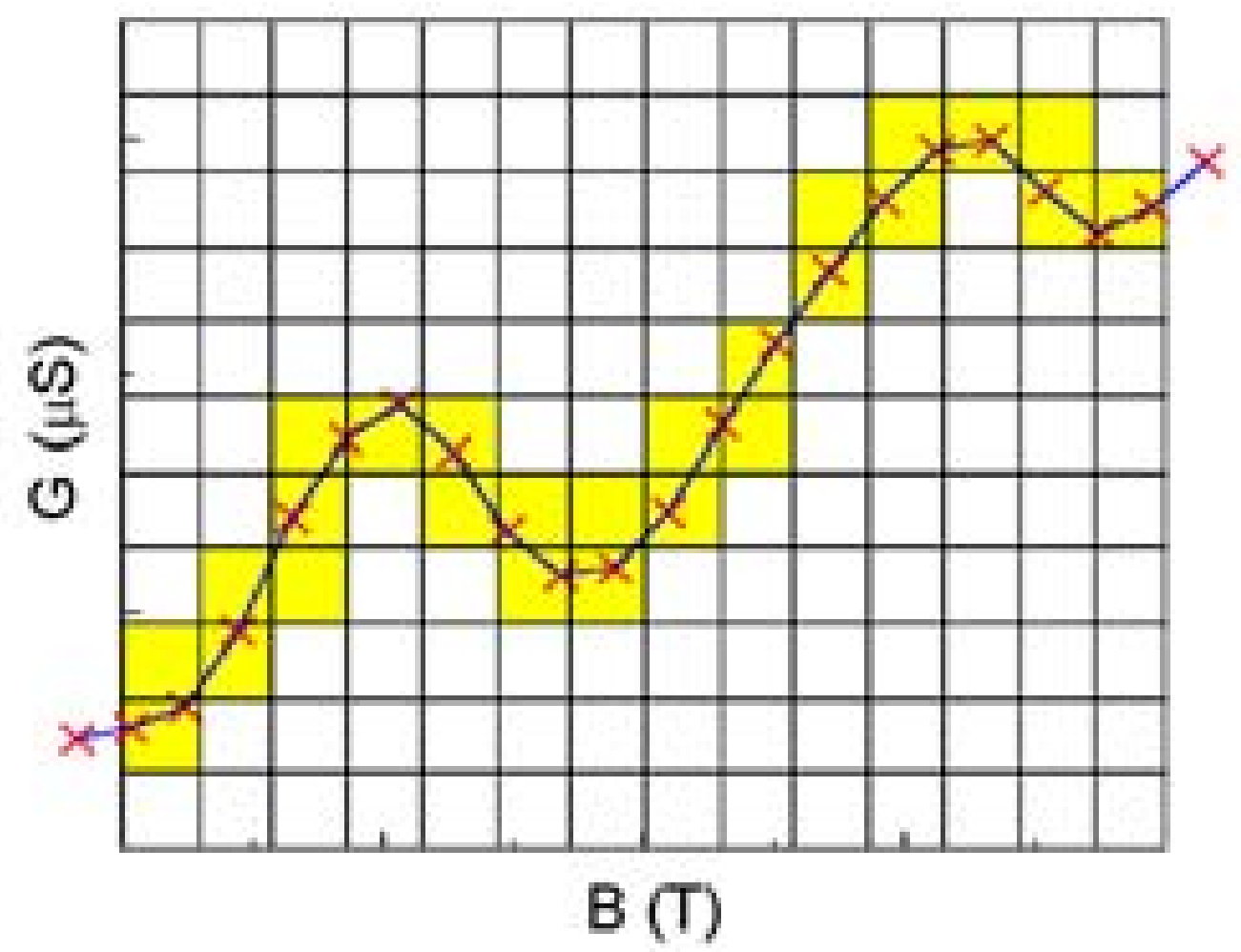

Figure 6.1: Schematic of the box-counting technique. Yellow squares contain some part of the data-trace and hence contribute to the box-count $N(\Delta B)$.

This counting procedure is repeated for successive integers $i$ within the range $1 \leq i \leq j$, where $j$ is the largest $i$ such that the resulting $\Delta B$ value is larger than the resolution of the data-trace $B_{res}$ (i.e. $\Delta B = R$, $R/2$, $R/3$, $R/4$, etc). If the data-trace is fractal over some range of $\Delta B$ values, then $N(\Delta B)$ will be related to $\Delta B$ by the power-law:

$$N(\Delta B) \propto \Delta B^{-D_F} \tag{6.1}$$

Such power-law behaviour appears linear in a plot[14] of $-\log N(\Delta B)$ versus $\log \Delta B$, as shown by the black crosses in Fig. 6.2(a). The fractal dimension $D_F$ is obtained as the gradient of the linear region in this plot. Note that it is possible to generate the $\Delta B$ values in other ways, such as by dividing the range $0 \rightarrow R$ by successive powers of two (i.e. $\Delta B = R$, $R/2$, $R/4$, $R/8$, etc). This produces the red triangles in Fig. 6.2(b). Whilst obtaining the $\Delta B$ values by dividing the range by powers of two results in a much faster analysis, dividing by integers is superior as it provides a significant increase in the number of points in the log-log plot. This allows a better assessment of deviations from linear behaviour and more reliable gradients to be obtained.

[14] Hereafter, plots of $-\log N(\Delta B)$ versus $\log \Delta B$ will also be referred to as log-log plots for convenience.

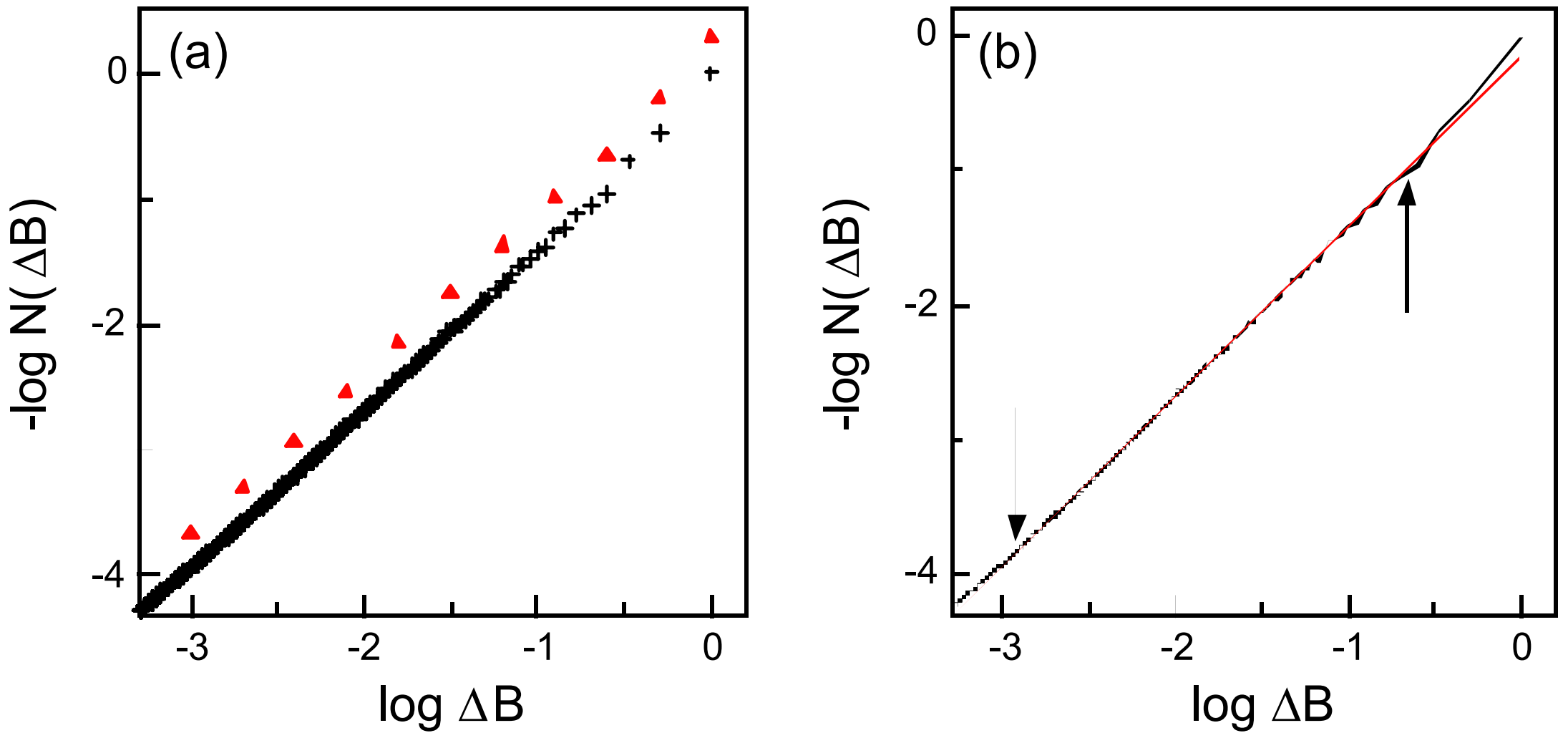

Figure 6.2: (a) Plot of –log $N(\Delta B)$ versus log $\Delta B$ obtained using the box-counting technique on a model fractal trace. The red triangles and black crosses correspond to obtaining $\Delta B$ values by dividing the data range by successive powers of two and successive integers respectively. The red triangles are vertically offset by +0.3 for clarity. (b) Linear fit (red) to the linear interpolation (black) of the black crosses in (a). The arrows indicate the upper and lower analytical cut-offs discussed in the text.

As discussed in §2.3.3, fractals in physical systems are generally observed over a finite range between upper and lower cut-offs. These cut-offs may have one of two origins – analytical and physical. The upper and lower cut-offs are determined by the overlap of the range between the physical limits and the range between the analytical limits (see §2.3.3). The physical limits depend upon the system under investigation and, in the case of MCF, are discussed in §6.3. The analytical cut-offs however, are a property of the specific analysis method employed and the data-trace itself. In the box-counting method, the upper analytical limit is due to the limited ability of large $\Delta B$ squares to assess features in the data-trace. This can be seen by considering that the largest $\Delta B$ is equal to the total width of the data-trace $R$ and hence there is a single square that must be occupied. The next $\Delta B$ is equal to $R/2$ (i.e. four squares covering the data-trace) and $N(\Delta B)$ in this case must be either 4, 3, or 2. Hence, the line joining these points in the log-log plot can only have a possible gradient of 2, 1.5 or 1 respectively. For $\Delta B = R/3$, between 3 and 9 squares are occupied, increasing the number of possible values of the gradient $D_F$ of the line joining $\Delta B = R/2$ and $\Delta B = R/3$ on the log-log plot, and so on. Typically, once $\Delta B < R/7$ occurs (corresponding to 49 squares in the mesh) there is a sufficient number of possible $D_F$ values between subsequent points on the log-log plot that one of them is equal to the real $D_F$ of the data-trace. However, the precise upper analytical limit can only be determined on a case-by-case basis by observing the location of the deviation from linear behaviour in the log-log plot for large $\Delta B$.

The lower analytical cut-off is due to the interpolation between data-points. For $\Delta B$ values smaller than the data resolution, the technique detects the linear interpolation between data points. Since this is non-fractal, the fractal dimension $D_F$ becomes equal to one in this limit. It is important to note that interpolation can have an effect before $\Delta B$ becomes smaller than the data resolution. Hence it is not possible simply to assign the location of the lower analytical cut-off as $\Delta B = B_{res}$. As with the upper analysis cut-off, the location of the lower analysis cut-off can only be accurately determined by observing the $\Delta B$ value where deviations from linear behaviour begin to occur. Note that the analysis cut-offs can be distinguished from physical cut-offs by changing the range of the analysis or the resolution of the data, since the location of the physical cut-offs is fixed by the physical properties of the system generating the fractal behaviour. The upper and lower analytical cut-offs obtained by analysing a mathematically-generated fractal trace (with no physical cut-offs) are indicated by arrows in Fig. 6.2(b).

There are two major disadvantages in the implementation of the box-counting method for analysis of data-traces. The first disadvantage is related to the exact definition of the fractal dimension. Strictly, the definition of the box-counting fractal dimension is the *minimum* number of non-overlapping boxes required to cover the trace as a function of $\Delta B$. Note, that while the boxes in this method are not overlapping, the rigidity provided by the mesh leads to occasional over-counting of the number of boxes required to cover the data-trace. This over-counting problem is demonstrated in Fig. 6.3. Whilst one might simply expect that over-counting leads to an over-estimate of the fractal dimension, it can also lead to concavity in the log-log plot for fractal behaviour since the over-counting occurs to different extents for different $\Delta B$ values [173].

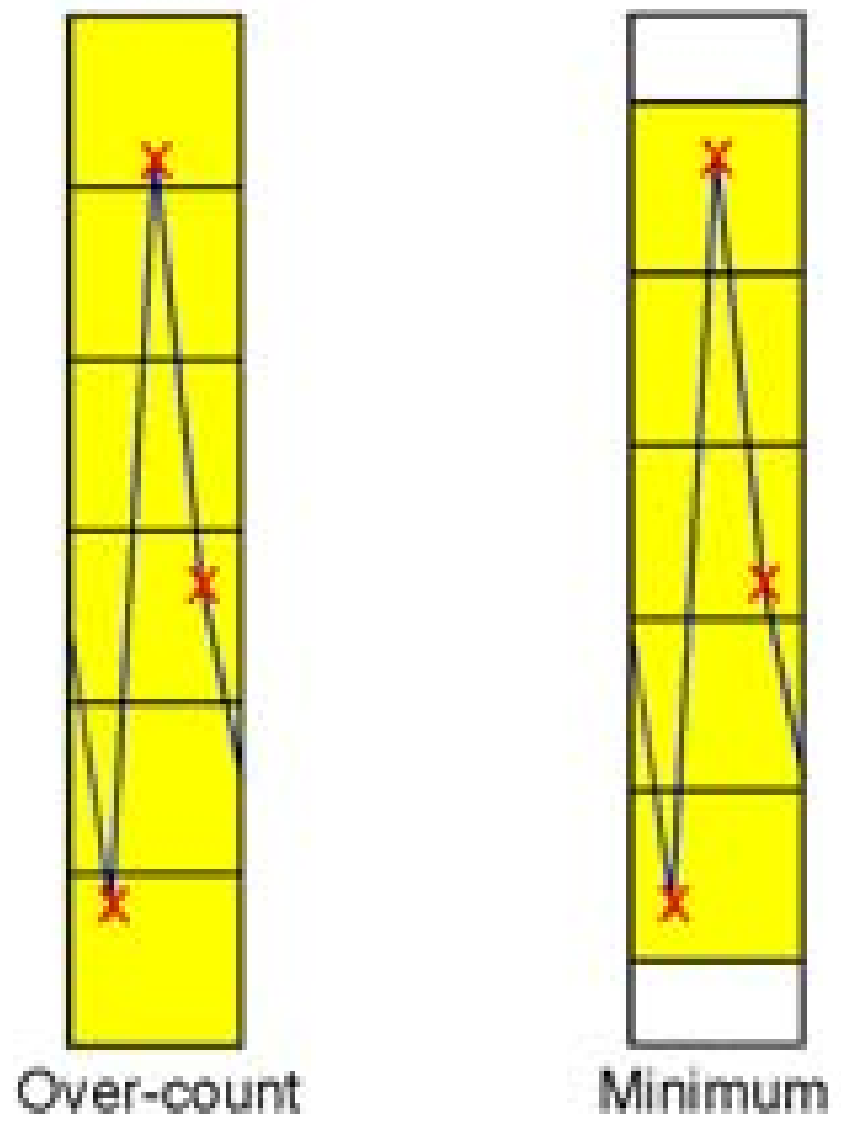

Figure 6.3: Schematic demonstrating the problem of over-counting due to the use of a rigid mesh in the box-counting method. Whilst the minimum number of boxes required to cover the data in this example is five, the requirement for boxes to be aligned horizontally means that six, rather than five boxes, are counted.

The second disadvantage is the computational difficulties in the method. Calculating the number of boxes containing some part of the data-trace entails examining each box and determining not just whether one of the data points is contained within it but also whether the linear interpolation between data points passes through the box. Note that the interpolation is an important part of the data-trace otherwise the data is simply a cluster of points in a two-dimensional space. This requires several million iterations of a routine searching for some part of the data-trace lying in a particular box, a process which is extremely time-intensive (taking up to days of computer time). Careful programming based on the fact that the magneto-conductance data is a mathematical function of magnetic field (i.e. there is a single $G$ value for each $B$ and that the graph of $G(B)$ is everywhere continuous) allows this situation to be improved slightly. The variation and HSE methods, discussed in §6.2.1 and §6.2.2, avoid both of these disadvantages.

# 6.2 – Fractal Analysis Techniques

Three techniques for the analysis of fractal behaviour have been used in this thesis. These are the variation method, the horizontal structured element (HSE) method and the Ketzmerick variance technique. Each of these techniques is discussed in detail in the following subsections, followed by a comparison of the methods and a brief discussion of their application to the analysis of the data in §6.2.4.

## 6.2.1 – The Variation Method

The variation method [173,174] is an improvement on the box-counting method for obtaining the fractal dimension of a data-trace. Unlike the box-counting method, this method relies on the fact that the magneto-conductance is a mathematical function of magnetic field. In this method the magnetic field range of the data is divided into $i$ non-overlapping columns of width $\Delta B$ centred at $B_i$ as shown in Fig. 6.4(a). For each column, the column cover $\nu(B_i,\Delta B)$ is:

$$\nu(B_i,\Delta B) = \max G(B') - \min G(B') \tag{6.2}$$

where $B'$ lies between $B_i - ½\Delta B$ and $B_i + ½\Delta B$ (i.e. within the particular column). The total area of $\Delta B \times \Delta B$ boxes covering the trace is the cover $V(\Delta B)$:

$$V(\Delta B) = \Delta B \sum_i \nu(B_i,\Delta B) \tag{6.3}$$

Hence, the total number of boxes is $N(\Delta B) = \mathrm{Int}(V(\Delta B)/(\Delta B)^2) + 1$. This is the minimum number of non-overlapping boxes required to cover the data-trace and hence it overcomes the over-counting problem raised with the box-counting method in §6.1. $D_F$ is obtained by a linear fit to $-\log N(\Delta B)$ versus $\log \Delta B$ as shown in Fig. 6.4(b).

In terms of computational requirements, the variation method is simpler than the box-counting method because instead of having to perform $1/(\Delta B)^2$ iterations to obtain $N(\Delta B)$ at each $\Delta B$ value, the variation method only has to perform $1/\Delta B$ iterations.

There is also a significant reduction in computation time since the assessment of box occupation in the box-counting method requires a large number of operations whilst the variation method only needs to calculate the maximum and minimum value of each column. In reality, what takes many hours for the box-counting method requires only seconds for the variation method.

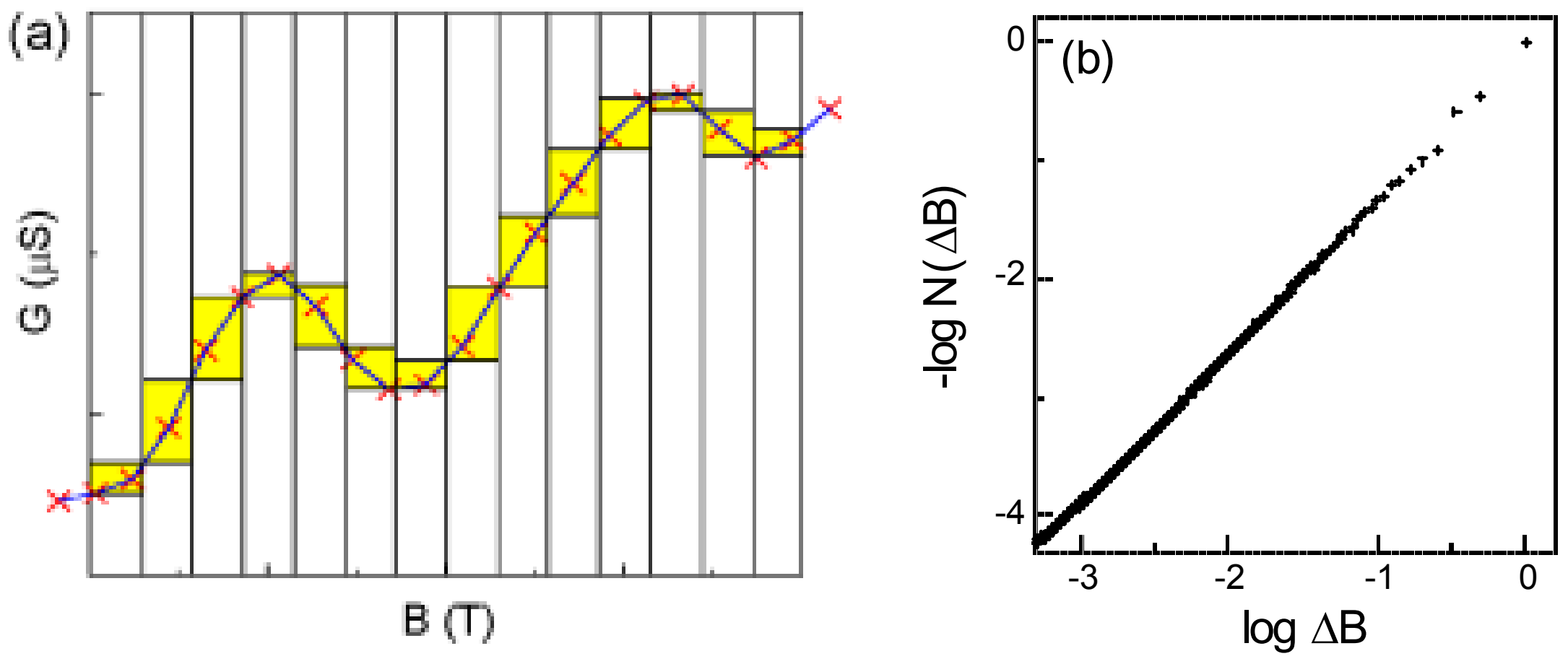

Figure 6.4: (a) Schematic of the variation method. The yellow regions indicate the contribution from each column to the $\Delta B$-variation. (b) Plot of $-\log N(\Delta B)$ versus log $\Delta B$ for the variation method applied to a model fractal trace. The fractal dimension $D_F$ is obtained as the gradient of a linear fit to this data.

An important consideration in the variation method, however, is the real maximum and minimum conductance value in the columns. Initially, one would expect to be able to simply search the data in a particular column and take $v(B_i,\Delta B)$ as the difference between the maximum and minimum conductance values within that column. However, this ignores the fact that the interpolation between data points can extend beyond the maximum/minimum data point conductance values within a particular column, as shown in Fig. 6.5(a). Ignoring any interpolation that extends beyond the minimum and maximum conductance points in the column, as shown in Fig. 6.5(a – left), produces the red trace shown in the log-log plot of Fig. 6.5(b). In comparison, the blue trace in Fig. 6.5(b) was obtained accounting for the interpolation as shown in Fig. 6.5(a – right). Note that there is little effect for large $\Delta B$. However, as $\Delta B$ decreases the line produced by ignoring the interpolation deviates, eventually reaching a minimum and heading rapidly upwards again. The deviation of this trace from the trace where the interpolation is accounted for corresponds to under-counting of the number of boxes required to cover the data-trace. The minimum location in the line produced by ignoring the interpolation coincides with the $\Delta B$ value where columns begin to contain only a single data point, causing $\nu(B_i,\Delta B)$ for such columns to have a value of zero.

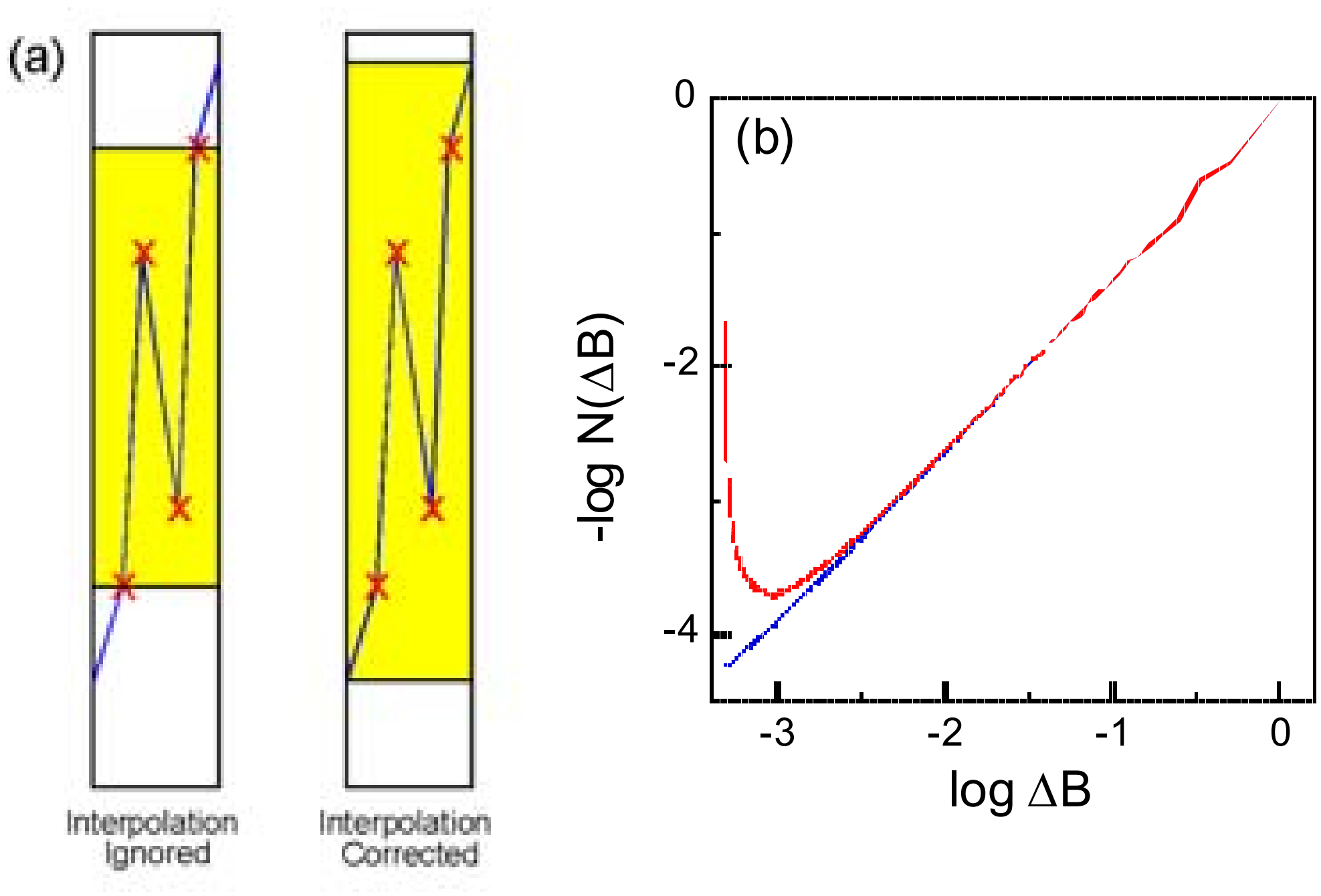

Figure 6.5: (a) Schematic illustrating the difference in calculating the column cover for the variation method by ignoring the interpolation between data points (left) and accounting for the interpolation between data points (right). (b) Plot of –log $N(\Delta B)$ versus log $\Delta B$ for the variation method where the interpolation is ignored (red line) and the interpolation is accounted for (blue line).

As $\Delta B$ continues to decrease, the number of columns with zero contribution to $N(\Delta B)$ increases, causing $N(\Delta B)$ to decrease rapidly and leading to the steep upward section at low $\Delta B$ in the log-log plot. Note that, whilst the main effect of ignoring the interpolation occurs for small $\Delta B$, there are significant deviations in the linear region of the log-log plot. For this reason, the interpolation must be considered in the variation method and the fractal dimension is always obtained by fitting to the blue line rather than the red line. However, the method where the interpolation is ignored is still useful in determining the location of the lower analytical cut-off since the effect of the limited resolution is considerably stronger in this method.

## 6.2.2 – Horizontal Structured Elements (HSE) Method

Whilst the HSE method still assesses the fractal dimension of a data-trace, it does so in a different way to the box-counting and variance methods. The HSE method commences by placing a mesh of squares of width $\varepsilon$ over the data-trace. Typically $\varepsilon$ is chosen to be close to the data resolution of the trace. Squares in the mesh that contain some part of the data-trace are assigned a value of zero. Values are assigned to other squares based on their horizontal distance to the nearest square of value zero. In other words, if a square is next to a square of value zero it is assigned a value of one, if it is the next square along it is assigned a value of two and so on, as demonstrated in Fig. 6.6(a). The distances are assigned entirely in the horizontal direction, even though a zero-valued square may be closer in some other direction. This is in order to avoid concavities in the log-log plot due to overlapping cover elements (see [173] for details), which produce a similar effect to the over-counting problem in the box-counting method. Once the distance values have been assigned, a count is made, firstly of the number of squares of value zero, $N(0)$, then of the number of squares of value zero or one, $N(0,1) = N(0) + N(1)$, then zero or one or two, $N(0,1,2) = N(0) + N(1) + N(2)$, and so on. This is analogous to counting boxes of width $\Delta B = \varepsilon, 2\varepsilon, 3\varepsilon$, etc. However, these $\Delta B \times \Delta B$ boxes are larger than those in the HSE mesh by a factor of $\Delta B^2/\varepsilon^2$. Hence the relationship between counts is $N(\Delta B) = \varepsilon^2 N(0,1,2,\ldots,k)/\Delta B^2$ where $\Delta B = (k - 1)\varepsilon$ (i.e. $N(\varepsilon) = N(0)$, $N(2\varepsilon) = N(0,1)/4$, $N(3\varepsilon) = N(0,1,2)/9$, etc). The fractal dimension is then obtained by a linear fit to a plot of $-\log N(\Delta B)$ versus $\log \Delta B$ as shown in Fig. 6.6(b).

An interesting feature of the HSE method is the point distribution in the log-log plot. Instead of having increasing point density in the low $\Delta B$ direction, as found in the box-counting and variation methods, the point density is highest in the high $\Delta B$ end of the log-log plot. The difference in point distribution between the variation and HSE methods is exploited in §6.3. In terms of computational requirements, the HSE method is better than the two preceding methods because the mesh does not need to be redefined for each $\Delta B$ value. A simple recursive routine can be used to establish the horizontal distance values. However, since the mesh of horizontal distance values needs to be established prior to counting, this program is considerably more memory-intensive than the two preceding methods.

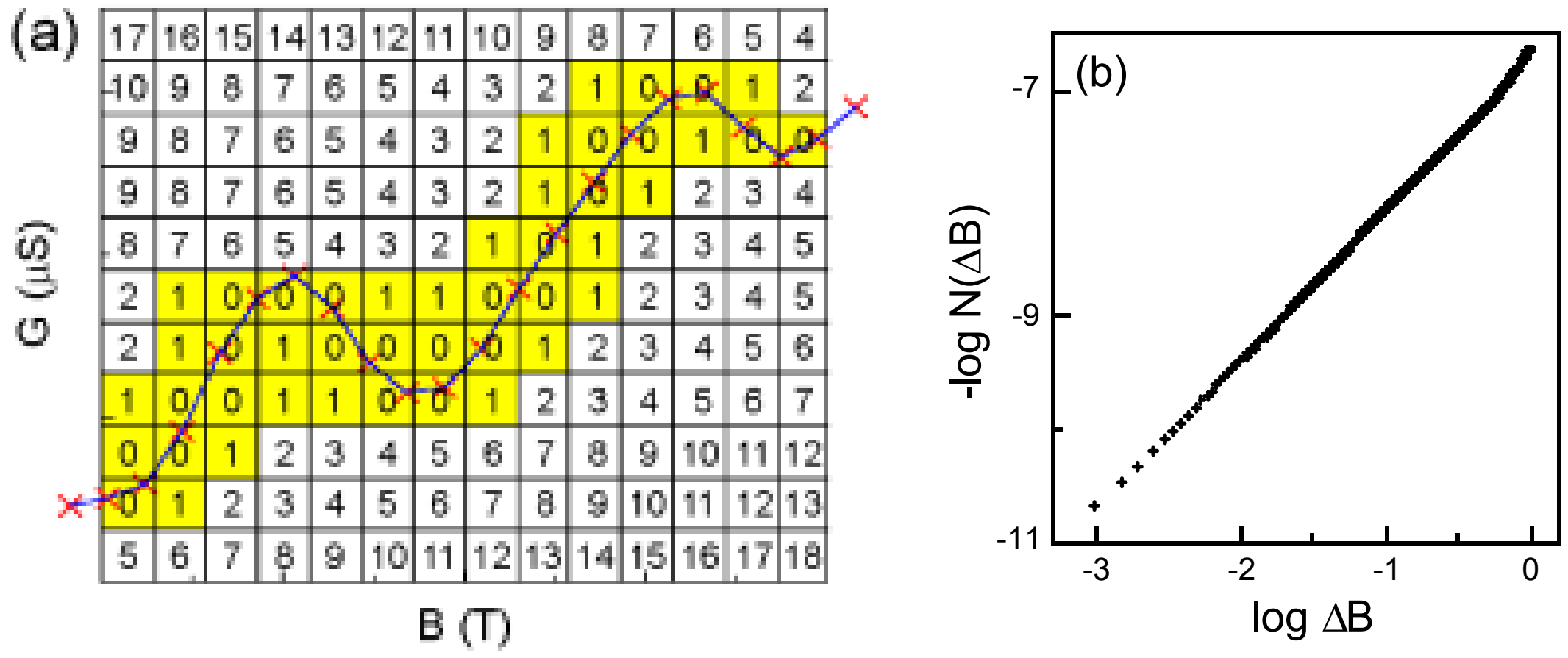

Figure 6.6: (a) Schematic for the HSE method. Numbers in the squares are assigned based on their horizontal distance from the nearest zero-valued square. Squares are assigned value zero if they contain some part of the data-trace. Yellow squares indicate those counted for $N(0,1)$. (b) Plot of $-\log N(\Delta B)$ versus $\log \Delta B$ for the HSE method applied to a model fractal trace. The fractal dimension $D_F$ is the gradient of a linear fit to this data.

Indeed, the data-traces examined in this thesis contain several thousand data points and require a mesh of several million squares, each containing an integer horizontal distance value ranging from zero up to the number of data points. To prevent the program exceeding the memory-limits of the computer it was necessary to implement a virtual-memory routine (i.e. save the mesh to file with direct-access). Ultimately, the HSE method is slower than the variation method but faster than the box-counting method, requiring approximately 30 minutes compared to 1 minute and several hours respectively.

## 6.2.3 – The Ketzmerick Variance Method

This method follows that used by Ketzmerick for the calculation of the variance of the MCF in [12]. It is important to note that this method does not obtain the covering fractal dimension for the data-trace like the two preceding methods. Instead it only extracts the fractal dimension in limited cases where the data-trace is a mathematical function and obeys fractional Brownian statistics [2,12]. It operates by comparing the statistics of the data-trace, namely the variance of the MCF as a function of field increment $\Delta B$, with those expected for fractional Brownian statistics, in order to extract $D_F$ (see §2.4.3). Hence, whilst this method is simple and efficient in a computational sense, its use is severely limited by the requirement for fractional Brownian statistics to hold in order for a meaningful fractal dimension value to be obtained. In this method, the variance is defined as:

$$\left\langle (\Delta G(B,\Delta B))^2 \right\rangle_B = \left\langle [G(B) - G(B+\Delta B)]^2 \right\rangle_B \tag{6.4}$$

and is calculated as a function of $\Delta B$ where the average $\langle\ \rangle_B$ is taken over the maximum possible magnetic field range for a particular $\Delta B$ value (i.e. for all $B$ such that $B+\Delta B < B_{max}$ where $B_{max}$ is the end of the data-trace). If $G(B)$ obeys fractional Brownian statistics, the variance scales with magnetic field increment $\Delta B$ as $\langle(\Delta G(B,\Delta B))^2\rangle_B \propto (\Delta B)^\gamma$ where the fractal dimension $D_F$ is given by $D_F = 2 - \gamma/2$. The exponent $\gamma$ is obtained as the gradient of the linear region in a plot of log $\langle(\Delta G(B,\Delta B))^2\rangle_B$ versus log $\Delta B$, as shown in Fig. 6.7. The fractal behaviour observed in the MCF for the variance method is predicted to be restricted to $\Delta B < B_{Corr}$ where $B_{Corr}$ is the correlation field defined by the auto-correlation function $F(\Delta B) = \langle[G(B)-\langle G(B)\rangle_B][G(B+\Delta B)-\langle G(B)\rangle_B]\rangle_B$ as $F(B_{Corr}) = F(0)/2$ [110]. However, in the data I have analysed I have seen no strong correlation between the upper $\Delta B$ limit of linear behaviour in the log-log plot and the calculated correlation field for the data. This is because in the majority of cases, poor averaging statistics (due to the restriction of $\langle\ \rangle_B$ to $B+\Delta B < B_{max}$), which lead to fluctuations from linear behaviour for large $\Delta B$, take effect at $\Delta B$ values well below $B_{Corr}$. Note that this is even the case for the analysis of a model fractal trace, as shown in Fig. 6.7.

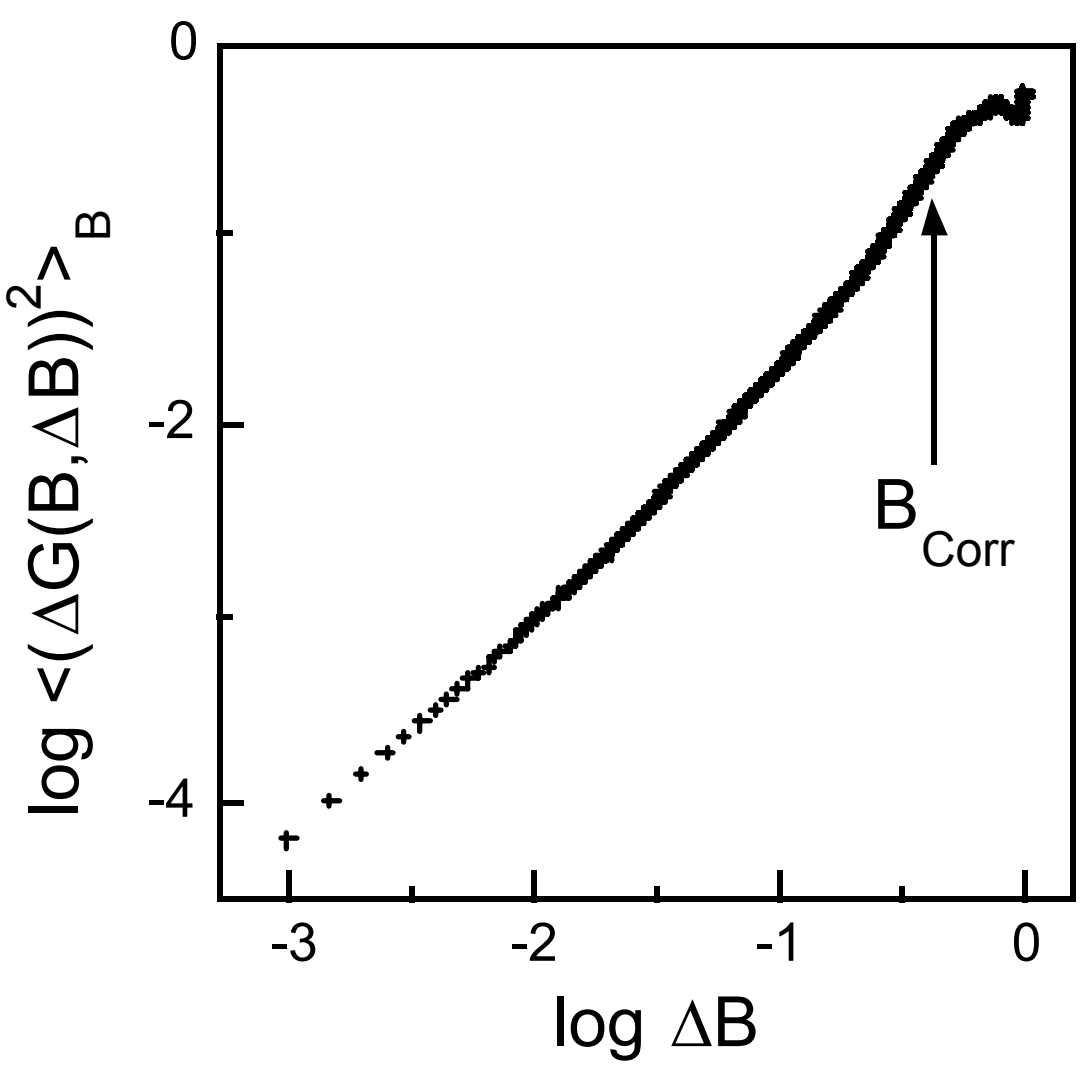

Figure 6.7: Plot of log $\langle(\Delta G(B,\Delta B))^2\rangle_B$ versus log $\Delta B$ for a model fractal trace generated using fractional Brownian statistics. Slight deviations from linear behaviour are due to finite sampling of the generating statistics. $\Delta B = B_{Corr}$ is indicated by the arrow.

An important issue relating to the implementation of the variance technique is the overlap of $\Delta B$ increments. Methods that assess the covering dimension of a data-trace strictly require the cover to be determined using non-overlapping cover elements (e.g. $\Delta B \times \Delta B$ squares in the preceding three methods). In contrast, however, it is not necessary to have 'statistically-independent' (i.e. non-overlapping) $\Delta B$ increments in the variance technique as asserted in [114]. Indeed, there should be no difference (aside from improved averaging) between the results obtained from using overlapping and non-overlapping $\Delta B$ increments. The reason for this is that the variance increments $\Delta G(B,\Delta B) = G(B) - G(B+\Delta B)$ have a certain distribution in $B$, which is being sampled in either case. Taking samples at $B_1$ and $B_2$ such that $B_1+\Delta B$ and $B_2+\Delta B$ overlap will not affect this distribution, even though $\Delta G(B_1,\Delta B)$ is likely to be more correlated with $\Delta G(B_1+1,\Delta B)$ than $\Delta G(B_1+5,\Delta B)$ for example. This means that although the method with the higher averaging will give a more accurate result for the average of this distribution and hence the overall variance, all methods should converge to the same result given that a reasonable average is taken in calculating the variance. To confirm this, three versions of the variance method have been applied to a data-trace to verify that they do indeed produce identical results. The first method operates with non-overlapping $\Delta B$ increments, and with $\Delta B$ values that are obtained by dividing the range of the data-trace $R$ by successive powers of two (i.e. $\Delta B = R$, $R/2$, $R/4$, $R/8$, etc) as used in [15]. Note that this requires that the data-trace has a total number of data points equal to $2^i+1$, where $i$ is a positive integer – a restrictive requirement in conducting experiments. The second method again operates with non-overlapping $\Delta B$ intervals, this time with $\Delta B$ values that are obtained by dividing the range of the data-trace by successive integers. The third method is as used in [113] and in this thesis, and operates with overlapping $\Delta B$ increments with $\Delta B$ values obtained again by integer division of the data-trace range. These methods will be called the non-overlapping power method, the non-overlapping integer method and the overlapping integer method hereafter.

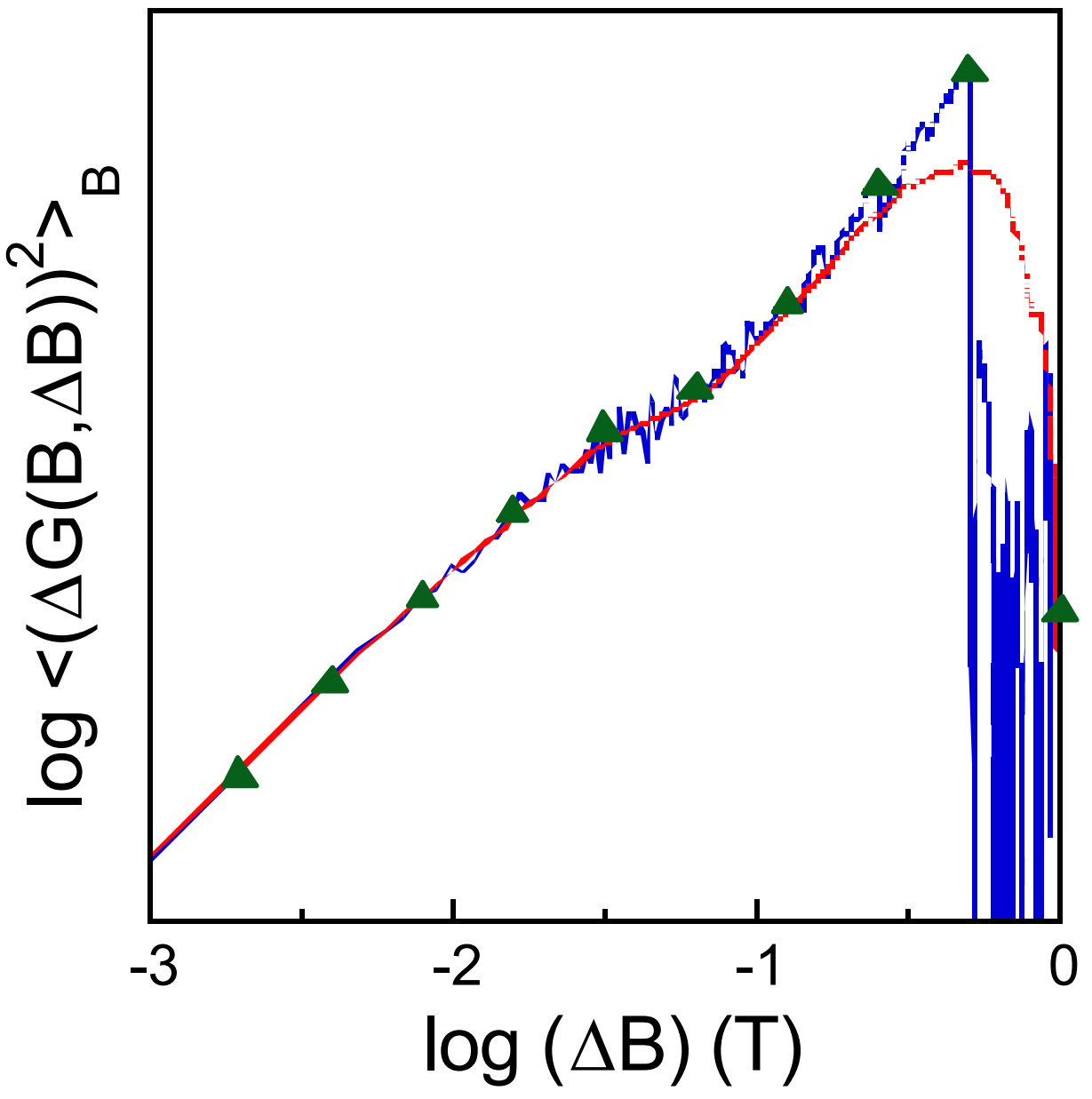

Figure 6.8: A plot of log $\langle(\Delta G(B,\Delta B))^2\rangle_B$ versus log $\Delta B$ for three methods of calculating the variance of a model data-trace: non-overlapping $\Delta B$ intervals with $\Delta B$ determined by power of two division of the data range (green triangles), non-overlapping $\Delta B$ intervals determined by integer division of the data range (blue line) and overlapping $\Delta B$ intervals determined by integer division of the data range (red line). Deviations between the overlapping and non-overlapping data for large $\Delta B$ are due to a break-down in averaging in the non-overlapping methods (see text).

The results of these three methods are presented in Fig. 6.8. Note that linear behaviour is not necessarily expected in the analyses in Fig. 6.8 because the model data-trace used does not obey fractional Brownian statistics. In the low $\Delta B$ limit, the three methods for calculating the variance closely agree. Moving towards larger $\Delta B$ values, the non-overlapping integer method develops structure compared to the other methods. This structure is due to a combination of averaging and point-resolution effects, which are explained further below. No structure appears in the non-overlapping power method simply because there are so few points. However, the non-overlapping integer method, which calculates the variance in the same way but for more $\Delta B$ values, has the resolution to reveal structure (same point-distribution as shown in Fig. 6.7). Note that every green triangle lies precisely on the blue line confirming that these two non-overlapping methods calculate the variance identically. Considering the two integer methods (red and blue lines), the first item to note is that the blue line has considerably more structure than the red line. The non-overlapping integer and overlapping integer methods have matching point distributions but the overlapping integer method has a substantially larger number of variance increments $\Delta G(B,\Delta B)$ contributing to the average in the variance. To be specific, for a data-trace with magnetic field resolution $\Delta B_{res}$ there are $(R - \Delta B)/\Delta B_{res}$ variance elements for the overlapping integer method compared to $R/\Delta B$ for the non-overlapping integer method. Hence the additional structure in the blue line (non-overlapping integer method) is actually statistical noise, which is produced by the significantly smaller number of variance increments $\Delta G(B,\Delta B) = G(B) - G(B+\Delta B)$ that contribute to the average. This averaging effect becomes most significant for $\log(\Delta B) > -1$, where the variance in the non-overlapping integer method is calculated by averaging over four or less variance increments. This leads to unstable behaviour as a function of $\Delta B$ and the significant deviation between the two methods. Due to such statistical problems, in the high $\Delta B$ limit, only the overlapping integer method still gives a statistically accurate assessment of the variance. Ultimately, however, for the purposes of the analysis performed in this thesis, only the low $\Delta B$ range is important and in this region the three methods are equivalent, as expected from the preceding discussion and demonstrated in Fig. 6.8. The overlapping integer method is expected to give the most sensitive and accurate indication of the behaviour of the variance as a function of $\Delta B$ and hence it will be used for calculating the variance hereafter.

An interesting feature of the variance method is its response to the presence of an underlying period in a data-trace, as demonstrated in Fig. 6.9. This feature is exploited later in §6.5. The bottom trace in Fig. 6.9(b) shows the variance method results for the pure cosine curve shown in Fig. 6.9(a) – bottom trace. For small $\Delta B$, $\log\langle(\Delta G(B,\Delta B))^2\rangle_B$ versus $\log\Delta B$ is linear with a gradient $\gamma = 2$, corresponding to a fractal dimension $D_F = 1$, as expected for a pure cosine. For larger $\Delta B$ however, the variance method results develop a series of singularities where $\log\langle(\Delta G(B,\Delta B))^2\rangle_B = -\infty$. For these singularities to occur, $\langle(\Delta G(B,\Delta B))^2\rangle_B$ must equal zero, and for the pure cosine curve this will only take place when $\Delta B$ is equal to a positive integer multiple of the period, so that each variance increment $\Delta G(B,\Delta B)$ is equal to zero.

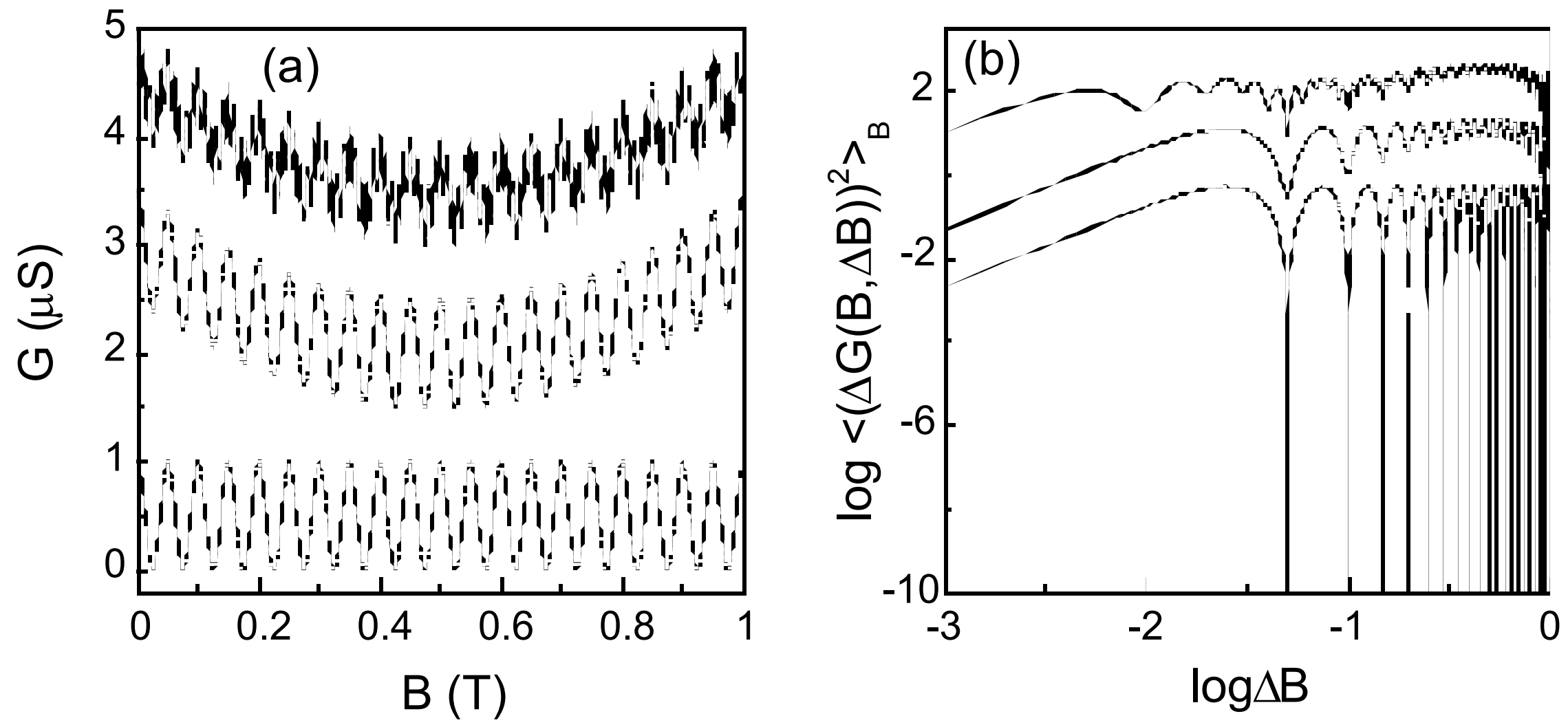

Figure 6.9: (a) Model data-traces used to demonstrate the detection of periodic behaviour in the variance method: (bottom) a pure cosine with a period of 0.05T, (middle) cosine with a period of 0.05T superimposed on a parabolic background and (bottom) a pair of cosines of equal magnitude and periods of 0.05T and 0.01T superimposed on a parabolic background. (b) Plots of $\log\langle(\Delta G(B,\Delta B))^2\rangle_B$ versus $\log\Delta B$ for the model data-traces in (a). The top/middle/bottom trace in (b) corresponds to the top/middle/bottom trace in (a). The top and middle traces in (a) are vertically offset by +3.5μS and +5.0μS and in (b) by +2.0 and +3.5 for clarity.

Examining the bottom trace in Fig. 6.9(b), the first singularity (smallest $\Delta B$) occurs at $\log \Delta B = -1.3$. The location of the first singularity corresponds exactly to a $\Delta B$ equal to the period of the cosine, confirming the statement made above. Subsequent singularities are found to coincide with $\Delta B$ values that are equal to positive integer multiples of the cosine period until $\Delta B$ becomes equal to the total width of the data-trace. To examine the effect of introducing a background and multiple periodicities into the data, the variance method has been applied to the middle and top traces in Fig. 6.9(a). These traces are a single cosine of period 0.05T superimposed upon a parabolic background, and the superposition of a pair of cosines with equal amplitudes and periods of 0.05T and 0.01T and the same parabolic background, respectively. The addition of a smooth background causes the singularities found for the pure cosine to become finite-valued minima. These minima also coincide with $\Delta B$ being a positive integer multiple of the cosine period. The singularities are no longer present at these positions because the background prevents the individual variance increments $\Delta G(B,\Delta B)$ from being equal to zero, giving a non-zero minimum variance. In this trivial case, the exact value of the minima is determined entirely by the background, however for a general data-trace this is not necessarily the case. The introduction of multiple periodicities leads to a significant increase in the number of minima as demonstrated in the top trace of Fig. 6.9(b). In this case, there are a number of 'cascades' of minima, one cascade for each period present in the data-trace. Within each cascade there are minima for all positive integer multiples of the period until $\Delta B$ becomes equal to the total width of the data-trace.

## 6.2.4 – Method Comparison and Application of Methods to the MCF Data

Comparisons of the accuracy of the box-counting, variation and HSE methods have been performed by Dubuc *et al.* [173]. This comparison used three types of standard fractal traces – fractional Brownian, Weierstrass-Mandelbrot and Kiesswetter curves – each of which can be generated with a specified fractal dimension. In this way it is possible to investigate the accuracy of the methods as a function of fractal dimension. It was shown in [173] that the most accurate method was the variation method, followed by the HSE method. In the early stages of this work I investigated a number of methods for assessing the fractal dimension of data-traces (extra to the methods presented in this thesis) with the view of establishing the most appropriate method for the assessment of MCF data. Methods investigated include: the coastline method [2,14], the correlation method [3,14], the Minkowski-sausage method [173], the power spectrum method [173,175], as well as numerous minor variations of the box-counting technique. My comparison between the methods qualitatively agreed with that of [173].

Interestingly, the variance method became quite important in this comparison. Due to time limitations, I restricted myself to using fractional Brownian traces generated using the random midpoint displacement method presented in [90]. Fractional Brownian (FB) traces were used in the calibration process for two reasons. The first is that based on the theory in [12], the MCF data was expected to obey fractional Brownian statistics and hence calibration of the methods using fractional Brownian traces would be ideal. The second is that the variance method can be used to obtain the fractal dimension of the FB-trace based on a comparison between the variance of the FB-trace and the variance expected for fractional Brownian statistics. This means that it is possible to check that the standard FB-trace really has the specified fractal dimension. It was generally found that there was a significant discrepancy between the specified fractal dimension for the statistics and the fractal dimension measured using the variance method for the model FB-traces. This was particularly the case as the specified fractal dimension approached two and for a low number of data points in the generated standard FB-trace. Note that the number of data points has to be a power of two in the method used for generation of the FB-traces [90]. This discrepancy appears to be due to poor sampling of the generating statistics, which produces aberrations from ideal statistics obtained in the final trace. Dubuc *et al.* [173] have not discussed the number of data points in their standard FB-traces so it is difficult to determine if their results are also affected by this problem. The only solution is to increase the number of data points in the standard FB-traces to achieve good sampling of the generating statistics. However, this makes the comparison process prohibitively long since the time required for fractal analysis scales with the number of data points in the data-trace. For the purpose of this work, however, a qualitative indication of the relative accuracy of the methods was sufficient to isolate the appropriate methods for further use. I will rely on the values obtained by Dubuc *et al.* for a rough estimate of the overall accuracy of the variation and HSE methods. These were found to be < 3% and ~5% respectively [173]. Since the variance method is not used for the assessment of the fractal dimension of MCF data in this thesis (see below), accurate calibration of this method was not pursued.

Of the three methods presented in the preceding subsections, only the variation method is used to obtain numerical values of the fractal dimension of MCF traces for three reasons. Firstly, the traditional method for assessing the fractal dimension of data-traces is the box-counting method. Hence to avoid breaking with tradition, the variance method is not used to obtain $D_F$ values. This is further supported by Sachrajda and Ketzmerick who claim “This fractal dimension and the range of fractal scaling can be determined only by a fractal analysis, e.g., by the box-counting algorithm … The reason for performing a variance analysis thus is to check the underlying theory and not to extract the scaling region of the fractal analysis” [114]. However, since the variation method is the more accurate than either the HSE or box-counting methods [173], it is the method of choice for obtaining $D_F$ in this thesis.

Secondly, the variation method is more robust over the total range of analysis. The HSE method remains slightly susceptible to the concavity present in its parent method (the Minkowski-sausage method [173]) causing the HSE method to occasionally deviate from linear behaviour for fractal traces. The third reason is computational. In conjunction with being the most accurate method, the variation method is fast and efficient, allowing the fractal analysis of a large number of data-traces to be performed in a considerably shorter period of time and with decreased computational resource requirements.

The HSE method is used to assist in determining the relevant cut-offs of the fractal behaviour obtained by the variation method – in particular the upper cut-offs. As discussed earlier, the variation method has increasing point density at lower $\Delta B$ values whilst the HSE method has increasing point density at higher $\Delta B$ values. Hence the upper cut-offs are difficult to see in the variation method due to the low point density in the high $\Delta B$ limit. The HSE method, which has maximum point density in this region, is used as a guide in locating the relevant upper cut-offs.

As discussed in §6.2.3, the variance method is highly sensitive to the presence of periodicity in a data-trace, and hence this method is used as a detector of periodic behaviour in a data-trace. Whilst Ketzmerick suggests the use of the variance method as confirmation of his theory, in this Chapter it will only be used as a method for examining the data for underlying periodicities.

# 6.3 – Fractal Behaviour in the MCF Data of the Square Billiard

At the conclusion of Chapter 5, the question of whether the MCF data produced by the square billiard remained fractal despite the absence of exact self-similarity remained unanswered. The methods established in the preceding section are now used to answer this question. A typical MCF data-trace obtained from a square billiard is shown in Fig. 6.10. This data was obtained from the Cambridge double-2DEG device (see §3.6 and §3.7.5) measuring the top 2DEG only, with the lower 2DEG fully pinched-off by the back-gates. This is to avoid any remnant of the Sinai diffuser that might be found in the NRC Sinai billiard at $V_I$ = +0.7V (see §6.5) and because MCF data from this device has been obtained over the largest range (over 3.5 orders of magnitude at continuous resolution). This device is lithographically identical to the NRC Sinai billiard except that the central circular surface-gate and bridging interconnect have not been fabricated, and will be called the Cambridge square billiard for the remainder of this chapter. Note that the MCF data shown in Fig. 6.10 is visually dissimilar to the typical data obtained from the Sinai billiard in a number of ways. The most important difference is the absence of clustering of structure on distinct scales as observed in the Sinai billiard. Instead, for the data presented in Fig. 6.10 there appears to be structure at all scales, suggesting that the exact self-similarity is no longer present, as demonstrated using the correlation function analysis in Chapter 5. Fractal analysis of a typical data-trace obtained from the Cambridge square billiard is presented in Fig. 6.11. Since at this point the focus is simply on demonstrating the presence of fractal behaviour rather than obtaining an accurate value for the fractal dimension of the data, the analysis presented in Fig. 6.11 combines both the variation and HSE method results into a single line. The objective of this is to maximise the point density in the log-log plot in order to detect any deviations from fractal behaviour and allow accurate assessment of the position of the upper and lower cut-offs. Note however, that it is not uncommon for the gradient of the HSE method to differ slightly from that of the variation method, and hence any fractal dimension values are extracted from the variation method results alone.

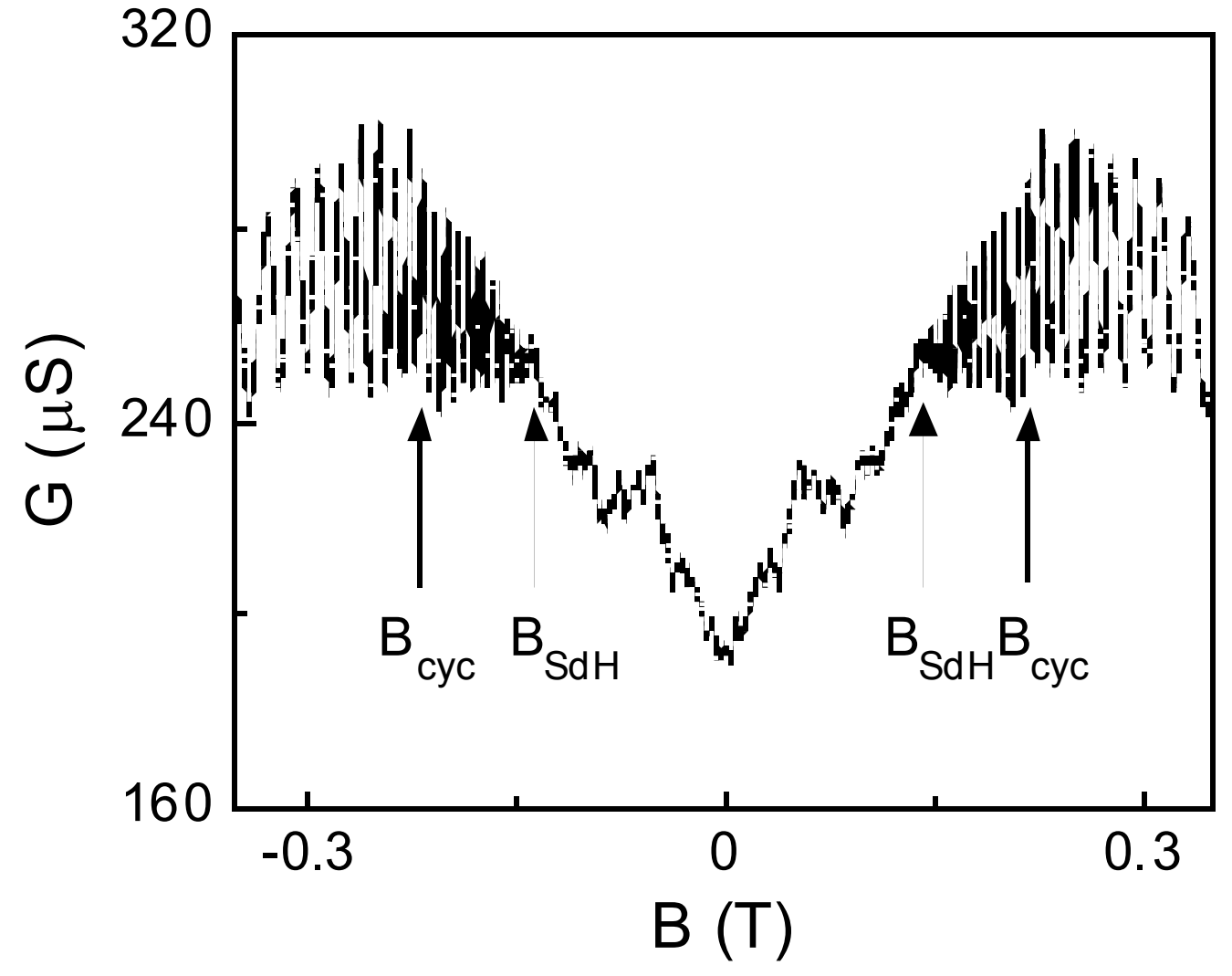

Figure 6.10: Typical MCF data obtained from the Cambridge square billiard for $n$ = 5 and $T$ = 30mK. $B_{cyc}$ and $B_{SdH}$ are discussed in the text.

As discussed in §6.1, fractal behaviour extends between upper and lower cut-offs that are determined by limitations in the analysis and physical considerations of the system under investigation. The locations of these limits are indicated by the arrows marked $\Delta B_1$ through to $\Delta B_5$, above and below the data in Fig. 6.11. The limit marked $\Delta B_1$ is due to finite resolution in the data-trace, as discussed in §6.1, and corresponds to a $\Delta B$ value less than double the magnetic field resolution of the data-trace. For $\Delta B < \Delta B_1$, it is possible for $\Delta B$-width columns in the variation method to contain only a single data point. Hence the analysis method 'sees' two straight lines in that particular column. As $\Delta B$ decreases below $\Delta B_1$, the occurrence of single point columns increases, causing $D_F$ = 1 for $\Delta B < \Delta B_1$ and the loss of fractal behaviour in this limit. $\Delta B_3$ and $\Delta B_4$ are also analytical limits and are due to the inability of the fractal analysis to assess the structure in the data, as also discussed in §6.1.

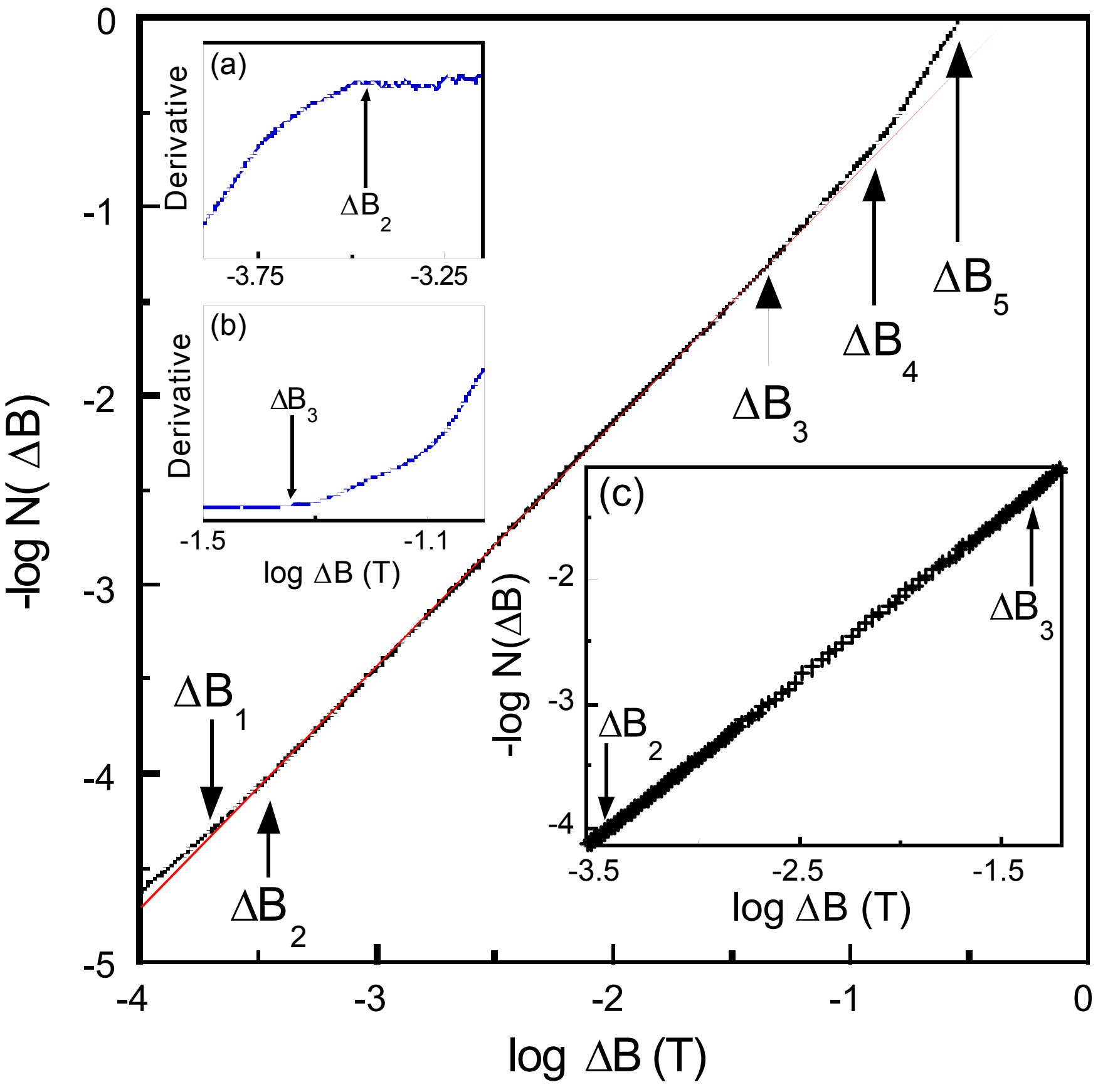

Figure 6.11: Fractal analysis of typical MCF data obtained from the Cambridge square billiard. The data (combined variation and HSE method results) is shown in black. The red line represents a linear fit to the data. The limits $\Delta B_1$ through to $\Delta B_5$ are indicated by the arrows and discussed in the text. Insets (a) and (b) are plots of the derivative of –log $N(\Delta B)$ with respect to log $\Delta B$ in the vicinity of $\Delta B_2$ and $\Delta B_3$ respectively. Inset (c) shows a close-up of the fractal scaling region. Only every $10^{th}$ data point in inset (c) is shown.

The location of $\Delta B_3$ corresponds to 49 squares covering the data in the variation method and for $\Delta B > \Delta B_3$ the analysis begins to break down. In the HSE method, $\Delta B_3$ corresponds to $N(0,1,\ldots,48)$ and for $\Delta B > \Delta B_3$ the cover produced by the HSE method becomes large enough that the structure in the data can no longer have a significant effect on the cover between subsequent counts of $N$ (i.e. $N(0,1,\ldots,48+i)$ and $N(0,1,\ldots,48+i+1)$), where $i$ is an integer greater than or equal to zero. $\Delta B_4$ corresponds to the more extreme situation where there are only four squares covering the data-trace. Between $\Delta B_4$ and $\Delta B_5$ all of the squares remain filled leading to $D_F = 2$ for $\Delta B > \Delta B_4$. In the HSE method, $\Delta B_4$ corresponds to $N(0,1,\ldots,136)$ and for $\Delta B > \Delta B_4$, the cover produced by the HSE method contains the full set of horizontal structured elements. No further changes in the HSE method cover are possible in this limit and $D_F = 2$ is also obtained for this method when $\Delta B > \Delta B_4$.

The remaining two limits, $\Delta B_2$ and $\Delta B_5$ have a physical origin. The location of the upper physical limit $\Delta B_5$ is determined by the minimum magnetic field where one of two processes begins to affect the MCF. The first, and most common, is the change in billiard transport regime that occurs once the electron cyclotron radius $r_{cyc}$ becomes smaller than half the width $W$ of the billiard. For $r_{cyc} < W/2$, the electrons traverse the billiard in skipping orbits along the walls as discussed in §2.2.4, and are no longer expected to produce fractal MCF. The second process occurs for billiards defined in 2DEGs with high electron mobilities, where bulk Shubnikov-de Haas oscillations can appear in the MCF data well before $r_{cyc}$ becomes smaller than $W/2$. Note that the fractal analysis techniques used in this Chapter give an equal weight in the sum to each $\Delta B$ square or column. This means that if non-fractal segments are present, their behaviour as a function of $\Delta B$ will obscure that of the fractal regions at lower fields. For this reason the data-trace is truncated at the highest magnetic field possible such that there are no bulk Shubnikov-de Haas oscillations present and $r_{cyc} > W/2$ for the whole truncated data-trace. MCF data for both positive and negative $B$ values within this range are used in the analysis. As a result $\Delta B_5$ coincides with the full-width of the truncated data-trace. Hence it is not possible to extend the analysis so that the upper physical limit $\Delta B_5$ lies inside the upper analytical limit $\Delta B_3$.

The limit $\Delta B_2$ is due to noise in the experimental data. Note that the MCF should be symmetric about $B = 0$ due to the Onsager relations [62,63]. Hence noise appears as asymmetry in the MCF about $B = 0$ and was identified as the cause of $\Delta B_2$ by a comparitive analysis of the MCF for $B > 0$, $B < 0$, and the average and difference of these two sub-traces for positive and negative $B$. This noise is small compared to the MCF and varies on a scale comparable to the data resolution. Hence the noise contribution to the log-log plot is only significant at small $\Delta B$ and leads to deviations from fractal behaviour in the small $\Delta B$ limit. Noise contributions to the experimental data can be removed in one of two ways. The first is by preventing it completely, in other words, by shielding the experimental system from all sources of electromagnetic and vibrational noise. Whilst in theory this is possible, in practice it is impossible to completely prevent noise from leaking into the system. However, even if all external noise sources were entirely eliminated, there are still noise contributions from the electronics (Johnson noise ($V = (4kTR\Delta f)^{1/2}$) from the resistors, shot noise, amplifier noise (>8nV/Hz$^{1/2}$), component instability, etc [152]) and from the device itself (e.g. telegraph noise, leakage currents, bias instabilities, etc [125]) that it is not possible to eliminate. The second option for removing noise is considerably simpler, and involves averaging a number of MCF traces obtained under a single set of experimental conditions. Since the MCF is identical for data-traces measured under identical experimental conditions, averaging should have no effect upon the MCF. However, the noise contribution is entirely random and should instead average to zero. The averaging process is time-consuming however, particularly when MCF traces under a set of different conditions are required.

As discussed in §2.3.3, the actual range of fractal behaviour is defined as the range common to the span between the physical limits and the analytical limits. Hence fractal behaviour extends only between $\Delta B_2$ and $\Delta B_3$. The precise location of the cut-offs on fractal behaviour, which are the two limits $\Delta B_2$ and $\Delta B_3$, is not entirely clear simply by comparing the linear fit to the log-log data. Insets (a) and (b) in Fig. 6.11 present the derivative of the log-log data in the regions close to $\Delta B_2$ and $\Delta B_3$. A horizontal line in the derivative plot corresponds to a straight line in the log-log plot and hence deviations from the horizontal line in the derivatives correspond to deviations from fractal behaviour. The precise locations of $\Delta B_2$ and $\Delta B_3$ become clear upon considering these derivatives and are marked by arrows in insets (a) and (b). The positions of these cut-offs ($\Delta B_2$ and $\Delta B_3$) in the main figure and inset (c) are based on the derivatives presented in insets (a) and (b). Inset (c) of Fig. 6.11 shows the region between $\Delta B_2$ and $\Delta B_3$ in greater detail. The log-log data between the upper and lower cut-offs ($\Delta B_2$ and $\Delta B_3$) in inset (c) clearly fall onto a good straight line. Any small deviations from exactly linear behaviour in these points are expected to be due to numerical/statistical errors in the calculation methods rather than true deviations from fractal behaviour. Taking a linear fit between $\Delta B_2$ and $\Delta B_3$ for the variation method results yields a fractal dimension $D_F$ = 1.29 for the particular data-trace analysed.

Fractal behaviour is hence observed over a total of 2.1 orders of magnitude rather than the 2.6 orders obtained by a simplistic examination of the linear fit alone. This highlights the importance of a careful analysis of the cut-off locations in assessing the presence of fractal behaviour. Particularly following recent debate [93-98] regarding the presence of fractal behaviour in various physical systems (see §2.3.3), where it is clear that whilst fractals observed over a limited range are no less fractal than those observed over an infinite range, the certainty of fractal behaviour increases with range of observation [95]. Indeed, following this debate, it is clear that merely obtaining some straight line behaviour in a log-log plot with a non-integer gradient is not enough to justify a claim of fractal behaviour. A significant range of observation of the fractal behaviour is also highly important. The 2.1 order of magnitude range of observation of fractal behaviour in the MCF of square billiards presented here is significantly larger than the average reported range of fractal observations in experimental systems [93,94]. It is also larger than the one order of magnitude lower limit on reliably fitting a power-law to data [15], and an order of magnitude larger than observations [14,75,76] made prior to embarking on the work presented in this thesis. Considering also that fractal behaviour has been observed in the Sinai billiard over 3.7 orders of magnitude, I believe there is little doubt that fractal behaviour does exist in the MCF of semiconductor billiards. However it will be the aim of future experiments to extend the range of observation of this fractal behaviour further. This will be achieved by improving the resolution of the data-traces in order to shift the lower cut-off to smaller $\Delta B$ values.

## 6.4 – Properties of Fractal MCF in the Square Billiard: Statistical Self-similarity

Returning to Fig. 6.11, the plot of log $-N(\Delta B)$ versus log $\Delta B$ is a graph of a statistical property of the MCF data – the minimum cover for the data – as a function of a scaling parameter – the covering element $\Delta B$. The presence of linear behaviour in this graph not only indicates the presence of fractal behaviour (since it gives $1 < D_F = 1.29 < 2$), it demonstrates that the data follow the same statistical relationship (i.e. the power law covering relationship in Eqn. 6.1) for all scales $\Delta B$ within the range of linear behaviour. As discussed in §2.3.2, this is a property frequently described as statistical self-similarity. It is important to note that there are two forms of the statistical self-similarity presented in §2.3.2. The first form shows statistical self-similarity not only under scaling but also under displacement (i.e. at a constant scale, the statistics of different sections of the data-trace are similar), whilst the second form only shows self-similarity under scaling.

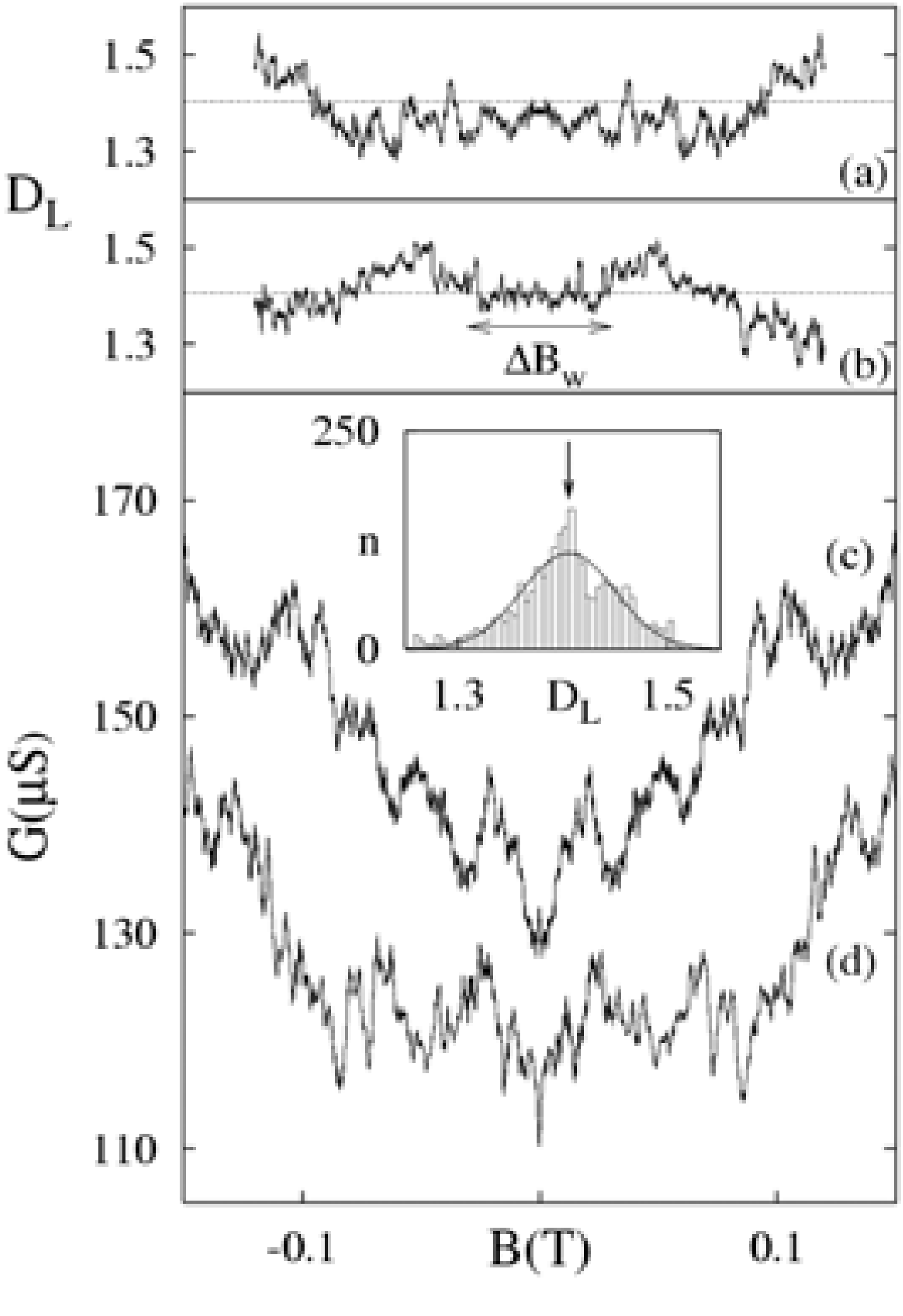

Figure 6.12: (a) and (b) are plots of local fractal dimension $D_L$ as a function of magnetic field for a model fractal trace (c) and typical experimental data from the Cambridge square billiard (d). Details of the model trace are discussed in the text. The width of the window $\Delta B_w$ used to obtain $D_L$ is shown in (b). The inset is a histogram of the number of windows giving a particular $D_L$ versus $D_L$ for the data shown in (d). A Gaussian fit to the histogram is shown and the arrow indicates the total fractal dimension for the two traces $D_F \sim 1.4$

To determine whether the statistics remain similar under displacement, the local fractal dimension $D_L$ has been assessed as a function of magnetic field as shown in Figs. 6.12(a) and (b). These two traces correspond to a model fractal trace (Fig. 6.12(c)) and typical experimental data obtained from the Cambridge square billiard (Fig. 6.12(d)), respectively. The model trace is generated using fractional Brownian statistics with a specified fractal dimension of $D_F$ = 1.4 to match closely the $D_F$ obtained for the experimental data shown in Fig. 6.12(c). The model trace has the same magnetic field resolution, magnetic field range, total amplitude and background as the experimental data. The plot of $D_L$ versus $B$ is obtained by sliding a window of width $\Delta B_w$ across the data-trace. This window is stepped forward by one data point at a time and at each stop the fractal dimension of the data contained in the window is assessed using the variation method to obtain $D_L$. In the particular case presented here, $\Delta B_w$ is equal to one fifth of the total magnetic field range of the data-trace and the magnetic field value associated with a particular $D_L$ value is taken as the magnetic field value at the centre of the window used to obtain $D_L$. Note that in both Figs. 6.12(a) and (b), $D_L$ varies about the total fractal dimension $D_F \sim 1.4$ for both data-traces. The scatter about $D_F$ in both Figs. 6.12(a) and (b) appears on two scales – a larger amplitude, low frequency background variation with small amplitude, high frequency variations superimposed upon it. The small amplitude, high frequency variation is expected to be due to random error in determining the local fractal dimension in each window. This is because within the window, which is five times smaller than the overall data-trace, the cut-offs $\Delta B_2$ and $\Delta B_3$ are closer together, giving a linear region smaller than one order of magnitude (rather than the 2.1 orders obtained for the whole trace) over which to fit for $D_L$.

The cause of the large scale background variation is currently not known. However, the smooth behaviour it displays suggests that it is a real variation in the statistics with displacement. This is further confirmed by the histogram of the number of windows that yield a particular $D_L$ value shown in Fig. 6.12(inset). In the case where deviations of $D_L$ from $D_F$ are entirely due to random error in fitting for $D_L$, the distribution in the histogram would be expected to be Gaussian with a mean value of $D_F$. However, the Gaussian fit shown in Fig. 6.12(inset) is not a perfect fit to the distribution, suggesting that there may be non-random contributions to the variation in $D_L$ about $D_F$. Note that while in this case the mean $D_L$ roughly coincides with $D_F$, this is not the case for other data-traces investigated in this manner. Statistical self-similarity under displacement is still under investigation, in particular to establish whether the large-scale background variation is a real effect and how the background variation originates. However, in order to improve the statistics and extend the range over which the variation can be examined it is necessary to obtain fractal MCF which extend over a significantly wider range than that presented in this thesis. A modified version of the correlation function approach presented in §5.3 is currently being employed to further investigate statistical self-similarity under displacement.

In summary of what has been established to this point, it was clear at the end of Chapter 5 that the Sinai billiard exhibits exact self-similarity, and fractal behaviour over 3.7 orders of magnitude. Using the correlation function analysis (see §5.3 and §5.6), it was established that by removing the Sinai diffuser, the exact self-similarity in the MCF was suppressed. Hence at the close of Chapter 5, it was suggested that the loss of exact self-similarity observed with the removal of the Sinai diffuser may correspond to a loss of fractal behaviour altogether. However, based on the results in this Chapter it is clear that fractal behaviour is present irrespective of the presence/absence of the Sinai diffuser. This suggests that the introduction of the Sinai diffuser instead leads to a change in the form of fractal behaviour. I will now return to the Sinai-square transition, this time examining the presence of statistical self-similarity rather than exact self-similarity in the MCF as the Sinai diffuser is removed.

## 6.5 – Absence of Statistical Self-similarity in the Sinai Billiard

In 1998 Sachrajda *et al.* [15] reported an observation of statistical self-similarity in the Sinai billiard over a range of two orders of magnitude. However, following a re-analysis of this data, it was clear that there were a number of shortcomings in this work. These shortcomings are discussed here and ultimately demonstrate that the Sinai billiard data does not exhibit statistical self-similarity. In fact, the re-analysis further confirms the analysis of the Sinai billiard data presented in Chapter 5. The top three traces in Fig. 6.13(a) show data obtained in the Sinai billiard experiment for Sinai diffuser biases $V_I$ = +0.7V (top), −0.1V and −2.9V (second from bottom) at $V_O$ = −0.51V. The data-trace where $V_I$ = –0.1V is the same used in the analysis of Sachrajda *et al.* [15] and whilst this data-trace does not correspond to those in which exact self-similarity was observed in Chapter 5, it is not that of an empty square ($V_I$ = +0.7V) either. For this reason the data-traces for $V_I$ = –2.9V and +0.7V (which correspond to the Sinai and square billiards) have been analysed to provide a comparison for the data-trace analysed in [15]. For further comparison, the bottom trace in Fig. 6.13(a) is a model oscillatory function constructed from a cosine with the same period used in the initial Weierstrass model to generate the fine structure (see §5.5). This cosine is superimposed on a smooth parabolic background similar to that of the experimental data. Note that this trace is qualitatively the same as the middle trace in Fig. 6.9(a), and analysed in §6.2.3. The analysis presented in [15] is performed using the variation method and confirmed using the variance method. In both analysis techniques employed by Sachrajda *et al.*, the $\Delta B$ values were determined by dividing the data range by successive powers of two (see §6.1 and §6.2.3). Sachrajda *et al.* used the non-overlapping power method (see §6.2.3) for their calculations of the variance.

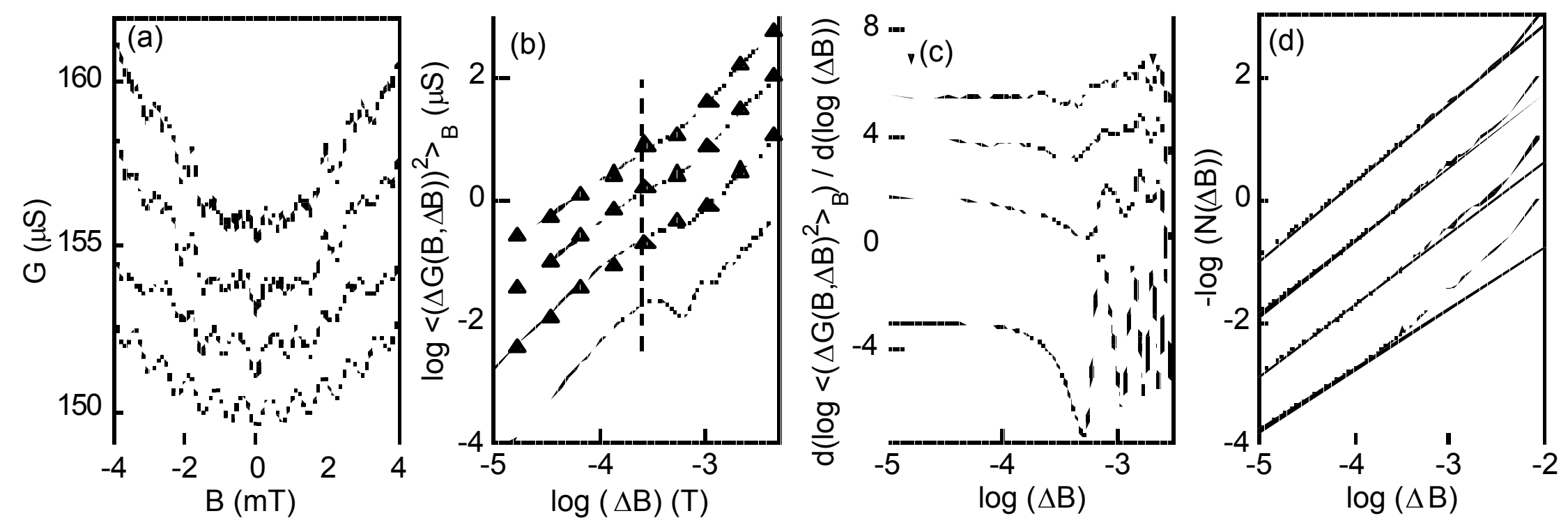

Figure 6.13: (a) MCF data-traces for the NRC Sinai billiard at $V_I$ = +0.7V (top), –0.1V, –2.9V, and a model trace constructed from a cosine superimposed on a parabolic background (bottom). (b) $\log\langle(\Delta G(B,\Delta B))^2\rangle_B$ versus $\log \Delta B$ for the data-traces in (a). (c) the derivative of the traces in (b) with respect to $\log \Delta B$ versus $\log \Delta B$. (d) $-\log N(\Delta B)$ versus $\log \Delta B$ for the data-traces in (a). Traces are vertically offset. Intercepts ($B$ = 0 or $\log \Delta B$ = –5) are as follows: (a) (top) 264, 218, 151 and 152μS; (b) –2.20, –2.52, –2.91 and –3.28; (c) 1.28, 1.53, 1.72 and 2; (d) –3.91, –3.92, –4.00 and –3.85.

The lack of resolution provided in the resulting log-log plots using the method employed by Sachrajda *et al.* for determining $\Delta B$ values means that although their analysis presents a reasonable linear fit to over two orders of magnitude, they are actually fitting to at most 11 points in the log-log plot. In contrast, by using $\Delta B$ values determined by dividing the data range by successive integers and overlapping $\Delta B$ increments, it is possible to obtain more than 1000 points over an equivalent range for the same data-traces. This allows a much more thorough investigation of the presence of linearity in the log-log plots generated by the fractal analysis methods. Although the variation method was preferred to the variance method for assessing the presence of fractal behaviour in [15], I will discuss the results of the variance method first. The reason for this is that the findings obtained from the variance method play an important part in the interpretation of the results obtained using the variation method analysis.

Plots of log $\langle(\Delta G(B,\Delta B))^2\rangle_B$ versus log $\Delta B$, and the derivative of log $\langle(\Delta G(B,\Delta B))^2\rangle_B$ with respect to log $\Delta B$ versus log $\Delta B$, for the data in Fig. 6.13(a) are shown in Figs. 6.13(b) and (c) respectively. Variance method results using both the non-overlapping power method (triangles) and the overlapping integer method (line) are presented in Fig. 6.13(a). According to the theory presented by Ketzmerick in [12] and again by Sachrajda *et al.* in [15], over the range that fractal conductance fluctuations exist, the traces in Fig. 6.13(b) should be linear with gradient $\gamma = 4 - 2D_F$, where $D_F$ lies in the range $1 < D_F < 2$. The corresponding situation in the derivative traces presented in Fig. 6.13(c) is a horizontal line with ordinate $4 - 2D_F$, again with $1 < D_F < 2$. As expected, the model trace is not fractal. At small $\Delta B$, for the bottom trace in Fig. 6.13(c), the horizontal line gives $D_F = 1$, whilst at larger $\Delta B$ the derivative breaks into oscillations. The oscillations observed in the derivative correspond to the series of oscillations present in the variance method results for the model trace (bottom trace – Fig. 6.13(b)). In turn, the oscillation sequence in the variance method results is due to periodicity in the data-trace, as discussed in §6.2.3. Note that the first minimum in the bottom trace of Fig 6.13(b) is equal to the period of the cosine structure in the model trace as found in §6.2.3. Similar oscillations in the other traces in Figs. 6.13(b) and (c) indicate a dominant period in the MCF data, particularly in the case of the data for $V_I = -2.9$V (trace second from bottom). This observation is confirmed by an inspection of the raw data in Fig. 6.13(a), where a clear period develops as the Sinai diffuser is introduced. The variation method results for the four traces in Fig. 6.13(a) are presented in Fig. 6.13(d). The gradient obtained from the variation analysis of the model trace must be one, since it is non-fractal (i.e. has $D_F = 1$). The gradient of the fit assigned to the corresponding trace in Fig. 6.12(d) is fixed to a value of one based on the previous statement. This fit is satisfactory over approximately one order of magnitude and limited by upper and lower cut-offs. The lower cut-off is analytical and due to the limited resolution of the data-trace. The upper cut-off location coincides with the emergence of the underlying period in the variance method results, as indicated by the dashed vertical lines in Fig. 6.13(b) and (d). For higher $\Delta B$ values, the variation method results deviate significantly from the fit. For the Sinai billiard, the fit, which in this case does not have a restricted gradient, also extends between these ranges and yields a fractal dimension $D_F = 1.1$, consistent with the value obtained from the gradient of the linear range of the corresponding variance trace in Fig. 6.13(c). As the Sinai diffuser is removed, the upper cut-off extends to slightly higher values indicating the decreased presence of the underlying period. This is also confirmed in the variance method and the derivative analysis.

Based on the non-overlapping power method (triangles in Fig. 6.13(c)), Sachrajda *et al.* claim that fractal behaviour is present over two orders of magnitude in the data-traces second from top in Fig. 6.13(a). The arrows at the top of Fig. 6.13(c) indicate the range over which this claim is made. The deviation corresponding to the period present in the experimental data is apparent in the variance method results presented in [15]. However, by employing the overlapping integer variance method combined with an analysis of its derivative, the presence of the underlying period in the Sinai billiard data is considerably more obvious. Initially, the re-analysis of the data demonstrates that the claim of fractal MCF made by Sachrajda *et al.* should be restricted to only one order of magnitude. However, there are four important conclusions that can be drawn from the re-analysis. The first conclusion relates to the methods themselves. Whilst my analysis of the data using the variation method shows the break-down in fractal behaviour due to the emergence of an underlying period in the data, it is not as obvious as it is in the variance method, particularly when combined with the derivative analysis. This indicates that the variation method is less sensitive to the emergence of periodicity and hence caution must be exercised in using this method alone. The discussion presented above also highlights the need for a careful treatment of the cut-offs. Although a simple fit that can be performed with a ruler might give a good fit over an extended range [114], such a fit is not necessarily a true indicator of the properties of the data.

The second conclusion is that the analysis presented here confirms that there are distinct periods in the Sinai billiard data. Note that the Sinai billiard data presented in the second from bottom data-trace in Fig. 6.13(a) matches the fine scale data presented in Fig. 5.12. The analysis performed by Sachrajda *et al.* was not of the complete range of data including the ultra-coarse, coarse and ultra-fine levels. Indeed, such an analysis has not been presented in this thesis either. The reason is that each individual scaling level is a separate data-trace obtained at different field resolutions. Hence compiling these traces into a single set of data for investigation using the analysis presented in this section is not possible. However, an analysis of each level reveals the presence of underlying periodic structure. This is consistent with the presence of exact self-similarity in the Sinai billiard data, and provides further justification for the Weierstrass model, which imitates the experimental data using a series of scaled periodic functions.

The third conclusion is that statistical self-similarity is not observed in the MCF of the Sinai billiard. Only the four points corresponding to the ultra-coarse, coarse, fine and ultra-fine levels lie on the power-law line shown in Fig. 5.13(a). The fourth and final conclusion is that some remnant of the Sinai diffuser remains when $V_I$ = +0.7V. This conclusion is based on the fact that the underlying periodic structure has not been completely removed at $V_I$ = +0.7V, as demonstrated by the small minima remaining in the variance (top trace – Fig. 6.13(b)), and the corresponding small oscillations in the derivative (top trace – Fig. 6.13(c)). This remnant may be due either to the inability of the applied positive bias $V_I$ = +0.7V to fully remove the illumination-shadow depletion region beneath the circular surface-gate (see §5.1), or the presence of damage due to EBL beneath the circular surface-gate. However, the latter is unlikely as such damage should significantly increase phase breaking and lead to suppression of the quantum interference processes that generate the MCF. Further justification for the third conclusion is provided by applying the variance method and derivative analysis to the MCF data obtained from the Cambridge square billiard. The results of the variance method and derivative analysis are presented in the top traces in Figs. 6.14(a) and 6.14(b) respectively. The four traces displayed in Figs. 6.13(b) and 6.13(c) are included as the bottom four traces (and in the same order) in Figs. 6.14(a) and 6.14(b) for comparison. It is clear in Fig. 6.14(a) that the minima due to the underlying period in the model and Sinai billiard traces (indicated by the dashed vertical line) are absent in the variance of the empty square billiard.

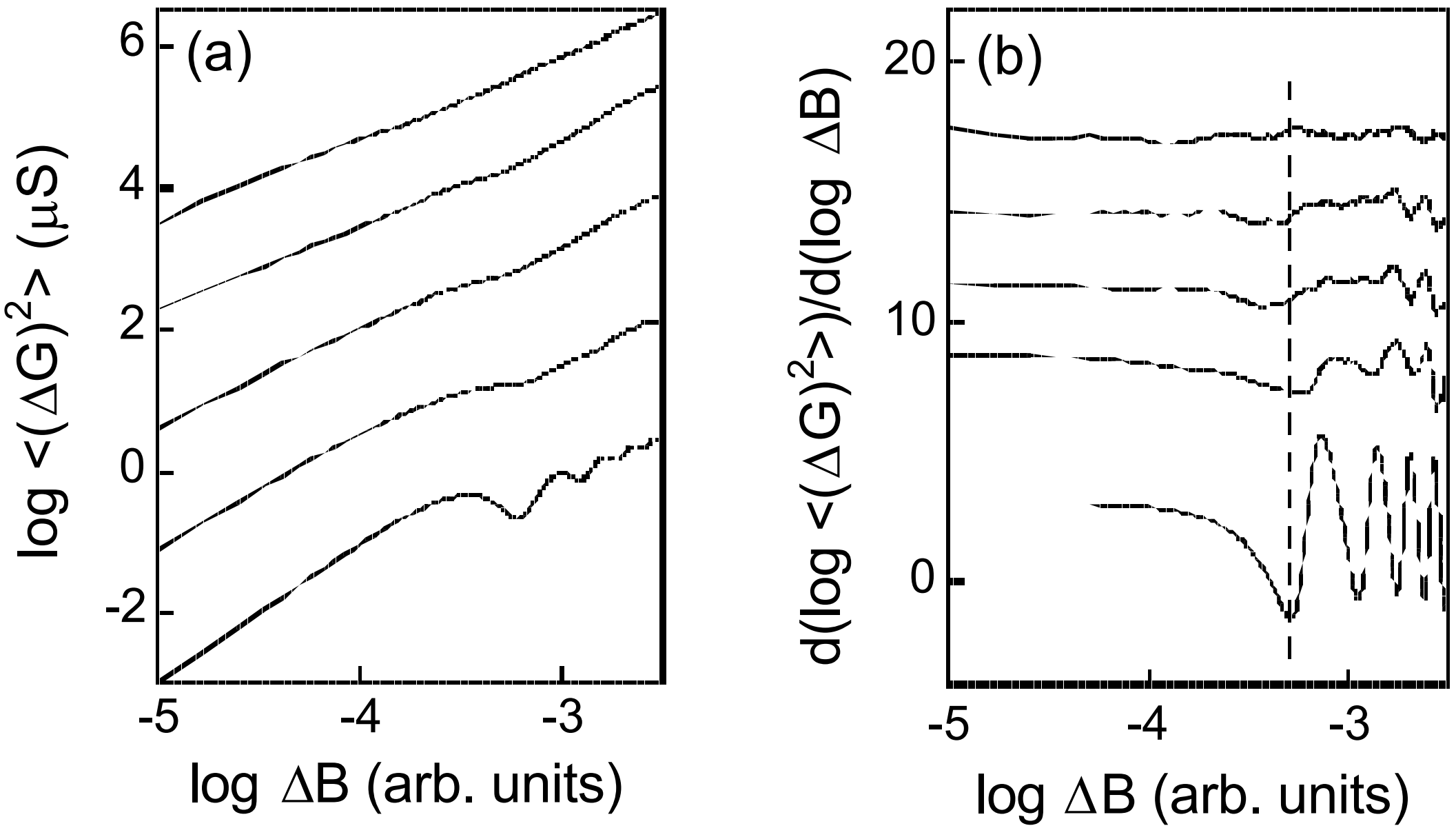

Figure 6.14: (a) Plots of log $\langle(\Delta G(B,\Delta B))^2\rangle_B$ versus log $\Delta B$ for: (top) Cambridge square billiard, NRC Sinai billiard for $V_I$ = +0.7V, −0.1V, −2.9V, and a model trace constructed from a cosine superimposed on a parabolic background (bottom). (b) The derivative of the traces in (a) with respect to log $\Delta B$ versus log $\Delta B$.

This is supported by the derivative analysis in Fig. 6.14(b) where the derivative minimum that corresponds to the drop into the first variance minimum from the small $\Delta B$ side (indicated by the dashed vertical line) is also absent. To summarise, this means that the Sinai-square transition not only corresponds to a change in fractal behaviour rather than a total loss of fractality, but that it also corresponds to a transition between exact and statistical self-similarity. I now return to the Sinai-square transition with this in mind, to present a proposal for a unified theory explaining the behaviour of exact and statistical self-similarity in semiconductor billiards.

# 6.6 – Exact and Statistical Self-similarity: A Unified Model

Returning to the discussion of the scaling properties of exactly self-similar MCF in the NRC Sinai billiard in §5.4, a set of magnetic field and conductance scaling factors $\lambda_B$ and $\lambda_G$ were defined based on the ratio of structural scales in subsequent structural levels. That is, $\lambda_B = \Delta B_i/\Delta B_{i+1}$ and $\lambda_G = \Delta G_i/\Delta G_{i+1}$, where $\Delta B_i$ was the characteristic period (obtained as the full-width at half maximum of the central feature) and $\Delta G_i$ was the characteristic amplitude (obtained as the full height of the central feature) for level *i*. As required for fractal behaviour, the scaling factors are related by the power law $\lambda_G = (\lambda_B)^\beta$, where $D_F = 2 - \beta$ and $1 < D_F < 2$. Performing an equivalent analysis, replacing $\Delta G_i$ with $\langle \{\delta G(B) - \delta G(B+\Delta B_i)\}^2 \rangle_B$, the variance of $\delta G(B)$ with $\Delta B_i$ for the *i*th structural level (i.e. analysing all features in the MCF), yields an analogous power-law expression:

$$\frac{\left\langle [\delta G(B) - \delta G(B+\Delta B_i)]^2 \right\rangle_B}{\left\langle [\delta G(B) - \delta G(B+\Delta B_{i+1})]^2 \right\rangle_B} = \left( \frac{\Delta B_i}{\Delta B_{i+1}} \right)^\gamma \tag{6.5}$$

where $\lambda_B = \Delta B_i/\Delta B_{i+1}$ and $D_F = 2 - \gamma/2$. Note that $\gamma = 2\beta$, accounting for the square in the variance. On the other hand, the theory presented for statistically self-similar MCF in [12] gives, for an arbitrary increment $\Delta B$ in the range of fractal scaling:

$$\left\langle [\delta G(B) - \delta G(B+\Delta B)]^2 \right\rangle_B \propto \Delta B^\gamma \tag{6.6}$$

where $D_F = 2 - \gamma/2$ also holds. By selecting two arbitrary scales $\Delta B_1$ and $\Delta B_2$, however, Eqn. 6.6 can be written in the same form as Eqn. 6.5 to give:

$$\frac{\left\langle [\delta G(B) - \delta G(B+\Delta B_1)]^2 \right\rangle_B}{\left\langle [\delta G(B) - \delta G(B+\Delta B_2)]^2 \right\rangle_B} = \left( \frac{\Delta B_1}{\Delta B_2} \right)^\gamma \tag{6.7}$$

where $\lambda_B = \Delta B_1/\Delta B_2$. The difference between Eqns. 6.5 and 6.7 is that the latter holds for a continuous range of $\lambda_B$ while the former is only true for a single value. For both exact and statistical self-similarity, the MCF are described by a unique $D_F$ that links each ($\lambda_B, \lambda_G$) pair by $\lambda_G = (\lambda_B)^{\gamma/2}$ and $D_F = 2 - \gamma/2$.

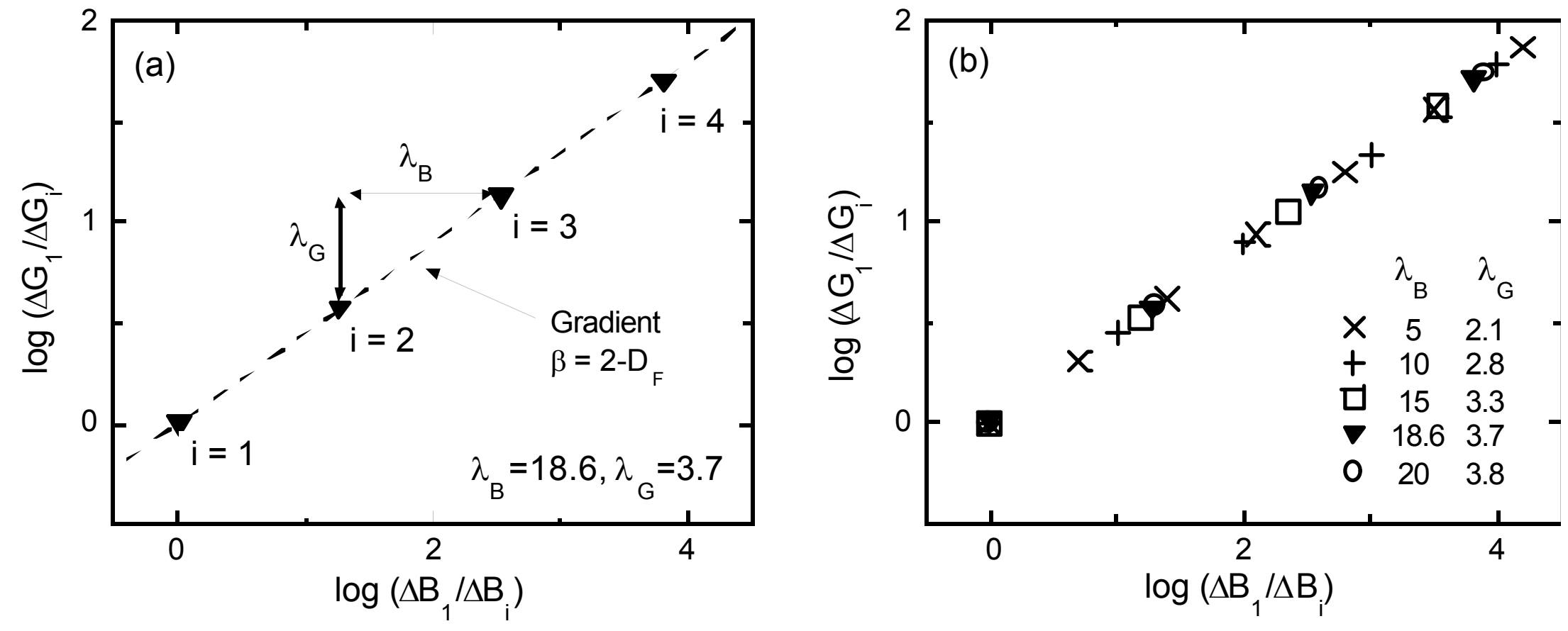

Figure 6.15: Schematic representation of the scaling properties of the proposed unified model for exact and statistical self-similarity. Exact self-similarity (a) is generated by a single hierarchy of structural levels determined by a single scaling factor pair ($\lambda_B, \lambda_G$). In contrast, statistical self-similarity (b) is generated by an infinite number of hierarchies determined by a continuum of ($\lambda_B, \lambda_G$). Only five hierarchies have been shown in (b) for clarity.

Based on this, I propose a model where activation of the Sinai diffuser selects one ($\lambda_B, \lambda_G$) pair from the continuum of ($\lambda_B, \lambda_G$) pairs that lead to statistical self-similarity, in order to produce the transition from statistical to exact self-similarity. This is summarised in the schematic of Fig. 6.15(a), where the Sinai diffuser is activated and Fig. 6.15(b), where the Sinai diffuser is de-activated. In Fig. 6.15(a), the exact self-similarity is generated by a single hierarchy of points ($i$ = 1, 2, 3, 4,...) located at equal spacings, as determined by a single pair of scaling factors ($\lambda_B, \lambda_G$) = (18.6, 3.7) on a line with gradient $\gamma = 4 - 2D_F$. In Fig. 6.15(b), de-activation of the Sinai diffuser allows additional hierarchies. The spacing between adjacent levels in these hierarchies is determined by the particular value of ($\lambda_B, \lambda_G$) for the hierarchy.

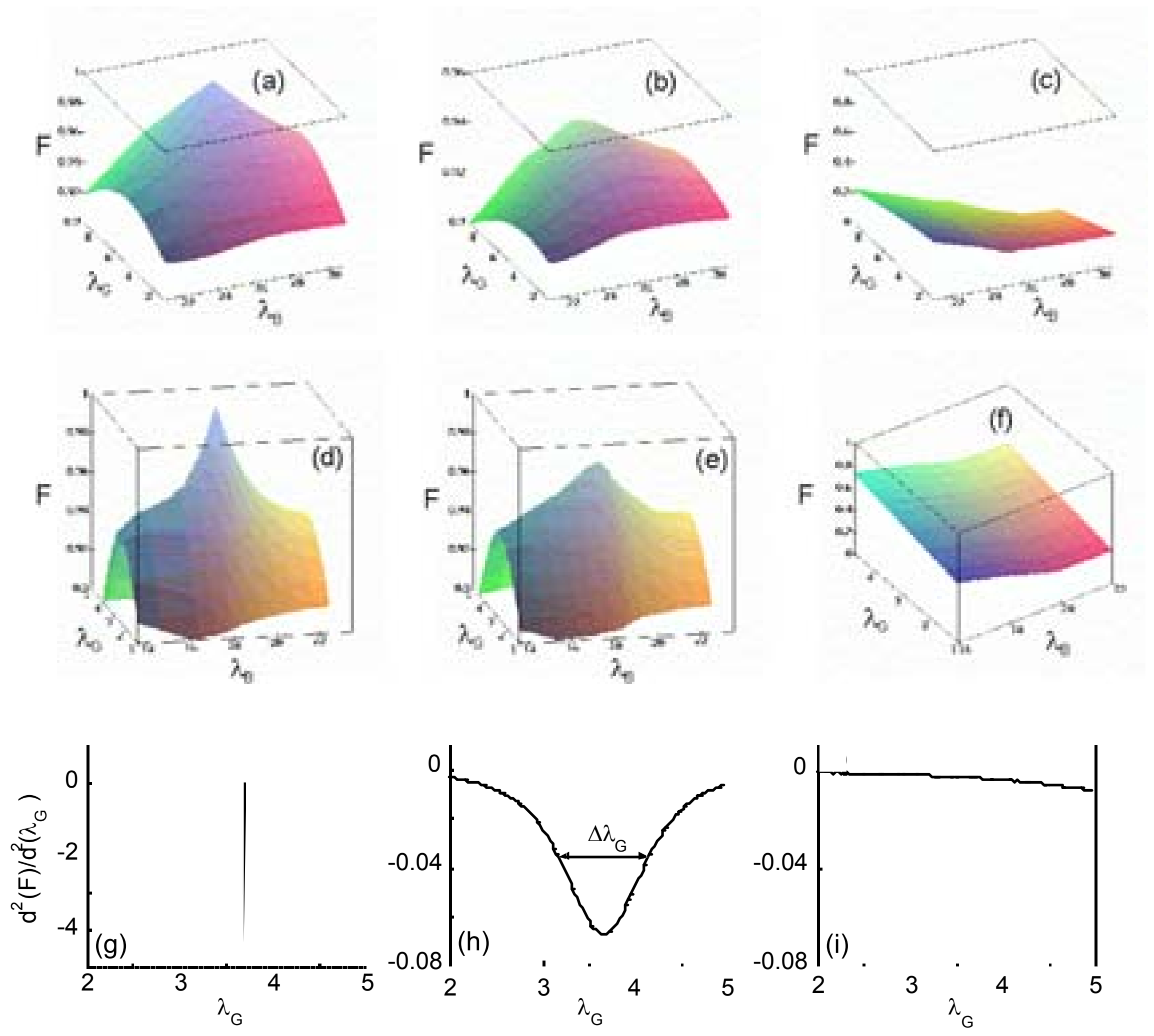

Figure 6.16: Scaling factor maps for the NRC Sinai device at (a,b,c) $V_O$ = −0.55V and (d,e,f) $V_O$ = −0.51V. (a,d) Sinai diffuser activated – Ideal case, (b,e) Sinai diffuser activated – Experimental case and (c,f) Sinai diffuser deactivated – Experimental case. (g,h,i) are second derivatives of $F$ with respect to $\lambda_G$ versus $\lambda_G$ for the scaling factor maps in (d,e,f).

For simplicity, only four additional hierarchies have been presented in Fig. 6.15(b) although a continuum of possible ($\lambda_B$, $\lambda_G$) pairs is expected for ($\lambda_B$, $\lambda_G$) such that $\lambda_G = (\lambda_B)^{\gamma/2}$ and $D_F = 2 - \gamma/2$ hold, and $\gamma$ has a single value. Note that although the gradient shown in the schematics in Figs. 6.15(a) and (b) is identical, corresponding to the same fractal dimension for both exact and statistical self-similarity, in reality this is not necessarily the case. As an example, the MCF of the Sinai billiard has $D_F$ = 1.55 with the Sinai diffuser activated ($V_I$ = −2.9V) and $D_F$ = 1.20 with the Sinai diffuser de-activated ($V_I$ = +0.7V). A mechanism for such a change in $D_F$ in the transition between exact and statistical self-similarity (i.e. Figs. 6.15(a) and (b)) is not provided in this model and currently not understood.

The proposed model for the transition between exact and statistical self-similarity is confirmed by the scaling factor map analysis in Fig. 6.16. Figures 6.16(a) and (d) show scaling factor maps for the ideal case of exact self-similarity, where the cascade has been generated mathematically using the coarse structure (Figs. 5.9(a) and 5.12(b) respectively) as the initiator. That is, by using the correlation function to compare the coarse scale structure to a scaled version of itself. The peak rises to a maximum of $F = 1$ at a single point in the scaling factor map. This is demonstrated in Fig. 6.16(g), where the second derivative of $F$ with respect to one of the scaling factors features a $\delta$-function that coincides with the location of the peak in the scaling factor map. Figures 6.16(b) and (e) show the experimentally observed exact self-similarity. Although it is centred on the same ($\lambda_{B,}$ $\lambda_G$) point, the peak has a slightly lower maximum of $F = 0.97$ (0.94 in (e)) and, as demonstrated by the second derivative in Fig. 6.16(h), this peak is not as sharp. Ideal exact self-similarity is not quite achieved in the experiment due to contributions from a narrow range of scaling factors rather than from a unique pair, and the full widths at half maximum in the second derivative plots quantify this range as ($\Delta\lambda_{B,}$ $\Delta\lambda_G$) = (2.7, 1.0). Note that this is consistent with the refined Weierstrass model (see §5.5). Each pair of scaling factors within this range generates its own fine structure from the common coarse scale initiator, producing the observed small reduction in $F$ and also the emergence of statistical self-similarity at very fine scales ($\Delta B << \Delta B_F$). Note, however, that unless the coarse structure corresponds to $i$ = 1 (i.e. is the initiator for the self-similar sequence) then there will be a set of coarse levels, each of which has a set of fine structure levels. Hence, while the argument above still holds, it is a slightly more complicated situation than before in that there are far more levels to consider. When the Sinai diffuser is de-activated, the range of scaling factors increases dramatically and this generates statistical self-similarity over the full range of magnetic field scales. For statistical self-similarity, the large peak in Figs. 6.16(a), (b), (d) and (e) will be absent. Instead, for an ideal statistically self-similar system, with a continuous range of $\lambda_B$ values, $F$ will condense onto a smooth background with a significantly lower $F$ value than that of Figs. 6.16(a), (b), (d) and (e). This is demonstrated in Figs. 6.16(c) and (f), which show the same section of the scaling factor map as Figs. 6.16(a), (b), (d) and (e), but with the Sinai diffuser de-activated. If selected scaling factors still dominate the spectrum, this should result in small, distinct peaks in $F$ at ($\lambda_B$, $\lambda_G$) positions set by $D_F$. Such small remnant peaks are not expected or observed. However, it is not clear if this is because they do not exist or because they are not resolved. The smooth contours of Figs. 6.16(c), (f) and (i) are consistent with the large spectrum of scaling factors predicted in the proposed model.

As a final point, it is interesting to compare the description of the Sinai-square transition in the model proposed above with the Weierstrass model. In the Weierstrass model (see §5.6), widening the narrow distribution of areas contributing to the individual levels led to fine structure that appeared remarkably similar to that observed in the experiment once the Sinai diffuser is de-activated (see Fig. 5.19). Widening the area distribution has a similar effect to increasing the number of possible scaling factors in the model proposed here. The quality of the match provided by widening the area distribution in the Weierstrass model further suggests the viability of the model proposed here in understanding the transition between exact and statistical self-similarity. It is certain that increasing the number of possible scaling factors in the Weierstrass model removes the exact self-similarity, simply because this violates the basis of the model. However, it is not yet known if this produces statistical self-similarity as observed in the experimental data and suggested by the model proposed in this section. Unfortunately, whilst introducing a distribution of possible scaling factors to the Weierstrass model seems simple initially, there are a large number of possible ways of doing this, none of which may be totally correct, particularly in the case of the refined Weierstrass model. Due to this uncertainty, I have not actively pursued this addition to the Weierstrass model. However, I think this would be an interesting avenue for further theoretical studies of self-similarity in the MCF of semiconductor billiards.

## 6.7 – Summary and Conclusions

In summary, three techniques for fractal analysis of MCF data have been developed and optimised. Two of these – the variation and HSE methods – are based upon the well-known box-counting method for obtaining the covering fractal dimension of objects. The third method is the variance method developed by Ketzmerick in [12]. The variance method only assesses the fractal dimension of data obeying fractional Brownian statistics and is highly sensitive to underlying periodicity in the data. Hence the variance method is only used to detect underlying periodicity in the data in this Chapter. It has been established that the variation method is the most accurate for assessments of the fractal dimension and this method will be used for obtaining $D_F$ values for the remainder of this thesis (i.e. in Chapter 7). The remaining two methods are useful in establishing the presence of fractal behaviour in the MCF.

Using a combination of the variation and HSE methods, fractal behaviour extending over 2.1 orders of magnitude was observed in the MCF data of a square billiard. A rigorous study of the upper and lower fractal cut-offs was presented in relation to the analysis of this data and demonstrated that the observed fractal behaviour was limited in the lower end by noise, and in the upper end by limitations of the analysis technique. The upper physical cut-off lies outside the fractal scaling range and is generally due to a changeover to skipping orbit transport once the electron cyclotron radius becomes smaller than half the width of the billiard. However, for billiards in high electron mobility 2DEGs, the presence of bulk Shubnikov-de Hass oscillations can obscure the MCF before this occurs. Both of these effects lead to non-fractal effects and since these interfere with the observation of fractal behaviour, the data-traces are truncated so that only MCF data obtained for $r_{cyc} > W/2$, and not containing Shubnikov-de Haas oscillations, is used. The fractal analysis demonstrates that the data obtained from the square billiard is not only fractal, but also statistically self-similar.

In 1998, an observation of statistical self-similarity in the Sinai billiard, claimed to extend over a range exceeding two orders of magnitude, was reported in the literature [15]. Since it is not expected that both statistical and exact self-similarity can be observed concurrently, a re-analysis of this data was performed. Using the variance method and a derivative analysis it was demonstrated that there was an underlying period present in the Sinai billiard data, restricting the presence of linear behaviour in the variance method results to one order of magnitude rather than the reported two orders. The re-analysis was further confirmed by the variation method, which highlighted the need for considerable care in assessing fractal behaviour, particularly in assigning cut-offs, indicated by recent debate about fractal behaviour in physical systems [93-98]. The re-analysis led to two key conclusions. The first is that statistical self-similarity is not observed in the Sinai billiard. Indeed, the analysis confirms the original claim of exact self-similarity made in [13]. The analysis reported in [15] was only for the fine structure level discussed in Chapter 5. Analysis of the other levels revealed a similar underlying period in each level and these periods are expected to closely match the scaling behaviour presented in Chapter 5. The second conclusion is that the Sinai diffuser is not entirely removed at $V_I$ = +0.7V since some remnant of the underlying periodic behaviour observed in the Sinai billiard ($V_I$ = −2.9V) is present. This is confirmed by analysing a lithographically identical billiard where the central circular surface-gate and bridging interconnect have not been fabricated. The periodic remnant is found to be absent in this data.

Finally, based on the findings up to this point, a model describing a unified picture of exact and statistical self-similarity in semiconductor billiards was proposed. In this model, exact self-similarity is generated by a hierarchy of levels constructed from an initiator pattern using a single pair of scaling factors in magnetic field and conductance. In contrast, statistical self-similarity is generated in the same way, but with a continuum of possible scaling factor pairs rather than only one. In both cases a power-law, where the exponent is directly related to the fractal dimension, relates the two scaling factors. Hence it is proposed that the transition from the square to the Sinai billiard is due to the elimination of all but one of the scaling factor pairs. However, the exact physical mechanism for this has not yet been established. The proposed model is supported using the scaling factor map analysis discussed in Chapter 5. Further support is provided by the behaviour observed in the refined Weierstrass model as the narrow area distributions contributing to each level are expanded.

Having now established the presence of fractal behaviour in semiconductor billiards, the next step is to examine the various physical dependencies of this fractal behaviour. In terms of exactly self-similar fractal behaviour, this is difficult since it has only been observed in one sample to date, however the physical dependencies of this behaviour that were established in the Sinai experiment have already been presented in Chapter 5. The prospects of making a study of the physical dependence of the statistically self-similar fractal behaviour are better, largely because the billiards in which it can be studied are easier to fabricate and investigate. The results of such a study are presented in Chapter 7.

# Chapter 7 – Physical Dependencies of Fractal MCF

There are a number of parameters that determine the properties of semiconductor billiards and the behaviour of electrons traversing them. These parameters can be changed reversibly, with precision, and in many cases independently of the other parameters of the billiard. Hence, semiconductor billiards provide a controllable environment for investigations of electron transport and quantum interference phenomena. Fractal behaviour is observed in many physical systems. However, in a large majority of these physical systems it is difficult and sometimes impossible to vary the parameters of the system and investigate the effect that these parameters have upon the fractal behaviour. The ability to tune the parameters so easily and so controllably in the semiconductor billiard also makes it an ideal system for studying fractal behaviour.

In the preceding two Chapters, the presence of fractal behaviour in the MCF of semiconductor billiards was established. This fractal behaviour took on one of two forms – exact self-similarity and statistical self-similarity as discussed in Chapters 5 and 6 respectively. This Chapter commences by investigating the origin of exact self-similarity further, focussing on the role played by the location of curvature in the billiard geometry. §7.1 presents the results performed using a side-wall Sinai billiard, where the circular obstacle at the centre of the billiard is replaced with a curved side-wall. This experiment confirms that it is the presence of an obstacle at the centre of the billiard, rather than merely the presence of curvature in the billiard that causes the exact self-similarity observed in Chapter 5. Since exact self-similarity has only been observed in the Sinai billiard device discussed in Chapter 5, attention is turned instead to statistically self-similar fractal behaviour, which has been observed in a number of samples. An extensive study of the physical dependencies of statistically self-similar fractal MCF is presented in the remainder of this Chapter. §7.2 discusses the various devices and parameters involved in this investigation. In §7.3 it is established that modifications in the billiard parameters act only to change the fractal dimension, not the locations of the upper and lower cut-offs on the fractal behaviour. Thereafter, the physical dependencies of the fractal dimension are discussed. In §7.4, the various dependencies are combined into a unified picture where the fractal dimension is found to depend only upon the ratio of the average energy level spacing to the total broadening of the energy levels. This unified picture confirms that the fractal behaviour is a semiclassical effect, as proposed by Ketzmerick [12], and demonstrates that fractality is suppressed in the fully classical and fully quantum mechanical limits. Finally, §7.5 presents the results of an investigation into the effect of the billiard potential profile on the fractal MCF. The potential profile of the billiard is dependent upon the separation between the surface-gates and the 2DEG. Hence this investigation was performed using a double-2DEG heterostructure, constituting the first double-2DEG billiard experiment performed. A summary of the results and discussion of the conclusions obtained in this Chapter is presented in §7.6.

# 7.1 – The Dependence of the Type of Self-similarity on the Billiard Geometry

In 1997, I performed an experiment aimed at establishing the cause of the exact self-similarity observed in the MCF of the NRC Sinai billiard (Fig. 7.1(c)). This experiment involved two devices, the empty square billiard shown in Fig. 7.1(a) and the side-wall Sinai billiard shown in Fig. 7.1(b). Both of these devices were fabricated at RIKEN, Japan and are referred to as the RIKEN square billiard and the RIKEN Sinai billiard hereafter. Further details of these devices are presented in §3.7.2 and §3.7.3 respectively.

Curvature is present in both the NRC and RIKEN Sinai billiard geometries. Hence if exact self-similarity in the MCF is simply caused by the presence of curvature, then it is expected that exact self-similarity should be present in the RIKEN Sinai billiard data also. On the other hand, if the presence of an obstacle at the centre of the billiard is the cause of the exact self-similarity, as proposed at the conclusion of Chapter 5, then something other than exact self-similarity is expected to be observed in the MCF of the RIKEN Sinai billiard. The RIKEN square billiard was used as a comparison device, firstly because it lacks curvature and secondly because it had already been shown to exhibit statistical self-similarity [14,75,76], albeit over only one order of magnitude at that stage.

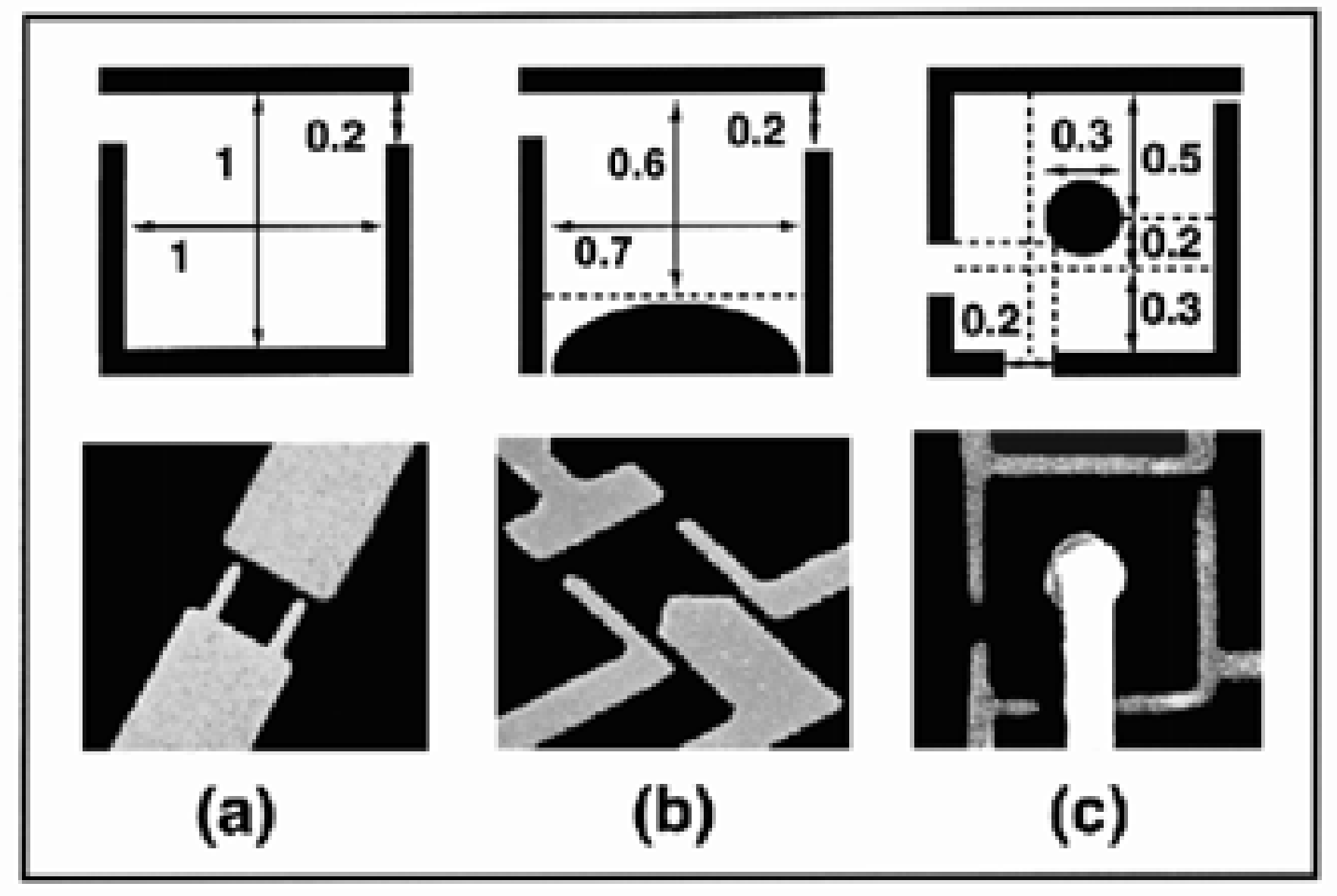

Figure 7.1: Schematic diagrams and corresponding scanning electron micrographs for the devices discussed in this section: (a) RIKEN 1μm square billiard (§3.7.2), (b) RIKEN side-wall Sinai billiard (§3.7.3) and (c) NRC Sinai billiard (§3.7.1). All dimensions in the schematics are in μm.

Typical MCF data obtained from the three devices is shown in Fig. 7.2. The MCF obtained for the RIKEN square (Fig. 7.2(a)) and Sinai (Fig. 7.2(b)) billiards is considerably different to that obtained from the NRC Sinai billiard with the Sinai diffuser activated (Fig. 7.2(c)). In particular, the clustering on distinct magnetic field scales observed in the NRC Sinai billiard (see §5.2) is not observed in either the RIKEN Sinai or square billiard MCF data. Subsequent fractal analysis using the variation method demonstrated that the MCF for the RIKEN Sinai and square billiards instead displayed statistical self-similarity over at least one order of magnitude. Note that the limited range of observation of this behaviour is largely due to the lack of magnetic field resolution possible at the time these experiments were performed ($B_{res}$ = 0.5–1mT compared to $B_{res}$ = 0.1mT for the data in §6.3). The presence of statistical self-similarity in the MCF of both the RIKEN Sinai and square billiards confirms the earlier conclusion that exact self-similarity in the NRC Sinai billiard MCF data is due to the presence of an obstacle (i.e. the Sinai diffuser) at the centre of the billiard. Furthermore, the presence of statistical self-similarity in both the RIKEN Sinai and square billiards suggests that fractal behaviour in semiconductor billiards has little to do with the presence of curvature in the billiard geometry.

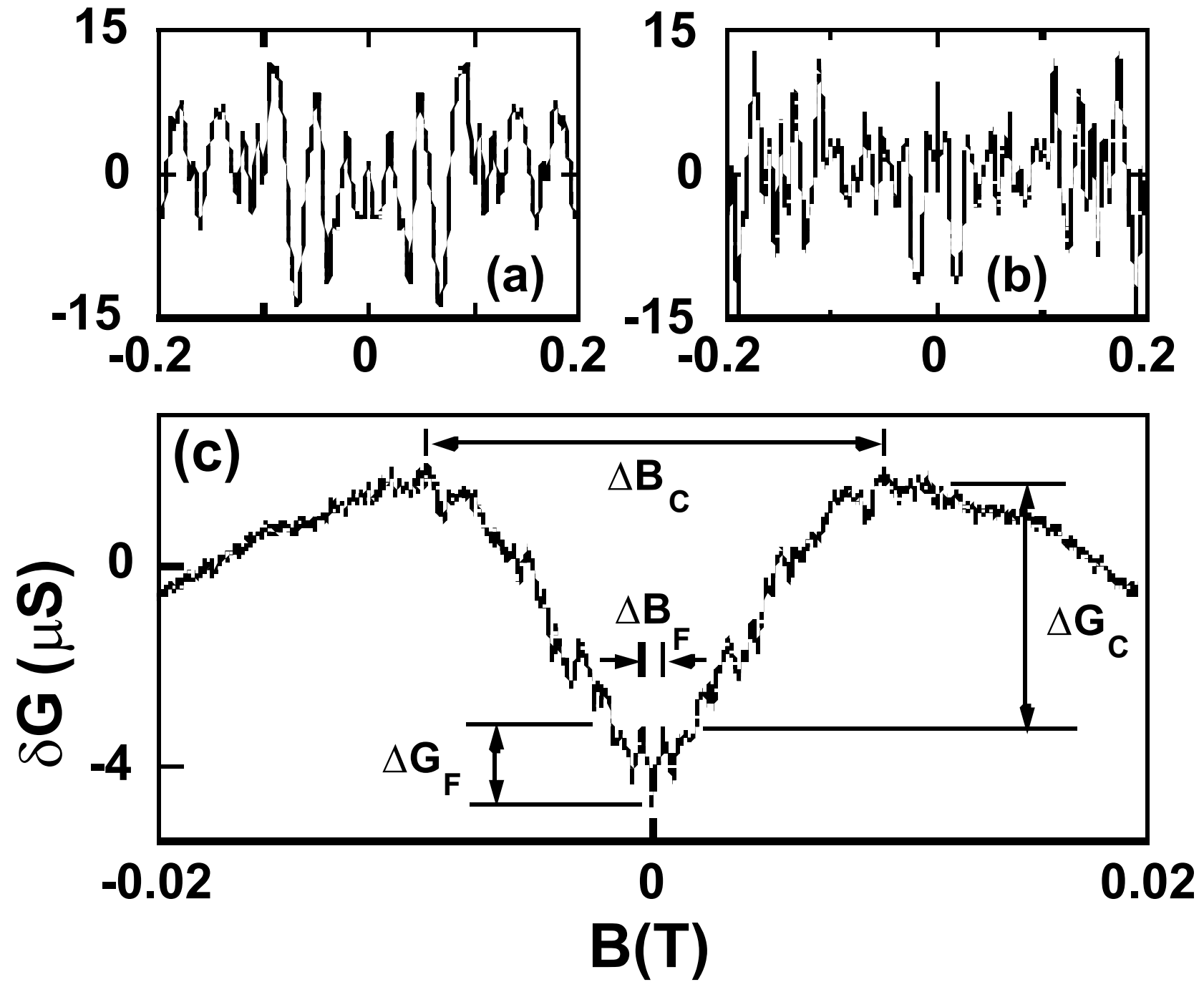

Figure 7.2: $\delta G(B) = G(B) - \langle G(B)\rangle_B$ versus $B$ for (a) the RIKEN 1μm square billiard, $V_O = -0.95$V, $G(B = 0) = 52$μS; (b) the RIKEN side-wall Sinai billiard, $V_O = -0.47$V, $G(B = 0) = 155$μS; (c) the NRC Sinai billiard, $V_O = -0.52$V, $G(B = 0) = 147$μS.

Indeed, all of the billiards investigated in this thesis, aside from the NRC Sinai billiard, are 'empty' billiards (i.e. no obstacle at the centre of the billiard) and exhibit statistically self-similar fractal behaviour over at least one order of magnitude. Over this set of billiards there are various geometries and lead positions confirming the suggestion by Ketzmerick [12] that the soft-wall potential profile, rather than the billiard geometry is responsible for fractal behaviour. The most important item to note is that the NRC Sinai billiard is the only 'non-empty' billiard investigated and only this billiard exhibits exact self-similar MCF, highlighting the fact that the presence of an obstacle inside the billiard is responsible for the exact self-similarity. Further investigations are planned to examine the link between the obstacle and exact self-similarity (see §8.2).

# 7.2 – Physical Dependencies of Statistically Self-similar MCF: Parameters and Devices

Details of the various devices used in the first part of the investigation of the dependence of fractal behaviour on billiard parameters are presented in Table 7.1. Schematics and scanning electron micrographs of these devices are presented in Fig. 7.3. Note that whilst seven devices are presented in Table 7.1, only four devices are shown in Fig 7.3. This is because devices B, F, G and H have the same nominal billiard geometry (shown in Fig. 7.1(B)) with lithographic dimensions of 2μm × 2μm, 1μm × 1μm, 0.6μm × 0.6μm and 0.4μm × 0.4μm respectively.

| Device | Origin | Area ($m^2$) | Temperature (K) | Mobility ($m^2$/Vs) | Port Modes |
|---|---|---|---|---|---|
| A | NRC, Canada | $6.0\times10^{-11}$ | 0.03 | 45 | 4 |
| B | RIKEN, Japan | $4.0\times10^{-12}$ | 0.03 | 40 | 2 |
| C (*n*) | Cambridge, UK | $1.0\times10^{-12}$ | 0.03 | 130 | 2-6 |
| C (*T*) | Cambridge, UK | $1.0\times10^{-12}$ | 0.03 – 4.2 | 130 | 5 |
| D | RIKEN, Japan | $2.2\times10^{-13}$ – $9.0\times10^{-14}$ | 0.03 | 50 | 2 |
| E | RIKEN, Japan | $1.0\times10^{-12}$ | 0.03 – 4.2 | 40 | 2 |
| F | RIKEN, Japan | $3.6\times10^{-13}$ | 0.03 – 2.5 | 40 | 2 |
| G | RIKEN, Japan | $1.6\times10^{-13}$ | 0.03 | 40 | 2 |

Table 7.1: Details of the various devices investigated in §7.3 and §7.4. Device C appears twice because both the temperature and mode dependence of this device were investigated. The area listed corresponds to the lithographic area in all devices except for D (see text).

The scanning electron micrograph of device A shows a metal stripe down the middle of the billiard that is not present in the corresponding schematic. This stripe was used for another experiment [79]. For the data obtained from device A for this study, the stripe was biased at +0.7V to remove its presence in the 2DEG, in a similar way to the removal of the Sinai diffuser in Chapter 5 (see [79] for details). The top and bottom walls of the schematic of device A are formed by the edges of the 30μm wide Hall-bar on which this billiard was fabricated. The metal gates shown in the scanning electron micrograph of device A extend beyond the Hall-bar edge, forming a 30μm × 2μm billiard, as shown in the schematic for device A.

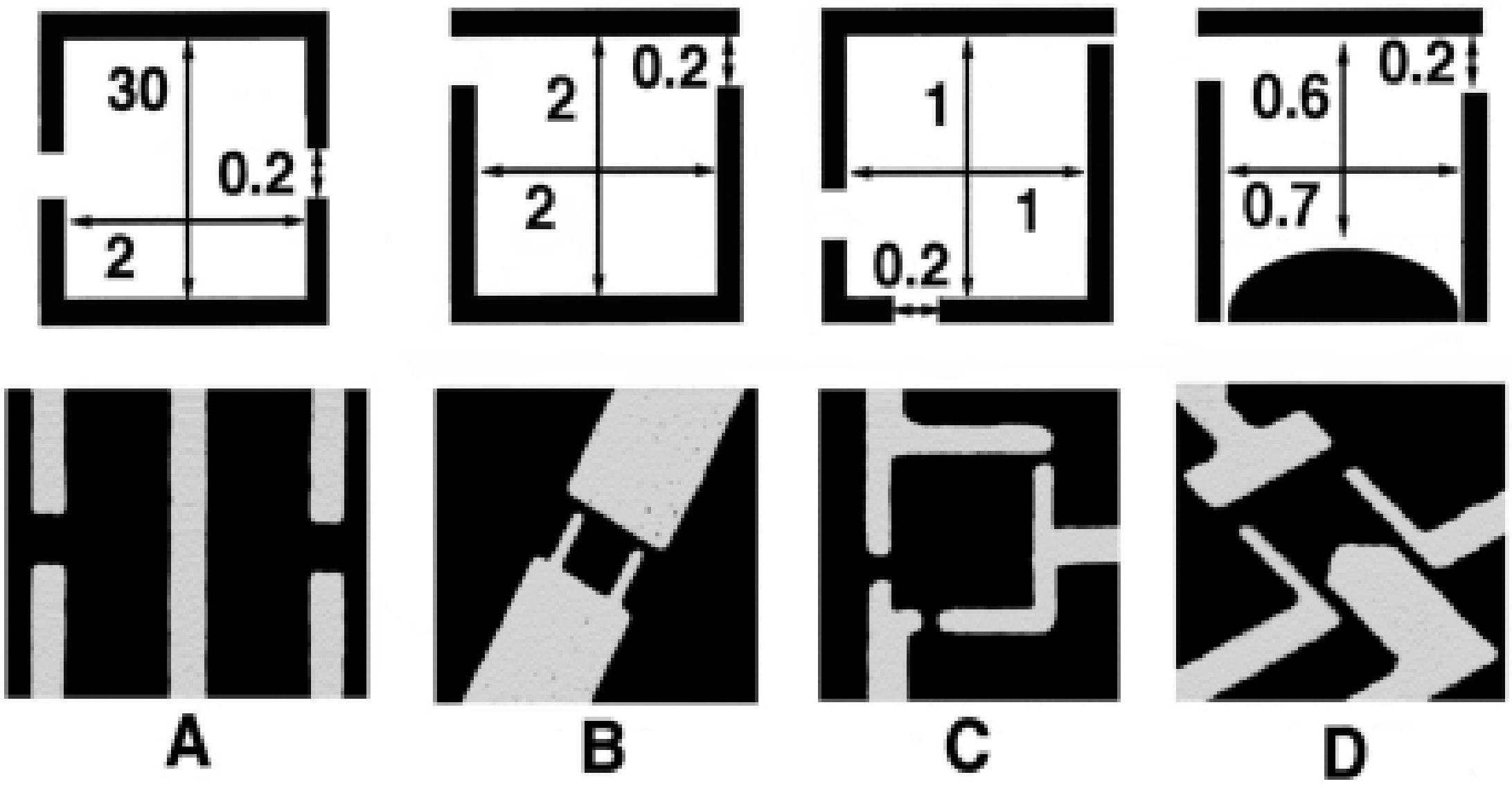

Figure 7.3: Schematics (top) and scanning electron micrographs (bottom) of devices investigated in §7.3 and §7.4. Details for devices A-D are given in Table 7.1. Devices E-G have an identical geometry to device B but have dimensions of 1μm × 1μm, 0.6μm × 0.6μm and 0.4μm × 0.4μm respectively and their details are also presented in Table 7.1. Further details of each of these devices are also discussed in §3.7.

Note that device C is the Cambridge square billiard referred to in Chapter 6. This device is lithographically identical to the NRC Sinai billiard except that the central circular gate and the bridging interconnect have not been fabricated. Device C is formed in the upper of the two 2DEGs in the heterostructure with the lower 2DEG isolated using the isolation surface-gates discussed in §3.6. The area presented in Table 7.1 is the lithographic area of the billiard with exception of device D where it is the area measured using the edge-state Aharonov-Bohm effect (see §4.2.4). The reason for this is that edge-state Aharonov-Bohm oscillations are not present for all of the devices, and rather than making assumptions about depletion edges that may be incorrect, it is more consistent to simply assume that the billiard is equal to the lithographic area instead. In the case of device D, however, the plunger-gate (rounded feature at the bottom in the schematic and bottom right corner of micrograph) is used to reduce the area well below the lithographic area of the device. Hence the measured area of this device is necessary instead. The ability to achieve such small billiard areas in device D is a desirable feature in §7.4. Further details and fabrication/experimental credits on the devices presented in Table 7.1 are to be found in §3.7.1 – §3.7.5.

The parameters investigated in §7.3 and §7.4 are billiard geometry, the position of the entrance and exit ports, the number of modes $n$ in these ports, the billiard area $A_B$, the temperature $T$ and the electron mobility $\mu$ in the 2DEG. The first two of these parameters are fixed by the lithography of the surface-gates that define the billiard. Hence the geometries and port locations are limited to those shown in Fig. 7.3. Aside from device D (discussed above), the billiard area is also fixed by the lithography of the surface-gates. The number of modes in the entrance and exit ports is adjusted by changing the bias applied to the surface-gates. Note that whilst this will also lead to changes in $A_B$, these changes are expected to be minimal over the range of $n$ investigated in device C. The reduction in $A_B$ below the lithographic area in device D is achieved by increasing the negative bias applied to the plunger-gate. Since the plunger-gate can be biased independently of the surface-gates that define the entrance and exit ports, $A_B$ can be adjusted with minimal effect on $n$ in this device.

# 7.3 – Dependence of Fractal Behaviour on Billiard Parameters

Electron quantum interference processes in the billiard produce the fractal MCF presented in Chapters 5 and 6. Hence I will begin by investigating the dependence of fractal behaviour on temperature. The reason for this is that by increasing the temperature, the phase coherence length of the electron decreases, which in turn suppresses the quantum interference processes that generate fractal behaviour. In particular, as the temperature is raised, quantum interference contributions from long trajectory, large area loops are reduced relative to the short trajectory, small area loops formed within the billiard. Considering the Aharonov-Bohm relationship (see §2.1.4 and §2.2.3), this means that smaller period MCF components are damped relative to larger period MCF components. This is clearly seen in the MCF data obtained from devices F and G (shown in Fig. 7.4) as the temperature is increased.

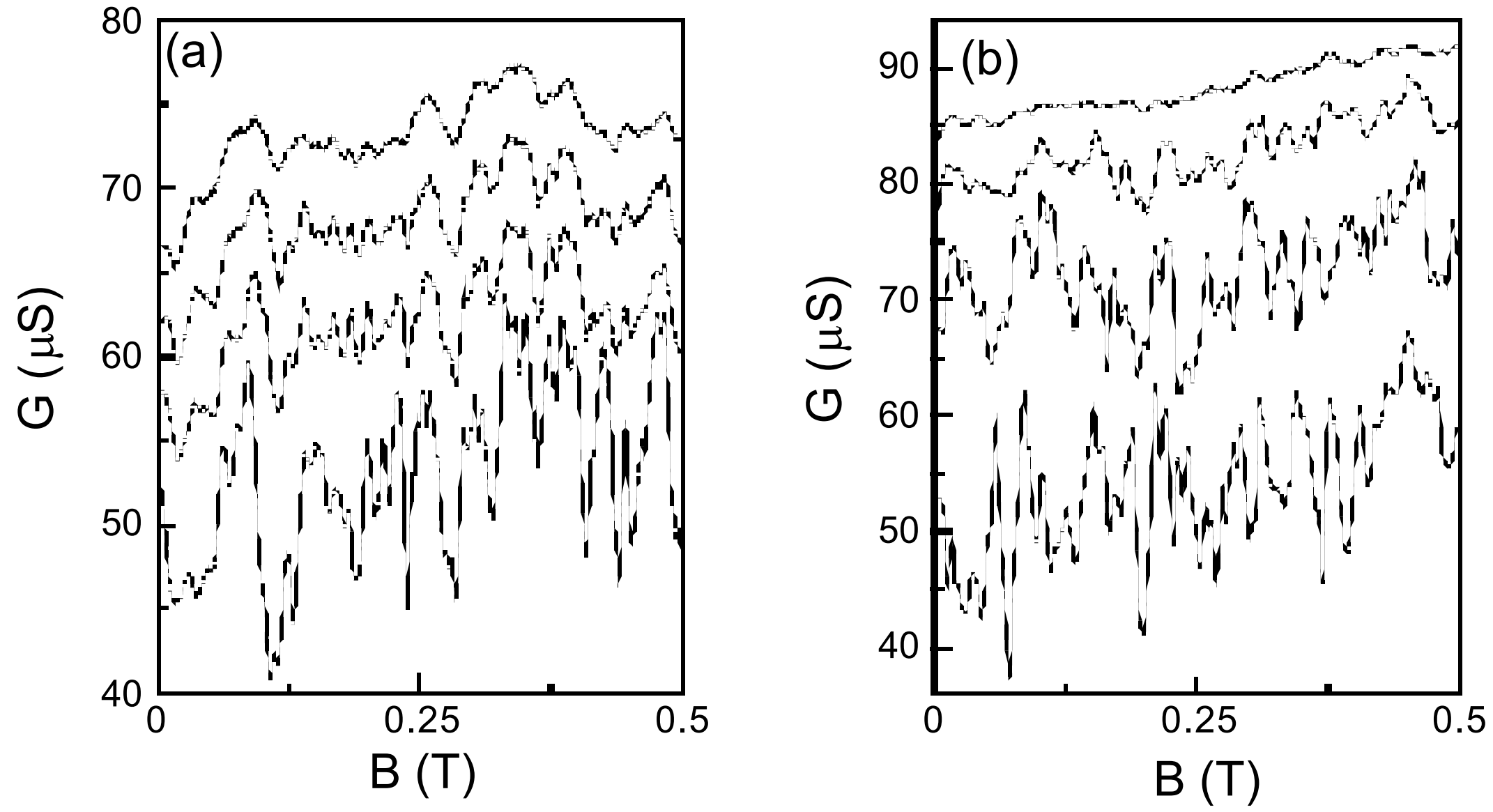

Figure 7.4: MCF data obtained from (a) Device G at temperatures (from top): 2.5K, 1.3K, 900mK and 220mK, and (b) Device F at temperatures (from top): 4.2K, 1.4K, 480mK and 80mK.

For the MCF to be fractal, the distribution of loop areas contributing to the MCF must obey a power-law. At zero temperature, this power-law is induced by the soft-wall potential profile of the billiard geometry (see §2.4.3). As the temperature is increased there are two possibilities for how the temperature-induced redistribution of loop areas that are able to contribute to the MCF affects the fractal behaviour. If the redistribution of contributing loops leads to deviations from a power-law, then fractal behaviour will either be totally removed or observed over a diminished range. However, if the redistribution results in a new power-law then the fractal nature of the MCF will be retained but described instead by a different fractal dimension. Figure 7.5 demonstrates that increasing the temperature produces a change in gradient rather than a deviation from fractal behaviour. That is, the power-law behaviour is retained, the locations of the upper and lower cut-offs (indicated by the arrows) remain unchanged but the fractal dimension changes with temperature. Note that the lower cut-off in (b) is an order of magnitude higher than $\Delta B_2$ in Fig. 6.11 due to the order of magnitude decrease in magnetic field resolution in the MCF data obtained for device G compared to device C, which was investigated in §6.3. In Fig. 7.5, $D_F$ changes from 1.47 to 1.35 as the temperature is raised from 0.22K to 2.5K. Note that $D_F$ decreases towards one as the temperature is increased, suggesting that at some temperature $D_F = 1$ will occur and the MCF will become non-fractal. This loss of fractal behaviour is expected because in the high temperature limit the phase coherence length becomes much smaller than the length of the trajectories, eliminating the quantum interference contributions to the magneto-conductance necessary for fractal MCF to be observed.

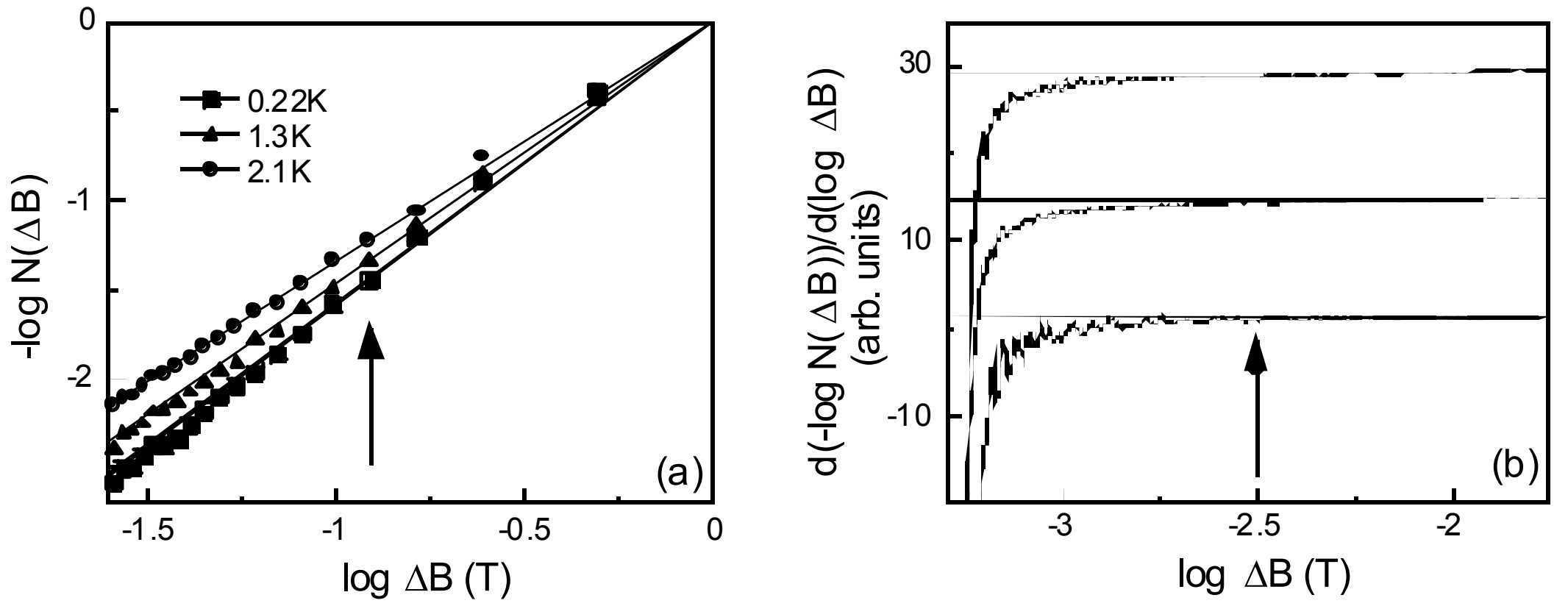

Figure 7.5: (a) Plot of $-\log N(\Delta B)$ versus $\log \Delta B$ for the MCF data presented in Fig. 7.4(a) (Device G) for temperatures of 220mK, 1.3K and 2.5K in the high $\Delta B$ limit, demonstrating that the linear behaviour is maintained and the location of the upper cut-off (indicated by the arrow) is unchanged as the temperature is varied. (b) derivative of $-\log N(\Delta B)$ with respect to $\log \Delta B$ versus $\log \Delta B$ in the low $\Delta B$ limit demonstrating the location of the lower cut-off (indicated by the arrow) also remains unchanged as the temperature is varied.

The behaviour observed as a function of temperature is also observed under changes in each of the parameters listed in the preceding section over the ranges given in Table 7.1. That is, for changes in each of the parameters, the power-law behaviour is retained, the location of the upper and lower cut-offs remain fixed and the fractal dimension changes as the parameter is varied. Note, however, that the range of fractal behaviour and the locations of the upper and lower cut-offs vary between individual samples and experiments due to differences in the cyclotron field, data resolution and noise. Following the above observation that only the fractal dimension changes with the parameters investigated in this section, discussion will be restricted to the behaviour of the fractal dimension as a function of the various parameters for the remainder of §7.3 and §7.4. Furthermore, since fractal MCF are generated by electron quantum interference effects, the quantum lifetime $\tau_Q$, which is the average time that an electron can contribute to quantum interference effects, is an important parameter. The quantum lifetime and the method employed for measuring it in billiards is discussed in §4.2.3. Due to the importance of the quantum lifetime in considering quantum interference effects, the dependence of $\tau_Q$ on the various parameters is discussed together with the fractal dimension for the remainder of this section. Figure 7.6 shows plots of the dependence of $D_F$ and $\tau_Q$ upon $T$, $n$, $A_B$ and $\mu$. Note that in each case the remaining parameters are kept constant at the values listed in each plot. Whilst an initial inspection of Fig. 7.6(a)–(c) suggests that $\tau_Q$ and $D_F$ are directly related, later analysis (§7.4) will show that the relationship is far more subtle and also far more significant.

As the temperature is reduced in Fig. 7.6(a), $\tau_Q$ is observed to initially increase and then saturate for temperatures below 150mK. Figure 7.7(a) shows that the root-mean-square fluctuation amplitude continues to increase as $T$ becomes less than 150mK, indicating that the saturation observed in Fig. 7.6(a) does not correspond to a saturation in the sample temperature. The saturation behaviour observed in $\tau_Q$ is due to a saturation in $\tau_\phi$ since the characteristic dwell-time $\tau_D$ should remain constant as a function of $T$. The saturation of $\tau_\phi$ at low temperature has previously been reported in a number of mesoscopic systems [176] including semiconductor billiards [43,44,46,47]. Although the exact cause of this saturation in $\tau_\phi$ is not currently known, in billiards, it has been established experimentally that the onset of $\tau_\phi$ saturation coincides with the thermal smearing $kT$ becoming smaller than the average energy level spacing $\Delta = 2\pi\hbar^2/m^*A_B$ (see §7.4 for details) of the billiard [43,44,46]. This is also the case in Fig. 7.7(b) where the saturation in $\tau_Q$ for both devices E and F occurs for $\Delta > kT$. $D_F$ follows a similar behaviour to $\tau_Q$ in Fig. 7.6(a). The onset of saturation in $D_F$ also coincides with the temperature where $kT$ becomes smaller than $\Delta$. Note that $\Delta$ depends upon the size of the billiard and hence the saturation temperature should increase as the billiard size is reduced. This is demonstrated in Figs. 7.7(b) and (c) where in device F (0.6μm × 0.6μm), $\tau_Q$ and $D_F$ saturate at a higher temperature than in device E (1.0μm × 1.0μm). In both cases, the saturation occurs at a temperature that coincides with $\Delta \approx kT$.

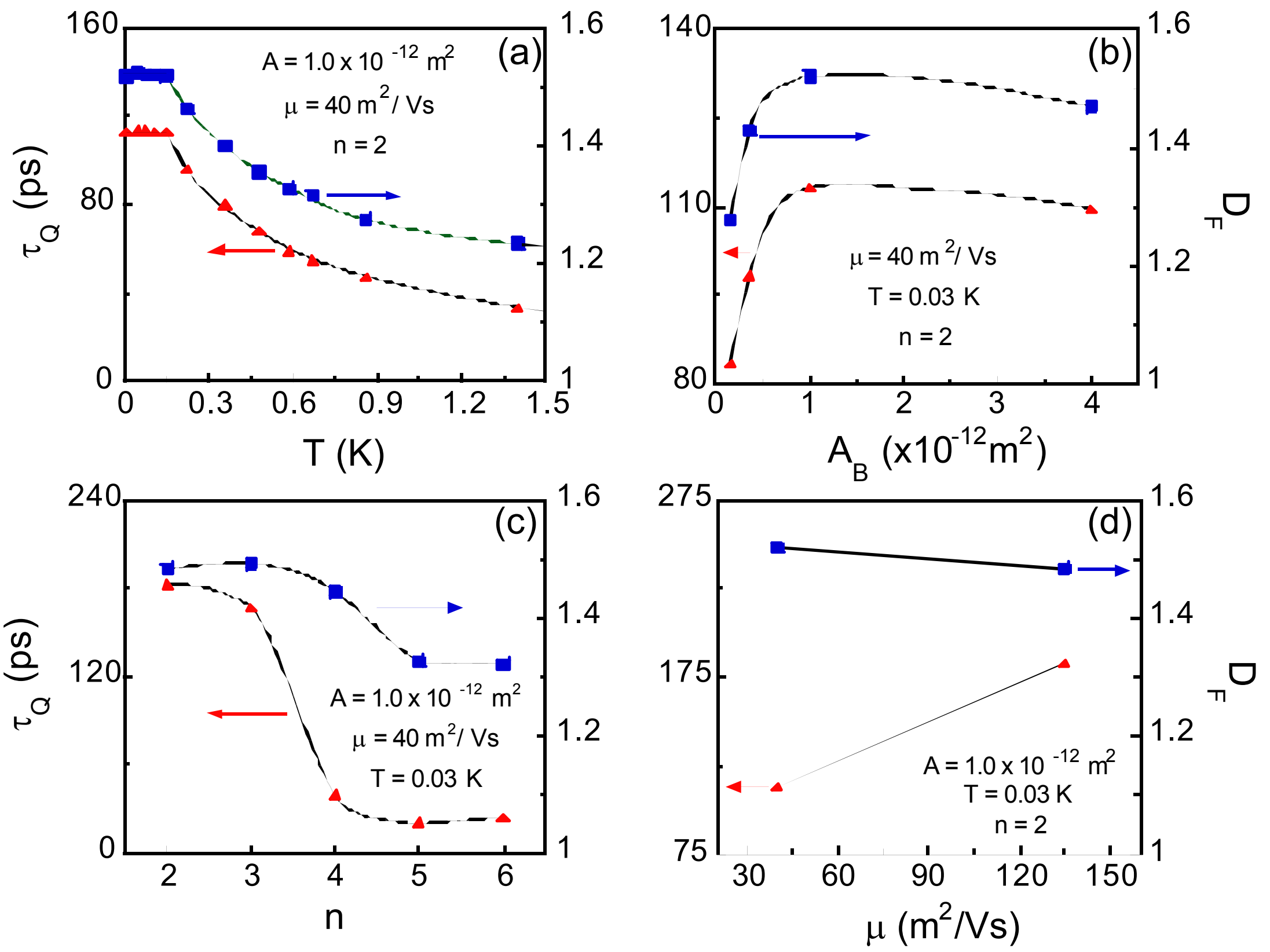

Figure 7.6: Quantum lifetime $\tau_Q$ (red triangles) and fractal dimension $D_F$ (blue squares) as a function of: (a) temperature, (b) billiard area, (c) number of modes in the entrance and exit ports and (d) 2DEG electron mobility with other parameters constant. The remaining parameter values are quoted in each plot.

Figure 7.6(b) shows the dependence of both $\tau_Q$ and $D_F$ on $A_B$. Again $\tau_Q$ and $D_F$ exhibit remarkably similar behaviour. It is important to note in Fig. 7.6(b) that $\Delta > kT$ for the smallest three $A_B$ values presented, whilst $\Delta < kT$ for the largest $A_B$ value. This means that $\tau_Q$ has saturated for the three smallest $A_B$ values but not for the largest $A_B$ value. Unfortunately it was not possible to establish the saturation $\tau_Q$ value for $A_B$ = $4.0\times10^{-12}$m$^2$ because saturation in this sample is expected to occur at 20mK, which is below the base temperature (30mK) obtainable in the experiment where device B was examined. However, the results in Fig. 7.6(b) suggest that the saturation values of $\tau_Q$

increase with $A_B$. The expected saturation value of $\tau_Q$ and the reason for any dependence of $\tau_Q$ on $A_B$ is currently not known and is not pursued further in this thesis.

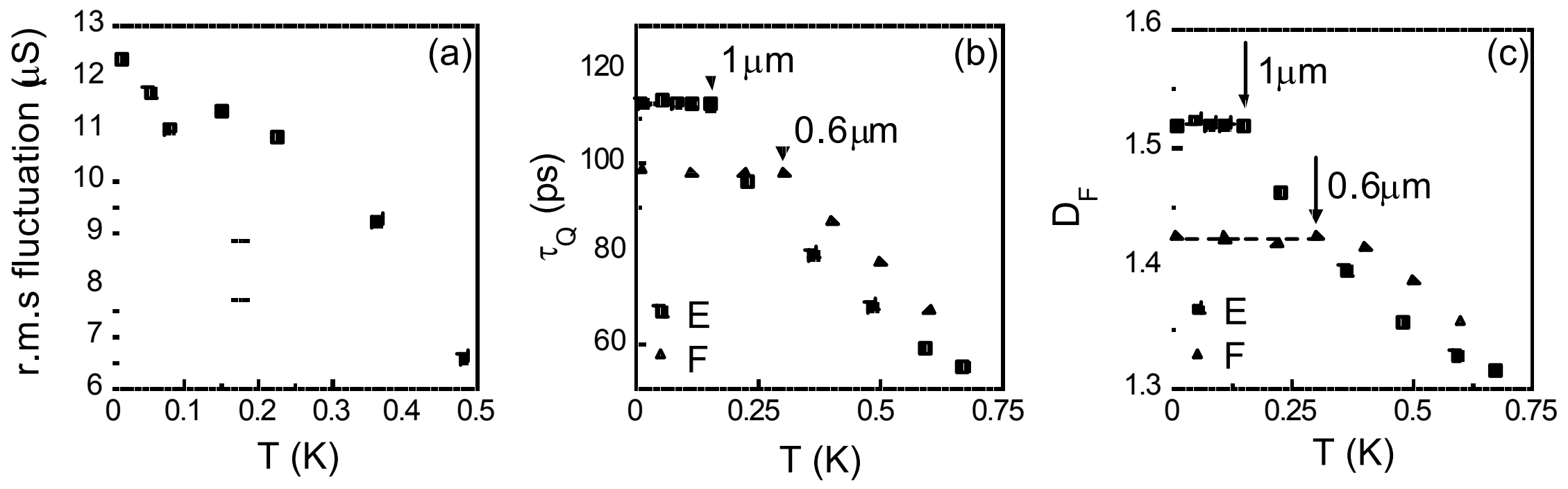

Figure 7.7: (a) Root-mean-square fluctuation amplitude versus temperature for the data presented in Fig. 7.6(a). Continued increase in amplitude below 150mK indicates that the saturation of $\tau_Q$ in Fig. 7.6(a) is a real effect and does not correspond to saturation of the actual electron temperature. (b) and (c) are plots of $\tau_Q$ and $D_F$ versus $T$ for devices E and F respectively, demonstrating that the temperature at which $\tau_Q$ saturates increases as the billiard size is reduced. This result is consistent with the saturation corresponding to $\Delta > kT$.

Figure 7.6(c) shows $\tau_Q$ and $D_F$ versus $n$ for device C. Again $D_F$ and $\tau_Q$ display remarkably similar behaviour. $\tau_Q$ has a high value for low mode number ($n$ = 2,3) and a low value for higher mode number ($n$ = 4-6), with a sudden 'step' occurring between $n$ = 3 and $n$ = 4. This behaviour has been previously reported in [45,177] using the skipping orbit analysis (albeit in terms of $\tau_\phi$ rather than $\tau_Q$) although the transition occurred at a lower $n$ (between $n$ = 1 and $n$ = 2) in these reports. It was suggested in [45,177] that this behaviour in $\tau_\phi$ was due to changing the 'environmental coupling' with $n$. That is, for large $n$ the ports are open and coupling between the external environment and electrons within the billiard leads to an increase in phase breaking. As $n$ is reduced and the ports are gradually closed, coupling to the environment is suppressed, which leads to a sudden increase in $\tau_\phi$ as $n$ is reduced below some threshold value. In contrast, whilst the dwell-time increases with decreasing $n$, the change in $\tau_D$ should be continuous rather than step-like. Hence the explanation for the sudden 'step' in $\tau_Q$ as a function of $n$ should not be affected by considering that the time measured by the skipping orbit method is a combined dwell and phase coherence time rather than simply a phase coherence time (see §4.2.3). The transition between high $\tau_Q$ and low $\tau_Q$ with increasing $n$ is coincident with a similar transition from high $D_F$ (~1.5) to low $D_F$ (~1.3). However, the transition appears to be smoother with $D_F$ than it is with $\tau_Q$.

Finally, Fig. 7.6(d) shows the dependence of $\tau_Q$ and $D_F$ upon $\mu$. Unfortunately only two different $\mu$ values with common $A_B$, $T$ and $n$ values were available for investigation. The quantum lifetime is observed to increase by ~60% with the roughly threefold increase in mobility. Whilst it would be surprising if $\tau_Q$ decreased with increasing $\mu$, a direct relationship between $\tau_Q$ and $\mu$ is not expected because the billiard is much smaller than the elastic mean free path $l_{el}$ of the 2DEG. It is interesting to note that $D_F$ behaves oppositely to $\tau_Q$ as a function of $\mu$ suggesting that the relationship between $D_F$ and $\tau_Q$ may not be as simple as Figs. 7.6(a)–(c) might otherwise suggest.

## 7.4 – Unified Picture of the Dependencies of the Fractal Dimension on Billiard Parameters

In order to understand the various dependencies of $D_F$ presented in §7.3 within a single framework, it is necessary to return to the picture of a semiconductor billiard as a two-dimensional potential well discussed in §2.4.2. In this picture there are three important parameters – the energy level spacing, the broadening of the individual energy levels and the thermal smearing of the Fermi energy. The precise energy level spectrum of the billiard is determined by the exact form of the potential well (called the potential profile hereafter). In turn, the potential profile is dependent upon the specific details of the billiard. Determining the precise energy level spectrum is time-consuming and difficult and in most cases an average measure of the energy level spacing is sufficient to interpret experimental results. The average energy level spacing $\Delta$ is defined by assuming that all energy levels up to the Fermi energy contain two electrons (assuming spin degeneracy) and that there are $n_sA_B$ electrons in the billiard. Hence there are $n_sA_B/2$ levels below the Fermi energy and:

$$\Delta = \frac{2E_F}{n_s A_B} = \frac{2\pi\hbar^2}{m^* A_B} \tag{7.1}$$

Note that this definition assumes that there is no further degeneracy. The validity of $\Delta$ is confirmed by the results in Fig. 7.6(a) and Figs. 7.7(b) and (c), where significant changes in the behaviour in both $\tau_Q$ and $D_F$ occur at the temperature where $\Delta \approx kT$. For the same reasons, the energy broadening of the individual energy levels is also replaced with an average energy broadening which is defined via the uncertainty principle as $\Delta E_Q = \hbar/\tau_Q$, where $\tau_Q$ is the quantum lifetime. It is important to note that the quantum lifetime draws together both processes that contribute to energy level broadening – loss of phase coherence and escape from the billiard – as discussed in §4.2.3. The Fermi-Dirac distribution determines the occupancy of energy levels in the billiard. At zero temperature all levels up to the Fermi energy are occupied, whilst those above remain unoccupied. At a finite temperature $T$, a range of energy levels about the Fermi energy can be partially occupied. The typical width of this range of partial occupation (known as the thermal smearing) is approximately $\Delta E_T = kT$ [178]. In this study, the energy broadening $\Delta E_Q$ and the thermal smearing $\Delta E_T$ are combined into a single 'total broadening' parameter $\Delta E$, which is defined differently depending upon the relationship between $\Delta$ and $kT$. In the case where $\Delta > kT$, the thermal smearing is no longer significant and the total broadening is simply the energy level broadening $\Delta E = \Delta E_Q$. For $\Delta < kT$ however, the thermal smearing is significant and the total broadening is *assumed* to be the sum of the energy level broadening and the thermal smearing $\Delta E = \Delta E_Q + \Delta E_T$. The validity of this definition of $\Delta E$ is demonstrated by the results discussed below.

It is possible to combine the average energy level spacing and the total broadening into a single parameter $Q$ that quantifies the overlap of the energy levels:

$$Q = \frac{\Delta}{\Delta E} \tag{7.2}$$

and which lies in the range $0 < Q < \infty$. In the limit where $Q \rightarrow 0$, the total broadening becomes much larger than the level spacing and the energy spectrum can effectively be viewed as a continuum. A fully classical picture is sufficient to adequately describe the billiard in this limit. As $Q$ increases, the overlap is reduced and this can occur by one of two mechanisms – either by increasing the level spacing or by reducing the total broadening. At $Q$ = 1 the total broadening is equal to the level spacing. As $Q$ increases above 1 the discrete nature of the energy level spectrum becomes significant and, ultimately, in the limit where $Q >> 1$, the total broadening is small compared to the level spacing, the energy level spectrum is strongly quantised and a fully quantum mechanical picture is required. Hence, $Q$ can be viewed as a parameter that quantifies the degree to which the billiard can be described by a quantum mechanical or classical picture.

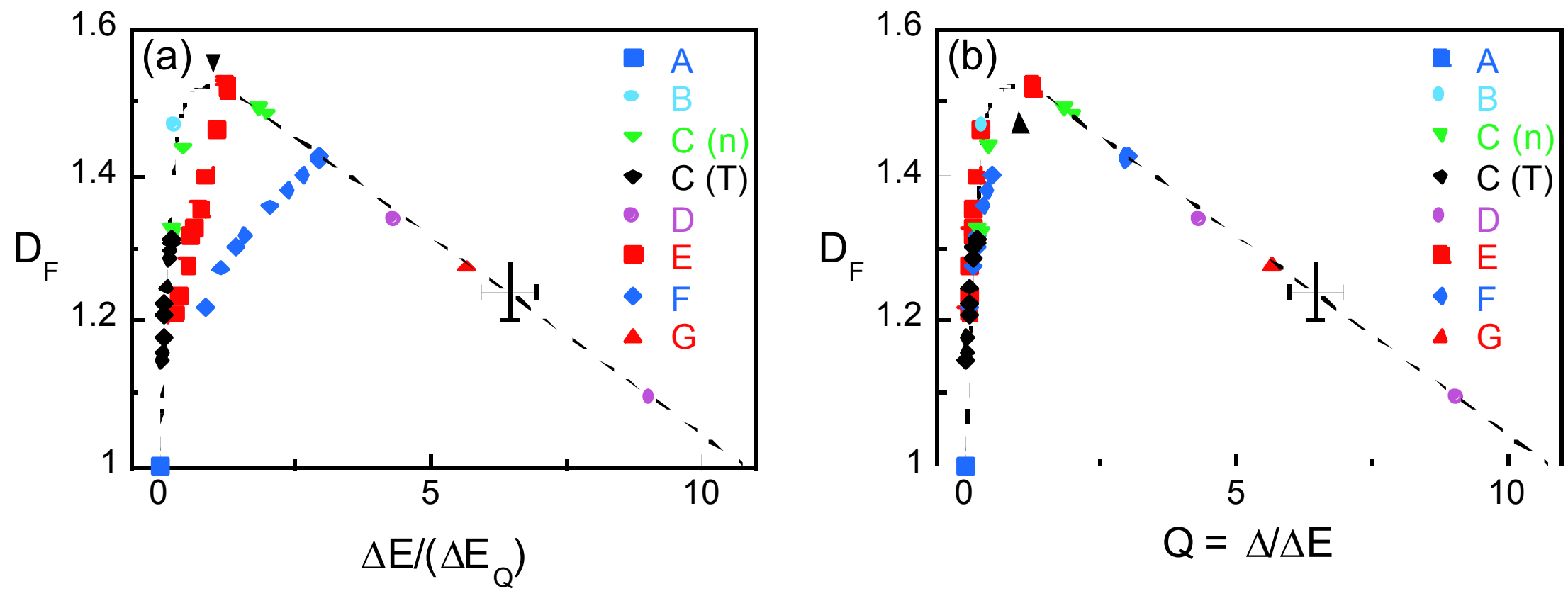

Figure 7.8: (a) Plot of $D_F$ versus the ratio of average energy level spacing Δ to energy level broadening $\Delta E_Q$ for devices A-G and the range of parameters specified in Table 7.1. (b) Plot of $D_F$ versus the ratio of Δ to the total broadening $\Delta E$, which is defined in the text. The dashed lines in (a) and (b) serve as guides to the eye only.

Since Δ is inversely proportional to $A_B$, as the billiard size is decreased, Δ increases and consequently, so does $Q$, allowing investigations in the fully quantum mechanical limit ($Q >> 1$). In contrast, increasing both $T$ and $n$ leads to a decrease in $\tau_Q$, an increase in $\Delta E_Q$ (and $\Delta E_T$ in the case of increasing $T$) and consequently a decrease in $Q$, allowing investigations in the fully classical limit ($Q \rightarrow 0$). Using all of the results obtained from examining the various devices and dependencies in Table 7.1, it is possible to investigate the relationship between $D_F$ and $Q$ for $Q$ values ranging from 0 up to ~9. However, before discussing the relationship between $D_F$ and $Q$, it is necessary to demonstrate the validity of the definition of $\Delta E$. In order to do this I will first examine the relationship between $D_F$ and a more simplistic definition of the ratio of the level spacing to the level broadening $\Delta/\Delta E_Q$ (i.e. ignore the thermal smearing). Figure 7.8(a) shows $D_F$ as a function of $\Delta/\Delta E_Q$ for devices A-G over the range of parameters quoted in Table 7.1. The data in this plot can be divided into two groups. The first group of data lies on a single curve (indicated by the dashed line) in Fig. 7.8(a). The second group of data lies off this curve, crossing instead between the segments of the curve that lay on opposite sides of $\Delta/\Delta E_Q = 1$. It is significant to note that the first group corresponds to data where $\Delta > kT$ whilst the second group corresponds to $\Delta < kT$, irrespective of device geometry, port location, $A_B$, $T$, $n$ or $\mu$. Hence, it is clear from Fig. 7.8(a) that $D_F$ behaves differently with $\Delta/\Delta E_Q$ depending on the relationship between $\Delta$ and $kT$ (i.e. whether or not the thermal smearing is significant). Based on this result it seems obvious that the thermal smearing needs to be accounted for in some way once $\Delta$ becomes smaller than $kT$. This is achieved using the definition of total broadening $\Delta E$ presented earlier.

Figure 7.8(b) shows the data presented in Fig. 7.8(a), plotted as a function of $Q$ rather than simply $\Delta/\Delta E_Q$, hence including the thermal smearing. All of the data now condenses onto the single curve in Fig. 7.8(a), irrespective of device or parameter, demonstrating the validity of the definition of the total broadening $\Delta E$ given earlier. However, the significance of this result goes well beyond proving the validity of $\Delta E$. In total, in Fig. 7.8(b) there are 48 individual measurements of $D_F$ from seven different devices with different geometries and port locations, under changes in four separate parameters, all of which lay on a single curve as a function of $Q$. This is especially significant considering that obtaining repeatable results in mesoscopic physics experiments is quite difficult. The single curve in Fig. 7.8(b) demonstrates that the fractal MCF obtained using the various devices and parameters listed in Table 7.1 can be viewed under a single unified picture where the fractal dimension depends only upon the ratio of the average energy level spacing to the total broadening. However, the full significance of this single curve becomes apparent on considering the behaviour of $D_F$ with $Q$.

Commencing with $Q = 0$, which corresponds to the extreme classical limit, $D_F$ is equal to 1. This is not surprising since in this limit, there is no quantum interference and hence no MCF. As $Q$ is gradually increased towards 1 and phase coherent effects gradually become significant, $D_F$ is observed to also rise gradually, peaking at a value of 1.52 at $Q \sim 1$ (indicated by the arrow in Figs. 7.8(a) and (b)). Note that $Q = 1$ corresponds to an intermediate regime that lies between the fully classical and fully quantum mechanical descriptions. As $Q$ is increased beyond 1, the overlap of the energy levels continues to decrease and $D_F$ begins to decrease linearly with $Q$. Ultimately, as $Q$ becomes much larger than 1, the billiard moves into the fully quantum regime and $D_F$ heads towards 1. Note that the point with highest $Q$ in Fig. 7.8(b) is from device D ($Q$ = 9) with $A_B = 9.0\times10^{-14}\text{m}^2$. This corresponds to a 0.3μm × 0.3μm square where $W/\lambda_F \cong 8$. Extrapolating the linear trend in $D_F$ for $Q > 1$, $D_F = 1$ occurs at value of $Q \sim 11$, which corresponds to a QPC with $n \cong 5$ – a truly quantum system. As discussed in §2.4.3, under the current theory presented by Ketzmerick [12], the presence of statistically self-similar fractal MCF is a semiclassical effect relying not only on $Q$ being close to 1 but also on $E_F/\Delta$ being large. This is the case for the samples near $Q = 1$ in Fig. 7.8(b). Hence it is quite significant that the maximum $D_F$ is obtained at $Q \sim 1$, as this provides experimental confirmation of the suggestion that fractal MCF is a semiclassical phenomenon. It is interesting to note however, that no discussion of how the fractal behaviour is affected by the break-down of the semiclassical approximation is discussed in [12]. Rather than a sudden loss of fractal behaviour as the semiclassical approximation becomes invalid, the results in Fig. 7.8(a) show that $D_F$ instead decreases to 1 in a smooth continuous manner in both the fully quantum mechanical and fully classical limits. What is most surprising is that fully non-fractal MCF (i.e. $D_F = 1$) is not obtained until the fully quantum mechanical, and more particularly the fully classical limit ($Q = 0$) is reached. This demonstrates that the semiclassical approximation needs to be completely invalid (i.e. no quantum interference whatsoever or $E_F/\Delta$ very small), not just an inadequate description, before total loss of fractal behaviour occurs. Future investigations will focus on this aspect further.

# 7.5 – Investigating the Effect of Potential Profile: The Double-2DEG Billiard

The one parameter that was not considered in the preceding sections was the softness of the potential profile of the billiard. Whilst this has no effect upon the average level spacing, which is entirely dependent upon $A_B$, or the total broadening, which is dependent upon $\tau_Q$ and also $T$ for $\Delta < kT$, the precise energy level spectrum of the billiard is highly dependent upon the exact details of the potential profile. Indeed, it is important to note that Ketzmerick also emphasises the importance of the exact form of the potential profile in determining the expected value of $D_F$ [12,15]. In Ketzmerick's theory (see §2.4.3), it is the soft-walled nature of the potential profile that generates the mixed phase-space that is responsible for fractal MCF. In this section, I discuss the effect of potential profile on the fractal dimension of statistically self-similar fractal MCF.

The billiards investigated in this thesis are defined using negatively biased surface-gates. Hence it would be expected that the potential profile will vary with the depth of the 2DEG beneath the heterostructure surface because the strength of the Coulomb interaction depends upon the distance between two charges. However, the situation is more complicated than this because the intervening semiconductor layers affect the electrostatics between the surface-gates and the 2DEG. Furthermore, the precise details of the surface-gates should also have some effect upon the potential profile and the presence of ionised donors will impose an added random disorder potential upon the 2DEG. These issues are highlighted by Ketzmerick as major problems in obtaining accurate predictions of $D_F$ from theoretical models [12]. The results of the preceding section suggest that these problems may not be as significant as suggested in [12]. Devices A-G have 2DEG depths ranging from 60-90nm, with differing material structures between the surface-gates and the 2DEG, differing doping profiles and different surface-gate geometries and thicknesses, yet $D_F$ only appears to depend upon $Q$, with the results from each of these devices lying on a single curve. The issue here then, is whether the exact form of the potential profile does not effect $D_F$ or whether its effect is too small to resolve in the results of devices A-G. Since the 2DEG depth alters the potential profile, and the 2DEG depth is a readily controllable parameter, it is clear that this is the most appropriate parameter to investigate in order to answer this question.

A system for performing this experiment is the double-2DEG billiard. Using such a device, two billiards can be formed using the same set of surface-gates. This prevents any differences in the precise details of the surface-gates from causing a difference in potential profile between the billiards. This would be a problem if our investigation was performed using two separate single-2DEG billiards instead. The barrier between these devices is 70nm thick (20nm GaAs, 30nm AlGaAs, 20nm GaAs). Hence the 2DEGs are well separated, preventing interaction effects (e.g. Coulomb drag [179] and interlayer tunnelling [180]) from affecting the results.

The fabrication and details of the double-2DEG billiard studied here are presented in §3.6 and §3.7.5. A schematic and scanning electron micrograph of the surface-gates for this device are shown in Fig. 7.3(c). Note that the top-2DEG in this sample has already been investigated as device C in §7.2 and §7.3. The results from device C were presented in Fig. 7.8(b). Results from the bottom-2DEG have not been presented thus far, but have been obtained for the same range of parameters as device C (see Table 7.1). The most important detail in this section is the depth of the bottom-2DEG, which is 160nm, between 1.8 and 2.7 times the depth of the 2DEG in devices A-G. Hence if there is any effect due to changes in potential profile it is expected they will be observed in the billiard formed in the bottom-2DEG.

Since this was the first experiment on a double-2DEG billiard, a careful characterisation of the device was performed to establish how similar the two billiards are and the suitability of such a device for the intended investigation of fractal behaviour as a function of potential profile. Characterisation commenced with an investigation of the properties of the two 2DEGs in order to establish whether they had similar electron density/mobility, and that the two 2DEGs could be measured independently. Following this, an attempt was made using the edge-state Aharonov-Bohm effect to establish which billiard has the softer potential profile. The techniques employed in these characterisations have been presented in §4.2.1 and §4.2.4. However, given that the presence of two 2DEGs complicates the results and that this is the first use of a double-2DEG heterostructure in a billiard experiment, the characterisations are discussed here rather than in Chapter 4.

As discussed in §3.6, the ohmic contacts in this device are fabricated so that the 2DEGs can be measured in parallel or independently by depleting the Hall-bar arms of one of the 2DEGs using the isolation surface- or back-gates. Shubnikov-de Haas oscillations and the Hall effect for the bulk-2DEGs are shown in Figs. 7.9(a) and (b) respectively. The black traces in Figs. 7.9(a) and (b) are for the two 2DEGs measured in parallel. The Shubnikov-de Haas oscillations exhibit a beating effect indicating the presence of two 2DEGs with slightly differing electron density. Note that the beating is not present in the single-2DEG traces confirming that the 2DEGs can be measured independently. The Hall effect trace gives an $R(B)$ exactly half that for the single-2DEGs since the two 2DEGs have roughly equal resistance and add in parallel. In the black trace only even-numbered Hall plateaux are observed. Note that this effect is again due to the parallel addition of resistance, which halves the resistance of the single-2DEG plateaux, doubling the apparent filling factor $\nu$ of plateaux at the same $B$ in the double-2DEG trace, as indicated by the dashed lines in Fig. 7.9(b).

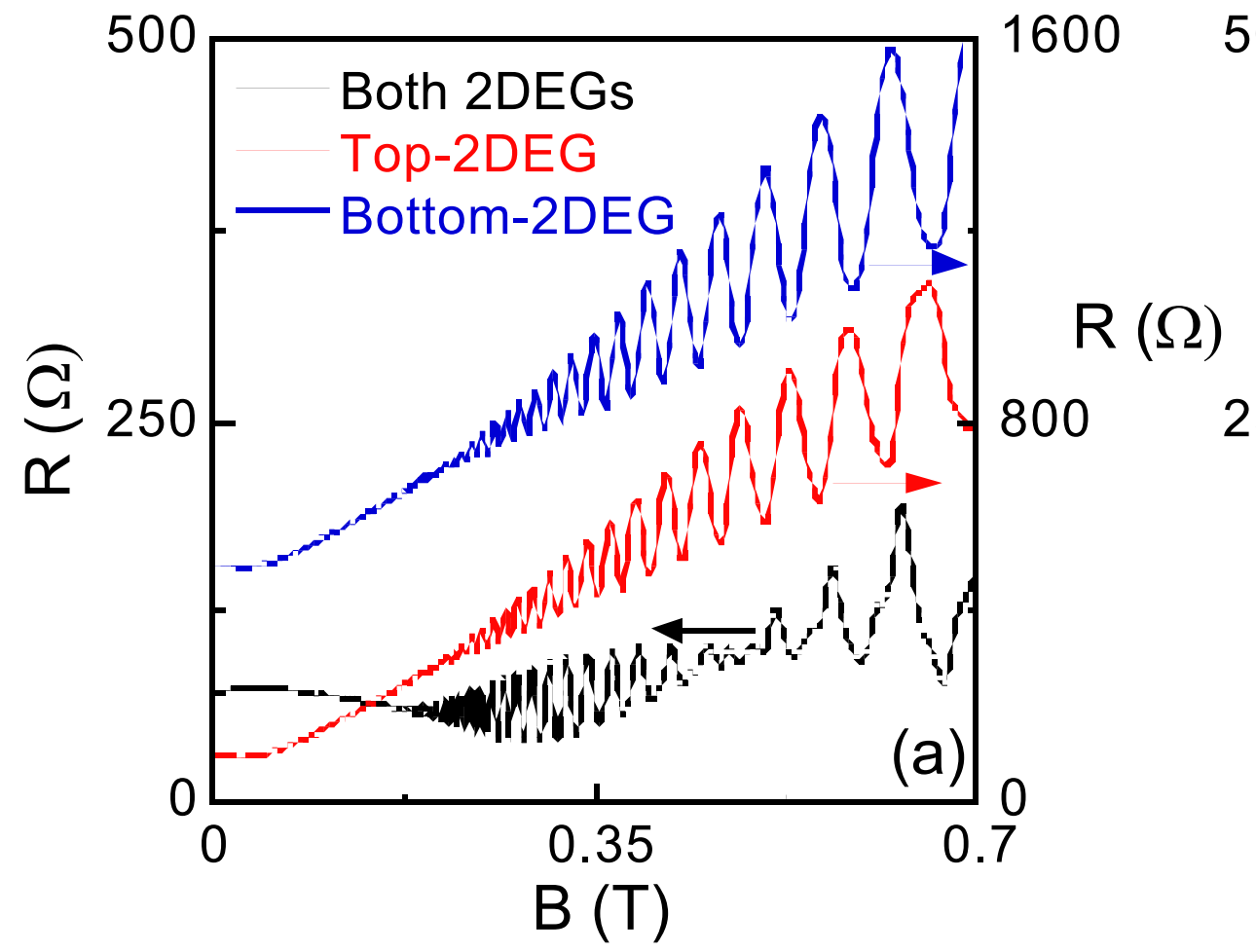

Figure 7.9: Characterisation of the double-2DEG billiard. (a) Shubnikov-de Haas oscillations for the top- and bottom-2DEGs (red, blue – right axis) and both 2DEGs (black – left axis) measured in parallel. Traces have been offset for clarity. Resistances at zero magnetic field are 102Ω, 130Ω and 74Ω respectively. (b) Hall effect traces corresponding to those in (a). Dashed lines indicate plateaux lying at common $B$ values and the numbers correspond to the filling factor $\nu$ of the plateaux. Traces have been offset for clarity. Resistances at zero magnetic field are all ~20Ω.

Note that short odd-numbered plateaux are sometimes observed in the double-2DEG Hall trace and these are due to slight misalignment of the plateaux in $B$ in the individual 2DEGs. The electron densities and mobilities of the two-2DEGs differ by ~3% and

~20% respectively with the top-2DEG having the higher electron density and mobility, as indicated by the lower resistance at zero magnetic field in the top-2DEG.

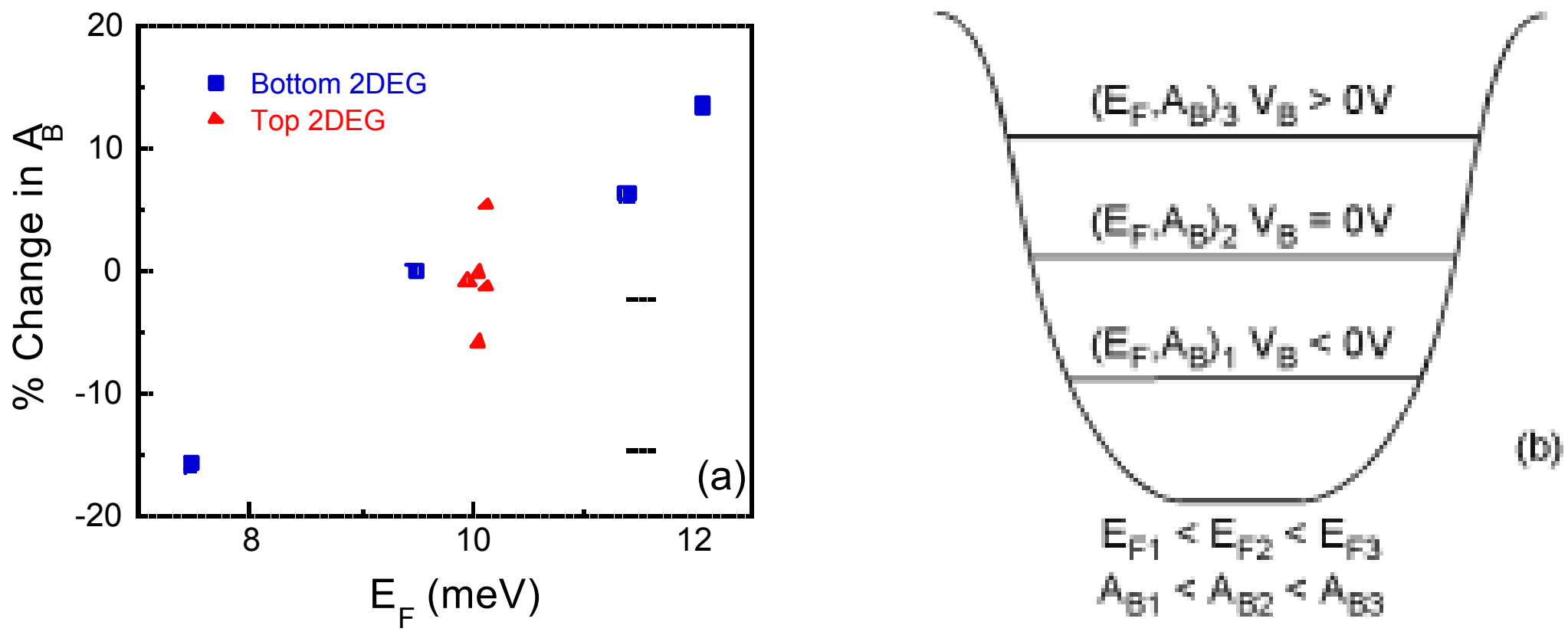

Figure 7.10: (a) Results of an attempt to assess whether the billiard in the top- or bottom-2DEG has the softer potential profile. Change in billiard area $A_B$ from that found with the full back-gate unbiased is measured as a function of Fermi energy $E_F$, which is directly dependent on electron density in the 2DEG. Unfortunately it was not possible to adjust the electron density in the top-2DEG using the back-gate, however these results serve as an estimate of the error in this characterisation. $A_B$ was measured using the edge-state Aharonov-Bohm oscillations discussed in §4.2.4. (b) Schematic illustrating the method employed in obtaining (a) and demonstrating that $A_B$ should increase with $E_F$ for a soft-wall potential as observed in (a).

Figure 7.10(a) presents the results of an investigation to determine which 2DEG has a softer potential profile, and were obtained using a separate characterisation device with nominally identical surface-gate lithography. This investigation was achieved by adjusting the Fermi energy $E_F$ and measuring the billiard area $A_B$ via the edge-state Aharonov-Bohm effect, as illustrated in the schematic in Fig. 7.10(b). In a soft-wall potential well, $A_B$ should increase with $E_F$. However, the change in $A_B$ as a function of $E_F$ should be larger for softer potential profiles. Adjustments in $E_F$ are achieved by adjusting the electron density $n_s$, since $E_F = \hbar^2 \pi n_s / m^*$, using the full back-gate (see §3.6). Unfortunately, using the full back-gate, significant changes in $E_F$ could only be achieved in the bottom-2DEG. Hence it was only possible to establish that a soft-wall potential exists for the billiard in the bottom-2DEG. In order to achieve the intended

assessment (i.e. determining which 2DEG has the billiard with the softer potential profile) a full top-gate over the surface-gates defining the two billiards is necessary. Such a device is planned for future investigations.

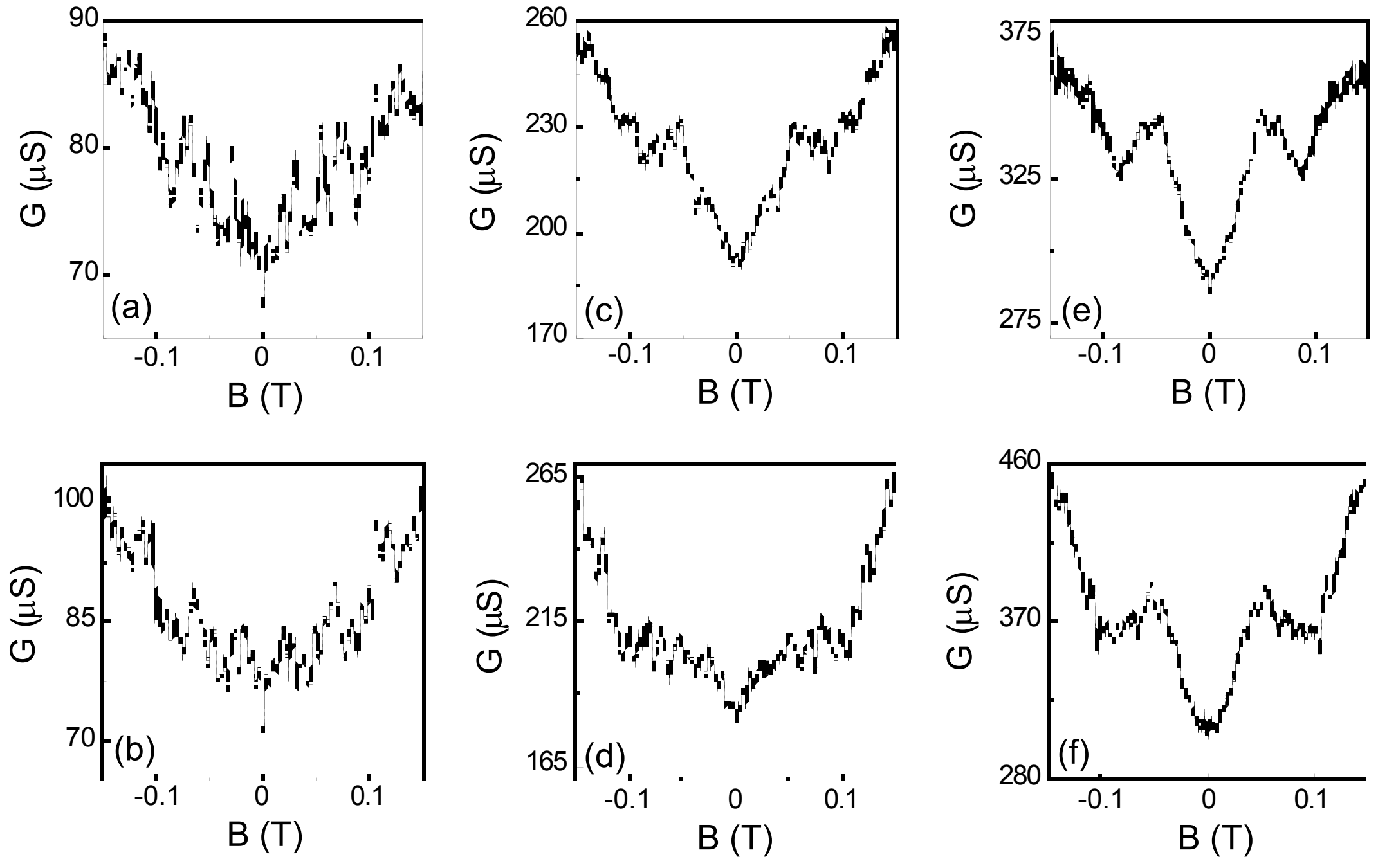

Figure 7.11: Magneto-conductance traces from the billiard in (a,c,e) the top-2DEG and (b,d,f) the bottom-2DEG. Traces have been obtained for $n$ values of (a,b) 2, (c,d) 5, and (e,f) 8 at $T = 30$mK.

All data from the double-2DEG billiard presented hereafter is obtained from the billiard in either the top- or the bottom-2DEG with the other 2DEG disconnected from all four measurement leads using the isolation surface- or back-gates (see §3.6 for details). Typical magneto-conductance data obtained from the billiards formed in the top- and bottom-2DEGs for $n = 2$, 5, and 8 is presented in Figs. 7.11(a,b), (c,d) and (e,f) respectively. By comparing the data between the billiards formed in the top- and bottom-2DEGs, it is clear that the two billiards are different since the MCF obtained for the top and bottom billiards at each $n$ value are not entirely identical. However, it is clear that the two billiards have a similar geometry since their classical backgrounds are similar and the background of both billiards changes in a similar way as a function of $n$. These background similarities are expected because both billiards are defined by the same set of surface-gates. The differences in the MCF are due to differences in the quantum interference processes that occur in the two billiards. This suggests that the energy level spectrum of the two billiards is different, as would be expected if they have different potential profiles. Note, however, that this may also be due to differences in $A_B$ and $\tau_Q$ between the two billiards. It is important to note that characterisation studies demonstrate that the two billiards do in fact have slightly different areas, and that the billiard formed in the bottom-2DEG is actually larger than the billiard formed in the top-2DEG. The difference in area in the characterisations was found to be ~$9\times10^{-14}$m$^2$ (i.e. approximately 9% of the total billiard area). However, to remain consistent with the earlier analysis in §7.3 and §7.4, where measured values of $A_B$ are not available for all billiards, I will assume here that they both have an area matching that defined by the surface-gates ($A_B$ = $1.0\times10^{-12}$m$^2$). I note that the absence of accurate $A_B$ values is a weakness in this result. Further experiments are planned to address this issue now that the behaviour in this chapter has been discovered. Furthermore, it is possible that differences between the billiards may be due to material properties since the two 2DEGs have separate n-AlGaAs donor layers. This may occur in one of two ways: via the random 'eggshell' potential of the ionised donors or the presence of an impurity within the billiard itself. The former effect is expected to be minimal whilst the latter should affect the electron dynamics through the billiard. The remarkable similarity between the classical backgrounds of the top- and bottom-billiards in Fig. 7.11 suggest that the latter effect is not occurring in this device. Measurements of $D_F$ versus $Q$ for the billiard in the bottom-2DEG were obtained for the same range of $n$ and $T$ as the billiard formed in the top-2DEG (device C) in Table 7.1. These results have been added to those obtained in Fig. 7.8(b), which is shown again as Fig. 7.12(a) for comparison, as Fig. 7.12(b).

In Fig. 7.12(b), the results obtained from the billiard in the bottom-2DEG lay on a curve (red dashed/dotted line) of similar form to that in Fig. 7.12(a) (black dashed line). The $D_F$ values for the billiard in the bottom-2DEG clearly head towards one for $Q \to 0$. Unfortunately, high $Q$ results could not be achieved for the billiard in the bottom-2DEG. Further experiments using double-2DEG billiards with smaller billiard areas are planned to address this. Despite the lack of results for high $Q$, the small number of results that have been obtained for $Q > 1$ appear consistent with the linear decrease in $D_F$ with increasing $Q$ towards $D_F = 1$ at $Q \sim 11$, as observed in Fig. 7.12(a) (indicated by the red dotted line). The peak $D_F$ value still occurs at $Q \sim 1$ for the billiard in the

bottom-2DEG. However, the peak $D_F$ value is lower than that obtained for the other billiards (1.45 compared to 1.52). Note that this behaviour is not expected to be due to billiard area or quantum lifetime differences between the billiards formed in the top- and bottom-2DEGs since these parameters simply act to shift the data along the curve, not onto another curve, as discussed in §7.4. Hence it seems clear that the effect of the potential profile is to adjust the peak $D_F$ value of the curve.

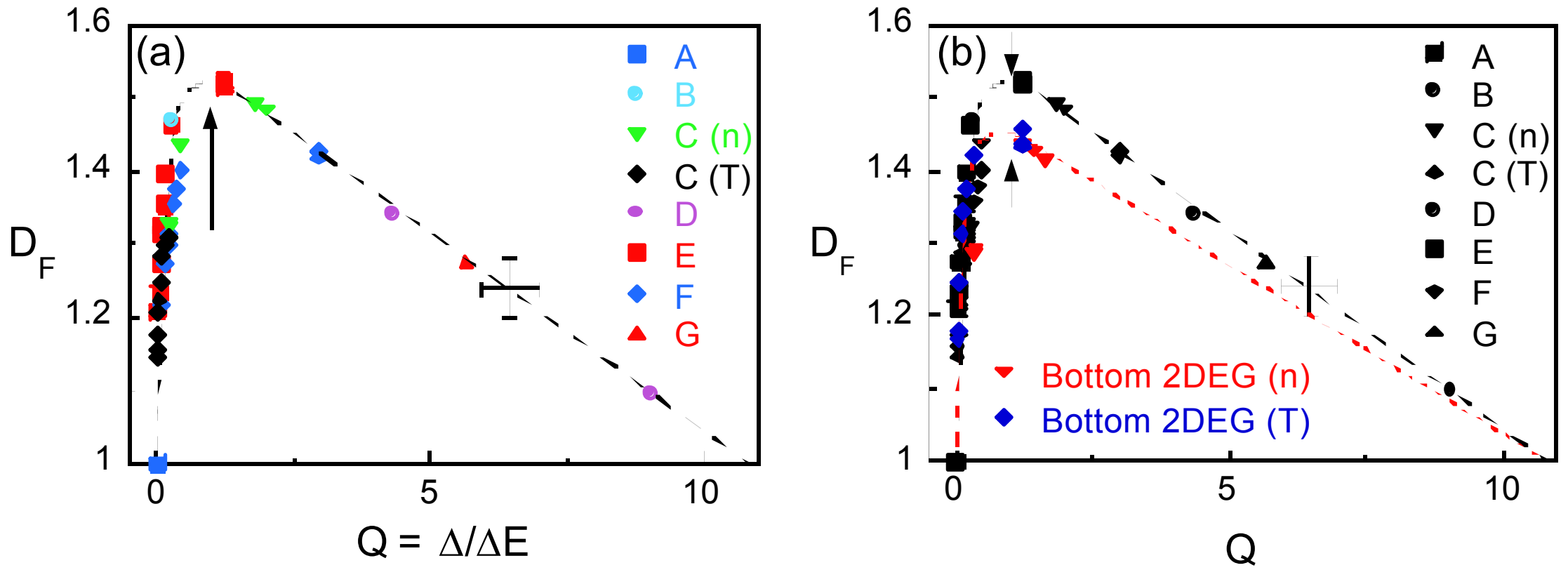

Figure 7.12: (a) The results obtained in §7.4 for the devices and parameters listed in Table 7.1. (b) $D_F$ versus $Q$ for the bottom billiard with adjustments in both $n$ and $T$. The data in (a) is shown in black for comparison. The arrows in both plots indicate $Q = 1$. Dashed/dotted lines serve as guides to the eye.

At present, however, it is unclear how much the potential profile changes between the billiards formed in the top- and bottom-2DEGs. Hence, it is not possible to say whether the effect of changing the potential profile is less significant than Ketzmerick suggests [12], or that the effect of changing the potential profile is significant but that the potential profile has only changed by a small amount between the two billiards. This will be the subject of further investigations; however, the results from this first experiment are promising.

# 7.6 – Summary and Conclusions

The physical dependencies of the fractal behaviour discussed in the preceding two Chapters were investigated in this Chapter. Further experiments were performed in order to establish the cause of the exact self-similarity reported in Chapter 5. These focussed on a side-wall Sinai billiard where the curvature is on one of the walls of the square rather than the obstacle at the centre of the billiard. The MCF of the side-wall Sinai billiard was found to exhibit statistical self-similarity rather than exact self-similarity, confirming that the presence of the obstacle is most likely responsible for the exact self-similarity observed in the Sinai billiard discussed in Chapter 5. Indeed, all of the 'empty' billiard geometries (i.e. without an obstacle at the centre) investigated in this thesis exhibit statistically self-similar MCF. Exact self-similarity has only been observed in the Sinai billiard discussed in Chapter 5 to date.

The remainder of this Chapter examined the physical dependencies of this type of fractal MCF. Seven devices with differing billiard geometry, port locations, billiard area and 2DEG electron mobility were investigated as a function of temperature and the number of modes transmitting in the ports. Based on Chapter 6, it is clear that fractal behaviour can change in one of two ways as the billiard parameters are modified. The first is a change in the range of observation of fractal behaviour due to shifts in the upper and lower cut-offs. The second is a change in the fractal dimension $D_F$ observed over the range of fractal behaviour. This was initially investigated with temperature since the phase coherence time is reduced by increasing the temperature, gradually eliminating the quantum interference processes that generate the fractal MCF observed in Chapter 6. It was found that the range of fractal observation was unaffected by changes in $T$, contrary to the suggestion by Ketzmerick in [12]. Instead $D_F$ was observed to decrease as $T$ was increased. Similar behaviour was observed with changes in the other billiard parameters.

The fractal dimension was then investigated as a function of the four quantifiable parameters – temperature, billiard area, number of modes in the QPCs and 2DEG electron mobility. Whilst $D_F$ exhibits a dependence upon each of these parameters, the most significant result is that the various dependencies can be drawn into a unified picture where $D_F$ relies on a single parameter $Q$. This parameter $Q$ is the ratio of the average energy level spacing to a total broadening parameter that includes the effect of finite lifetime in a single quantum state and thermal smearing of the Fermi-Dirac distribution. $D_F$ is found to lay on a single curve as a function of $Q$, irrespective of the billiard geometry, port location, billiard area, temperature, number of modes in the QPCs or electron mobility. This curve passes through $D_F = 1$ at $Q = 0$ and $Q \cong 11$ (corresponding to the fully classical and fully quantum regimes respectively) with a single peak of $D_F = 1.52$ at $Q = 1$. Providing the average level spacing is small compared to the Fermi energy, the semiclassical approximation for transport through the billiard is expected to be most valid at $Q \sim 1$. Over the entire range $0 < Q < 11$ the curve is smooth and continuous and extends linearly between $Q \sim 1$ and $Q \sim 11$. This trend is significant as it provides the first investigation of the effect of the break-down of the semiclassical approximation on statistically self-similar fractal MCF in both the classical and quantum limits. The MCF only becomes non-fractal ($D_F = 1$) in the extreme classical and quantum limits where the semiclassical approximation is completely invalid. The suppression of fractal behaviour occurs in a smooth continuous way as these limits are approached.

The fractal behaviour in billiards is expected to be strongly dependent upon the exact form of the potential profile, which in turn is expected to be dependent upon the depth of the 2DEG beneath the heterostructure surface. An investigation of the dependence of fractal behaviour on 2DEG depth was conducted using a double-2DEG billiard. The 2DEG depth in the single-2DEG samples (used for the investigation discussed in the preceding paragraphs) varied between 60 and 90nm. The 2DEG depths in the double-2DEG sample were 90nm and 160nm. Hence the billiard formed in the bottom-2DEG was between 1.8 and 2.7 times as deep as the 2DEGs investigated in the single 2DEG samples. Data was obtained for the bottom billiard as a function of $T$ and $n$ and the dependence of $D_F$ on $Q$ was again investigated. For the billiard in the bottom-2DEG, $D_F$ follows a similar behaviour to that observed in the other billiards. However, the maximum $D_F$ value, whilst still occurring at $Q = 1$, is significantly lower for the billiard in the bottom-2DEG and it is expected that this is due to the difference in potential profile. The data from the single-2DEG billiards and the billiard in the top-2DEG (60-90nm deep) all lay on a single curve while the data from the billiard in the bottom-2DEG (160nm deep) lay on a curve with the same form as that above but with a different peak $D_F$ value. This suggests either that the change in potential profile with depth is small, or that the effect of potential profile on $D_F$ is not as significant as suggested in [12].

Further investigations of the results presented in this Chapter are planned. There are two main aims of these further investigations. The first is to understand why $D_F$ depends upon $Q$ in the way that it does. The second is to further understand how the peak $D_F$ value is determined, and how it depends upon the depth of the 2DEG beneath the heterostructure surface.

# Chapter 8 – Future Directions

At the conclusion of this thesis, it is clear that the study of fractal MCF in billiards is far from complete. Whilst theoretical investigations of fractal MCF are currently in progress, the need for further experiments is also obvious. In this final Chapter, I will discuss some suggestions for future experiments. For clarity, this Chapter is divided into three sections. Future experiments to further examine exact and statistical self-similarity in billiards are discussed in §8.1 and §8.2 respectively. The future for double-2DEG billiards is discussed in §8.3.

## 8.1 – Exact Self-similarity

To date, exact self-similarity has only been observed in one Sinai billiard sample discussed in Chapter 5 of this thesis. As discussed in Chapter 3, attempts were made as part of my research to fabricate another Sinai billiard of the type measured by Taylor *et al.* [13]. Unfortunately, these attempts were unsuccessful and hence a second observation of exact self-similarity remains to be made. Reproducing the results of [13] is an important goal for future experiments. Not only for experimental verification of the existence of exact self-similarity in the MCF of billiards, but also because by doing this, the technology required for further investigations – namely the capability to fabricate bridging interconnects – will be re-established.

Following a successful second observation of exact self-similarity in a Sinai billiard, there are a number of further experiments that would enable the cause of exact self-similarity to be investigated. Figure 8.1 shows schematics of a few possible variations to the original Sinai billiard design. Note that in each case, the same bridging interconnect technology employed in the original Sinai billiard would be required in order for independent biasing of the obstacles to be possible. The first possible variation involves changing the outer gate geometry (Fig. 8.1(a)). This would allow the role of the confinement in the observation of exact self-similar MCF to be examined. The second possible variation is to move the locations of the ports (Fig. 8.1(b)). The locations of the ports play an important role in electron transport through the billiard as recently highlighted by Bird *et al.* [181]. Thirdly, the role of the obstacle in generating exactly self-similar MCF could be established by changing its geometry (Fig. 8.1(c)) or its location/orientation in the billiard (Fig. 8.1(d)). Such an investigation could also be performed using the double-obstacle billiard shown in Fig. 8.1(e). In this device, it is possible to 'move' the obstacle in a single billiard, hence eliminating small differences in outer-gate lithography and material properties as causes for any effects observed. Furthermore, by biasing both obstacles at once, it would be possible to create an extended obstacle that could be continuously modified in shape, again with the advantage of eliminating other variables as causes for any effects observed. Note that this final device would require two bridging interconnects. However, the largest hurdle is going to be re-establishing the fabrication process for the bridging interconnect in the first place. The fabrication of two or more interconnects should follow comparatively easily.

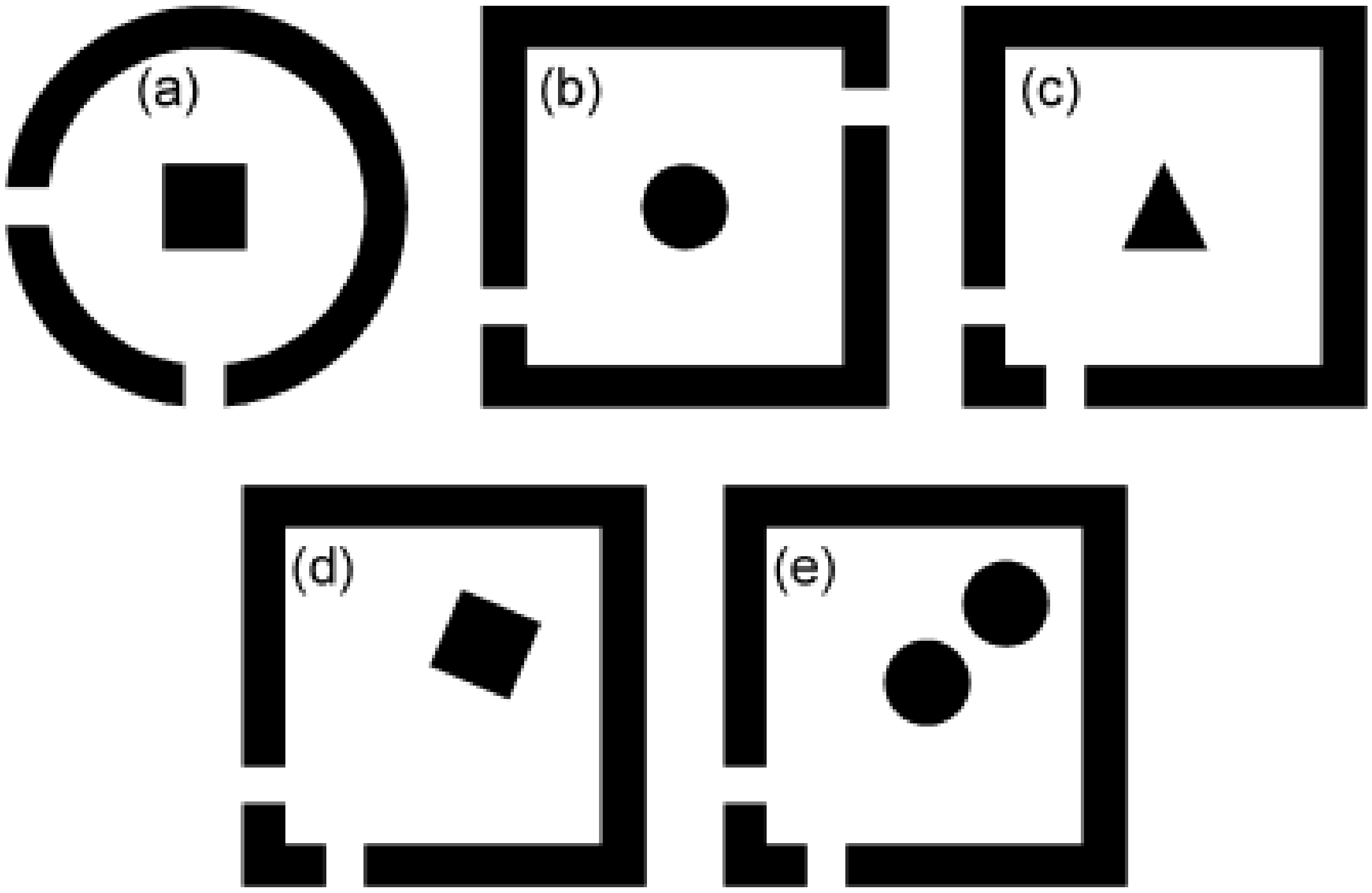

Figure 8.1: Possible modifications to the original Sinai billiard design for future investigations of exact self-similar MCF: (a) different outer gate geometry, (b) different port positions, (c) different obstacle geometry, (d) different obstacle location and orientation, and (e) double obstacle device.

Another possible device, which is also relevant to the further study of statistical self-similarity (see §8.2), is shown in Fig. 8.2(a). The Quantum Flow Disruptor (QFD) device is designed to investigate the role of periodic orbits in electron transport by interfering with the injection onto the orbit (left billiard), or the orbit itself (right billiard). The central billiard in this device is present as a control. Note that only one billiard in this device is active at a time to prevent coupled billiard effects and the formation of sub-devices by the gate-leads of the different billiards. The small gap at the top of each billiard in Fig. 8.2(a) is to allow independent control of the two QPCs in the bottom left-hand corner of each billiard (as in the Sinai billiard – see §5.1). This gap is expected to be fully pinched-off during operation of each particular billiard. Whilst this particular device is unlikely to present any major revelations in terms of exact self-similar MCF, extensions of this device into those suggested in Fig 8.1 may prove interesting. Firstly, such a device will allow an investigation of whether exactly self-similar MCF requires that the obstacle is detached from the outer gates (i.e. whether it is necessary for electrons to be able to orbit the obstacle). Secondly, recent theoretical investigations [52] suggest that periodic orbits play an important role in the generation of exactly self-similar MCF, and such a device would allow the importance of these periodic orbits to be examined experimentally.

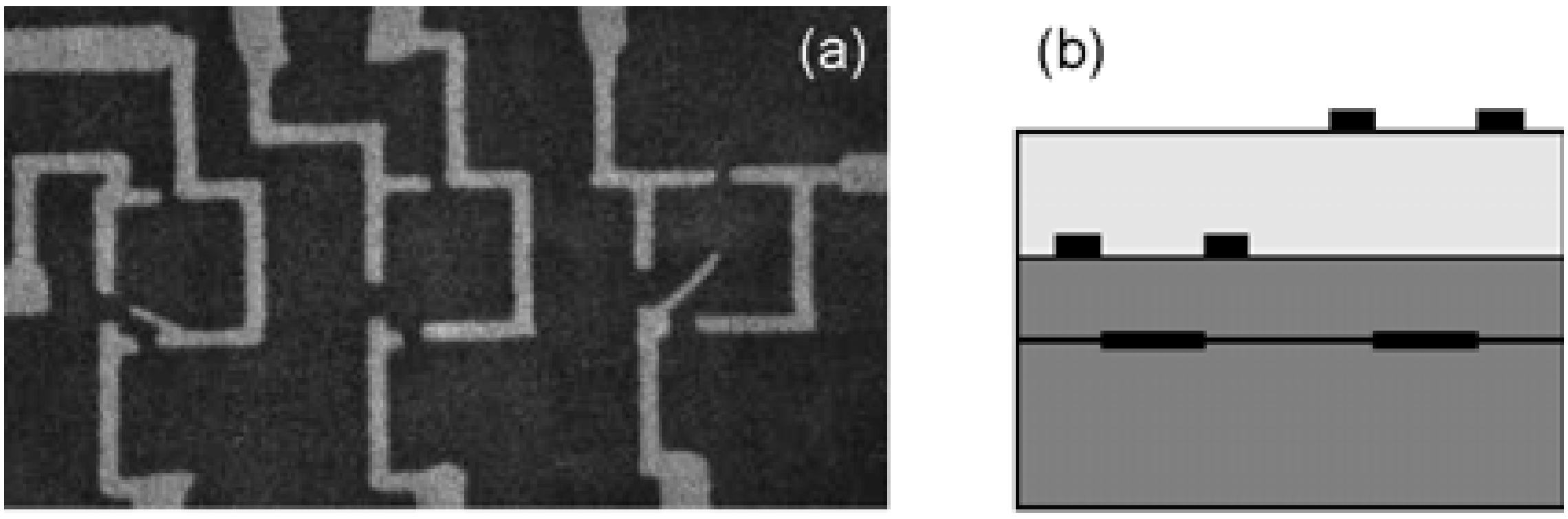

Figure 8.2: (a) Scanning electron micrograph of the Quantum Flow Disruptor (QFD) device. (b) Schematic of the double-gate billiard: an alternative to the double-2DEG billiard for studies of the influence of soft-wall potentials on fractal MCF

Finally, the fabrication of a Sinai billiard on a double-2DEG heterostructure would allow an investigation of the role of the soft-wall potential profile on exactly self-similar MCF, similar to the investigation conducted for statistical self-similarity in Chapter 7.

# 8.2 – Statistical Self-similarity

In contrast to exactly self-similar MCF, the study of statistically self-similar MCF in this thesis is more complete. However, the need for further investigations of statistically self-similar MCF is also significant. Firstly, there is clearly a need for smaller billiards to further investigate statistically self-similar MCF for large $Q$ values. It would also be interesting to examine larger billiards ($W = 3 - 10\mu m$) to establish, for a fixed number of modes and at base temperature, that the curve in Fig. 7.8(b) is followed solely on the basis of billiard area. This will allow the role of the semiclassical approximation in the generation of statistically self-similar MCF to become better established.

Further investigations of the influence of the soft-wall potential profile on statistically self-similar MCF are also required. More experiments using the double-2DEG billiards are planned including extra characterisation devices and billiards with a range of sizes to more adequately probe the fractal behaviour as a function of $Q$. Other experiments on double-2DEG billiards that are not related to the soft-wall profile are also planned, as discussed in §8.3. A possible improvement for studying soft-wall profile effects is the billiard shown schematically in Fig. 8.2(b). The double-gated billiard is fabricated on a single 2DEG heterostructure. In this device there are actually two billiards on one chip defined by lithographically identical surface-gates. For one of these billiards the surface-gates are on the heterostructure surface (Fig. 8.2(b) – left) and for the other the surface-gates are on top of an insulating layer (PMMA or polyimide). In this way, the distance between the gates and the 2DEG is different between the two billiards so that they have different potential profiles. The advantage of this device is that the two billiards are defined in the same 2DEG which makes the measurements simpler than in the 2DEG device. The profile characterisation attempted in §7.5 would also be far simpler in this device since only a single back-gate would be required, rather than a back- and top-gate for the double-2DEG devices. Note that it is not possible to have the upper set of gates directly above the lower set of gates in the double-gate device. The reason is that the lower gates would act to screen the bias on the upper gates, preventing proper formation of the billiard defined by the upper gates.

Finally, the Quantum Flow Disruptor device (Fig. 8.2(a)) is already fabricated and experiments on this device are planned for late 2000. The presence of a diamond periodic orbit in square billiards has been established in theoretical simulations by Akis and Ferry and demonstrated experimentally by Bird *et al.* [51,115]. This device will allow the role of periodic orbits in the generation of statistically self-similar MCF to be investigated.

# 8.3 – Double-2DEG Billiards: A Promising Future

This thesis reports the results of the first billiard experiment performed using a double-2DEG heterostructure. Aside from the intended purpose of this experiment – the investigation of the influence of soft-wall profile on fractal MCF – this experiment can also be viewed more strategically as a feasibility study of double-2DEG billiard experiments in particular. Investigations of laterally coupled billiards have become popular in recent years [182-185]. However, double-2DEG billiards now allow the investigation of vertically coupled billiards. It is important to note that the latter are likely to exhibit vastly different behaviour to the former because the nature of the coupling between the billiards is different. In particular, Coulomb drag [179] and inter-layer tunnelling [180] effects may play an important role in electron transport through vertically coupled billiards. Studies of coupling effects on fractal MCF in double-2DEG billiards are planned, and it is hoped that some of the results and conclusions obtained in this thesis will be useful as a tool in understanding the physics of double-2DEG billiards.

# Appendix A

This Appendix discusses in further detail some parts of the experimental set-up presented in Chapter 4. §A.1 deals with the wiring refit I performed in 1999, with particular emphasis on the new dilution unit wiring, the tri-strut cold finger and the dual sample mounts that were installed. §A.2 discusses the LC cold filters mounted on the $^4$He pot and the choice of components for these filters.

## A.1 – Dilution Unit Wiring Refit 1999

In the summer of 1998/99 I performed a complete overhaul of the electronic measurement set-up used in the experiments performed at UNSW. The objective of this refit was twofold – to improve the signal-to-noise ratio in electrical measurements and to reduce the minimum electron temperature in the billiard samples. It was also desirable to increase the number of samples that could be investigated in a single cool-down to reduce the down-time that occurs in changing samples. Note that doubling the number of sample wires has the potential to double the possible heat path from the environment into the dilution unit, hence considerable care was required to ensure that the extra heat load of the additional wiring was kept to a minimum. Ultimately, the refit was very successful. Beforehand, the minimum mixing chamber temperature was 30mK on a single sample with up to 12 wires available for measurements. Due to poor thermal contact, eddy current heating in the cold finger, and Joule heating due to RF pick-up, the minimum sample temperature was probably ~100mK. Afterwards however, the base mixing chamber temperature was < 20mK with the capability for two samples, each with 18 wires. The noise was lower than had previously been achieved on this system, eddy current heating was significantly reduced and the minimum sample temperature (<50mK) much closer to the base mixing chamber temperature, due to improved thermal contact between the sample and the mixing chamber.

Most of the gains in terms of signal-to-noise ratio were made by changing the grounding and shielding of the cryostat. Although there are basic principles for correct grounding and shielding [152], even Ott [152] (the ‘standard’ reference on RF interference) points out that there is no set solution for minimum noise in every system. In other words, what works in one system, may not work in another and vice versa. Due to this, and the excellent discussion of noise reduction techniques in [152], I will not pursue this further, aside from mentioning one salient piece of advice for reducing noise in low-temperature electrical measurements. Cool down the cryostat so that you have turned off all possible instruments, except the thermometry and then start to turn on other necessary pieces of equipment, and connect their cables one by one. The temperature will rise as you approach the final configuration and each instrument contributes some noise to the system. The most desirable set-up is likely that which leads to the smallest rise in temperature [186]. This advice proved most useful in achieving low-noise, low-temperature measurements following the refit. Although, one extension to this would be to connect the sample measuring equipment first to monitor the effect of connecting the other instruments on both the temperature and the signal itself.

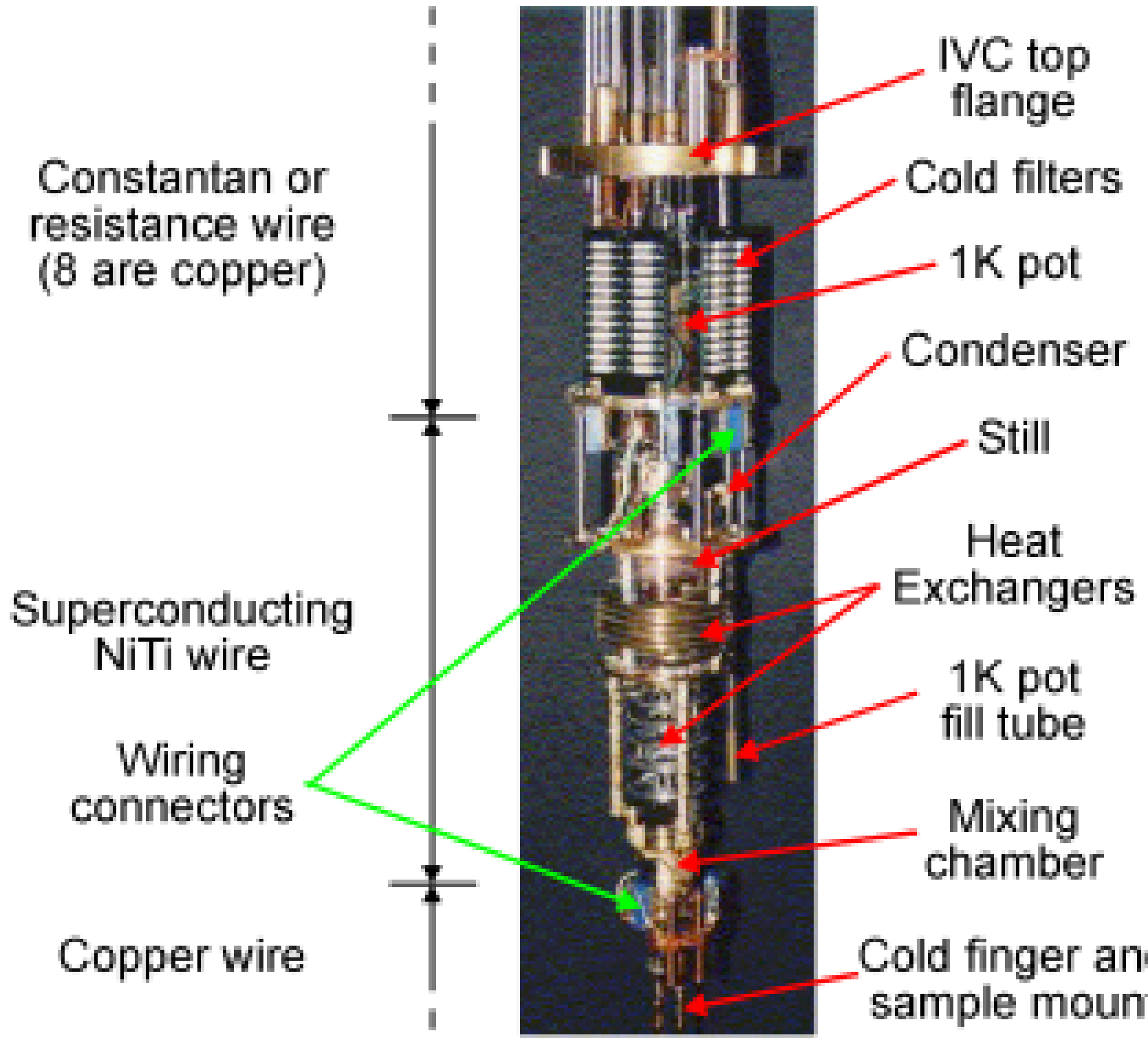

Figure A.1: Photograph showing the three different wiring sections that extend down the dilution unit.

The reduction in minimum sample temperature was achieved in four ways – careful choice of wire material and thickness, good thermal anchoring of the sample wires, the design of the cold finger, and the use of LC cold filtering. The last of these is the subject of the next section. The sample wiring from the room temperature connectors at the top of the dilution unit down to the sample mounting at the bottom of the cold finger, consists of three sections as shown in Fig. A.1. The top section runs from the room temperature connectors to the $^4$He pot and is made using 52SWG Eureka resistance wire (12Ω/m). Note: some of the original wires were left intact in this section, of these 8 are 42SWG Cu and 10 are 44SWG Constantan. The middle section extends between the $^4$He pot and the mixing chamber and is made using 0.1mm ǁ Cu clad $Nb_{0.48}Ti_{0.52}$ monofilament superconducting wire. The bottom section extends from the mixing chamber down to the sample mounts and is made using 0.2mm ǁ Cu wire. In explaining the choice of wire materials it is important to bear in mind the Wiedemann-Franz law which provides the relationship between the thermal conductivity and the electrical conductivity of a conductor [178]. The aim of the top section is to maximise the thermal resistance in order to prevent heat from the external environment entering the dilution unit. The middle section should also have a high thermal resistance to avoid providing a thermal short between the $^4$He pot and the mixing chamber. While resistance wire would achieve this in the middle section, superconducting wire has the added advantage that while the thermal resistance is high, the electrical resistance is zero, reducing the Joule heating ($I^2R$) in the wires. Note that the use of superconducting wire in the top section is pointless as the temperature is not low enough above the $^4$He pot for superconduction to take place. The aim of the bottom section is to provide a good thermal link between the sample and the mixing chamber, hence the choice of relatively thick Cu wire. The wires are thermally anchored at temperatures of 4.2K, 1.2K, ~600mK and ~30mK at the IVC plate, $^4$He pot, still and mixing chamber respectively using copper 'heatsink' bobbins as shown in Fig. A.2(a). Extra care is taken in thermal anchoring at the mixing chamber, as shown in Fig. A.2(b). The incoming superconducting wires are thermally anchored before prior to connecting to the Cu wires, which are thermally anchored again before extending down the cold finger. The aim was to achieve maximum contact between the copper wires and the mixing chamber to ensure the best possible thermal contact between the sample and the mixing chamber.

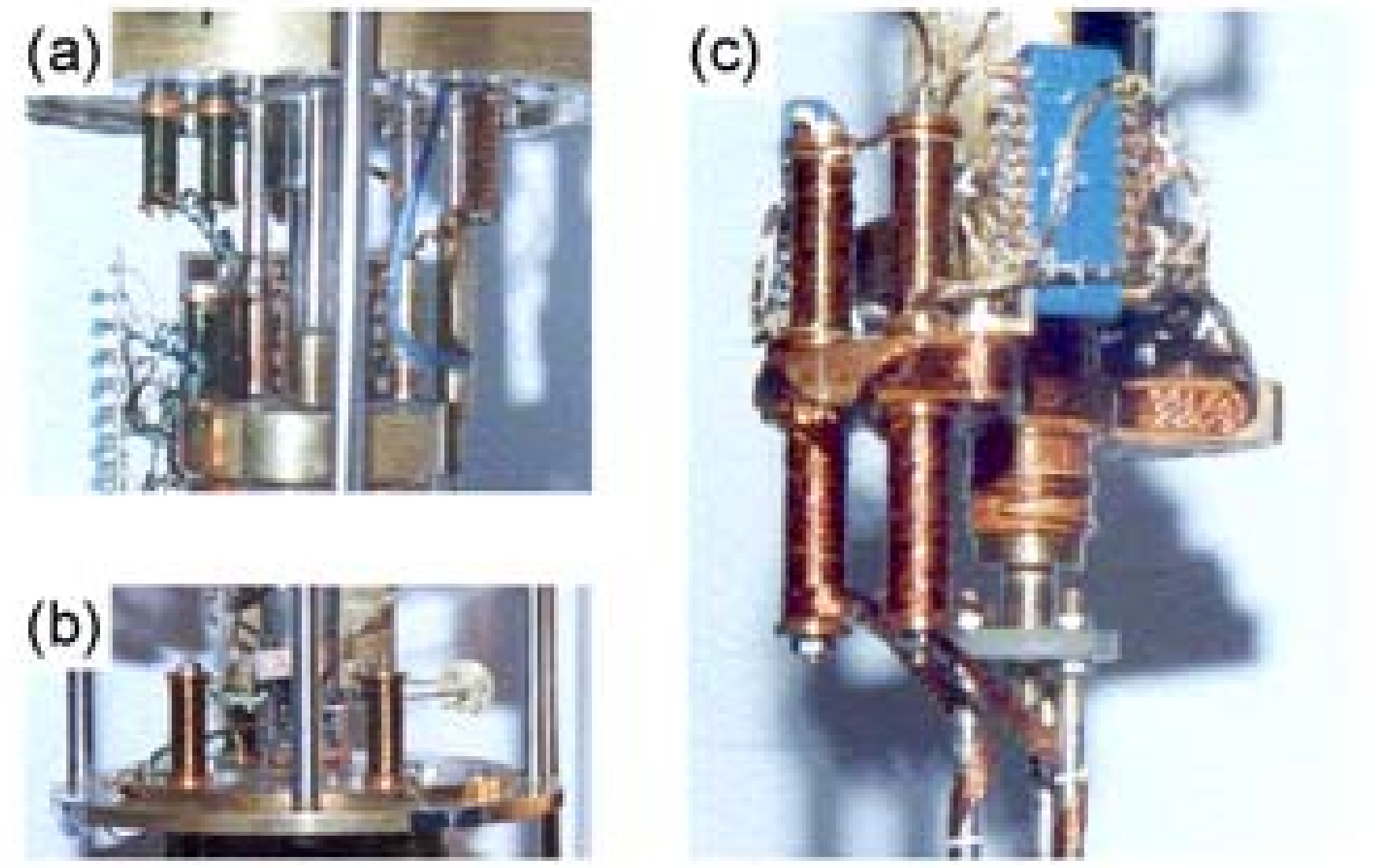

Figure A.2: Thermal anchoring of the sample wires at (a) the IVC plate and $^4$He pot, (b) still and (c) mixing chamber. Extra-care is taken to provide good thermal anchoring at the mixing chamber to ensure that the sample temperature is as close as possible to the mixing chamber temperature.

The original cold-finger was made of 12mm ǁ stainless steel tube with a wall thickness of 1mm. Substantial heating of the sample was observed as the magnetic field was swept due to eddy current heating in the cold finger. The aim was to minimise the diameter and wall thickness of the cold finger in order to reduce the effect of eddy current heating. This is particularly important since billiard measurements are obtained as a function of magnetic field. The new cold finger has a tri-strut design as shown in Fig. A.3. The advantage of this is twofold. Firstly, it is possible to use smaller diameter tubing without losing strength in the cold finger. Secondly, it allows better adjustment of the alignment of the cold finger in the bore of the superconducting solenoid. A variable length extension is also provided allowing the sample position along the bore to be adjusted, this is important in changing sample mountings. Note also that this extension can be removed to allow for a hydraulic pressure cell capable of up to 8kbar at millikelvin temperatures. Each strut is made of 4mm ǁ stainless steel tube with a wall thickness of 0.3mm. The extension piece is made of similar tube of 8mm ǁ. 3mm thick G10 fibreglass plates are used as spacers and held in place with stycast. The cold finger alignment is adjusted using brass bolts mounted at the ends of the struts. Sample mounting was constructed to accommodate samples in 12 to 18 pin dual-in-line packages (Fig. A.3(b)) or 20 pin Charntec packages (Fig. A.3(c)). Sample mounts are arranged so that the samples are facing each other and lie either side of the central field position within the solenoid bore and within the maximum field homogeneity region.

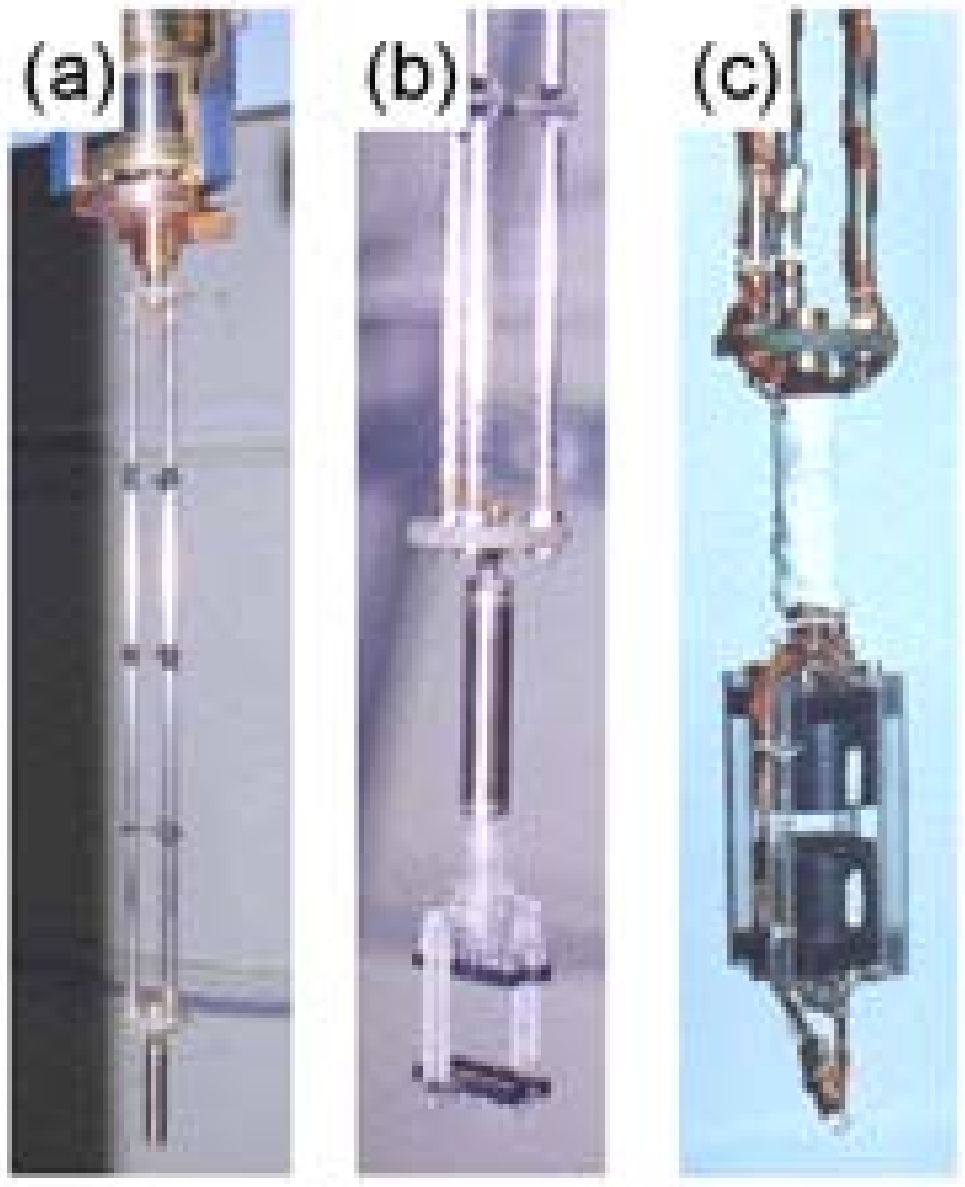

Figure A.3: Photographs of the new cold finger (a) prior to wiring, (b) with the 18 pin dual-inline sample mount and (c) fully wired with the 20 pin Charntec sample mounts.

Optical fibre access through the back of the Charntec mounts is provided to allow independent illumination of the samples. Unfortunately, independent illumination is not possible with devices in dual-in-line packages.

# A.2 – LC Cold Filters

As discussed in Chapter 4, each sample/thermometry wire extending down the dilution unit passes through a LC cold filter mounted in thermal contact with the $^4$He pot. Hence these LC filters are at a temperature of 1.2K during the experiment. The circuit diagram for the LC filters is shown in Fig. 4.7. Electrical details of the LC filters are also discussed in Chapter 4. Figure A.4(a) shows the original set of 48 LC cold filters installed in January 1999. The components for the original filters were Radio-Spares 1mH wire-wound ferrite-core inductors with a polypropylene outer casing (RS

cat:228-208) and Panasonic 2.2μF surface-mount polyphenylene sulphide (PPS) film capacitors (RS cat:175-019).

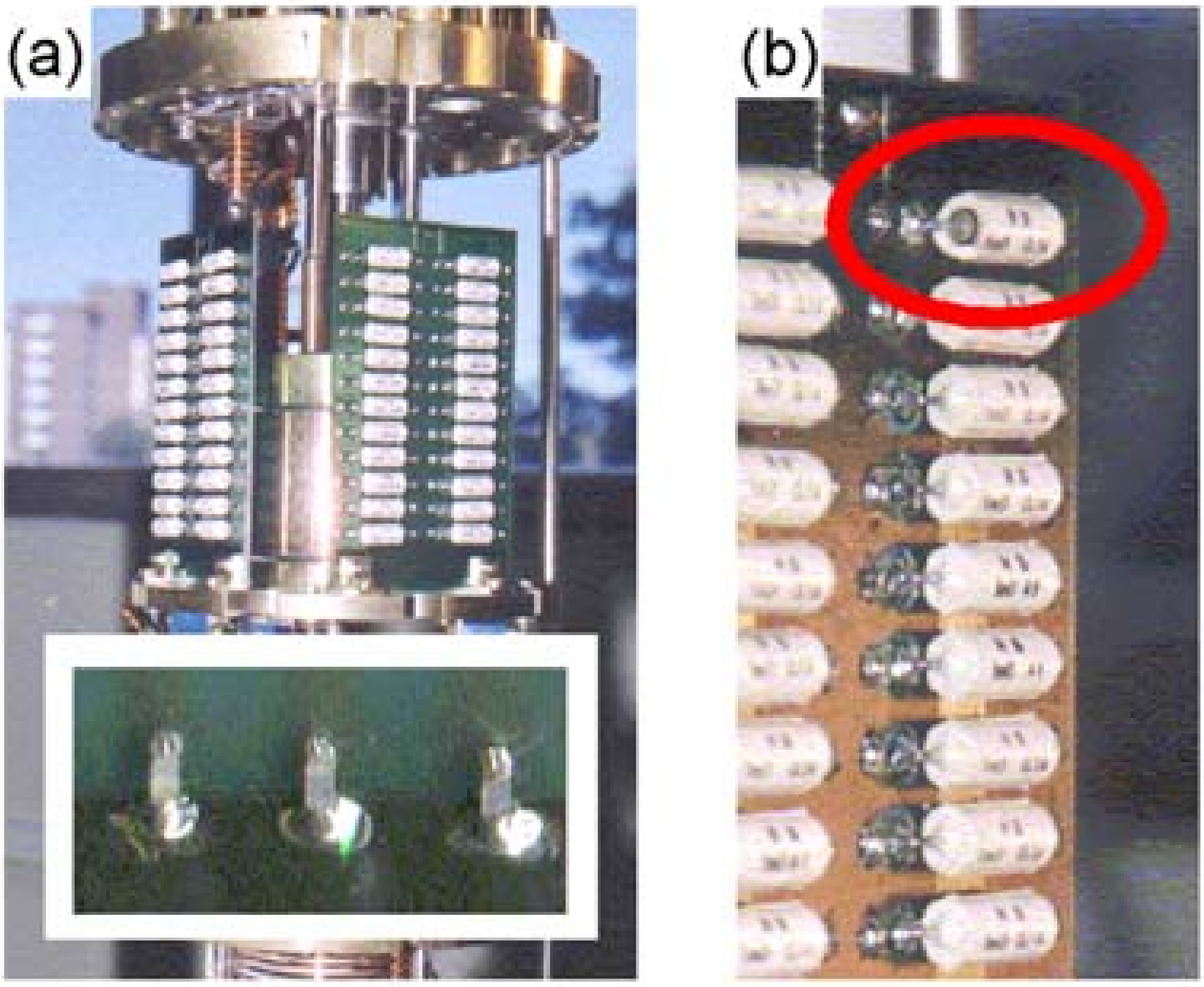

Figure A.4: (a) The original cold filters installed in January 1999 mounted on the $^4$He pot plate. Inset is a photograph of the surface-mount capacitors, which are located on the back of the PCB. (b) Photograph of the damaged inductors following repeated thermal cycling. The top inductor shows the most dramatic example of the damage.

The capacitors are mounted on the back of the LC filter PCBs as shown in the inset to Fig. A.4(a). The capacitors are soldered directly to the inductor lead to save space, as shown schematically in Fig. A.5. The LC filters have a common earth that is connected to the lock-in earth via a twisted pair that leads back up the dilution unit from the top of the LC filter PCB. Extra metallisation is provided on the PCB and is connected to the $^4$He pot via the PCB mounts to facilitate more rapid cooling of the PCBs.

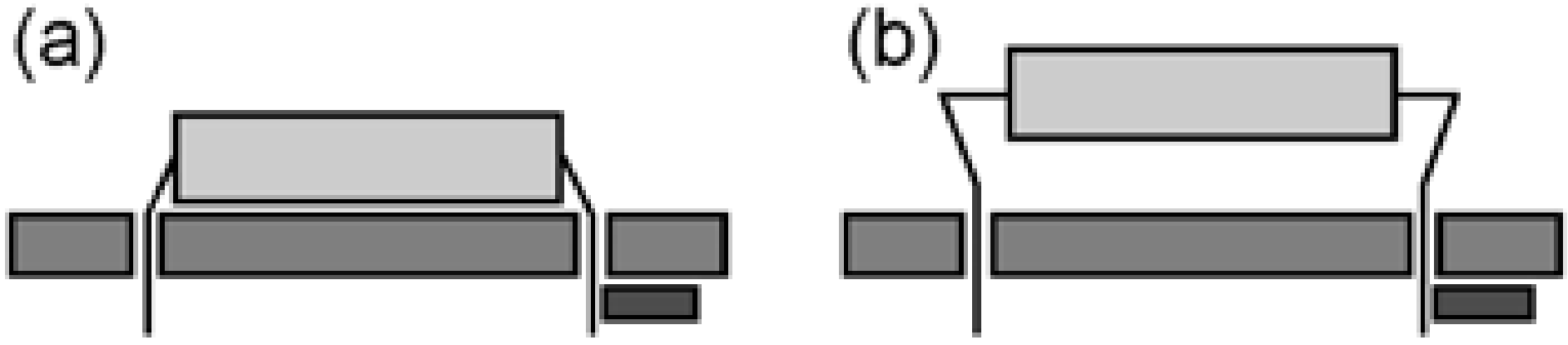

Figure A.5: (a) Old and (b) new mounting technique for the inductors on the PCB. The new technique allows for movement under thermal contraction preventing damage to the inductor. The capacitor is mounted on the back of the PCB and is soldered directly to the inductor lead to save space.

The inductors on the original cold filters were problematic. Whilst initial thermal cycling prior to construction of the filters presented no problems, after several thermal cycles of the dilution unit the outer casing began to break away as shown in Fig. A.4(b). Whilst the broken casing should present no problems to the operation of the inductors, later examination showed that the wire is bonded to the casing and that the cracks break the thin wires of the inductor. Further examination later found that the real problem is the mounting of the inductors on the PCB rather than the casing material of the inductor. As shown in Fig. A.5(a), the original inductors were mounted with the leads tight through the PCB. During cool-down, thermal contraction placed large stresses upon the inductor casing causing it to fracture. In the second series of cold filters, the inductors were mounted with more room for thermal contraction (Fig. A.5(b)). Since replacing the cold filters is an exhaustive process (plenty of very tricky soldering) I decided to replace the inductors. The new cold filters have 1mH wire-wound ferrite core inductors, this time with a resin coating (RS cat:240-551). This resin coating can crack (and be removed) without breaking the wire in the inductor, unlike the packaging of the original inductors.

Photographs of the new inductors and inductor mounting are shown in Figs. A.6(a) and (b). The capacitors have proven quite robust to thermal cycling and have presented no problems, however they are limited to a maximum voltage of 50V DC preventing the use of ‘back-of-the-wafer’ back-gates which often require biases above 50V DC.

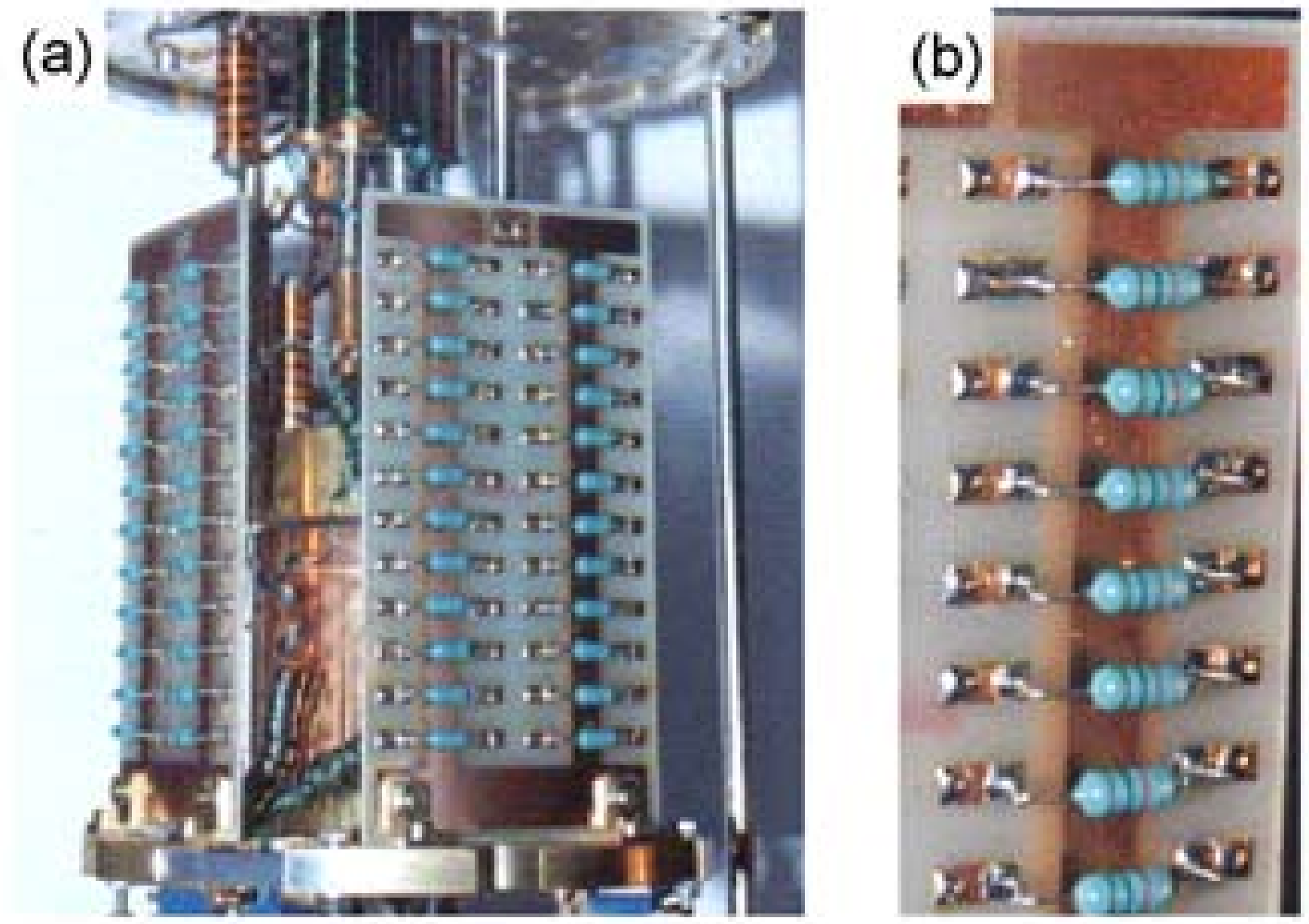

Figure A.6: (a) Photograph of the new cold filters. (b) Close-up highlighting the new inductor mounting technique shown in Fig. A.5(b).